\documentclass[prb,twocolumn,notitlepage,superscriptaddress,nofootinbib]{revtex4-2}
\usepackage{amsmath,amssymb}
\usepackage[hidelinks,colorlinks,linkcolor=blue,
citecolor=blue,urlcolor=blue]{hyperref}
\usepackage{graphicx,siunitx}
\usepackage[dvipsnames]{xcolor}

\usepackage{pifont}

\usepackage[normalem]{ulem}

\usepackage{xcolor}
\DeclareUnicodeCharacter{3000}{\textcolor{red}{BAD!!}}

\newcommand{\be}{\begin{equation}}
\newcommand{\ee}{\end{equation}}
\newcommand{\bea}{\begin{equation} \begin{aligned}}
\newcommand{\eea}{\end{aligned} \end{equation} }
\newcommand{\bi}{\begin{itemize}}
\newcommand{\ei}{\end{itemize}}

\renewcommand{\be}{\beta}
\newcommand{\al}{\alpha}
\newcommand{\bpm}{\begin{pmatrix}}
\newcommand{\epm}{\end{pmatrix}}
\newcommand{\eps}{\epsilon}

\renewcommand{\th}{\theta}

\newcommand{\lp}{\left(}
\newcommand{\rp}{\right)}

\newcommand{\Tr}{\text{Tr} \ }

\DeclareRobustCommand{\App}[1]{\cref{#1}}

\DeclareRobustCommand{\Fig}[1]{Fig.~\ref{#1}}

\DeclareRobustCommand{\Eq}[1]{Eq.~(\ref{#1})}

\usepackage{amsmath,amsfonts,amssymb,amsthm,epsfig,array}
\usepackage{dsfont}
\usepackage{slashed}
\usepackage{graphics}
\usepackage{float}
\usepackage{verbatim}
\usepackage{color}
\usepackage{tabularx}
\usepackage[mathscr]{euscript}
\usepackage{mathtools}
\usepackage{braket}

\newcommand{\bsl}[1]{\boldsymbol{#1}}
\newcommand{\mbf}[1]{\boldsymbol{#1}}

\renewcommand{\mod}{\,\mathrm{mod}\,}

\newcommand{\ii}{\mathrm{i}}

\newcommand{\dsR}{\mathbb{R}}

\newcommand{\eqnref}[1]{Eq.\,\eqref{#1}}
\newcommand{\figref}[1]{Fig.\,\ref{#1}}

\newcommand{\refcite}[1]{Ref.\,\cite{#1}}
\newcommand{\refscite}[1]{Refs.\,\cite{#1}}

\newcommand{\eq}[1]{\begin{equation} #1 \end{equation}}

\newcommand{\eqa}[1]{\begin{align}\begin{split} #1 \end{split}\end{align}}

\usepackage{environ}
\NewEnviron{eqs}{%
\begin{equation}\begin{split}
\BODY
\end{split}\end{equation}
}

\let\oldAA\AA
\renewcommand{\AA}{\text{\normalfont\oldAA}}

\newcommand{\ie}{{\emph{i.e.}}}

\newcommand{\V}{\mathcal{V}}

\newcommand{\Ch}{\text{Ch}}
\newcommand{\K}{\text{K}}

\newcommand{\Ebias}{E_{\text{bias}}}
\newcommand{\lfock}{\lambda_{\text{Fock}}}
\newcommand{\meV}{\text{meV}}

\usepackage{cleveref}
\crefname{appendix}{App.}{Apps.}
\crefname{equation}{Eq.}{Eqs.}
\crefname{figure}{Fig.}{Figs.}
\crefname{table}{Tab.}{Tabs.}
\crefname{section}{Sec.}{Secs.}
\creflabelformat{appendix}{#2#1#3}

\usepackage{rotating}

\begin{document}

\title{Moir\'e Fractional Chern Insulators IV: Fluctuation-Driven Collapse of FCIs in Multi-Band Exact Diagonalization Calculations on Rhombohedral Graphene}

\author{Jiabin Yu}
\thanks{These authors contributed equally.}
\affiliation{Department of Physics, Princeton University, Princeton, New Jersey 08544, USA}

\author{Jonah Herzog-Arbeitman}
\thanks{These authors contributed equally.}
\affiliation{Department of Physics, Princeton University, Princeton, New Jersey 08544, USA}

\author{Yves H. Kwan}
\affiliation{Princeton Center for Theoretical Science, Princeton University, Princeton, NJ 08544}

\author{Nicolas Regnault}
\email{regnault@princeton.edu}
\affiliation{Laboratoire de Physique de l’Ecole normale sup\'erieure,
ENS, Universit\'e PSL, CNRS, Sorbonne Universit\'e,
Universit\'e Paris-Diderot, Sorbonne Paris Cit\'e, 75005 Paris, France}
\affiliation{Department of Physics, Princeton University, Princeton, New Jersey 08544, USA}

\author{B. Andrei Bernevig}
\email{bernevig@princeton.edu}
\affiliation{Department of Physics, Princeton University, Princeton, New Jersey 08544, USA}
\affiliation{Donostia International Physics Center, P. Manuel de Lardizabal 4, 20018 Donostia-San Sebastian, Spain}
\affiliation{IKERBASQUE, Basque Foundation for Science, Bilbao, Spain}

\date{\today}

\begin{abstract}
The fractional Chern insulators (FCIs) observed in pentalayer rhombohedral graphene/hexagonal boron nitride superlattices have a unique origin contrary to theoretical expectations: their non-interacting band structure is gapless, unlike standard FCIs and the Landau level. Hartree-Fock (HF) calculations at filling $\nu=1$ yield a gapped ground state with Chern number $1$ through band mixing, identifying a possible parent state. However, many-body calculations restricted to the occupied HF band predispose the system towards FCIs and are essentially uncontrolled. In this work, we use unbiased multi-band exact diagonalization (ED) to allow fluctuations into the gapless bands for two normal-ordering schemes. In the ``charge neutrality" scheme, the weak moiré potential leads to theoretical proposals based on Wigner crystal-like states. However, we find that FCIs seen in 1-band ED calculations are destroyed by band mixing, becoming gapless as fluctuations are included. In the ``average" scheme, the Coulomb interaction with the periodic valence charge background sets up a stronger moir\'e potential. On small systems, FCIs at $\nu=1/3$ are destroyed in multi-band calculations, while those at $\nu=2/3$ are initially strengthened. However we do not converge to a stable FCI at $\nu=2/3$ even on the largest accessible systems. These findings question prior results obtained within projection to a single HF band. They suggest that current models do not support FCIs with correlation length small enough to be converged in accessible, unbiased ED calculations, or do not support FCIs at all.
\end{abstract}

\maketitle

\section{Introduction}

\begin{figure}
\centering
\includegraphics[width=\linewidth]{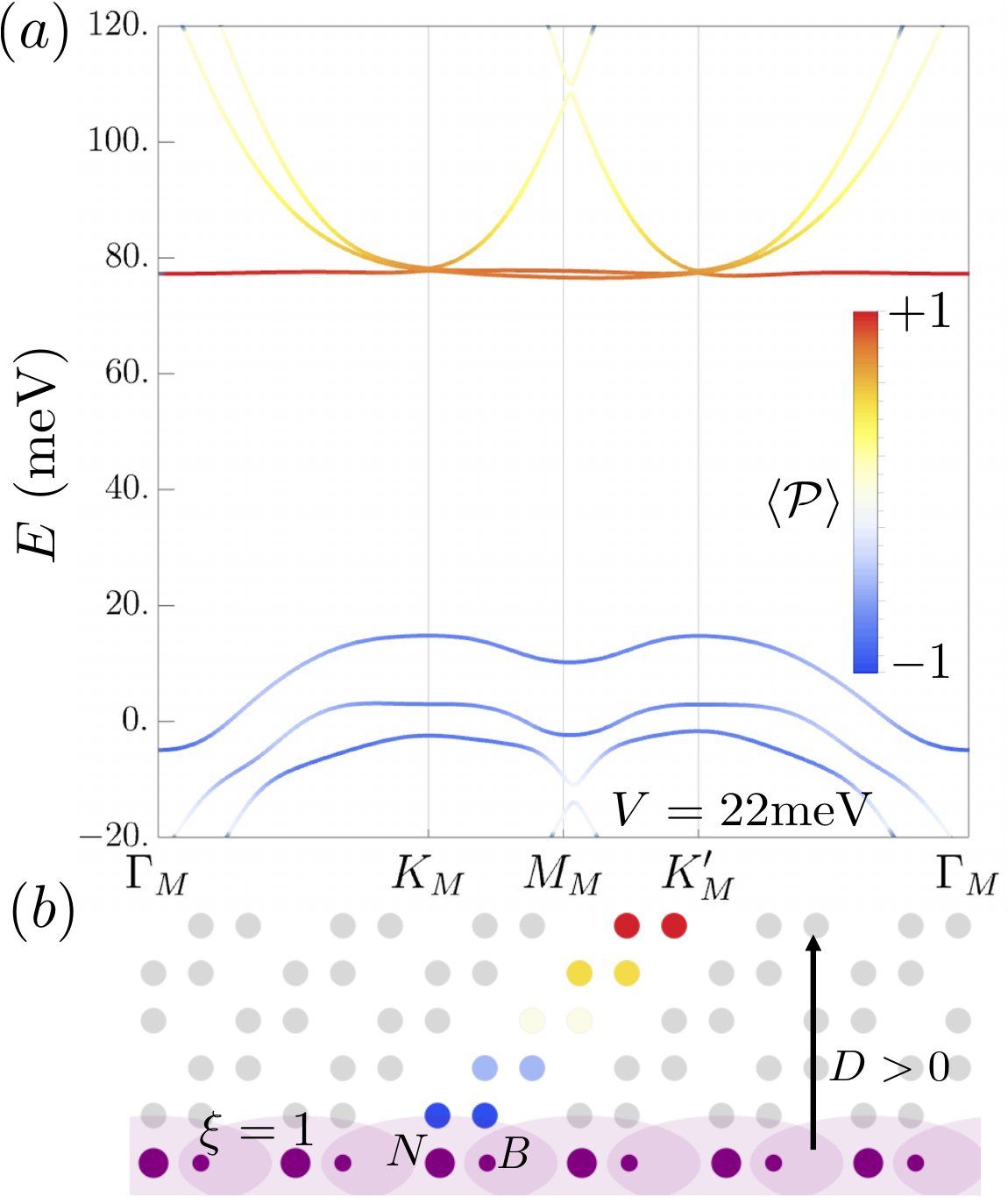}
\caption{$(a)$ Non-interacting band structure of R5G/hBN at $\theta=0.77^\circ$. Applied displacement field (incorporated with interlayer potential $V=22$\,meV) separates the bands into a gas of nearly gapless, low-energy topological conduction electrons on the top layers (red) and valence charges creating an electrostatic background bound to the bottom layers (blue). States are colored according to their layer polarization $\braket{\mathcal{P}}$ where $\mathcal{P} \ket{\mbf{R},\al,l} = \frac{l-2}{2}\ket{\mbf{R},\al,l}$ for states with sublattice $\al$ and layer index $l = 0,\dots,4$. $(b)$ Schematic of the $\xi = 1$ stacking configuration where the N atom is under the carbon $A$ site. For $\xi=0$ stacking, the hBN layer would be rotated by $180^\circ$ so that N atom would be under the carbon $B$ site.}
\label{fig:BS}
\end{figure}

Fractional Chern insulators (FCIs) are now experimentally reported in twisted bilayer MoTe$_2$~\cite{cai2023signatures,zeng2023thermodynamic,Park2023observation,Xu2023observation} and rhombohedral pentalayer~\cite{Lu2024fractional} and hexalayer~\cite{xie2024even} graphene/hexagonal boron nitride superlattices (R5G/hBN). The theoretical prediction of FCIs~\cite{neupert,regnaultbernevig,sheng} originated from extracting the key physics of the lowest Landau level, specifically, the fractional filling of a nearly flat, isolated band with a Chern number of $\Ch=1$. This provides important intuition for understanding twisted MoTe$_2$ \cite{cai2023signatures,zeng2023thermodynamic,Park2023observation,Xu2023observation,2024arXiv240407157J,2024arXiv240510269R,2024arXiv240519308T,2024arXiv240609591P,2023PhRvB.108h5117R,wang2023fractional}, where at integer filling $\nu=-1$, the flat Chern band valley polarizes and spontaneously breaks time-reversal symmetry, setting the stage for FCIs at fractional fillings. However, despite a $10$\,meV gap around this parent Chern band, Coulomb-driven band mixing~\cite{Yu2024MFCI0,Xu2024maximally,Abouelkomsan2024mixing} is essential in order to reproduce the phase diagram seen in experiments.

The robust Jain sequence of FCIs reported in $\theta\simeq 0.77^\circ$ R5G/hBN under large displacement field appears to exhibit the standard FCI phase diagram~\cite{Lu2024fractional}.  However, closer inspection of the single-particle band structure, expected to be reliable in graphene systems \cite{2023Natur.614..682I,2023arXiv231213637Z}, reveals a more complex situation (see \cref{fig:BS}). The large displacement field, required to obtain the Chern insulator (CI) at filling $\nu=1$ and the multiple FCIs at $\nu<1$ in experiment~\cite{Lu2024fractional}, pushes conduction electrons \emph{away} from the moir\'e pattern. Thus, while the lowest conduction band is significantly flattened by the displacement field, the weak moir\'e potential is almost incapable of opening a single-particle gap at the moir\'e Brillouin zone (BZ) edge \cite{herzog2024MFCI2,kwan2023MFCI3,dong2023anomalous,zhou2023fractional,dong2023theory,guo2023theory}. The tiny gap of $<.1$\,meV that it does open yields a Chern number $\Ch = 5$ at $\nu=1$ ~\cite{herzog2024MFCI2}, in contradiction with experimental reports \cite{Lu2024fractional} of $\Ch=1$ and $0.5-1$\,meV gaps~\cite{2024arXiv240402192X}.
As we will show, the essentially gapless flat band poses a fundamentally different starting point for FCIs~\cite{ju2024fractional}, preventing projection to a single band. Neglecting fluctuations into all available low energy states can prejudice calculations towards FCIs. A \emph{controlled} theory of if, how, and why FCIs appear in theoretical studies of R5G/hBN (and hexalayer graphene/hBN superlattices~\cite{xie2024even}) is still lacking.

Existing HF calculations at filling $\nu=1$ and twist angle $\theta\simeq 0.77^\circ$ predict a spin-valley-polarized $\Ch = 1$ ground state with a charge gap~\cite{kwan2023MFCI3,dong2023anomalous,zhou2023fractional,dong2023theory,guo2023theory,2024arXiv240708661H}, seemingly in agreement with experiment. Within a single spin-valley flavor, the interaction couples the lowest three nearly gapless conduction bands (\cref{fig:BS}a) and significantly reconstructs the single-particle bands.
The HF ground state has been shown to acquire a large charge gap of \mbox{$15-25$\,meV} for two proposed normal-orderings of the Coulomb interaction, the charge-neutrality (CN) scheme~\cite{kwan2023MFCI3,dong2023anomalous,zhou2023fractional,dong2023theory} and the average (AVE) scheme~\cite{kwan2023MFCI3}, which will be elaborated on in \cref{sec:schemes}.

The CN scheme neglects the valence electrons, such that the low-energy conduction electrons only feel an extremely weak moir\'e potential. This leads to the interpretation of the $\Ch=1$ HF state as a topological ``Wigner crystal"-like state~\cite{kwan2023MFCI3,dong2023anomalous,zhou2023fractional,dong2023theory,tan2024parent,soejima2024anomalous,2024arXiv240307873D,zeng2024sublattice,2024arXiv240512294S,crepel2024efficient,2024arXiv240614354K} which has been dubbed the anomalous Hall crystal (AHC) \cite{dong2023anomalous} since it persists in HF even as the moir\'e potential is artificially turned off and continuous translation symmetry emerges.  However, it is well known that HF overestimates the tendency to spontaneously break symmetries and open charge gaps, and fails to accurately capture phase boundaries when compared to unbiased numerical methods. For example, in the standard 2D electron gas, HF predicts the Wigner crystal phase to emerge at $r_s\gtrsim 1.2$~\cite{Trail2003HF,Bernu2011HF,2024arXiv240307873D}\footnote{Interestingly, the translation-invariant Fermi liquid is never the ground state in HF for \emph{any} $r_s$~\cite{bernu2008metal,Bernu2011HF}, since at high density (small $r_s$) it yields to various gapless incommensurate phases.}, while quantum Monte Carlo finds that it only beats the Fermi liquid for much stronger interactions $r_s\gtrsim 31$~\cite{tanatar1989_2deg,attaccalite2002_2deg,drummond2009_2deg}. The phase diagram is also potentially complicated by several possible intermediate phases~\cite{spivak2004phases,reza2005universal,falakshahi2005hybrid,kim2024dynamical,valenti2024nematic}. It appears so far that in pentalayer samples that are not nearly aligned to hBN and hence experience no moir\'e potential, FCIs are absent \cite{han2024electron}. This raises the question of what minimum moir\'e strength is required to pin the putative AHC, and protect quantized transport signatures that are centered around both integer and fractional fillings of the moir\'e unit cell~\cite{Lu2024fractional}.

Alternatively, the AVE scheme captures the effect of the moir\'e-bound valence electrons, which act on the conduction electrons through the Coulomb interactions and magnify the effect of the moir\'e potential. A similar scenario was proposed in \refcite{Lu2024fractional} at a phenomenological Hartree level. Crucially, while the density matrices of the $\nu=1$ ground states obtained for both schemes are similar in HF, their excitations are quantitatively  distinct~\cite{kwan2023MFCI3}. In particular, the moir\'e (pseudo-)phonons, which are related to the gapless Goldstone modes of the spontaneously-broken continuous translation symmetry in the moir\'e-less limit, have a small $<1$\,meV gap at $\mbf{q}=0$ in the CN scheme, but develop a more substantial gap that is $3-5$ times larger in the AVE scheme~\cite{kwan2023MFCI3}.

At fractional fillings, earlier theoretical work has resorted to projection onto the occupied gapped HF band at $\nu=1$~\cite{dong2023anomalous,zhou2023fractional,dong2023theory,guo2023theory}, despite the fact that the bare single particle spectrum is gapless. Restriction to 1-band calculations is an \emph{a priori} unjustified approximation, and severely biases the system towards FCIs. In this work, we implement multi-band exact diagonalization (ED) calculations which aim to preserve all low-energy degrees of freedom in order to critically assess this approach. To systematically improve upon 1-band ED while managing the computational demand, we implement approximations limiting the number of particles and available states within the multi-band Hilbert space.

We find that strong fluctuations enabled by band mixing destroy the gapped FCI state at both $\nu=1/3$ and $\nu=2/3$ in the CN scheme.  The $\nu=1/3$ FCI is initially more robust than at $\nu=2/3$ in the projected 1-band limit, but both many-body gaps quickly collapse as the nearby HF bands are re-introduced. In the AVE scheme, we also find the eventual collapse of FCIs at both $\nu=1/3$ and $\nu=2/3$, but the $\nu=2/3$ state is much more robust to band mixing, which seems to follow the same trend as in experiment~\cite{Lu2024fractional}. However, on accessible systems, band mixing does eventually destroy the FCI ground state.
Furthermore, we consider integer filling in ED. Our combined multi-band HF and ED calculations at $\nu=1$ lean towards the possibility of having a $\Ch=1$ gapped ground state in the AVE scheme in the thermodynamic limit, which is consistent with the experiments.
This is in contrast to the CN scheme, where we find that band mixing in ED leads to the collapse of the gap due to continuum excitations at $\nu=1$, and hence no sign of a ``pinning" gap of the collective mdoes.
These findings underscore the pivotal role of band mixing, questioning the validity of 1-band projected ED calculations and suggesting caution be taken when interpreting the HF calculation in the CN scheme for $\nu=1$.

Assuming that a gapped FCI is the ground state of R5G/hBN at $\nu=2/3$ in the thermodynamic limit, our results suggest the following two possibilities: (1) the AVE scheme is the appropriate scheme for the study of FCIs, since the experiment~\cite{Lu2024fractional} does show FCIs at $\nu=2/3$ and not $\nu=1/3$, but strong finite size effects in the calculation prevent us from converging to the right states; (2) the AVE scheme, or the model, does not correctly capture the stability of the FCIs, and an important piece of the physics is missing.

\section{Hamiltonians and Interacting Schemes}
\label{sec:schemes}

Since two distinct interaction schemes, the CN and AVE schemes, have been proposed to describe R5G/hBN, we review their motivations and introduce their explicit Hamiltonians. In \cref{sec:MBED}, we will discuss their contrasting behavior in 1-band and multi-band ED.

\subsection{Interaction Schemes}

For the CN scheme~\cite{dong2023anomalous,zhou2023fractional,dong2023theory,kwan2023MFCI3}, the interaction is normal-ordered with respect to the single-particle charge neutrality gap.
This does not include any effects from the valence electrons, which are pushed by the displacement field towards the aligned hBN substrate and rendered inert.
As a result, the conduction electrons only feel the effect of the moir\'e at the single-particle level, which is weak since the action of the moir\'e potential on the low-energy conduction bands is exponentially suppressed by their layer polarization~\cite{PhysRevB.81.125304,herzog2024MFCI2}. For instance, at an interlayer potential of $V>20$\,meV, the moir\'e gap above the lowest conduction band is $\leq 0.1$\,meV at $K_M'$. Thus the moir\'e potential can be neglected entirely in the HF calculation while barely altering the HF ground state and charge excitation spectrum \cite{dong2023anomalous,zhou2023fractional,kwan2023MFCI3,2024arXiv240307873D}.

In the AVE scheme, the densities are measured symmetrically with respect to charge neutrality, and the valence electrons bound to the moir\'e potential create considerable moir\'e-periodic background terms on the conduction electrons through the Coulomb interaction (such as the potential generated from the moir\'e modulated charge density). This effect dramatically broadens the non-interacting bandwidth of $1.5$\,meV to more than $15$\,meV at $V=22$\,meV, and opens an $8$\,meV gap at the $K_M$ point.

The extremely weak effect of the moir\'e in the CN scheme leads to a description of the $\nu=1$ HF ground state in terms of spontaneous breaking of a continuous translation symmetry, leading to the AHC as a topological Wigner crystal. As such, at zero moir\'e strength, Goldstone's theorem ensures two ``moir\'e phonons" that are gapless at $\mbf{q}=0$ appear in the collective mode spectrum, which are gapped by a small amount $<1$\,\text{meV} when the moir\'e is realistically incorporated~\cite{kwan2023MFCI3}. In the AVE scheme, the moir\'e \mbox{(pseudo-)phonons}, which are gapped because of the strong moir\'e background potential, can also be identified, but we find them to not be the lowest energy collective mode at $\mbf{q}=0$. The moir\'e pseudo-phonon gap is $3-5$ times larger in the AVE scheme compared to the CN scheme~\cite{kwan2023MFCI3}.

If there is a sufficiently large gap between the valence and conduction bands, then it becomes reasonable to consider the CN scheme, which does not account for any effects from the occupied valence subspace on the conduction bands. For zero displacement field, the validity of the CN scheme is undermined by the absence of a sizable gap. However, for moderate displacement fields relevant for experiments, it is not \textit{a priori} clear that the influence of the valence bands can be neglected.  In particular, the CN scheme eliminates the moir\'e potential induced by the valence background, which may lead to an underestimate of the strength of moir\'e effects on the low-energy conduction bands. This is not the case in the AVE scheme where the background effect of the valence bands is not neglected for any $V$. Another feature of the AVE scheme is that the interacting part of the Hamiltonian does not depend on parameters, such as the displacement field, that could be tuned experimentally \textit{in situ}. As discussed in Ref.~\cite{kwan2023MFCI3,2024arXiv240708661H}, a subtlety of the AVE scheme is that the background term depends weakly on the momentum cutoff of the single-particle model, even for large cutoff radius. However, the continuum Hamiltonian ceases to be a reliable model at $\sim$\,eV energies, so the background contribution from states at or beyond this scale is unphysical. Hence, we impose a finite momentum cutoff when specifying the Hamiltonian in the next subsection.

Ideally, \textit{ab initio} studies that explicitly include the hBN and the interlayer potential would be able to shed light on the correct interaction scheme. A more quantitative treatment of the interactions at low energies can be obtained with renormalization group approaches as used by Refs.~\cite{vafek2020RG,guo2023theory}.

\refcite{kwan2023MFCI3} showed that the AVE scheme leads to differing phase diagrams between the two hBN stackings in HF calculations. In particular at $\nu=1$, for displacement fields that bias the conduction electrons away from the hBN, the $\xi=1$ stacking favors the $\Ch=1$ state over the $\Ch=0$ state compared to the $\xi=0$ stacking (see \Fig{fig:BS}b). On the other hand, the results for the two stackings are nearly identical for the CN scheme since the lowest conduction bands feel minimal moir\'e effects. \emph{If} HF is valid, this difference has been proposed to help distinguish the two schemes experimentally by comparing the Chern number of the $\nu=1$ correlated insulator in the two stackings~\cite{kwan2023MFCI3}.
One possible way to perform this comparison is to cut and rotate half of a pentalayer sample by $60^\circ$ on a large hBN flake.

\subsection{Hamiltonians}

We introduce the Hamiltonians studied in this work, starting with $H_{0,\eta}$, the single-particle Hamiltonian, which reads~\cite{herzog2024MFCI2}
\bea
\label{eq:SP}
H_{0,\eta} &= H_{R5G}^{\eta} + H_{\text{moir\'e},\xi}^{\eta}\ ,
\eea
where $H^\eta_{R5G}$ is the rhombohedral pentalayer graphene Hamiltonian that contains the effect of the displacement field (modeled as a linear interlayer potential $V$), and $H^\eta_{\text{moir\'e}, \xi}$ captures the non-interacting moir\'e potential arising from the hBN twisted by an angle $\theta$. $\xi=0,1$ distinguishes the two distinct stackings [see \cref{fig:BS}b for $\xi=1$], and
$\eta=\pm\K$ is the valley index. Detailed expressions for these Hamiltonians and a summary of parameters are provided in \App{app:SP}.
\cref{fig:BS}(a) shows the single-particle band structure at $V=22$\,meV. In this work, we focus on the $\xi=1$ stacking configuration at $\theta=0.77^\circ$, which yields $\Ch=1$ HF bands for both CN and AVE schemes~\cite{kwan2023MFCI3}. The moir\'e potential obtained from our relaxation calculations is weaker (about $1/3$ the strength) than other non-relaxed estimates \cite{PhysRevB.90.155406} using the two-center approximation \cite{2011PNAS..10812233B} and about half as strong as estimates based on untwisted calculations in different stackings \cite{PhysRevB.89.205414,Park2023RMGhBNChernFlatBands}.

Next, we consider the interaction.
We will discuss in the main text the 2D limit which neglects the thickness of the pentalayer structure, whereas the 3D interaction that screens the displacement field~\cite{kwan2023MFCI3} is described in \App{app:3Dint}. We show in \App{app:3Dint} that similar behavior is found for the 3D interaction as for the 2D interaction, but at a larger displacement field which more closely matches the values relevant for the experiment~\cite{Lu2024fractional}.
We study both CN and AVE schemes.

The Hamiltonian in the CN scheme with a 2D gate-screened Coulomb interaction $V(\mbf{q})$ is \cite{kwan2023MFCI3}
\bea
H_{\text{CN}} &= \sum_{\eta} H_{0,\eta} + \frac{1}{2\mathcal{V}}\sum_{\mbf{q}, \mbf{G}} V(\mbf{q}+\mbf{G}) : \rho_{\mbf{q}+\mbf{G}}\rho_{-\mbf{q}-\mbf{G}}:
\eea
where $\mbf{q}$ is in the moir\'e BZ, and $\mbf{G}$ is a moir\'e reciprocal lattice vector. The normal-ordering notation $:\hat{O}:$ places all annihilation (creation) operators on the right for conduction (valence) electrons in $\hat{O}$, keeping track of minus signs.
Note that we place the creation operators on the right for valence electrons because we always consider the case where the valence bands are fully filled.
The projected density operator is
\bea
\rho_{\mbf{q}+\mbf{G}} &= \sum_{\mbf{k} mn \eta s} M_{mn}^{\eta}(\mbf{k},\mbf{q}+\mbf{G}) c^\dag_{\eta,\mbf{k}+\mbf{q}, m,s} c_{\eta,\mbf{k}, n,s}
\eea
where $c^\dag_{\eta,\mbf{k},m,s}$ creates an electron in valley $\eta$, spin $s$, band $m$, and Bloch momentum $\mbf{k}$, and $M_{mn}^{\eta}(\mbf{k},\mbf{q}+\mbf{G}) $ is the form factor for the bands in valley $\eta$ (see \App{app:2Dint}).

The difference between the CN scheme and the AVE scheme is the presence of a one-body moir\'e background term from normal ordering the interaction in the AVE scheme. This term arises from the Hamiltonian~\cite{kwan2023MFCI3}
\bea
H_{\text{AVE}} &= \sum_{\eta} H_{0,\eta} + \frac{1}{2\mathcal{V}}\sum_{\mbf{q}, \mbf{G}} V(\mbf{q}+\mbf{G}) \, \delta \rho_{\mbf{q}+\mbf{G}}\delta\rho_{-\mbf{q}-\mbf{G}}
\eea
where $\delta \rho$ is the (projected) density measured relative to a uniform background at neutrality:
\bea
\delta \rho_{\mbf{q}+\mbf{G}} &= \!\!\!\!\sum_{\mbf{k} mn \eta s} \!\!\!\! M_{mn}^{\eta}(\mbf{k},\mbf{q}+\mbf{G})(c^\dag_{\eta,\mbf{k}+\mbf{q},m,s}c_{\eta,\mbf{k},n,s} - \frac{1}{2}\delta_{\mbf{q},0} \delta_{mn}) \ .
\eea
We recast the AVE scheme Hamiltonian into the form
\bea
H_{\text{AVE}} = H_{\text{CN}} + \sum_{\eta} H_{b}^{\eta}
\eea
where the one-body background term is
\eqa{
\label{eq:background}
\sum_{\eta} H_{b}^{\eta} & = \sum_{\mbf{q} \mbf{G}} \frac{V(\mbf{q}+\mbf{G})}{2\mathcal{V}} ( \delta \rho_{\mbf{q}+\mbf{G}}\delta\rho_{-\mbf{q}-\mbf{G}} - \!:\!\rho_{\mbf{q}+\mbf{G}}\rho_{-\mbf{q}-\mbf{G}}\!:)
}
corresponding physically to the moir\'e-periodic background including all spins and valleys (see \App{app:2Dint} for explicit expressions). $H_{b}^{\eta}$ is obtained by restricting the creation/annihilation operators in \cref{eq:background}, when expanded out, to valley $\eta$. The band summations implicit in \cref{eq:background} run over all bands in the full single-particle Hilbert space, which we take to encompass all plane waves with momenta $\leq 4|\mbf{q}_1|$, where $\mbf{q}_1$ connects the graphene and hBN Dirac momenta.

Throughout this work, we limit the active degrees of freedom to a single spin-valley flavor (explicitly $\eta=\K$ valley and $s=\uparrow$ spin) in order to achieve a manageable Hilbert space for multi-band ED calculations.
This approximation is justified at $\nu=1$ by HF calculations of `flat band' ferromagnetism that show full flavor polarization.  As a \emph{most favorable} scenario for FCI, we assume this polarization continues to hold at the fractional fillings $\nu=1/3$ and $2/3$ of interest. This assumption can potentially be verified in experiment \cite{2024arXiv240508074X}.
It is important to distinguish which aspects of the interacting physics HF is expected to capture correctly.
The HF approximation should accurately describe the polarizing of valley and spin at $\nu=1$; it would be exact if the flat band were separated from other bands~\cite{kang2019strong,TBG3,TBG4,2024arXiv240104163H} for projected interactions.  After spin and valley polarization, a nearly gapless single-particle band structure remains, which is further gapped by additional interaction effects in HF. This latter mechanism requires more careful consideration: in the CN scheme where the moir\'e potential is very weak, the opening of a gap suggests the formation of an AHC, but HF is known to give erroneous results for the crystallization transition and significantly overestimate the stability of the Wigner crystal for the 2D electron gas~\cite{tanatar1989_2deg,attaccalite2002_2deg,drummond2009_2deg,Trail2003HF,Bernu2011HF}. See \cref{eq:HFbasis} for further discussion.

We will restrict our calculations to the lowest \emph{three} conduction bands (see \cref{fig:BS}a). This is justified for interlayer potentials in the range of $V = 10 - 30$\,meV, where the other conduction bands are significantly higher in energy compared to the energy scale $\sim 20$\,meV of the interaction, and can therefore be assumed to be unoccupied.

To summarize, the Hamiltonian in the CN scheme reads
\eqa{
\label{eq:H_CN}
H^{\text{K}}_{\text{CN}} & = H_0 + \frac{1}{2\V} \sum_{\bsl{k}_1 \bsl{k}_2 \bsl{q} }  \sum_{n_1 n_2 n_3 n_4}  V_{n_1 n_2 n_3 n_4}^{\K \K}(\bsl{k}_1,\bsl{k}_2,\bsl{q})\\
& \quad \times c^\dagger_{\K,\bsl{k}_1+\bsl{q},n_1 ,\uparrow} c^\dagger_{\K,\bsl{k}_2-\bsl{q},n_2 ,\uparrow}  c_{\K,\bsl{k}_2,n_3 ,\uparrow}  c_{\K,\bsl{k}_1,n_4 ,\uparrow} \ ,
}
where $H_0$ is $H_{0,\K}$ (\cref{eq:SP}) restricted to lowest three conduction bands with spin $\uparrow$ in valley $\K$, $V_{n_1 n_2 n_3 n_4}^{\K \K}(\bsl{k}_1,\bsl{k}_2,\bsl{q})$ is defined in \App{appeq:Vetaeta} of \App{app:2Dint}, and the Hamiltonian in the AVE scheme reads
\eqa{
H^{\text{K}}_{\text{AVE}} = H_{\text{CN}}  + H_b \ ,
}
where $H_b$ is $H_{b}^{\K}$ (\cref{eq:background}) restricted to the lowest three conduction bands with spin $\uparrow$ in valley $\K$. Note that while we focus on a single spin-valley flavor, the background term $H_b$ takes into account contributions from \emph{all} spin-valley flavors.

\section{Multi-Band Exact Diagonalization Approaches}
\label{sec:MBED}

For finite systems with periodic boundary conditions, projecting onto the lowest three conduction bands and performing ED is possible on small systems at $\nu=1/3$. However, the 3-band Hilbert space grows rapidly, exceeding $10^9$ even for 15 sites at $\nu=2/3$. While as few as 12 sites sufficed for initial FCI studies on gapped toy models~\cite{regnaultbernevig}, we show that in this gapless system, finite size effects are too strong to make reliable statements about the phases on 12 sites. Thus we must implement additional approximations that reduce the Hilbert space in order access larger systems.

\subsection{Choice of Basis}
\label{sec:choice_of_basis}

In ED calculations, we have the freedom of choosing the orthonormal single-particle basis, which combined with further approximations can lead to reasonable truncation of the Hilbert space.
Our ED calculations in \cref{sec:CN_HFbasis,sec:AVE_HFbasis,sec:nu1_HFbasis} involve making a unitary transformation into the eigenbasis of the $\nu=1$ self-consistent HF Hamiltonian (see \cref{eq:HFbasis})
\bea
\label{eq:H_HF}
H_{\text{HF}} &= H_0 + H_b + H_{\text{HF,int}}[P] \ ,
\eea
which is projected to the 3-band Hilbert space. Here $H_{\text{HF,int}}[P]$ is the mean-field decoupled interaction defined by a Slater determinant with density matrix $P$, $H_0$ is defined below \cref{eq:H_CN}, and $H_b$ is the background term in valley $\K$ and spin $\uparrow$ defined below \cref{eq:background} for the AVE scheme~\cite{kwan2023MFCI3} ($H_b$ is omitted in the CN scheme).
The eigenbasis of $H_{\text{HF}}$ is represented by a unitary transformation $\tilde{U}_\alpha(\mbf{k})$ mixing the single-particle bands of $H_0$ at each $\mbf{k}$, where $\alpha=0,1,2,\dots$ orders the HF bands by increasing energy.
The HF basis operators and band basis operators are related by
\bea
\label{eq:unitary}
\gamma^\dag_{\mbf{k},\alpha} &= \sum_m c^\dag_{\K,\mbf{k},m,\uparrow} \tilde{U}_{m \alpha}(\mbf{k}) \ ,
\eea
where we have dropped the spin $s=\uparrow$ and valley $\eta=\text{K}$ indices. The 3-band projected many-body Hamiltonian can be rewritten in the HF basis using the unitary transformation in \Eq{eq:unitary}:
\bea
\label{eq:H_HF_basismain}
H &= \sum_{\bsl{k},\alpha\beta}t_{\alpha\beta}(\bsl{k}) \gamma^\dagger_{\bsl{k},\alpha} \gamma_{\bsl{k},\beta}  \\
& + \frac{1}{2}\sum_{\bsl{k}\bsl{k}'\bsl{q}} \sum_{\alpha\beta\gamma\delta} V_{\alpha\beta\gamma\delta}(\bsl{k},\bsl{k}',\bsl{q}) \gamma^\dagger_{\bsl{k}+\bsl{q},\alpha} \gamma^\dagger_{\bsl{k}'-\bsl{q},\beta} \gamma_{\bsl{k}',\gamma} \gamma_{\bsl{k},\delta}
\eea
where $t_{\alpha\beta}(\bsl{k})$ and $V_{\alpha\beta\gamma\delta}(\bsl{k},\bsl{k}',\bsl{q})$ are matrix elements of the one-body term and interaction in the HF basis.
We emphasize that eigenvalues of $t_{\alpha\beta}(\bsl{k})$ are \emph{not} the HF dispersion: $t_{\alpha\beta}(\bsl{k})$ has the same spectrum as the one-body term, which is $H_0 + H_b$ in \cref{eq:H_HF} for the AVE scheme or just $H_0$ in the CN scheme, and is gapless.

When generating $\tilde{U}_\alpha(\mbf{k})$, we perform translationally-invariant HF calculations at $\nu=1$ on sizes that are no smaller than $12\times 12$ (see \App{app:mommesh} for details),
select the lowest energy state from many random initial seeds, and choose $\tilde{U}_n(\mbf{k})$ for $\mbf{k}$ on the coarse ED mesh which is a subset of the HF momenta.
Performing the HF on these larger momentum meshes ensures that the lowest HF state has $\Ch=1$ (see \cref{sec:nu1_HFbasis} for a discussion on the competition between $\Ch=0$ and $\Ch=1$ states for small system sizes).
This unitary change of basis does \emph{not} modify the 3-band Hilbert space, but will allow us to \emph{systematically} study and track how the ground state evolves under band mixing.
We stress that calculations of the full 3-band projected Hamiltonian have no bias towards partial occupation of a single gapped HF band.

Besides the HF basis, our ED calculations in \cref{sec:one-body-diagonal} use a different single-particle basis, called the one-body diagonal basis.
Explicitly, we choose the single-particle basis to be
\bea
d^\dag_{\mbf{k},n} &= \sum_m c^\dag_{\mbf{k},m} U_{m n}(\mbf{k}),
\eea
where $U_{m n}(\mbf{k})$ diagonalizes the one-body Hamiltonian $H_{\text{one-body}}$ in \cref{eq:H_one_body} of \cref{sec:one-body-diagonal}. This choice of basis will be explained in \cref{sec:one-body-diagonal}. We emphasize again that in the absence of truncation, both the one-body diagonal basis and the HF basis lead to the exact same results.

\subsection{Multi-band ED Truncation}

The unitary transformation of the 3-band Hamiltonian into the HF basis does not reduce the computational complexity of the ED problem. Since the full 3-band ED problem at $\nu=2/3$ on 18 sites already has Hilbert space dimension of $\sim2 \times 10^{10}$ per momentum sector (beyond state-of-the-art capabilities), controlled approximations are required to reduce the Hilbert space in order to reach larger systems where finite size effects are diminished. We now propose and implement two such approximations, which we refer to as ``band maximum" and ``orbital restriction".

The Fock basis of the 3-band ED Hilbert space with $N_{s}$ sites and fixed particle number $N$ consists of the $\binom{3 N_{s}}{N}$ Slater determinants obtained by enumerating all combinations of momenta and band indices. In the band maximum approximation (abbreviated band-max $\{N_{\text{band1}},N_{\text{band2}}\}$ for short), we restrict each Slater determinant in the basis to have no more than $N_{\text{band1}}$ ($N_{\text{band2}}$) particles in band 1 (band 2). The number of particles allowed in band $0$ is unrestricted. This technique of band maximum truncation can be straightforwardly generalized to calculations involving more than 3 bands. Note that 1-band ED is equivalent to taking $N_{\text{band1}}=N_{\text{band2}}=0$.

If the true ground state only has a few particles in the higher bands, this approximation will be accurate for finite and small $\{N_{\text{band1}},N_{\text{band2}}\}$, and the ground state and many-body gap will converge quickly with respect to these band-max parameters. This would be the case at filling $\nu=1$ if band $0$ were separated from bands $1$ and $2$ by a large single-particle gap. However in R5G/hBN, the lowest three conduction bands are gapless, which suggests that significant band mixing is likely. \cref{eq:hilbertspacedim} contains the formula for the dimension of the Hilbert space as a function of $\{N_{\text{band1}},N_{\text{band2}}\}$. For 18 sites at $\nu=2/3$, the dimension interpolates between  $\sim 2 \times 10^{10}$ in the 3-band Hilbert space to $\sim 10^3$ in the 1-band $\{0,0\}$ Hilbert space. At $\{3,1\}$ for instance, the Hilbert space is $\sim 4.6\times 10^7$, which is tractable. This approximation has been used before~\cite{Rezayi2011breaking,2017PhRvL.119b6801R,kwan2024abelianFTI} to incorporate the effects of band-mixing in the case where the remote bands are separated from the active band by a considerable gap that is still smaller than the interaction scale, but has never been used in nearly gapless systems.

We also implement a new ``orbital restriction", where the basis of Fock states is restricted to exclude all Slater determinants containing a specified set of $[\mbf{k}_b,n_b]$. This approximation is justified when the removed orbitals $[\mbf{k}_b,n_b]$ have high energy, and are expected to be suppressed in the ground state. For instance, the energies of the  bare dispersive bands in R5G/hBN at $\Gamma_M$ are at least $\sim 40$\,meV higher than the flat band. Hence the occupation of this orbital will be very close to $0$ for an interaction of order $20$\,meV, and so we may remove it from the calculation. In particular, based on the non-interacting bare bands in Fig.~\ref{fig:BS}, we can see that the flat band is far away in energy from bands 1 and 2 for $\mbf{k}$ near $\Gamma_M$. The latter orbitals are not expected to be relevant for capturing the many-body states at $\nu\leq 1$.
Similar constraints can be imposed on any effective one-body spectrum, such as the HF dispersion.
\cref{eq:hilbertspacedim} contains the formula for the dimension of the Hilbert space with both band-max and orbital restriction. For instance, removing the 7 highest energy orbitals in both dispersive bands lowers the dimension of the band-max $\{3,1\}$ Hilbert space on 18 sites at $\nu=2/3$ by almost an order of magnitude to $\sim 7\times 10^6$. A list of calculations using orbital restriction is given \App{app:restrictedorb}.

Both these restrictions can be implemented in any basis and interaction scheme, and, if chosen carefully and with physical reasoning, can be leveraged to systematically improve on 1-band ED calculations by moving them towards the unbiased $3$-band calculation for sizes that cannot be reached otherwise. This is because the full 3-band ED problem contains the 1-band projected ED problem as a principal sub-matrix in the many-body Hilbert space. Increasing band-max and relaxing orbital restriction enlarges this sub-matrix towards the full many-body matrix representing the 3-band ED problem.

\subsection{Momentum Mesh}
\label{sec:mommesh}

The momentum meshes we use for our calculations are specified by the form
\eq{
\label{eq:momentum_mesh_main}
\bsl{k} =\frac{k_x}{N_x} \bsl{f}_1 + \frac{k_y}{N_y} \bsl{f}_2
}
with $k_x = 0, 1, 2, ..., N_x -1$ and $k_y = 0, 1, 2, ..., N_y -1$, and $N_s=N_xN_y$ the number of lattice sites. Here, $\mbf{f}_i$ are moir\'e reciprocal lattice vectors related by $SL(2,\mathds{Z})$ transformations to $\mbf{b}_{M,i} = \mbf{q}_3 - \mbf{q}_i$, which are defined in App.~\ref{app:SP}.
Explicitly, the $SL(2,\mathds{Z})$ transformations are specified by $\bsl{f}_1 = \widetilde{n}_{11} \bsl{b}_{M,1} + \widetilde{n}_{12} \bsl{b}_{M,2}$ and $\bsl{f}_2 = \widetilde{n}_{21} \bsl{b}_{M,1} + \widetilde{n}_{22} \bsl{b}_{M,2}$ with $\widetilde{n}_{11}$, $\widetilde{n}_{12}$, $\widetilde{n}_{21}$ and $\widetilde{n}_{22}$ being integers satisfying $|\widetilde{n}_{11} \widetilde{n}_{22} - \widetilde{n}_{21} \widetilde{n}_{12}|=1$.
We choose lattices that achieve the following properties: (1) the total number of sites is divisible by 3 in order to access $\nu=1/3$ and $2/3$;  (2) the $\K_M$ and $\K'_M$ points are included to resolve the near degeneracy of the bare bands at the moir\'e BZ corners; and (3) the FCI momenta are distinct, in order to minimize level repulsion.
For more detailed discussion, see \App{app:mommesh}.

\section{Relation to Lowest HF Band Projection and Summary of Results}

Thus far we have focused on the 3-band ED problem, where the unitary rotation between the band basis and the HF basis has no effect whatsoever on the spectrum. However, the basis choice is crucial when truncating the Hilbert space.
For example, when we restrict our Hamiltonian in the HF basis (\cref{eq:H_HF_basismain}) to $\gamma^\dagger_{\bsl{k},0}$ only (i.e.~orbitals belonging to the occupied HF band at $\nu=1$), we recover the approach thus far employed in \refscite{zhou2023fractional,dong2023anomalous,dong2023theory,guo2023theory} which, as we will show, and as is expected, biases the system towards FCI states.

Explicitly, we rewrite the Hamiltonian $H$ in \cref{eq:H_HF_basismain} so that the interaction is normal-ordered relative to the HF ground state. \App{app:HFhamiltonianbasis} shows that the appropriate identity is
\bea
\label{eq:HnormalorderHF}
H &= E_0^{\text{HF}} + \sum_{\mbf{k},\alpha} \eps^{\text{HF}}_\alpha(\mbf{k}) ::\gamma^\dag_{\mbf{k},\alpha} \gamma_{\mbf{k},\alpha}:: + :: H_{\text{int}} ::
\eea
where $E_0^{\text{HF}}$ is the energy of the self-consistent HF ground state and $\eps^{\text{HF}}_\alpha(\mbf{k})$, $\alpha=0,1,2\ldots $, is the dispersion of the self-consistent HF Hamiltonian. Crucially, in \Eq{eq:HnormalorderHF}, the double colons signify normal ordering against the $\nu=1$ HF ground state, \ie, the electron operator $\gamma^\dag_{\mbf{k},0}$ and hole operators $\gamma_{\mbf{k},1}, \gamma_{\mbf{k},2}$ are placed to the right (keeping track of the fermionic signs).
Explicitly, \cref{eq:HnormalorderHF} with $\alpha$ limited to $0$ is what was used for FCI calculations in \refscite{zhou2023fractional,dong2023anomalous,dong2023theory,guo2023theory}, which is equivalent to \cref{eq:H_HF_basismain} with $\alpha$ limited to $0$.
In other words, projecting $H$ to the lowest HF band can be performed in either representation of $H$, \Eq{eq:H_HF_basismain} or \Eq{eq:HnormalorderHF}, yielding equivalent 1-band ED problems, as verified in \App{app:HFhamiltonianbasis}.

One subtlety is that the equivalence in \cref{eq:HnormalorderHF} exactly holds only when the HF is done on the same momentum mesh as the ED calculations.
In practice, as discussed in \cref{sec:choice_of_basis}, we always choose the momentum mesh in HF to be larger than ED calculations.
Nevertheless, the equivalence in \cref{eq:HnormalorderHF} still approximately holds in this case to a very high precision, as discussed in \App{app:HFhamiltonianbasis}.
Thus to a very good approximation, our multi-band calculations reduce to the 1-band approximations employed in \refscite{zhou2023fractional,dong2023anomalous,dong2023theory,guo2023theory,2024arXiv240708661H} when we consider only one band. This provides the opportunity to assess the validity of existing results by including multiple bands.

It is tempting to conclude that projection to the lowest HF band is justified because of the $\sim 20$\,meV charge gap in $\eps^{\text{HF}}_\alpha(\mbf{k})$ in \Eq{eq:HnormalorderHF}. Even if one believes the HF approximation quantitatively captures the ground state at $\nu=1$, its gap is \emph{always} of the same order as the interaction (or smaller): it is an interaction-induced gap. This invalidates the projection into the HF band, as the interaction mixes further bands non-perturbatively. Thus there is no analogy to the case of an isolated flat band whose gap is determined by single-particle hoppings. Indeed, we will show that the 1-band ED results are not reliable, and bias the system heavily towards FCIs. Our multi-band calculations yield results contrary to those of existing work.

In \cref{sec:Displacement_Field}, we first motivate the choices of $V$ we will use in the subsequent sections. In \cref{sec:nu1_HFbasis}, we combine the HF and HF-basis ED calculations to provide an argument that the $\Ch=1$
state should be the ground state at $\nu=1$ in the thermodynamic limit for the AVE scheme, while in the CN scheme the continuum collapses.

In \cref{sec:CN_HFbasis} we demonstrate that the FCIs obtained in the CN scheme at $\nu=1/3$ and $\nu=2/3$ are destroyed as multiple bands are included. At $\nu=1/3$, we show this in 3-band ED at Hilbert spaces of dimension $1.6\times 10^6$ (Fig.~\ref{fig:CNrdm}). While 3-band ED is impossible at $\nu=2/3$, we employ band-max techniques to reach Hilbert spaces of $10^7$ and establish the demise of the FCI due to band-mixing effects (Fig.~\ref{fig:AHC2main}a). We then argue based on Goldstone's theorem that the many-body gap must vanish in the moir\'e-less limit for the 1-band FCI ground state since it spontaneously breaks continuous translation symmetry, and show that the 1-band approximation excludes fluctuations that would enforce gaplessness in the spectrum.

In \cref{sec:AVE_HFbasis}, we show that the AVE scheme displays markedly different behavior between $\nu=1/3$ and $\nu=2/3$. At $\nu=1/3$, the larger dispersion prevents a well-developed FCI even in 1-band ED, and the FCI gap quickly vanishes as fluctuations into the higher bands are allowed (Fig.~\ref{fig:AVE2main}b). At $\nu=2/3$, the FCI gap is actually \emph{enlarged} as higher bands are first included, and could be mistaken for a developed FCI state if calculations are not pushed to higher sizes. Ultimately, the gap starts to show decreasing behavior and  the continuum excitations collapse (Fig.~\ref{fig:AVE2main}a and Fig.~\ref{fig:AVEnk}). We reach Hilbert space dimensions of  $1.7\times 10^7$ and $5.5\times 10^8$ at $\nu=1/3$ and $\nu=2/3$ respectively.
In \cref{sec:one-body-diagonal}, we will show that the truncated ED results in the one-body diagonal basis are consistent with those performed with the HF basis.

\section{Displacement Field}
\label{sec:Displacement_Field}

\begin{figure*}
\centering
\includegraphics[width=1.8\columnwidth]{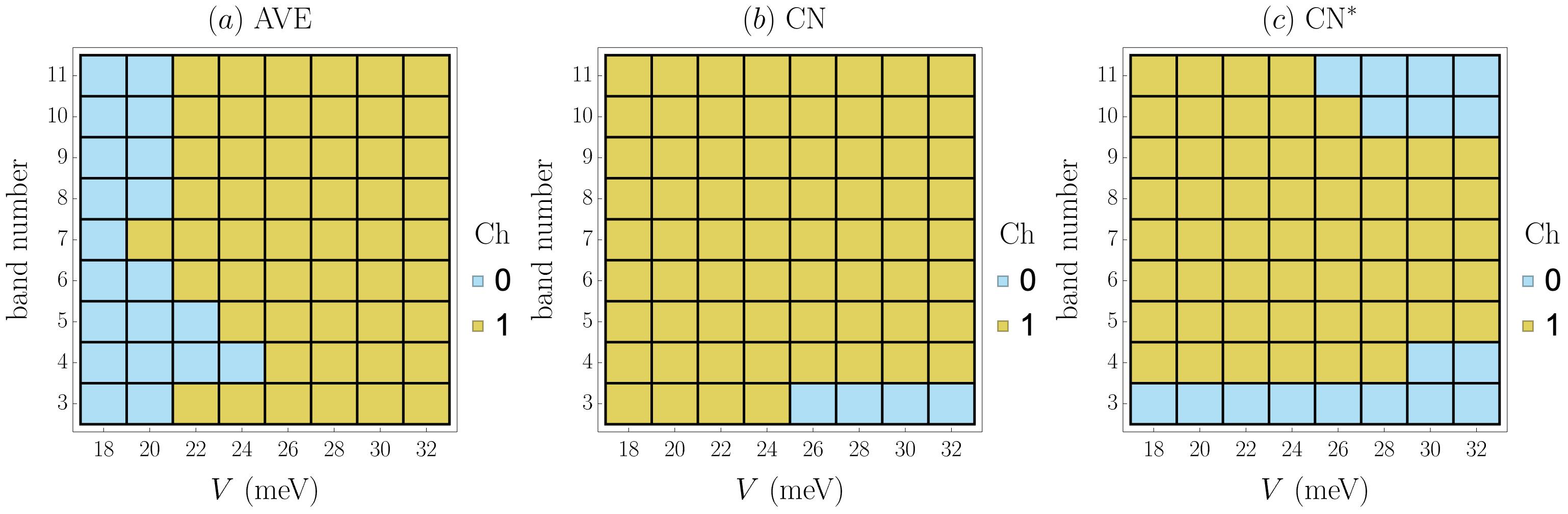}
\caption{
We show the HF phase diagram at $\nu=1$ in the (a) AVE, (b) CN and (c) CN$^*$ schemes as a function of the interlayer potential $V$ and the number of conduction bands (`band number') included in the HF calculation.
CN$^*$ in (c) stands for the moir\'e-less limit of the CN scheme.
Blue (green) indicates that the ground state has $\Ch=0$ ($\Ch=1$).
The momentum mesh  is $(N_x, N_y, \widetilde{n}_{11}, \widetilde{n}_{12}, \widetilde{n}_{21}, \widetilde{n}_{22}) = (12,12,1,0,0,1)$.
}
\label{fig:HFPDplot12x12}
\end{figure*}

\begin{figure}
\centering
\includegraphics[width=\columnwidth]{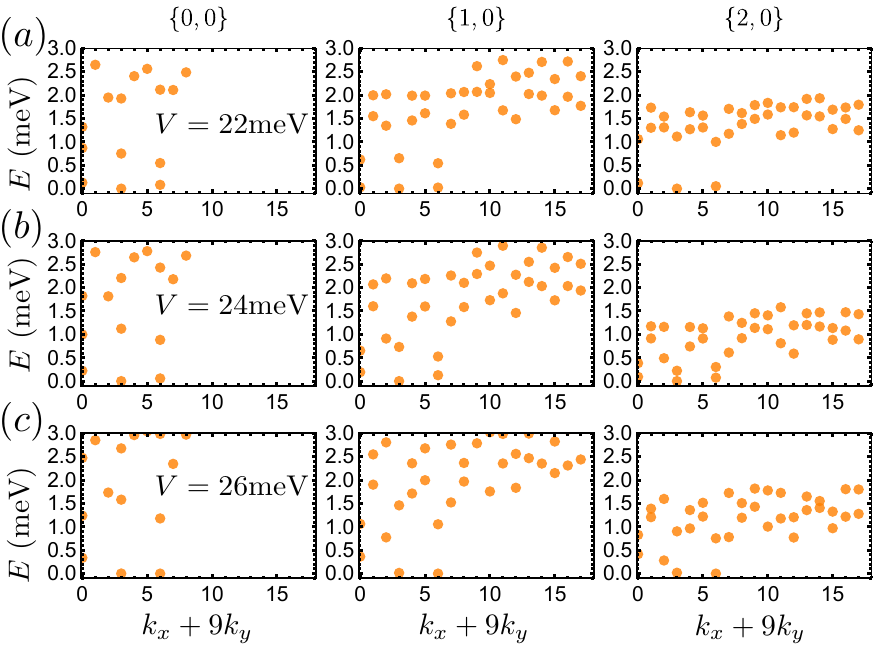}
\caption{
We show the band-max truncated 3-band HF-basis ED spectra at $\nu=2/3$ in the AVE scheme on the $9\times 2$ lattice for $V=22,24$ and $26$\,meV. The HF basis is computed on a larger $18\times 18$ system which ensures that the HF ground state has $\Ch=1$. Although $V=22$meV in $(a)$ shows the smallest FCI gap at band-max $\{0,0\}$, it exhibits the most robust FCI at band-max {2,0}.
}
\label{fig:AVE_V_9x2}
\end{figure}

Before embarking on extensive numerical calculations, we need to choose representative values of the displacement field.
To do so, we first perform moir\'e translation-invariant HF calculations  at $\nu=1$ on the momentum mesh $(N_x, N_y, \widetilde{n}_{11}, \widetilde{n}_{12}, \widetilde{n}_{21}, \widetilde{n}_{22}) = (12,12,1,0,0,1)$ (see conventions in \cref{eq:momentum_mesh_main}) at various values of $V$ for the AVE, CN and moir\'e-less CN (CN$^*$) schemes.
We consider HF calculations projected into the $3,4,5,6,7,8,9$ and $11$ lowest conduction bands in order to determine the values of $V$ where the 3-band calculation is expected to be valid and the HF ground state has $\Ch=1$.

As shown in \cref{fig:HFPDplot12x12}a, the ground state in the AVE scheme for $V\geq 22\,$meV has $\Ch=1$ for 3-band calculations, which is also true for the limit of large numbers of bands as demonstrated by the 6, 7, 8, 9, 10 and 11-band calculations.
(HF energy differences are shown in \cref{app:2D_Int_HF}.)
This consistency motivates the validity of 3-band calculations for $V\geq22\,$meV in the AVE scheme, despite the fact that the HF $\Ch=0$ phase extends to $V=24\,$meV ($V=22$\,meV) for the intermediate calculation with $4$ bands ($5$ bands).
To further select a value of $V$ from the range $V\geq22$\,meV, we perform 3-band HF-basis ED calculations at $\nu=2/3$ using the momentum mesh $(N_x, N_y, \widetilde{n}_{11}, \widetilde{n}_{12}, \widetilde{n}_{21}, \widetilde{n}_{22}) = (9,2,1,-2,0,1)$ for $V=22,24$ and $26$\,meV with either no or small band-mixing, \ie, band-max $\{N_\text{band1},N_\text{band2}\}=\{0,0\},\{1,0\}$ and $\{2,0\}$.
As shown in \cref{fig:AVE_V_9x2}, the FCI is most robust at $V=22$\,meV for small band-mixing, as indicated by the larger FCI gap at $\{2,0\}$ for $V=22$\,meV than for $V=24$\,meV, while the FCI gap already collapses at $\{2,0\}$ for $V=26$\,meV.
Therefore, we choose $V=22$\,meV for $3$-band calculations the AVE scheme.

For the CN scheme, $V=22$\,meV is also appropriate since the HF ground state has $\Ch=1$ for all $3,4,5,6,7,8,9,10$ and $11$-band calculations as long as $V\leq 24$\,meV, as shown in \cref{fig:HFPDplot12x12}b.
Therefore, we also choose $V=22$\,meV for 3-band calculations in the CN scheme, unless specified otherwise.

In the CN$^*$ limit, we do not yet observe signs of convergence even for relatively large numbers of bands, as shown in \cref{fig:HFPDplot12x12}c, suggesting that extra caution needs to be taken when doing HF calculations with the CN$^*$ scheme.
Nevertheless, within ED calculations, we find close similarity between the CN scheme and the CN$^*$ limit, as discussed in \cref{sec:CN_HFbasis}c.

\section{$\nu=1$ Competition in Hartree-Fock and Exact Diagonalization Studies}
\label{sec:nu1_HFbasis}

In this section, we address the $\nu=1$ state, which has so far only been approached at the level of HF~\cite{dong2023anomalous,zhou2023fractional,dong2023theory,kwan2023MFCI3,guo2023theory} despite reservations about its validity in the moir\'e-less limit \cite{2024arXiv240307873D}. With a thorough finite-size analysis, we show that there is a clear competition between the $\Ch=0$ and $\Ch=1$ states which is severely affected by system size even in HF, with the $\Ch=1$ state ultimately winning on large sizes. In ED, we also find and confirm the close competition between $\Ch=0$ and $\Ch=1$ states which is similarly seen in HF. We argue that the gap to the continuum excitations (at momenta different from the ground state) shows signs of convergence to a nonzero value ($\sim 1$\,meV) in the AVE scheme, in contrast to the CN scheme where the continuum can collapse with sufficient band-mixing.
In addition, we show that one-body correlation functions in the $\Ch=1$ and $\Ch=0$ states are Slater-like in the AVE scheme ED calculations, giving evidence that HF might give a reliable approximation of the $\nu=1$ GS in the AVE scheme (while likely over-estimating the charge gap). By combining the HF and ED calculations, we conjecture that the ground state (GS) in the thermodynamic limit has $\Ch=1$ in the AVE scheme.

\begin{figure}
\centering
\includegraphics[width=\columnwidth]{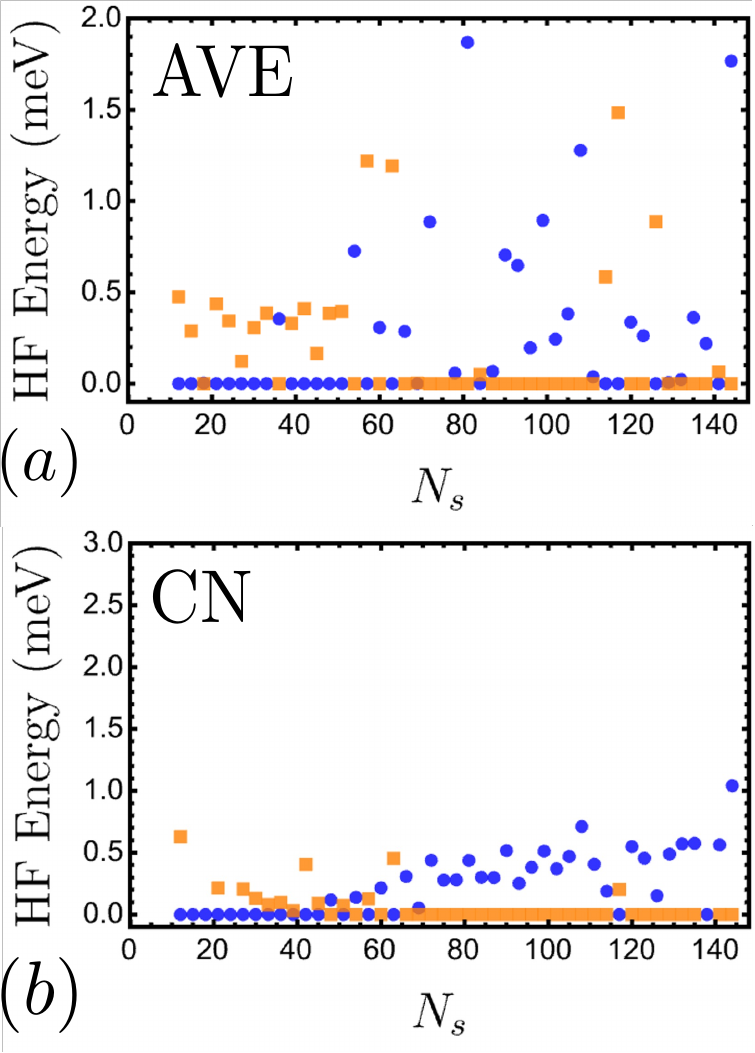}
\caption{Total HF energy versus the system size $N_s$ in the $(a)$ AVE and $(b)$ CN scheme at $V=22$\,meV and $\nu=1$.
The orange dots mark the $\Ch=1$ states, while the blue dots mark the $\Ch=0$ states.
Moir\'e translation invariance is enforced in the HF calculation, and we choose 20 random initial states to seed the self-consistent calculation for each system size $N_s$, leading to only two distinct phases.
At each size, we subtract the HF energy of the lowest state to show the gap.
}
\label{fig:HFscalingPlot}
\end{figure}

In both AVE and CN schemes, we perform moir\'e translation-invariant HF calculations on various system sizes with 20 initial random states for each system size at $V=22$\,meV.
We note that the convergence of the HF calculations in the CN scheme requires more iterations than that in the AVE scheme.
In \figref{fig:HFscalingPlot}, we summarize the HF results on meshes (see conventions in \cref{eq:momentum_mesh_main}) specified by $(N_x, N_y, \widetilde{n}_{11}, \widetilde{n}_{12}, \widetilde{n}_{21}, \widetilde{n}_{22}) = (2,6,1,0,1,1)$, $(15,1,1,-5,0,1)$, $(3,6,1,0,0,1)$, $(21,1,1,-5,0,1)$, $(4,6,1,0,1,1)$, $(9,3,1,-2,0,1)$, $(5,6,0,-1,-1,-1)$, $(33,1,-4,-1,-1,0)$, $(6,6,1,0,0,1)$, $(6,9,1,0,0,1)$, $(6,12,1,0,0,1)$, $(9,9,1,0,0,1)$, $(9,12,1,0,0,1)$, $(12,12,1,0,0,1)$ and $(x,1,1,10,0,1)$ with $x=39,42,45,...,141$ excluding $54,72,81,108$, which all contain the $K_M$ and $K_M'$ points.
We did not include $x=54,72,81,108$ for $(x,1,1,10,0,1)$ (see the HF results on those meshes in \cref{app:2D_Int_HF}), because they have the the same $N_s$ as $(6,9,1,0,0,1)$, $(6,12,1,0,0,1)$, $(9,9,1,0,0,1)$ and $(9,12,1,0,0,1)$.
As shown in \figref{fig:HFscalingPlot}, the HF ground state is mostly $\Ch=0$ ($\Ch=1$) for $N_s < 60$ ($N_s \geq 60$) in both AVE and CN schemes.
The Chern number is determined by $C_3$ eigenvalues at $\Gamma_M$, $\K_M$ and $\K_M'$ (up to mod 3), except $(N_x,N_y)=(6,6)$ and $(12,12)$ where the Chern number is determined by the integration of the Berry curvature.
The fact that the ground state is not consistently $\Ch=1$ for different $N_s$ indicates strong finite-size effects. This is also reflected by the finding that as we twist the boundary condition, the $\Ch=1$ state does not stay as the ground state even on systems as large as $N_s=144$~\cite{YZprivate}, indicating strong finite-size effects on the system (see \App{app:2D_Int_HF} for details).
Nevertheless, our data exhibits a clear trend in HF where small system sizes favor the $\Ch=0$ state, while the $\Ch=1$ state eventually wins as the system size increases.

\begin{figure*}
\centering
\includegraphics[width=\linewidth]{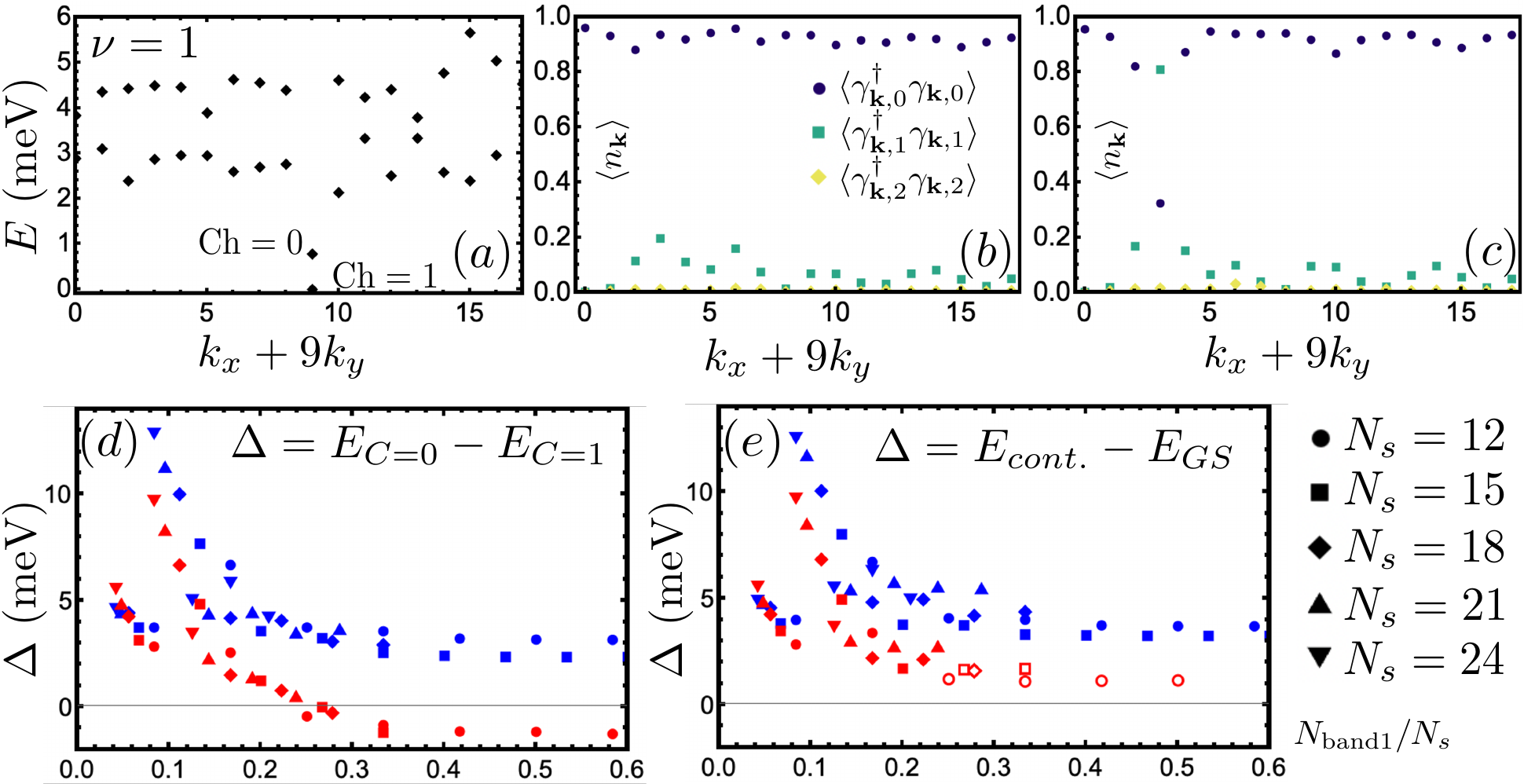}
\caption{ED calculations in the AVE scheme at $\nu=1$ and $V=22$\,meV.
$(a)$ Representative spectrum on the $9\times 2$ lattice at band-max $\{4,1\}$. The $\Ch=1$ and $\Ch=0$ ground states are labeled below the continuum. Corresponding band occupations are plotted in $(b)$, showing a nearly fully occupied $\Ch=1$ HF band, and $(c)$, showing a band inversion at the $K_M$ point leading to a $\Ch=0$ state. $(d)$ We show the scaling behavior of the many-body gap between the $\Ch=0$ and $\Ch=1$ states in the HF momentum sector with $N_{\text{band1}}/N_s$. Blue and red symbols denote $N_{\text{band2}} = 0$ and $1$ respectively. $(e)$ We show the scaling behavior of the many-body gap between the ground state and the lowest energy of the continuum (at a different momentum) with $N_{\text{band1}}/N_s$. Solid (open) symbol indicates that the ground state has $\Ch=1$ ($\Ch=0$). }
\label{fig:AVEinteger}
\end{figure*}

The fact that small sizes favor $\Ch=0$ states can also be seen in our ED results in the AVE scheme within multi-band calculations. These calculations enable an assessment of the accuracy of the proposed HF ground state as well as the stability against fluctuations.
We can separate the many-body spectrum into three parts: a $\Ch=1$ state, a $\Ch=0$ state, and a continuum of excitations (see \cref{fig:AVEinteger}a for an example). The Chern states are identifiable by their many-body momentum,$\sum_{\mbf{k}} \mbf{k}$ mod reciprocal lattice vectors, which is that of a fully-occupied Slater state,  and the occupation factors, shown for example in Figs.\,\ref{fig:AVEinteger}b,c. Specifically, for a given ED state, we can calculate its occupation factor in the HF basis, \ie, $\langle \gamma^\dagger_{\bsl{k},\alpha} \gamma_{\bsl{k},\alpha}\rangle$ for $\alpha=0,1,2$.
If the ED state is exactly the same as the HF state, we would have $\langle \gamma^\dagger_{\bsl{k},\alpha} \gamma_{\bsl{k},\alpha}\rangle = \delta_{\alpha 0}$ for all $\bsl{k}$.
Therefore, we identify a ED state as a $\Ch=1$ state if (1) its many-body momentum is equal to $\sum_{\bsl{k}} \bsl{k}$ mod reciprocal lattice vectors and (2) it has $\langle \gamma^\dagger_{\bsl{k},0} \gamma_{\bsl{k},0}\rangle>\langle \gamma^\dagger_{\bsl{k},\alpha} \gamma_{\bsl{k},\alpha}\rangle $ for $\alpha=1,2$.
The reasoning for (2) is that the HF calculation for generating the HF basis was done on a large enough size such that $\gamma^\dagger_{\bsl{k},0}$ corresponds to a $\Ch=1$ state.
\cref{fig:AVEinteger}d shows that the $\Ch=1$ state, which consists of fully occupying band 0 in the band-max $\{0,0\}$ limit, is stable to band mixing when $N_{\text{band1}}$ is increased (blue symbols). However, for the sizes achievable with our Hilbert space truncation techniques, the $\Ch=1$ state is out-competed by the $\Ch=0$ state when $N_{\text{band2}}$ is nonzero (red symbols). This is largely consistent with moir\'e translation-invariant HF performed directly on these system sizes\footnote{Ref.~\cite{kwan2023MFCI3} has found that the HF energy can be lowered by breaking moir\'e translation and doubling the unit cell along one axis. This is consistent with the presence of negative energy collective modes at non-zero $\mbf{q}$, including at the $M_M$ points, for the translation-invariant HF solution. Within the system sizes and band-max truncations accessed in our current ED calculations, we find no clear indications of a density wave ground state in the many-body spectrum.}.

To access the physics that is less afflicted by finite size effects, we then study the gap between the continuum and the ground state (either $\Ch=0$ and $\Ch=1$). From \cref{fig:AVEinteger}e, we see that this gap converges to a finite value when plotted as a function of $N_{\text{band1}}/N_s$, the maximum density of particles allowed in band 1, for both $N_{\text{band2}}=0$ and $N_{\text{band2}}=1$. This provides evidence that the continuum modes do not collapse, which is in sharp contrast to case of fractional fillings in the AVE scheme presented in \cref{sec:AVE_HFbasis}, where we will find that the continuum states at generic momenta eventually replace the FCI ground states.

Combined with the HF results showing that the $\Ch=1$ state wins over the $\Ch=0$ state and becomes the ground state as the size increases in the AVE scheme (Fig.~\ref{fig:HFscalingPlot}), we conjecture that in thermodynamic limit, the ground state of the full 3-band model in the AVE scheme is a gapped, moir\'e translation invariant $\Ch=1$ state.

\begin{figure*}
\centering
\includegraphics[width=\linewidth]{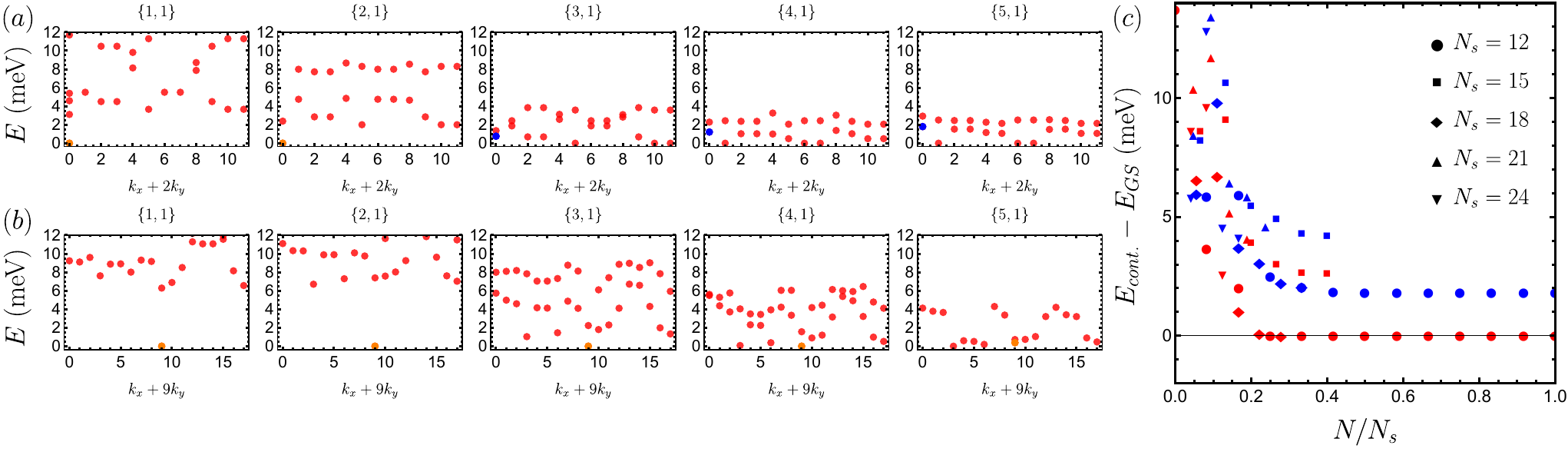}
\caption{ED calculations in the CN scheme at $\nu=1$ and $V=22$\,meV for $N_{\text{band2}}= 1$ on the $2\times 6$ mesh in $(a)$ and $9\times 2$ mesh in $(b)$.
In $(a)$ and $(b)$, the lowest-energy state is marked with $\Ch=1$ (orange) or $\Ch=0$ (blue). \App{app:result_2D_int_nu_1} contains plots of the correlation functions.
At band-max $\{3,1\}$ for 12 sites and $\{4,1\}$ for 18 sites, we observe the collapse of the continuum. $(c)$ We plot the continuum gap $E_{cont.}-E_{GS}$ on 12, 15, 18, 21 and 24 sites. Blue and red symbols denote $N_{\text{band2}} = 0$ and $1$ respectively.}
\label{fig:CNintegercollapse}
\end{figure*}

In contrast to the AVE scheme, the continuum does not consistently maintain a nonzero gap above the ground state in the CN scheme, \ie, it can drop down and destroy the ground state when sufficient band mixing is allowed. This can be seen, for example, at band-max $\{5,1\}$ for 12 and 18 sites in Figs.\,\ref{fig:CNintegercollapse}a,b.
More specifically, we find that the trend of the continuum dropping down is followed by the calculations on $12,18$ and $24$ sites, though the continuum appears to drop more slowly on $15$ and $21$ sites, as shown in \cref{fig:CNintegercollapse}c. One potential reason for this even/odd effect is that the 15- and 21-site meshes do not include any of the three $M_M$ points, where  band 0 and band 1 are nearly degenerate and band 0 and band 2 are close, as shown in \cref{fig:BS}.

\refscite{zhou2023fractional,dong2023anomalous,dong2023theory,kwan2023MFCI3} found a moir\'e-pinned AHC at $\nu=1$ in the CN scheme within HF calculations.
This state, whose periodicity coincides with the moir\'e lattice, appears in HF for a range of twist angles, which control the size of the moir\'e unit cell.
In ED calculations, evidence for the Wigner crystal order can be seen from the ED spectrum~\cite{Gorshkov2018WCinED,2021PhRvB.104h5107K} through the quasi-degenerate ground state momenta, which are consistent with the translation-breaking order, and through the lack of a clear gap indicating Goldstone modes and/or the Anderson tower of states on large enough systems~\cite{anderson1952antiferromagnetic,wietek2017studying}.

As shown in Figs.\,\ref{fig:CNintegercollapse}a,b, the band-mixing-induced collapse of the continuum can make the ground state momentum different from the many-body momentum expected for a $\nu=1$ moir\'e-periodic crystalline order, which is $(0,0)$ for 12 sites and $(0,1)$ for 18 sites. On 12 sites, the ground states occur at the 3 $M_M$ points for large band-max, while on 18 sites the $K_M$ point is the lowest energy, although there are many competing states at band-max $\{5,1\}$.
Hence, our ED results show that with enough band mixing, the ground state momentum on 12 and 18 sites is different from that of the moir\'e-pinned AHC at $\nu=1$, which shares the same ground state momentum as the $\nu=1$ moir\'e-periodic HF Slater state. This gives credence to the doubts articulated in \refcite{2024arXiv240307873D} about the suitability of HF for describing the many-body ground state at $\nu=1$ in the CN scheme. The apparent gapless phase we observe replacing the moir\'e-pinned AHC may be a gapless liquid or a distinct Wigner-like state with different spatial periodicity that out-competes the moir\'e-pinned AHC~\cite{kwan2023MFCI3}. A more detailed discussion of the potential roles played by different fluctuations is given in \cref{sec:CN_HFbasis}.

\section{CN Scheme Results at $\nu=1/3$ and $\nu=2/3$}
\label{sec:CN_HFbasis}

\begin{figure*}
\centering
\includegraphics[width=\linewidth]{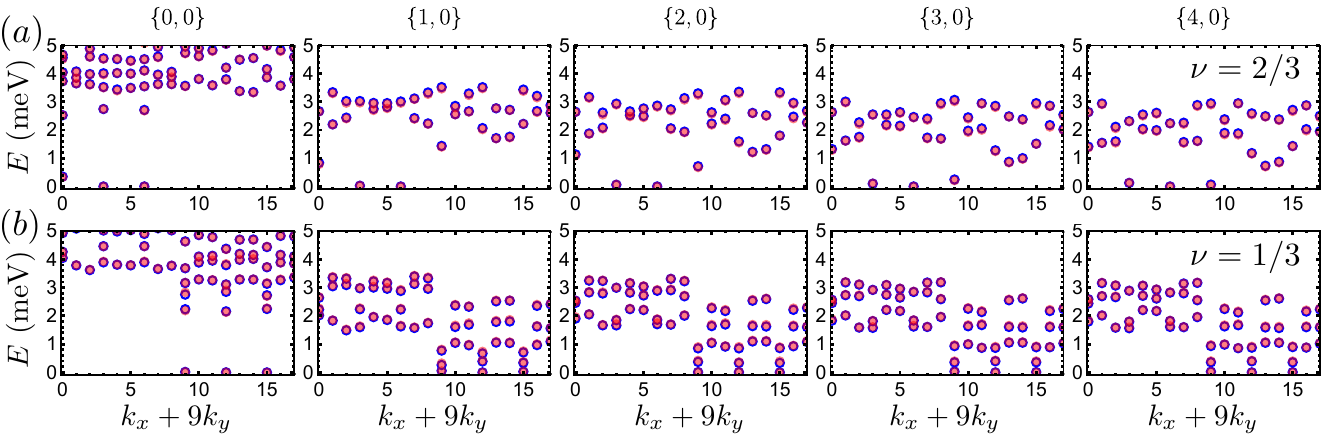}
\caption{Collapse of the CN scheme FCI with $N_{\text{band1}}$ for $N_{\text{band2}}=0$ at $(a)$ $\nu=2/3$ and $(b)$ $\nu=1/3$. The CN scheme spectrum (red filled dots) matches very closely with that in the moir\'e-less limit (blue  circles). Calculations are performed on the $9\times2$ tilted lattice at $V=28$\,meV. Similar results are seen for other sizes, values of $V$, and including $N_{\text{band}2}\neq 0$ (see \App{app:CNscheme}). The ground state at band-max $\{0,0\}$ has the momenta of the Laughlin state \cite{2012PhRvB..85g5128B} and shows a $\sim 2$\,meV gap in panel $(a)$ and $(b)$. On the other hand, the ground state in panel $(a)$ $\{4,0\}$ must be gapless by the LSM theorem \cite{LSM,PhysRevLett.84.1535,PhysRevLett.127.237204,PhysRevB.110.045107}, and the gap collapses in panel $(b)$. }
\label{fig:AHC2main}
\end{figure*}

In this section, we present ED results in the CN scheme (\Eq{eq:H_CN}) at $\nu=1/3$ and $2/3$. We perform both multi-band and 1-band ED calculations to investigate the stability of the FCIs obtained in the 1-band limit against band mixing. We also examine the proximity of the CN scheme to the moir\'e-less limit, and discuss the role of Goldstone modes and fluctuations for the putative FCI phases.

\subsection{ED Results}
We first note that at $V=22$\,meV, both $\nu=1/3$ and $\nu=2/3$ do not show FCIs at all band-max values considered, including even the 1-band $\{0,0\}$ limit (see \App{app:FCI_CN}).
To compare with Refs.~\cite{dong2023anomalous,zhou2023fractional,dong2023theory} which find that FCIs can exist at $\{0,0\}$ for the CN scheme with other band structure parameters, we resort to a larger value of $V=28$\,meV.
As the absence of a $\Ch=1$ HF ground state at $\nu=1$ in \cref{fig:HFPDplot12x12}b suggests that 3-band projection in the CN scheme at $V=28$\,meV may not be reliable, we perform 5-band HF and undertake ED calculations restricted to the Hilbert space of the lowest 3 HF bands. This is equivalent to imposing $N_\text{band3}=N_\text{band4}=0$ within the framework of 5-band ED. We will find that FCI states are destroyed even without allowing any particles in band 3 and band 4.

We begin with the 1-band results (left-most column of  \cref{fig:AHC2main}) at band-max $\{0,0\}$. This projection to the occupied HF band obtained using $\nu=1$ HF is expected to severely bias towards FCIs at fractional fillings, as it is known that projected Chern bands can host FCIs on lattices as small as 12 sites \cite{regnaultbernevig}.  Consistent with Refs.~\cite{dong2023anomalous,zhou2023fractional,dong2023theory,guo2023theory}, we observe FCIs at both $\nu=1/3$ and $\nu=2/3$ with a clear gap greater than 2\,meV.  The appearance of FCIs is unsurprising in this restricted calculation: the lowest HF band has $\Ch = 1$ and has a relatively flat 10\,meV bandwidth at $V=28$\,meV with somewhat ideal quantum geometry~\cite{dong2023anomalous,zhou2023fractional,dong2023theory,guo2023theory,kwan2023MFCI3} characterized by a dimensionless trace condition violation of $\sim 1.2$~\cite{Parameswaran13,Roy14} (exact FCI ground states can be obtained if the trace condition is 0 for flat bands with Haldane pseudo-potentials \cite{PhysRevLett.127.246403,PhysRevResearch.5.023167,2021arXiv210709039M,2022arXiv220915023L,2023PhRvR...5c2048E,2024arXiv240413455S}). This is precisely the setting where FCIs are typically seen. In the 1-band approximation, an FCI appears at $\nu=1/3$ (stronger than $\nu=2/3$) although no such state is observed in experiment~\cite{Lu2024fractional}.

We now incorporate band mixing by allowing particles to populate the second HF band in order to systematically remove the bias towards FCIs from 1-band truncation. \cref{fig:AHC2main}a shows the immediate decrease of the FCI gap at $\nu=2/3$ and the dramatic loss of topological degeneracy. When two or more particles are allowed in band 1, the ground states are no longer consistent with the Lieb-Schultz-Mattis (LSM) momenta of a gapped state~\cite{LSM,PhysRevLett.84.1535,PhysRevLett.127.237204}. Thus the FCI definitively ceases to be the ground state in ED, and is replaced by a gapless phase. At $\nu=1/3$, \cref{fig:AHC2main}b shows a steady decrease of the gap, although the ground state momenta do not change, and full 3-band ED at $\nu=1/3$ (Fig.~\ref{fig:CNrdm}a) shows the absence of a  clear gap in the spectrum. The rapid convergence in band-max to the 3-band result for $\nu=1/3$ is shown in \App{app:CNscheme}. From analyzing the ED spectra, we conclude that both the $\nu=1/3$ and $2/3$ FCIs do not survive band-mixing, inevitably present in the problem.

\begin{figure}
\centering
\includegraphics[width=\linewidth]{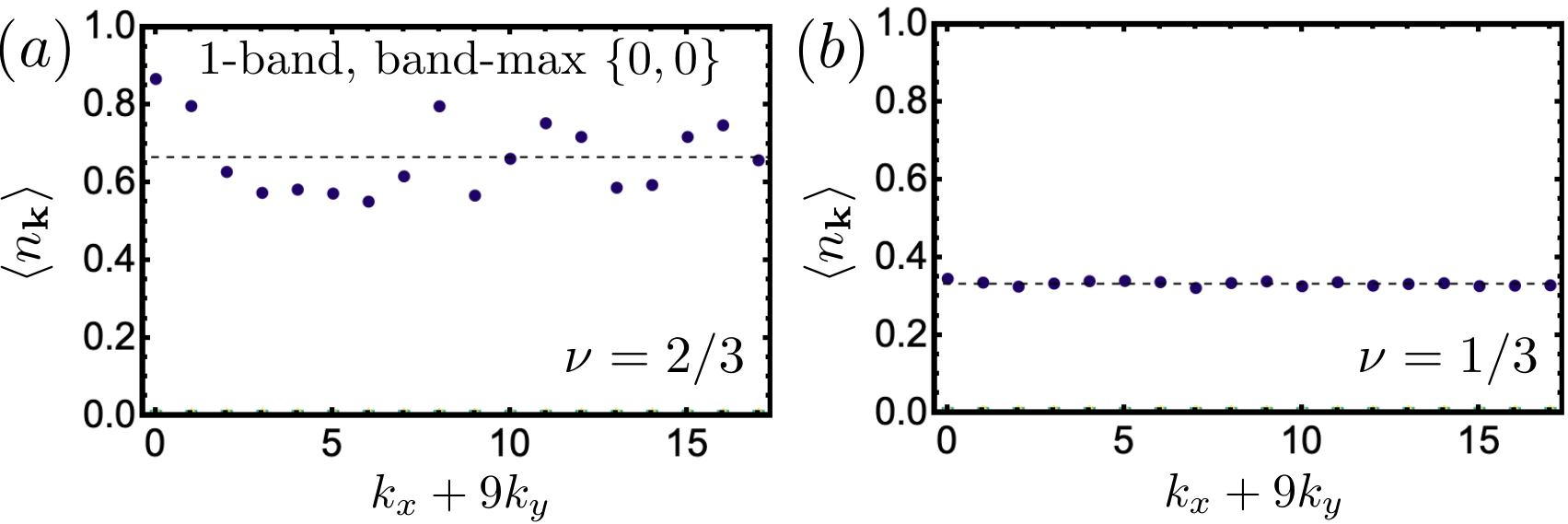}
\caption{Correlation functions in 1-band ED calculations (band-max $\{0,0\}$) in the CN scheme on the $9 \times 2$ lattice at $V=28$\,meV, averaged over the 3 quasi-degenerate ground states.
$(a)$ $\braket{n_{\mbf{k}}}$ shows no clear sign of a Fermi surface, indicating an FCI at $\nu=2/3$. $(b)$ The $\nu=1/3$ state also shows very flat $\braket{n_{\mbf{k}}}$, indicating a well-formed FCI. }
\label{fig:CNrdm_1band}
\end{figure}

To analyze the properties of the finite size ground states obtained above, we compute one-body correlators defined by
\bea
\braket{\gamma^\dag_{\mbf{k},\alpha}\gamma_{\mbf{k},\alpha}} &= \frac{1}{N_{GS}} \sum_{a=1}^{N_{GS}} \braket{a|\gamma^\dag_{\mbf{k},\alpha}\gamma_{\mbf{k},\alpha}|a}  \ ,\\
\braket{n_{\mbf{k}}} &= \sum_\alpha \braket{\gamma^\dag_{\mbf{k},\alpha}\gamma_{\mbf{k},\alpha}},
\eea
where $N_{GS} = 3$ is the number of quasi-degenerate ground states $\ket{a}$ at $\nu=1/3$ and $2/3$. In \cref{fig:CNrdm_1band}, we first compare the $\braket{n_{\mbf{k}}}$ fluctuations between $\nu=1/3$ and $\nu=2/3$ in the 1-band (band-max $\{0,0\}$) approximation, finding much more uniform occupation at $\nu=1/3$, indicative of a well-developed FCI, compared to $\nu=2/3$. This is consistent with the larger splitting of the topological degeneracy at $\nu=2/3$, leading to the rapid destruction of the FCI as additional bands are included. (This asymmetry between the quality of the FCI at $\nu=1/3$ versus $2/3$ in a 1-band calculation can also occur for perfectly ideal bands~\cite{2024arXiv240204303L}). These results are inconsistent with the experimental finding of an FCI at $\nu=2/3$ versus its absence at $\nu = 1/3$~\cite{Lu2024fractional}.
For the full 3-band ED calculations at $\nu=1/3$ (see \cref{fig:CNrdm}), $\braket{\gamma^\dag_{\mbf{k},\alpha}\gamma_{\mbf{k},\alpha}}$ shows enlarged fluctuations from populating the higher bands, further establishing the absence of an FCI here.

\subsection{Goldstone's Theorem}

In all cases, we find that the spectra computed with (CN scheme) and without (CN$^*$ scheme) the moir\'e potential are extremely similar (compare red and blue symbols in \cref{fig:AHC2main}, which differ by at most $0.1$\,meV), demonstrating that the moir\'e-less limit is approximately attained in the CN scheme calculation.
We conjecture that our numerical results demonstrating the destabilization of the FCIs to band mixing can be contextualized using the framework of Goldstone's theorem in the moir\'e-less limit of the Hamiltonian. Let us first consider the $\Ch=1$ AHC at $\nu=1$ obtained in HF~\cite{dong2023anomalous,zhou2023fractional,dong2023theory,guo2023theory,kwan2023MFCI3}. This ground state spontaneously breaks the continuous translation symmetry and, by Goldstone's theorem, must have gapless collective modes (the moir\'e phonons) which are indeed seen in the time-dependent HF collective mode spectrum~\cite{kwan2023MFCI3}. Although the HF spectrum exhibits a charge gap, the many-body spectrum must be gapless\footnote{In an ED calculation, the many-body spectrum corresponding to a Wigner crystalline ground state will be gapless not only because of the Goldstone modes, but also the Anderson tower of states~\cite{anderson1952antiferromagnetic,wietek2017studying}.}. At fractional filling, we pursue a similar argument. For the sake of contradiction, assume that the many-body ground state of the full multi-band Hamiltonian is an FCI which spontaneously breaks the continuous translation symmetry of the moir\'e-less Hamiltonian. This is the case, for instance, if the FCI ground state is fully captured within the Hilbert space of the lowest HF band. Such a state must therefore have gapless Goldstone modes. But by definition, the FCI is a topologically ordered ground state protected by a finite gap (within the same global symmetry sector).  Hence it is incompatible with the gapless Goldstones in the moir\'e-less limit.

\begin{figure}
\centering
\includegraphics[width=\linewidth]{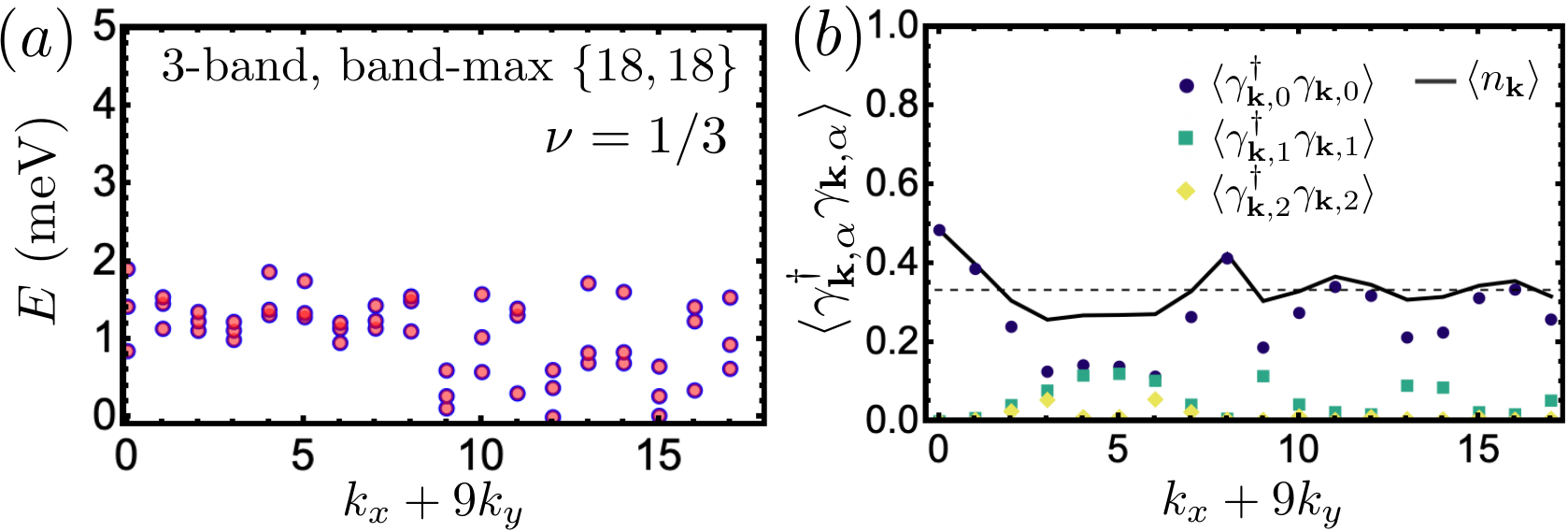}
\caption{Full 3-band ED calculations in the CN scheme on the $9 \times 2$ lattice for $\nu=1/3$ at $V=28$\,meV.
$(a)$ 3-band ED spectrum, showing no clear sign of a
gap. $(b)$ The ground states show enhanced fluctuations in the correlation functions, which are averaged over the 3 lowest states. Off-diagonal expectation values $\braket{\gamma^\dag_{\mbf{k},\alpha}\gamma_{\mbf{k},\beta}}$ are very small ($\lesssim 0.05$)
and can be neglected. The total occupations of bands 0, 1 and 2 are $79\%, 16\%$ and $5\%$ respectively.
}
\label{fig:CNrdm}
\end{figure}

The preceding discussion provides a candidate explanation for the observed gap collapse at $\nu=1/3$ and $2/3$ (\cref{fig:AHC2main,fig:CNrdm}), which is the enforced gaplessness of continuous translation-breaking phases in the full Hilbert space.
However, an alternative possibility is that the system does not actually break translation symmetry in the full multi-band limit, but is instead in some translation symmetry-preserving phase that is gapless. Within this scenario, the destruction of Wigner crystalline order present in the 1-band limit may be driven by some other mechanism entirely.
In our current finite-size calculations, we are not able to definitively distinguish between these interpretations.
Nevertheless, we emphasize that projection to the lowest HF band explicitly breaks the continuous translation symmetry, and thus guarantees that the gapped FCI states obtained in this method also break this symmetry.
If the symmetry-breaking survives to the full multi-band limit, then 1-band projection artificially removes the Goldstone modes that would have led to the gap collapse.
On the other hand, if the system does not actually spontaneously break the symmetry in the full multi-band limit, then the explicit breaking of continuous translation imposed by 1-band projection is invalid.
Either way, this suggests that 1-band projection calculations in the moir\'e-less limit using the HF basis are qualitatively inaccurate.
As our ED calculations in \cref{fig:AHC2main} suggest that the CN scheme spectrum is nearly identical with that of the moir\'e-less limit (at least for the system sizes studied here), this raises key quantitative questions regarding the importance of the moir\'e in the CN scheme, and highlights the importance of unbiased, beyond-HF methods of studying the Chern insulator at integer filling.

\section{AVE Scheme Results at $\nu=1/3$ and $\nu=2/3$}
\label{sec:AVE_HFbasis}

\begin{figure*}
\centering
\includegraphics[width=\linewidth]{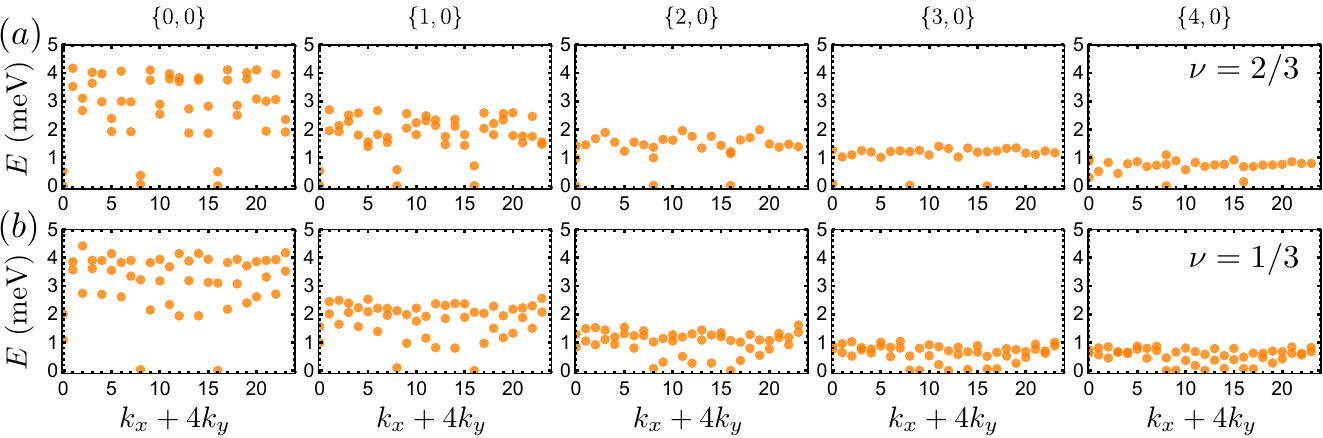}
\caption{Comparison of the FCIs at $\nu=2/3$ (upper panels) versus $\nu=1/3$ (lower panels) for $V=22$\,meV in the AVE scheme on the $4\times 6$ lattice. $(a)$ At $\nu=2/3$, 1-band ED (left-most column) shows two closely-competing sets of threefold quasi-degenerate ground states. As $N_{\text{band1}}$ is increased (indicated by the column heading $\{N_{\text{band1}},N_{\text{band2}}\}$), an FCI ground state emerges and improves up until  band-max $\{4,0\}$ when the topological degeneracy is destabilized and the continuum excitations close the gap. $(b)$ In contrast, $\nu=1/3$ shows no evidence of a stable FCI. }
\label{fig:AVE2main}
\end{figure*}

\begin{figure*}
\centering
\includegraphics[width=\linewidth]{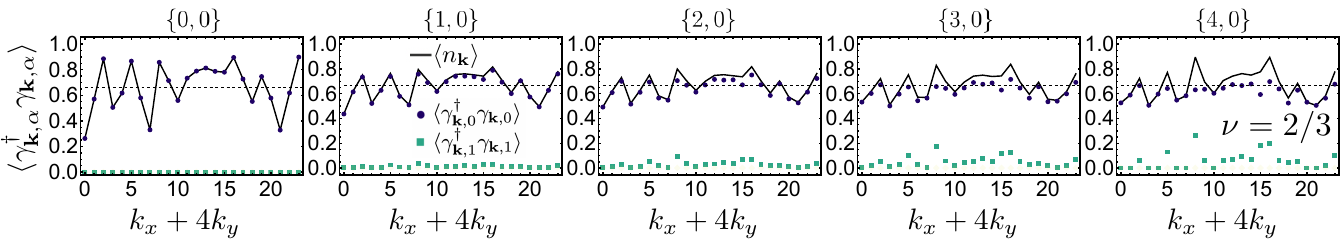}
\caption{Band occupation at $\nu=2/3$ in the AVE scheme at $V=22$\,meV  on the $4\times 6$ lattice. We compute $\braket{\gamma^\dag_{\mbf{k},\al}\gamma_{\mbf{k},\al}}$ averaged over the three-fold quasi-degenerate FCI ground states at $\nu=2/3$, which shows increasingly uniform $\braket{n_{\mbf{k}}}$ within the $\Ch=1$ HF band up to band-max $\{3,0\}$, before deteriorating at $\{4,0\}$ where \cref{fig:AVE2main}a shows the FCI gap closing.
}
\label{fig:AVE2mainnk}
\end{figure*}

Since the AVE scheme Hamiltonian includes a large coupling to the moir\'e through the valence charge background (see \Eq{eq:background}), we would expect that the low-lying pseudo-Goldstone modes that are present in the CN scheme can be gapped out. This indeed occurs in HF studies at $\nu=1$~\cite{kwan2023MFCI3}, with a moir\'e phonon gap $3-5$ times larger than that of the CN scheme.  To investigate whether the gapping~\cite{PhysRevB.33.2481,2014PhRvB..90d5114R,PhysRevLett.132.236502,2024arXiv240610709W} of such soft fluctuations leads to the survival of the putative FCI ground state in the presence of band-mixing, we turn to ED calculations.

\begin{figure}
\centering
\includegraphics[width=\linewidth]{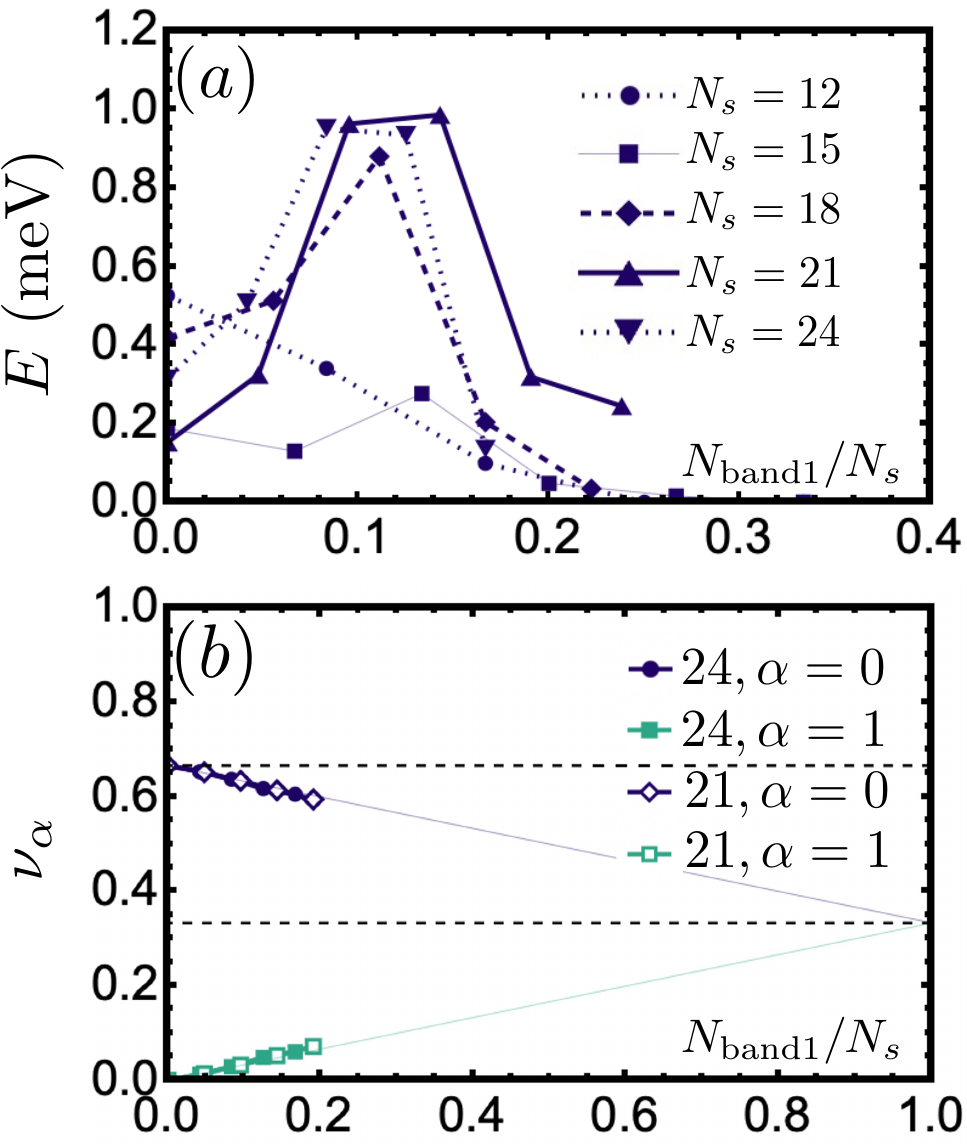}
\caption{$(a)$ Behavior of the $\nu=2/3$ FCI gap at $N_\text{band2}=0$ in the AVE scheme at $V=22$\,meV. On small systems $N_s \leq 18$ and $N_s = 24$, we observe the collapse of the gap. The $21$ site data shows a decrease in the gap, but does not appear to fall along the same scaling line. $(b)$ We demonstrate the importance of higher bands in the $\nu=2/3$ ground state, showing that their occupation linearly increases as the cutoff $N_{\text{band1}}$ is enlarged. This trend holds on $N_s = 21$ and $24$ sites.}
\label{fig:AVEnk}
\end{figure}

\cref{fig:AVE2main} shows a clear distinction between $\nu=2/3$ and $\nu=1/3$ (panels $(a)$ and $(b)$ respectively) in our multi-band ED calculations at $V=22$\,meV.
At $\nu = 2/3$ and band-max $\{0,0\}$, we see 3 nearly degenerate states appear at the FCI momenta, but with a very small gap to a competing set of 3 nearly degenerate states and highly fluctuating $\braket{n_{\mbf{k}}}$ spectrum as shown in \cref{fig:AVE2mainnk}. However, the gap \emph{grows} as the number of allowed particles in band 1 is increased. This is accompanied by a decrease in the  fluctuations in $\braket{n_{\mbf{k}}}$, indicating the stabilization of a gapped FCI ground state at $\nu=2/3$. However, \Fig{fig:AVE2main}a shows that at band-max $\{4,0\}$, the FCI gap closes as the higher bands are increasingly populated (see \Fig{fig:AVE2mainnk}). Additionally, \cref{fig:AVEnk}a shows that on smaller systems where larger ratios of $N_{\text{band1}}/N_s$ are reachable, the $\nu=2/3$ FCI is also destroyed by the collapse of the continuum.

In contrast, at $\nu=1/3$, the ground state manifold at band-max $\{0,0\}$ already exhibits a large energy spread that is greater than the gap to the next lowest energy state. Additionally, the gap immediately collapses  with the inclusion of the second band at band-max $\{1,0\}$, which rules out an FCI at $\nu=1/3$.

These features are qualitatively different from the CN scheme (Sec.~\ref{sec:CN_HFbasis}), where the $\nu=1/3$ FCI appears as the strongest state in 1-band ED, and both the $\nu=1/3$ and $2/3$ FCIs quickly collapse upon increasing band-max to $\{2,0\}$. In the AVE scheme there is a clear difference between the two fillings, with the $\nu=2/3$ FCI gap initially \emph{increasing} upon allowing for more particles in the second band. In this finite size calculation, the FCIs at $\nu=2/3$ are much more robust than at $\nu=1/3$.

We  also investigate the effect of the higher bands on the three-fold quasi-degenerate ground state itself. In \cref{fig:AVEnk}b,
we show the average occupation of each HF band
\bea
\nu_\alpha &= \frac{1}{N_s} \sum_{\mbf{k}} \braket{\gamma^\dag_{\mbf{k},\alpha}\gamma_{\mbf{k},\alpha}}
\eea
in the FCI ground state at $\nu=2/3$ as a function of $N_{\text{band1}}$ on two system sizes, $N_s = 21$ and $N_s = 24$, for the first and second band, i.e. $\alpha=0, 1$. Although only small ratios of $N_{\text{band1}}/N_s$ are reachable, all points show a clear trend extrapolating towards the $\alpha=0$ and $\alpha=1$ HF bands being almost \emph{equally} occupied in the 3-band limit. This is a clear indication that the properties of the ground state itself, as well as the gap set by the excited states, are dependent on band-mixing, and that the FCI states steadily lose weight in the lowest HF band as mixing with the higher bands is allowed to occur. \App{app:result_2D_int} contains additional results, consistent with this picture, on 18 and 21 sites where nonzero $N_\text{band2}$ is considered. At $\nu=2/3$, \cref{fig:AVE2main} shows that the gap increases and the $\braket{n_{\mbf{k}}}$ fluctuations decrease when mixing with band 1 is allowed, despite the fact that the occupation number of band $0$ diminishes. However, at band-max $\{4,0\}$, where the largest Hilbert space $> 5.5\times 10^8$, we see that the gap closes and the topological quasi-degeneracy between the lowest three states is lifted. From \cref{fig:AVE2mainnk}, we observe band 1 becoming increasingly populated at the $K_M$ and $K_M'$ points. This depletes the occupancy of band 0 which carries $\Ch=1$, and prevents the uniform $2/3$ filling of band $0$ typically desired in a robust FCI. The combined requirements of large systems and converged band-max approximations remain a challenge for definitively identifying the fate of the FCI at $\nu=2/3$, since \cref{fig:AVEnk}a also does not show obvious signs of a scaling behavior.

\section{Biased Phase diagram}
\label{sec:one-body-diagonal}
\begin{figure*}
\centering
\includegraphics[width=\linewidth]{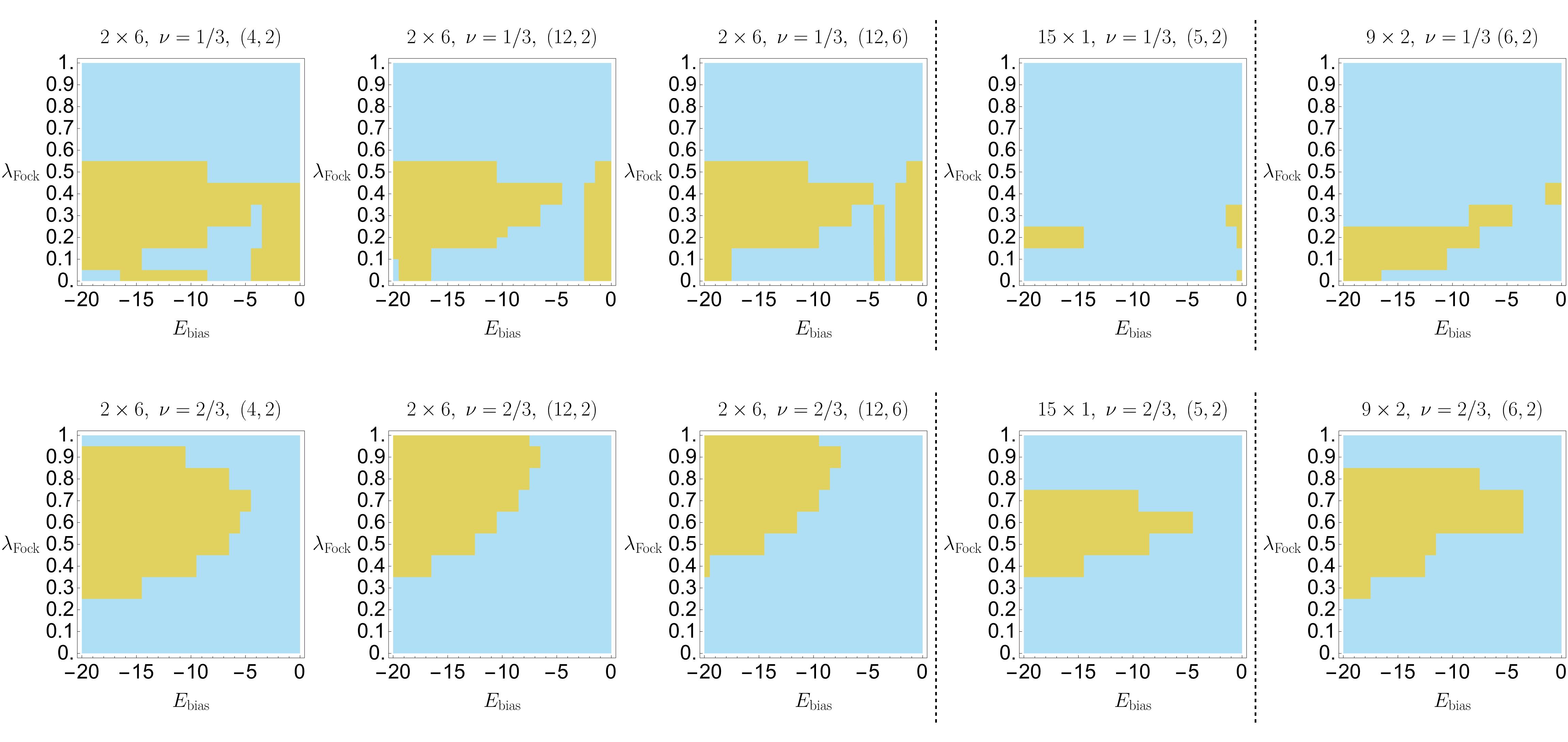}
\caption{FCI stability region as a function of biasing parameters $E_{\text{bias}}$ and $\lambda_{\text{Fock}}$ for the AVE scheme. The unbiased limit corresponds to $\Ebias=0$ and $\lfock=1$. The FCI region is shown in yellow for $\nu=1/3$ (top) and $\nu=2/3$ (bottom). We consider 12, 15 and 18 sites, and displacement field $V=22$\,meV.
The caption in each figure indicates the system size $N_x\times N_y$, the filling $\nu$, and the orbital restriction parameterized by ($N_\text{orb1},N_\text{orb2}$).
There is no orbital restriction on band 0.
}
\label{fig:Bias_Main_FCI_plots}
\end{figure*}

To further understand band mixing and finite size effects on the ground state at both fractional and integer fillings, we implement a complementary method in this section based on adding an engineered single-particle potential to bias the system towards the nearly-flat, isolated band limit within the 3-band Hilbert space. In this way we can achieve a converged 3-band ground state (within a certain range of orbital restriction of the Hilbert space), and investigate its fate as the bias is removed.

\App{app:biasing_method} contains a full account of this method.
In brief, we perform ED calculations on the following Hamiltonian
\eqa{
\label{eq:ED_H_2D_int_ave}
H & = H_{\text{one-body}} + : H_{\text{int},\K} : \ ,
}
where $: H_{\text{int},\K} :$ is the normal-ordered interaction restricted to valley $\K$ and spin $\uparrow$, and $H_{\text{one-body}} $ is the one-body term with the form
\eq{\label{eq:H_one_body}
H_{\text{one-body}} = \sum_{\bsl{k} n_1^c n_2^c } c^\dagger_{\K,\bsl{k},n_1^c, \uparrow} c_{\K,\bsl{k},n_2^c, \uparrow}  \left[ h_{\text{one-body}}(\bsl{k}) \right]_{n_1^c n_2^c}\ ,
}
where $n_1^c$ ranges over the conduction bands.
$h_{\text{one-body}}(\bsl{k})$ in the AVE scheme reads
\eq{
\label{eq:h_one-body_bias_AVE_main}
h_{\text{one-body}}(\bsl{k})  = h_0(\bsl{k}) + h_b^H(\bsl{k}) + \lambda_{\text{Fock}} h_b^F(\bsl{k}) + E_{\text{bias}} P_{\bsl{k}}^*\ ,
}
where $h_0(\bsl{k})$ is the bare kinetic term, $ h_b^H(\bsl{k})$ is the Hartree contribution to the background term, and $ h_b^F(\bsl{k})$ is the Fock contribution to the background term (see \App{app:2Dint} for definitions).
$P_{\bsl{k}}$ is the order parameter of the $\nu=1$ $\Ch=1$ state obtained in the self-consistent HF calculation, where the complex conjugation in \eqnref{eq:h_one-body_bias_AVE_main} is because $P_{\bsl{k}}=\tilde{U}_{\bsl{k},0}^*\tilde{U}_{\bsl{k},0}^T$ by definition ($\tilde{U}_{\bsl{k},0}$ is the eigenvector of the occupied HF band.)
To guarantee $P_{\bsl{k}}$ has $\Ch=1$, we always perform the HF calculation on a momentum mesh with no smaller than $12\times 12$ sites, and extract $P_{\bsl{k}}$ on the subset of momenta used for the ED calculation.
We will always choose $E_{\text{bias}}<0$ and $\lambda_{\text{Fock}}\in[0,1]$ with the unbiased limit being $\Ebias=0$ and $\lfock=1$

Physically, a large negative $E_{\text{bias}}$ will lead to the one-body spectrum having a well-isolated lowest band with wavefunctions identical to that of the $\Ch=1$ HF ground state, while $\lambda_{\text{Fock}}$ decreases the amplitude of the Fock contribution to the background term, flattening the lowest band. In the limit of large $E_{\text{bias}}\rightarrow -\infty$ and $\lambda_{\text{Fock}} \to 0$, the lowest band of the one-body spectrum is a well-isolated, nearly-flat $\Ch=1$ band, within which FCIs are expected at fractional fillings~\cite{zhou2023fractional,dong2023anomalous,dong2023theory,guo2023theory}.

We  perform ED with the biasing method for 2D interaction with the AVE scheme and $V=22$\,meV.
We focus on fillings $\nu=1/3,2/3$, and employ Hilbert space truncation techniques in order to perform calculations for 12, 15 and 18 sites. As discussed in \App{app:biasing_method}, we do so by limiting the number of single-particle eigenstates of the biased one-body term of the Hamiltonian. This is the orbital restriction approximation, with no band-max restriction, discussed in \cref{sec:MBED}. Specifically, we choose our single-particle basis to diagonalize the biased one-body term in \cref{eq:h_one-body_bias_AVE_main}.
At each momentum, we label the lowest, middle and highest energy bands at each momentum of the biased one-body term (shown in \App{app:biasing_method}) as  bands 0, 1 and 2.
Within the band $\alpha$ (with $\alpha=0,1,2$), we can also rank the single-particle states according to their corresponding eigenvalues of the biased one-body term in \cref{eq:h_one-body_bias_AVE_main}; then, we have the freedom to choose the $N_{\text{orb},\alpha}$ energetically lowest states for the $\alpha$th band to build up the Hilbert space.

To properly choose the values of $N_{\text{orb},\alpha}$, we first do benchmarks on 12 sites, and then extend the calculations to 15 and 18 sites. To perform the benchmarks, we need to first define concrete numerical criteria for identifying FCIs. 
At $\nu=1/3$ and $2/3$, the system is determined to be in the FCI phase if (1) its lowest three states are at the FCI momenta, (2) the spread of the $3$ lowest states is smaller than the gap between the 3rd and 4th lowest states, and (3) the averaged particle density $n_{\bsl{k}}$ for the 3 lowest states does not have a larger standard deviation than the 3 second lowest states at the FCI momenta.
The third condition is motivated from the fact that the fractional quantum Hall states have zero standard deviation in their $n_{\bsl{k}}$.
Practically, the third condition rules out cases where trivial charge density wave (CDW) states go below the FCI states at the same momenta, rendering the FCI states the excited states.
In this case, (3) is clearly violated since CDW states will have larger fluctuations in $n_{\bsl{k}}$ than FCI states.

Based on the above numerical criteria, we  show the FCI regions in \cref{fig:Bias_Main_FCI_plots} for 12, 15 and 18 sites.
As we increase the states included in band 1, the FCI region at $\nu=2/3$ requires less biasing on the Fock background term, eventually requiring no Fock biasing at all ($\lambda_{\text{Fock}}=1$), while there is barely any change in the FCI region at $\nu=1/3$.
Overall, the FCI at $\nu=2/3$ requires less biasing (especially in $\lambda_{\text{Fock}}$) than at $\nu=1/3$. This is an effect of the band mixing and  is consistent with our HF basis results (Sec.~\ref{sec:AVE_HFbasis}).
We also note that we do not observe considerable improvement by increasing the number of states in band 2 from $N_{\text{orb}2}=2$ to $N_{\text{orb}2}=6$, implying that including 2 states in band 2, at least for the system sizes considered here, is sufficient to achieve convergence for the FCI phase diagram.
We note that, on physical grounds, $N_{\text{orb}2}\ge 2$ is necessary since the two lowest states in band 2 of the biased one-body term in \cref{eq:h_one-body_bias_AVE_main} are at $K_M$ and $K_M'$ points where band 2 is close in energy to band 1.
Therefore, for 15 and 18 sites, we fix $N_{\text{orb}2}=2$.

In the calculations for 15 and 18 sites, we have chosen the ratio $N_{\text{orb}1}/(N_x N_y)\in [1/3,1/2)$ to be similar to that corresponding to $N_{\text{orb}1}=4,5$ for 12 sites, where $N_{\text{orb}1}=4$ for 12 sites is shown in Fig.~\ref{fig:Bias_Main_FCI_plots} and $N_{\text{orb}1}=5$ for 12 sites is shown in \App{app:biasing_method}.
In Fig.~\ref{fig:Bias_Main_FCI_plots}, we only show $N_{\text{orb}1}/(N_x N_y)=1/3$ for 15 and 18 sites; the results for larger $N_{\text{orb}1}$ within the considered range can be found in \App{app:biasing_method}, which are consistent with that for $N_{\text{orb}1}/(N_x N_y)=1/3$.
By comparing different sizes with $N_{\text{orb}1}/(N_x N_y)=1/3$ and $N_{\text{orb}2}=2$, we have not seen clear improvement of the FCI stability region by increasing the system size, which is consistent with the HF basis results that show the collapse of the FCI at the realistic $ E_{\text{bias}} = 0$ and $\lambda_{\text{Fock}} = 1 $.

\section{Conclusion}

By employing novel Hilbert space reduction techniques, we perform multi-band ED calculations for \mbox{$\theta=0.77^\circ$} R5G/hBN at fillings $\nu=1/3,2/3$ and $1$ on system sizes up to 24 sites. In typical FCIs within gapped band structures, these sizes would be more than large enough to show the appearance of stable FCIs.  Crucially, our methods enable a critical evaluation of 1-band calculations, where the physics is projected to the subspace of occupied HF orbitals at $\nu=1$ and fluctuations into the higher gapless bands are neglected.

We find the collapse of FCIs due to band mixing at $\nu=1/3$ and $2/3$ in both the AVE and CN schemes, the latter of which yields nearly identical results in the moir\'e-less limit where the hBN-induced potential is switched off.
As 1-band projected ED in the HF basis produces robustly gapped FCIs at both $\nu=1/3$ and $2/3$ in the CN scheme, our numerical results provide clear evidence of the inaccuracy of 1-band projected ED there.
The 1-band projected ED method is also questionable in the AVE scheme, as band mixing changes the ED spectrum dramatically.
The FCI states at $\nu=2/3$ are more robust against band mixing than those at $\nu=1/3$ in the AVE scheme, though they both eventually become destabilized as enough particles are allowed to populate the higher bands.
On the other hand, our combined ED and HF results at $\nu=1$ in the AVE scheme lead to the conjecture that the gapped $\Ch=1$ state is the ground state in the thermodynamic limit, which is consistent with the experiment~\cite{Lu2024fractional}.

While the practical limitations of finite size in ED have made it difficult to firmly establish to the nature of the multi-band ground states, we have accumulated strong evidence to invalidate a 1-band description of the FCIs in both the CN and AVE scheme. In the CN scheme, we have found that the principal challenge to a fractional AHC (an FCI state with spontaneous continuous translation symmetry breaking) description of the experiment is the absence of a many-body gap, as the continuum always collapses showing no indication of a ``pinning" gap. Further theoretical work may be required to investigate the moir\'e phonons in detail, and the effects of temperature, disorder, and the quantitative threshold of the hBN moir\'e. In the AVE scheme where one expects the moir\'e phonons to be gapped, we have still observed the continuum excitations collapse from band mixing at fractional fillings. It is potentially possible that this Hamiltonian does harbor an FCI ground state at $\nu=2/3$, but finite size effects obscure it (despite reaching system sizes where FCIs are routinely seen in ED on gapped bands). That said, it is also possible that there is no FCI ground state in the multi-band Hilbert space, and key physics is absent from the present Hamiltonian. What is however clear is the pressing need for a controllable theoretical approach to the multi-band problem posed by R5G/hBN, magnified by the observation of an even-denominator state in six-layer samples \cite{xie2024even}. It is likely that fractionalized phases in other systems \cite{2020Natur.579...56C,2024NatCo..15.2597Z,2024Sci...384..414S,2024arXiv240614289D,2024arXiv240501829Y,2024arXiv240509627S,2024arXiv240707894W,2024NatRP...6..349M,2024arXiv240513145Y}, thus far not considered as possible hosts of FCIs, can be predicted, engineered, and discovered once the fundamental mechanism in R5G/hBN is understood.

\section{Acknowledgements}
We thank Zhengguang Lu, Long Ju, Sid Parameswaran, Shivaji Sondhi, Daniel Arovas, Daniel Parker, Sankar Das Sarma, Senthil Todadri, Oskar Vafek, Akshat Pandey, Michael Zaletel, Liang Fu and Yang Zhang for helpful discussions.
J.Y. acknowledges support from the Gordon and Betty Moore Foundation through Grant No. GBMF8685 towards the Princeton theory program.
J. H.-A. is supported by a Hertz Fellowship, with additional support from DOE Grant No. DE-SC0016239.
Y.H.K is supported by a postdoctoral research fellowship
at the Princeton Center for Theoretical Science.
N.R. also acknowledges support from the QuantERA II Programme that has received funding from the European Union’s Horizon 2020 research and innovation programme under Grant Agreement No 101017733 and from the European Research Council (ERC) under the European Union’s Horizon 2020 Research and Innovation Programme (Grant Agreement No. 101020833).
B.A.B.’s work was primarily supported by the the Simons Investigator Grant No. 404513, by the Gordon and Betty Moore Foundation through Grant No. GBMF8685 towards the Princeton theory program, Office of Naval Research (ONR Grant No. N00014-20-1-2303), BSF Israel US foundation No. 2018226 and NSF-MERSEC DMR-2011750, Princeton Global Scholar and the European Union’s Horizon 2020 research and innovation program under Grant Agreement No 101017733 and from the European Research Council (ERC), as well  as from the Simons Collaboration on New Frontiers in Superconductivity.

\bibliography{fcigraphene}

\appendix
\onecolumngrid

\tableofcontents

\section{Review of the Model}

In this Appendix, we introduce the single-particle and interaction Hamiltonians used in the Main Text. \App{app:SP} summarizes the single-particle model, while \App{app:2Dint} and \App{app:2DintCNP} summarize the average (AVE) scheme and charge-neutrality (CN) scheme respectively.

\subsection{Single-Particle Hamiltonian}
\label{app:SP}

In the $\K$ valley, the matrix Hamiltonian for R$L$G is $[H_{\K}(\bsl{p})]_{l \sigma, l' \sigma'}$ with $l = 0,\dots, L-1$ denoting the layer index and $\sigma = A,B$ denoting the graphene sublattice index. In matrix notation, the Hamiltonian can be written~\cite{PhysRevB.90.155406,Park2023RMGhBNChernFlatBands,PhysRevB.89.205414,herzog2024MFCI2}
\eqa{
\label{eq:H_K}
H_{\K}(\bsl{p}) &= \bpm
v_F\mbf{p} \cdot \pmb{\sigma}  & t^\dag(\mbf{p}) & t'^\dagger &   &\\
t(\mbf{p}) & \ddots & \ddots & t'^\dagger \\
t' & \ddots & v_F\mbf{p} \cdot \pmb{\sigma} & t^\dagger(\mbf{p})\\
& t' & t(\mbf{p})  & v_F\mbf{p} \cdot \pmb{\sigma}
\epm + H_{ISP} + H_D,
}
where $\bsl{p}=-\ii \nabla$, and $\bsl{\sigma}=(\sigma_x,\sigma_y)$ are Pauli matrices in sublattice space. $t(\mathbf{p})$ and $t'$ are also $2\times 2$ matrices in sublattice space:
\eq{
t(\mbf{p}) = -\bpm v_4 p_+ & -t_1 \\ v_3 p_- &  v_4 p_+ \epm, \qquad  \qquad t' = \bpm 0 & 0 \\ t_2 & 0 \epm\ ,
}
where $p_\pm = p_x \pm \ii p_y$,  $v_F$ is the Fermi velocity, $t_1,v_3,v_4$ are parameters describing hopping between consecutive layers, and $t_2$ describes hopping between next-nearest layers. The inversion-symmetric potential describing the change in chemical potential of the interior layers is
\bea
\null [H_{ISP}]_{l \sigma,l' \sigma'}= V_{ISP} \left|l - \frac{n-1}{2} \right| \delta_{ll'} \delta_{\sigma \sigma'}\ ,  V_{ISP}  = 16.65\text{meV}
\eea
and the externally-applied displacement field is modeled as a linear interlayer potential
\bea
\label{eq:H_D}
\null [H_{D}]_{l \sigma,l' \sigma'}= V \left(l - \frac{n-1}{2} \right) \delta_{ll'} \delta_{\sigma \sigma'}\ .
\eea
The parameters of the graphene Hamiltonian have been computed in Ref.~\cite{herzog2024MFCI2} incorporating moir\'e relaxation in the presence of hBN:
\bea
v_F =542.1\text{meV nm}, \quad v_3 = 34\text{meV nm},\quad t_1 =  355.16\text{meV}, \quad t_2 = -7\text{meV} \ .
\eea
The signs of the parameters and the value of the Fermi velocity compare well with Ref.~\cite{PhysRevB.89.035405}, which was computed using Wannier orbitals in the absence any moir\'e relaxation.

Next, we integrate out the hBN substrate by perturbation theory at zero displacement field \cite{herzog2024MFCI2}, and obtain the following effective moir\'e potential:
\eqa{
\label{eq:Vxifinal}
V_\xi(\mbf{r}) &= V_0 + \left[V_1 e^{i\psi}\sum_{j=1}^3 e^{i \mbf{g}_j\cdot\mbf{r}}\bpm 1& \omega^{-j} \\ \omega^{j+1} &\omega \epm + h.c.\right]\ ,
}
which only acts on the bottom layer $l=0$ of graphene, and where $\mbf{g}_j = R(\frac{2\pi}{3}(j-1)) (\mbf{q}_2-\mbf{q}_3)$, for $j = 1,2,3$, are moir\'e reciprocal lattice vectors.
We define the $\mbf{q}_j$ via $\mbf{q}_{j+1} = R(\frac{2\pi}{3}) \mbf{q}_j$ and \eq{
\label{eq:qvecmain}
\mbf{q}_1 = \mbf{K}_G - \mbf{K}_{hBN} = \frac{4\pi}{3 a_G}\left(1 - \frac{R(-\th)}{1+0.01673} \right)\hat{x},
}
\noindent
where $\th$ is the twist angle, $R(\th)$ is the counterclockwise rotation matrix, $\mbf{K}_G$ and $\mbf{K}_{hBN}$ are the valley $\eta=K$ Dirac momenta of graphene and hBN, $a_G = 2.46\AA$ is the graphene lattice constant, and $(1+0.01673)a_G$ is the hBN lattice constant. We also define `standard' basis moir\'e reciprocal lattice vectors $\mbf{b}_{M,i}=\mbf{q}_3-\mbf{q}_i$ for $i=1,2$. The parameters of the moir\'e potential depend on the hBN stacking type $\xi$, corresponding to whether the carbon A site is aligned with the nitrogen ($\xi =1$) or boron ($\xi = 0$) atom. Throughout this work, we always consider $\xi=1$ since only this stacking reveals a robust $\Ch=1$ insulator at $\nu=1$ in HF studies in the AVE scheme \cite{kwan2023MFCI3}. We also focus on twist angle $\theta=0.77^\circ$. The values of the parameters in the moir\'e potential are $V_0 = 1.50$\,meV, $V_1 = 7.37$\,meV and $\psi=16.55^\circ$ for $\xi = 1$. Our parameters are obtained from best-fit optimization to Slater-Koster band structures generated from the relaxed structure.  The strength of the moir\'e in our relaxed calculations is about one-third of the estimates based on the Bistritzer-MacDonald two-center approximation \cite{PhysRevB.90.155406} which finds a $20-30$\,meV moir\'e scale, and about half as strong as estimates based on untwisted calculations for different stackings~\cite{PhysRevB.89.205414,Park2023RMGhBNChernFlatBands} which find a $12-15$\,meV moir\'e scale.

The full moir\'e Hamiltonian after integrating out hBN reads
\bea
\label{eq:H_K_nohBN}
H_{\K, \xi}(\mbf{r}) = H_{\K}(- i \pmb{\nabla}) + H_{\text{moir\'e},\xi}(\bsl{r})\ , \quad [H_{\text{moir\'e},\xi}(\bsl{r})]_{l \sigma,l' \sigma'} = \left[ V_\xi(\mbf{r}) \right]_{\sigma\sigma'} \delta_{l0}\delta_{ll'}
\eea
where
$V_\xi(\mbf{r})$ is in \eqnref{eq:Vxifinal}, and $H_{\K}(\bsl{p})$ is in \eqnref{eq:H_K}.

Finally we include both valleys $\eta=\pm\K$ and spins $s=\uparrow,\downarrow$ in second quantization so that the Hamiltonian reads (in $\eta$ valley)
\bea
H_{0,\eta} = H_{R5G}^{\eta} + H_{\text{moir\'e},\xi}^{\eta}\ ,
\eea
where
\eq{
H_{R5G}^{\K} = \sum_{s,l \sigma,l' \sigma'}\int d^2 r c^\dag_{\bsl{r}, l\sigma \eta s} \left[ H_{\K}(- i \pmb{\nabla})\right]_{l \sigma,l' \sigma'}  c_{\bsl{r}, l'\sigma' \eta s}\ ,
}
\eq{
H_{\text{moir\'e},\xi}^{\K} = \sum_{s,l \sigma,l' \sigma'}\int d^2 r c^\dag_{\bsl{r}, l\sigma \eta s} [H_{\text{moir\'e},\xi}(\bsl{r})]_{l \sigma,l' \sigma'}   c_{\bsl{r}, l'\sigma' \eta s}\ ,
}
$H_{R5G}^{-\K} $ and $H_{\text{moir\'e},\xi}^{-\K}$ are related to $H_{R5G}^{\K} $ and $H_{\text{moir\'e},\xi}^{\K}$ by spin-less  time-reversal symmetry, and $\bsl{r}=(x,y)$ is the continuum 2D position.
The Fourier transform of $H_{\eta, \xi=1}(\mbf{r})$ gives the single particle Hamiltonian $h_{0}^\eta(\bsl{k})$ in momentum space, whose basis is associated with creation operators
\eq{
\label{eq:plane_wave_basis}
c^\dagger_{\eta, \bsl{k},\bsl{G},l\sigma s} = \frac{1}{\sqrt{\V}} \int d^2 r \, e^{\ii (\bsl{k}+\bsl{G})\cdot\bsl{r}}
c^\dagger_{\bsl{r},l\sigma\eta s} \ ,
}
where $\V$ is the area of the whole sample, $\bsl{k}$ is in the first moir\'e Brillouin zone (BZ), and $\bsl{G}$ labels the reciprocal moir\'e lattice vectors. We assume the total system of area $\V$ consists of $N_s$ unit cells on periodic boundaries, so that $\mbf{k}$ takes $N_s$ distinct values. The reciprocal lattice $\mbf{G}$ is infinite but can be truncated since we are interested in low energies.

In the following subsections, we will discuss the interactions. There are multiple possible forms of the interaction that have been proposed in rhombohedral graphene~\cite{kwan2023MFCI3}. We will discuss 2D and 3D interactions, as well as two choices of normal ordering called the charge-neutrality (CN) and average (AVE) schemes.

\subsection{2D Coulomb Interaction}

In this section, we will review the Coulomb interaction without taking into account the thickness of the sample, which we refer to as the 2D interaction.

\subsubsection{Average Scheme}
\label{app:2Dint}

In the AVE scheme, the interaction Hamiltonian for the 2D interaction reads
\bea
\label{eq:H_2D_int}
H_{\text{int}} = \int d^2 r d^2 r' V(\bsl{r}-\bsl{r}) \delta \hat{\rho}_{\bsl{r}} \delta \hat{\rho}_{\bsl{r}'}\ , \quad \delta \hat{\rho}_{\bsl{r}} = \sum_{\eta,l,\sigma,s} \left[ c^\dagger_{\bsl{r},l\sigma\eta s} c_{\bsl{r},l\sigma\eta s} - \frac{1}{2}\delta(\bsl{r})\right], \quad \hat{\rho}_{\bsl{r}} = \sum_{\eta,l,\sigma,s} c^\dagger_{\bsl{r},l\sigma\eta s} c_{\bsl{r},l\sigma\eta s}.
\eea
The interaction potential can be Fourier-transformed:
\eq{
V(\bsl{r}) = \frac{1}{\V} \sum_{\mbf{q}} \sum_{\mbf{G}} e^{\ii (\bsl{q}+\bsl{G})\cdot\bsl{r}} V(\bsl{q}+\bsl{G})\ ,\qquad V(\mbf{q}) = \frac{e^2}{2 \eps} \frac{\tanh |\mbf{q}| d_{sc}}{|\mbf{q}|},
}
where we have assumed a dual-gate screened Coulomb interaction with dielectric constant $\eps = 5\eps_0$ and sample-to-gate distance $ d_{sc} = 10$\,nm.  We can rewrite $H_\text{int}$ in momentum space and obtain a normal-ordered two-body interaction as well as a residual one-body term:
\eq{\label{eq:Hint_Hintnormord}
H_{\text{int}} = :H_{\text{int}}: + H_{b}^{\text{full}} \ ,
}
where ``full" in $H_{b}^{\text{full}}$ means the term contains all bands.
To specify the form of these terms, in particular the meaning of the normal-ordering notation $:\hat{O}:$ used above, we must first discuss the single-particle basis.
We note that our ED calculations are not performed in the plane wave basis (\cref{eq:plane_wave_basis}), and transformation to a single-particle basis suitable for truncation/projection is important.
We denote the band basis by $c^\dagger_{\eta,\bsl{k},n ,s}$, where
\bea
c^\dagger_{\eta,\bsl{k},n ,s} =  \sum_{\bsl{G}l\sigma}c^\dagger_{\eta,\bsl{k},\bsl{G},l\sigma s} \left[ U_{n}^{\eta}(\bsl{k}) \right]_{\bsl{G}l\sigma}
\eea
and $U^\eta_n(\mbf{k})$ is a complete set of orthonormal eigenvectors of the non-interacting Hamiltonian (in valley $\eta$) labeled by the band number $n$.
In the momentum basis, the interaction term is given by
\eqa{
:H_{\text{int}}: & = \frac{1}{2\V} \sum_{\bsl{q}\bsl{G}} \sum_{\bsl{k}_1 \bsl{G}_1 \eta_1 l_1 \sigma_1 s_1} \sum_{\bsl{k}_2 \bsl{G}_2 \eta_2 l_2 \sigma_2 s_2} V(\bsl{q}+\bsl{G}) \\
& \qquad \times :c^\dagger_{\eta_1,\bsl{k}_1+\bsl{q},\bsl{G}_1+\bsl{G},l_1\sigma_1 s_1} c^\dagger_{\eta_2,\bsl{k}_2-\bsl{q},\bsl{G}_2-\bsl{G},l_2\sigma_2s_2} c_{\eta_2,\bsl{k}_2,\bsl{G}_2,l_2\sigma_2 s_2} c_{\eta_1,\bsl{k}_1,\bsl{G}_1,l_1\sigma_1  s_1}:
}
which in the band basis is equal to
\eqa{
:H_{\text{int}}: & = \frac{1}{2\V} \sum_{\bsl{k}_1 \bsl{k}_2 \bsl{q} } \sum_{\eta_1 \eta_2} \sum_{s_1 s_2} \sum_{n_1 n_2 n_3 n_4}  V_{n_1 n_2 n_3 n_4}^{\eta_1 \eta_2}(\bsl{k}_1,\bsl{k}_2,\bsl{q}):c^\dagger_{\eta_1,\bsl{k}_1+\bsl{q},n_1 ,s_1} c^\dagger_{\eta_2,\bsl{k}_2-\bsl{q},n_2 ,s_2}  c_{\eta_2,\bsl{k}_2,n_3 ,s_2}  c_{\eta_1,\bsl{k}_1,n_4 ,s_1}:  \ ,
}
where
\eqa{\label{appeq:Vetaeta}
V_{n_1 n_2 n_3 n_4}^{\eta_1 \eta_2}(\bsl{k}_1,\bsl{k}_2,\bsl{q})  & = \sum_{\bsl{G}_1 l_1 \sigma_1} \sum_{\bsl{G}_2 l_2 \sigma_2}  \sum_{\bsl{G}}
V(\bsl{q}+\bsl{G}) \left[ U_{n_1}^{\eta_1}(\bsl{k}_1+\bsl{q})\right]^*_{(\bsl{G}_1 + \bsl{G}) l_1 \sigma_1} \left[ U_{n_2}^{\eta_2}(\bsl{k}_2-\bsl{q})\right]^*_{(\bsl{G}_2 - \bsl{G}) l_2 \sigma_2} \\
& \qquad \times \left[ U_{n_3}^{\eta_2}(\bsl{k}_2)\right]_{ \bsl{G}_2 l_2 \sigma_2}  \left[ U_{n_4}^{\eta_1}(\bsl{k}_1)\right]_{ \bsl{G}_1 l_1 \sigma_1}\\
& = \sum_{\bsl{G}} V(\bsl{q}+\bsl{G}) M_{n_1 n_4}^{ \eta_1}(\bsl{k}_1,\bsl{q}+\bsl{G})  M_{n_2 n_3}^{ \eta_2}(\bsl{k}_2,-\bsl{q}-\bsl{G}) \\
& = \sum_{ \bsl{G}} V(\bsl{q}+\bsl{G}) M_{n_1 n_4}^{ \eta_1}(\bsl{k}_1,\bsl{q}+\bsl{G})  \left[ M_{n_3 n_2}^{ \eta_2}(\bsl{k}_2-\bsl{q},\bsl{q}+\bsl{G}) \right]^*\ ,
}
and
\eq{
M^{\eta}_{mn}(\bsl{k},\bsl{q}+\bsl{G}) = \sum_{\bsl{G}'l \sigma} \left[U^{\eta}_{m}(\bsl{k}+\bsl{q}+\bsl{G})\right]_{\bsl{G}'l\sigma}^* \left[U^{\eta}_{n}(\bsl{k})\right]_{\bsl{G}'l\sigma}\ .
}
In the expressions above, the normal-ordering operation $:\hat{O}:$ places all conduction band annihilation operators and valence band creation operators in $\hat{O}$ to the right (keeping track of fermionic minus signs).

When decomposing $H_{\text{int}}$ as in \cref{eq:Hint_Hintnormord}, we obtain an extra one-body term $H_{b}^{\text{full}}$.
We will perform ED calculations which project into a set of active degrees of freedom consisting of $c^\dagger_{\eta,\bsl{k},n^c,s}$ in the conduction band subspace with $n^c = 0,1,2$ labeling the lowest three conduction bands (all other bands are energetically far away from the flat $n^c=0$ band by a kinetic energy which is at least as large as the Coulomb energy), while assuming the valence bands are fully filled.
Projecting $H_{b}^{\text{full}}$ to the three conduction bands yields
\eq{
H_{b} = \sum_{\bsl{k} n_1^c n_2^c \eta s} c^\dagger_{\eta,\bsl{k},n_1^c, s} c_{\eta,\bsl{k},n_2^c, s}  h^{c,\eta}_{b,n_1^c n_2^c}(\bsl{k})\ ,
}
where
\bea
\label{eq:HbHF}
h^{c,\eta}_{b,n_1^c n_2^c}(\bsl{k})
& = - \frac{1}{\V} \sum_{ \bsl{G}} V(\bsl{G})  \sum_{\eta_1 \bsl{k}_1 \bsl{G}_1 l_1 \sigma_1} \left[ \tilde{P}_c^{\eta_1}(\bsl{k}_1) -  \tilde{P}_v^{\eta_1}(\bsl{k}_1) \right]_{\bsl{G}_1 l_1 \sigma_1,(\bsl{G}_1+\bsl{G})l_1 \sigma_1}^*  M^{ \eta}_{n_1^c n_2^c}(\bsl{k},\bsl{G}) \\
& \quad + \frac{1}{2\V} \sum_{\bsl{q}\bsl{G}} V(\bsl{q}+\bsl{G}) \sum_{\bsl{G}_1'\sigma_1'\bsl{G}_2'\sigma_2' l_1 l_2}\left[ U^{\eta}_{n_1^c}(\bsl{k}) \right]_{\bsl{G}_1'-\bsl{G}l_2 \sigma_1'}^*\left[ \tilde{P}_c^{\eta}(\bsl{k} + \bsl{q})- \tilde{P}_v^{\eta}(\bsl{k} + \bsl{q}) \right]_{\bsl{G}_1' l_2 \sigma_1', \bsl{G}_2' l_1 \sigma_2'} \left[ U^{\eta}_{n_2^c}(\bsl{k}) \right]_{\bsl{G}_2'-\bsl{G}l_1 \sigma_2'}\\
& = h^{H,\eta}_{b,n_1^c n_2^c}(\bsl{k}) + h^{F,\eta}_{b,n_1^c n_2^c}(\bsl{k}).
\eea
We have defined $\tilde{P}_c^\eta(\mbf{k}) = \sum_{n_c} U^\eta_{n_c}(\mbf{k})U^{\eta \dag}_{n_c}(\mbf{k})$ and $\tilde{P}_v^\eta(\mbf{k}) = \sum_{n_v} U^\eta_{n_v}(\mbf{k})U^{\eta \dag}_{n_v}(\mbf{k})$, and interpreted the first/second line in \Eq{eq:HbHF} as the Hartree/Fock term coming from bringing the interaction into normal-ordered form.

To summarize, the total many-body Hamiltonian within the subspace of the lowest three conduction bands reads
\eq{
H = \sum_{\bsl{k} n_1^c n_2^c \eta s} c^\dagger_{\eta,\bsl{k},n_1^c, s} c_{\eta,\bsl{k},n_2^c, s}  \left[h_{0,\eta}(\bsl{k})+ h^{H,\eta}_b(\bsl{k})  + h^{F,\eta}_b(\bsl{k}) \right]_{n_1^c n_2^c} + :H_{\text{int}}:\ ,
}
where $:H_{\text{int}}:$ above is understood to be restricted to the three conduction bands, and $h_{0,\eta}(\mbf{k})$ is the non-interacting Hamiltonian matrix in valley $\eta$ which is diagonal in the band basis.

In this work, we only consider the case where the conduction electrons are polarized in valley $\K$ and spin $\uparrow$. Thus we finally study the following Hamiltonian
\eqa{
H & = \sum_{\bsl{k} n_1^c n_2^c } c^\dagger_{\K,\bsl{k},n_1^c, \uparrow} c_{\K,\bsl{k},n_2^c, \uparrow}  \left[h_{0}(\bsl{k})+ h^{H}_b(\bsl{k})  + h^{F}_b(\bsl{k}) \right]_{n_1^c n_2^c} \\
& \quad + \frac{1}{2\V} \sum_{\bsl{k}_1 \bsl{k}_2 \bsl{q} }  \sum_{n_1 n_2 n_3 n_4}  V_{n_1 n_2 n_3 n_4}^{\K \K}(\bsl{k}_1,\bsl{k}_2,\bsl{q})c^\dagger_{K,\bsl{k}_1+\bsl{q},n_1 ,\uparrow} c^\dagger_{\K,\bsl{k}_2-\bsl{q},n_2 ,\uparrow}  c_{\K,\bsl{k}_2,n_3 ,\uparrow}  c_{\K,\bsl{k}_1,n_4 ,\uparrow}\\
& = H_0 + H_b + : H_{\text{int},\K\uparrow} : \ ,
}
where
\eq{
h_{0}(\bsl{k}) = h_{0,K}(\bsl{k})\ ;\ h^{H}_b(\bsl{k}) = h^{H,K}_b(\bsl{k})\ ;\ h^{F,K}_b(\bsl{k}) = h^{F}_b(\bsl{k}) \ ,
}
and
\eqa{
\label{eq:ED_H_2D_int_ave_terms}
& H_0 = \sum_{\bsl{k} n_1^c n_2^c } c^\dagger_{\K,\bsl{k},n_1^c, \uparrow} c_{\K,\bsl{k},n_2^c, \uparrow}  \left[h_{0}(\bsl{k}) \right]_{n_1^c n_2^c} \\
& H_b = \sum_{\bsl{k} n_1^c n_2^c } c^\dagger_{\K,\bsl{k},n_1^c, \uparrow} c_{\K,\bsl{k},n_2^c, \uparrow}  \left[ h^{H}_b(\bsl{k})  + h^{F}_b(\bsl{k}) \right]_{n_1^c n_2^c} \\
& : H_{\text{int},\K\uparrow} : = \frac{1}{2\V} \sum_{\bsl{k}_1 \bsl{k}_2 \bsl{q} }  \sum_{n_1 n_2 n_3 n_4}  V_{n_1 n_2 n_3 n_4}^{\K \K}(\bsl{k}_1,\bsl{k}_2,\bsl{q})c^\dagger_{K,\bsl{k}_1+\bsl{q},n_1 ,\uparrow} c^\dagger_{\K,\bsl{k}_2-\bsl{q},n_2 ,\uparrow}  c_{\K,\bsl{k}_2,n_3 ,\uparrow}  c_{\K,\bsl{k}_1,n_4 ,\uparrow}\ .
}

\subsubsection{Charge-Neutrality Scheme}
\label{app:2DintCNP}

For the charge neutrality (CN) scheme, we simply normal order the Coulomb interaction against the charge neutrality gap below the lowest conduction band. For $V>0$ where the conduction electrons are biased away from the hBN, this amounts to effectively treating the conduction bands as an electron gas (virtually unaffected by the valence bands bound to the moir\'e pattern). The total many-body Hamiltonian, projected to the lowest three conduction bands in valley $\K$ and spin $\uparrow$, reads
\eqa{
\label{eq:ED_H_2D_int_CN}
H & = \sum_{\bsl{k} n_1^c n_2^c } c^\dagger_{\K,\bsl{k},n_1^c, \uparrow} c_{\K,\bsl{k},n_2^c, \uparrow}  \left[h_{0}(\bsl{k})\right]_{n_1^c n_2^c} \\
& \quad + \frac{1}{2\V} \sum_{\bsl{k}_1 \bsl{k}_2 \bsl{q} }  \sum_{n_1 n_2 n_3 n_4}  V_{n_1 n_2 n_3 n_4}^{\K \K}(\bsl{k}_1,\bsl{k}_2,\bsl{q})c^\dagger_{\K,\bsl{k}_1+\bsl{q},n_1 ,\uparrow} c^\dagger_{\K,\bsl{k}_2-\bsl{q},n_2 ,\uparrow}  c_{\K,\bsl{k}_2,n_3 ,\uparrow}  c_{\K,\bsl{k}_1,n_4 ,\uparrow}\\
& = H_0 + : H_{\text{int},\K\uparrow} :\ ,
}
which differs from the Hamiltonian in the AVE scheme by the absence of one-body terms $H_b$ arising from the valence electron background.

The CN scheme is a natural scheme to consider if the non-interacting band structure possesses a large gap between the conduction and valence bands. However, the CN scheme is fine-tuned in such a way that when the electrons are biased away from the hBN, the conduction bands only feel the moir\'e extremely weakly. In the absence of \textit{ab initio} studies that compute the band structure in the presence of the hBN and a finite $V$, it is not clear if this fine-tuning is justified.  On the other hand, the charge background of the valence bands is not neglected in the AVE scheme. Another feature of the AVE scheme is that the part of the Hamiltonian that depends on the interaction potential does not depend on parameters, such as the displacement field, that could be tuned experimentally \textit{in situ}~\cite{kwan2023MFCI3}.

We note that projection into the lowest three conduction bands is not valid for significantly larger values of $V$ than those considered in this paper. This is because increasing $V$ further dramatically changes the band structure with more bands accumulating together, such that neglecting higher bands is not justified.

\begin{figure}[t]
\centering
\includegraphics[width=\columnwidth]{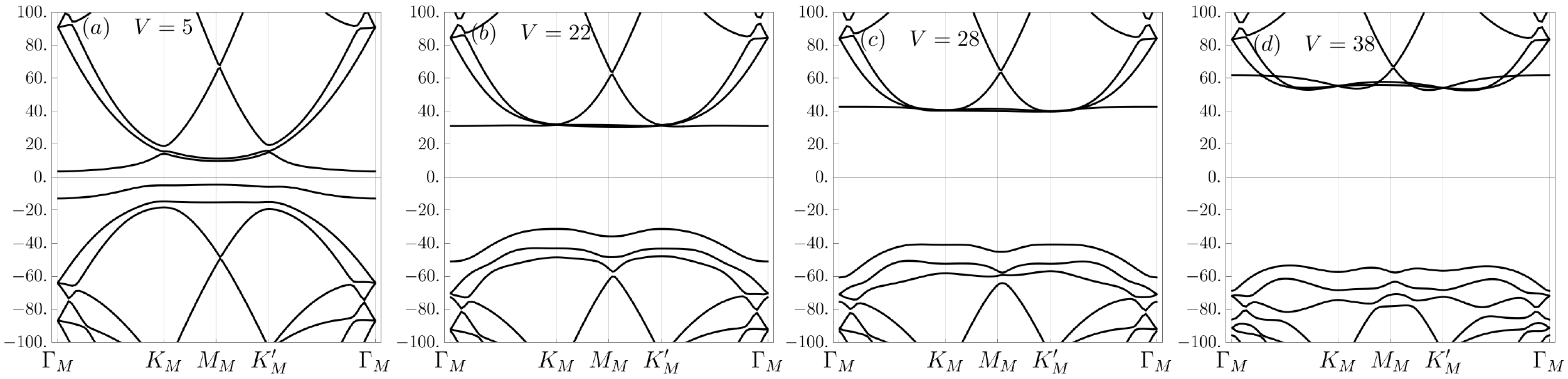}
\caption{We show the single-particle band structures of $\theta=0.77^\circ$ R5G/hBN in the $\xi=1$ stacking configuration for $V=5,22,28,38$\,meV in $(a)$ - $(d)$ respectively. At small $V$, the valence and conduction bands are close in energy. For $V = 22 -28$\,meV, the lowest conduction band ($\text{band 0}$) is quite flat but not isolated, and the first and second higher conduction bands are important for the physics at $\nu<1$. At large $V$, many bands accumulate in energy such that the lowest 4 or even 5 conduction bands may be appreciably coupled by the Coulomb interactions.
}
\label{fig:SPbands}
\end{figure}

\subsection{3D Interaction}
\label{app:3Dint}

In this section, we will review the Coulomb interaction that takes into account the thickness of the sample, which we refer to as the 3D interaction.
We will again discuss the AVE and the CN schemes.

\subsubsection{Average Scheme}
\label{app:3DintAVE}

In the AVE scheme, the 3D Coulomb interaction reads~\cite{kwan2023MFCI3}
\eq{
H_{\text{int}} = \frac{1}{2} \sum_{ll'} \int d^2 r d^2 r' V_{ll'}(\bsl{r}-\bsl{r}') \delta \hat{\rho}_{\bsl{r},l} \delta \hat{\rho}_{\bsl{r}',l'}\ ,
}
where
\eq{
\delta \hat{\rho}_{\bsl{r},l} = \sum_{\eta,\sigma,s} \left[ c^\dagger_{\bsl{r},l\sigma\eta s} c_{\bsl{r},l\sigma\eta s} - \frac{1}{2}\delta(\bsl{r}=0)\right]\ ,
}
and
\eq{
\hat{\rho}_{\bsl{r},l} = \sum_{\eta,\sigma,s} c^\dagger_{\bsl{r},l\sigma\eta s} c_{\bsl{r},l\sigma\eta s} \ .
}

The form of the interaction reads
\eq{
V_{ll'}(\bsl{r}) = \frac{1}{\V} \sum_{\bsl{p}\in\dsR^2} e^{\ii \bsl{p}\cdot\bsl{r}} V_{ll'}(\bsl{p})\ ,
}
where
\begin{equation}\label{eq:Vq0}
V_{ll'}(\mbf{p}=0)=-\frac{e^2|z_l-z_{l'}|}{2\epsilon}.
\end{equation}
and
\begin{equation}\label{eq:Vq_layer}
V_{ll'}(\mbf{p})=\frac{1}{2\epsilon p}\left[\frac{e^{-p(z_l+z_{l'})}\left(-e^{p(2d_{sc} + 2 z_l+ 2 z_{l'})}-e^{ 2p d_{sc} }+e^{2 p z_l}+e^{2 p z_{l'}}\right)}{e^{4 p d_{sc} }-1}+e^{-p|z_l-z_{l'}|}\right].
\end{equation}
Here $z_l=(l-2)0.333$nm is the $z$-component of the position of the $l$th layer with $l=0,\ldots,4$. We choose $\epsilon = 5\epsilon_0$ and $d_{sc}=10$\,nm.

Unlike the 2D interaction, the 3D interaction generate layer-dependent internal screening.
As a result, given a external displacement field $V$, the screened displacement field $U(V)$ should be used to generate the Hilbert space.
In other words, we should use the basis $c^\dagger_{\eta,\bsl{k},n,s}$ that diagonalizes $\left. H_0 \right|_{V\rightarrow U(V)}$ to generate the Hilbert space, \ie,
\bea
c^\dagger_{\eta,\bsl{k},n ,s} =  \sum_{\bsl{G}l\sigma}c^\dagger_{\eta,\bsl{k},\bsl{G},l\sigma s} \left[ U_{n}^{\eta}(\bsl{k}) \right]_{\bsl{G}l\sigma}
\eea
where $U^\eta_n(\mbf{k})$ is a complete set of orthonormal eigenvectors (in valley $\eta$) labeled by the band number $n$ for the screened displacmement field $U(V)$.
In this so-called screened basis, the interaction has the form
\eq{
H_{\text{int}} = :H_{\text{int}}: + H_{b}^{\text{full}} \ ,
}
where ``full" in $H_{b}^{\text{full}}$ means the term contains all bands of $\left. H_0 \right|_{V\rightarrow U(V)}$,
\eqa{
:H_{\text{int}}: & = \frac{1}{2\V} \sum_{\bsl{k}_1 \bsl{k}_2 \bsl{q} } \sum_{\eta_1 \eta_2} \sum_{s_1 s_2} \sum_{n_1 n_2 n_3 n_4}  V_{n_1 n_2 n_3 n_4}^{\eta_1 \eta_2}(\bsl{k}_1,\bsl{k}_2,\bsl{q}):c^\dagger_{\eta_1,\bsl{k}_1+\bsl{q},n_1 ,s_1} c^\dagger_{\eta_2,\bsl{k}_2-\bsl{q},n_2 ,s_2}  c_{\eta_2,\bsl{k}_2,n_3 ,s_2}  c_{\eta_1,\bsl{k}_1,n_4 ,s_1}:  \ ,
}
where
\eqa{
V_{n_1 n_2 n_3 n_4}^{\eta_1 \eta_2}(\bsl{k}_1,\bsl{k}_2,\bsl{q})  & = \sum_{\bsl{G}_1 l_1 \sigma_1} \sum_{\bsl{G}_2 l_2 \sigma_2}  \sum_{\bsl{G}}
V_{l_1 l_2}(\bsl{q}+\bsl{G}) \left[ U_{n_1}^{\eta_1}(\bsl{k}_1+\bsl{q})\right]^*_{(\bsl{G}_1 + \bsl{G}) l_1 \sigma_1} \left[ U_{n_2}^{\eta_2}(\bsl{k}_2-\bsl{q})\right]^*_{(\bsl{G}_2 - \bsl{G}) l_2 \sigma_2} \\
& \qquad \times \left[ U_{n_3}^{\eta_2}(\bsl{k}_2)\right]_{ \bsl{G}_2 l_2 \sigma_2}  \left[ U_{n_4}^{\eta_1}(\bsl{k}_1)\right]_{ \bsl{G}_1 l_1 \sigma_1}\\
& = \sum_{l_1 l_2 \bsl{G}} V_{l_1 l_2}(\bsl{q}+\bsl{G}) M_{n_1 n_4}^{l_1 \eta_1}(\bsl{k}_1,\bsl{q}+\bsl{G})  M_{n_2 n_3}^{l_2 \eta_2}(\bsl{k}_2,-\bsl{q}-\bsl{G}) \\
& = \sum_{l_1 l_2 \bsl{G}} V_{l_1 l_2}(\bsl{q}+\bsl{G}) M_{n_1 n_4}^{l_1 \eta_1}(\bsl{k}_1,\bsl{q}+\bsl{G})  \left[ M_{n_3 n_2}^{l_2 \eta_2}(\bsl{k}_2-\bsl{q},\bsl{q}+\bsl{G}) \right]^* \ ,
}
and
\eq{
M^{l\eta}_{mn}(\bsl{k},\bsl{q}+\bsl{G}) = \sum_{\bsl{G}'\sigma} \left[U^{\eta}_{m}(\bsl{k}+\bsl{q}+\bsl{G})\right]_{\bsl{G}'l\sigma}^* \left[U^{\eta}_{n}(\bsl{k})\right]_{\bsl{G}'l\sigma}  \ .
}
In the expressions above, the normal-ordering operation $:\hat{O}:$ places all conduction band annihilation operators and valence band creation operators in $\hat{O}$ to the right (keeping track of fermionic minus signs).
Hermiticity leads to
\eq{
\left[V_{n_1 n_2 n_3 n_4}^{\eta_1 \eta_2}(\bsl{k}_1,\bsl{k}_2,\bsl{q})\right]^* = V_{n_4 n_3 n_2 n_1}^{\eta_1 \eta_2}(\bsl{k}_1+\bsl{q},\bsl{k}_2-\bsl{q},-\bsl{q})\ ,
}
and fermionic statistics lead to
\eq{
V_{n_1 n_2 n_3 n_4}^{\eta_1 \eta_2}(\bsl{k}_1,\bsl{k}_2,\bsl{q})  = V_{n_2 n_1 n_4 n_3}^{\eta_2 \eta_1}(\bsl{k}_2,\bsl{k}_1,-\bsl{q}) \ .
}

We perform ED calculations which project into a set of active degrees of freedom consisting of $c^\dagger_{\eta,\bsl{k},n^c,s}$ in the conduction band subspace (of $\left. H_0 \right|_{V\rightarrow U(V)}$) with $n^c = 0,1,2$ labeling the lowest three conduction bands (all other bands are energetically far away from the flat $n^c=0$ band by a kinetic energy which is at least as large as the Coulomb energy), while assuming the valence bands (of $\left. H_0 \right|_{V\rightarrow U(V)}$) are fully filled.
With this constraint, we can restrict $H_{b}^{\text{full}}$ into the active conduction bands, and obtain
\eq{
H_{b} = \sum_{\bsl{k} n_1^c n_2^c \eta s} c^\dagger_{\eta,\bsl{k},n_1^c, s} c_{\eta,\bsl{k},n_2^c, s}  h^{c,\eta}_{b,n_1^c n_2^c}(\bsl{k})\ ,
}
where
\eqa{
h^{c,\eta}_{b,n_1^c n_2^c}(\bsl{k}) =h^{H,\eta}_{b,n_1^c n_2^c}(\bsl{k}) + h^{F,\eta}_{b,n_1^c n_2^c}(\bsl{k})
}
\eq{
h^{H,\eta}_{b,n_1^c n_2^c}(\bsl{k}) = - \frac{1}{\Omega} \sum_{l l' \bsl{G}} V_{l l'}(\bsl{G})   \Tr\left[ \frac{1}{N_s} \sum_{\eta_1 \bsl{k}_1 } \{ \tilde{P}_c^{\eta_1}(\bsl{k}_1) -  \tilde{P}_v^{\eta_1}(\bsl{k}_1) \} S_{\bsl{G},l'}\right]^*  M^{l \eta}_{n_1^c n_2^c}(\bsl{k},\bsl{G})\ ,
}
\eq{
h^{F,\eta}_{b,n_1^c n_2^c}(\bsl{k}) =  \frac{1}{2\V} \sum_{l l'\bsl{q}\bsl{G}} V_{l l'}(\bsl{q}+\bsl{G}) \left[ U^{\eta}_{n_1^c}(\bsl{k}) \right]^\dagger S^\dagger_{\bsl{G},l'} \left[ \tilde{P}_c^{\eta}(\bsl{k} + \bsl{q}) - \tilde{P}_v^{\eta}(\bsl{k} + \bsl{q}) \right] S_{\bsl{G},l} U^{\eta}_{n_2^c}(\bsl{k})  \ ,
}
\eq{
\left[ S_{\bsl{G},l} \right]_{\bsl{G}_1 l_1 \sigma_1, \bsl{G}_2 l_2 \sigma_2} = \delta_{\bsl{G}_1,\bsl{G}_2 + \bsl{G}} \delta_{l_1 l} \delta_{l_2 l} \delta_{\sigma_1 \sigma_2}\ ,
}
$\tilde{P}_c^\eta(\mbf{k}) = \sum_{n_c} U^\eta_{n_c}(\mbf{k})U^{\eta \dag}_{n_c}(\mbf{k})$, and $\tilde{P}_v^\eta(\mbf{k}) = \sum_{n_v} U^\eta_{n_v}(\mbf{k})U^{\eta \dag}_{n_v}(\mbf{k})$.

To summarize, the total many-body Hamiltonian reads
\eq{
H = \sum_{\bsl{k} n_1^c n_2^c \eta s} c^\dagger_{\eta,\bsl{k},n_1^c, s} c_{\eta,\bsl{k},n_2^c, s}  \left[h_{0,\eta}(\bsl{k})+ h^{H,\eta}_b(\bsl{k})  + h^{F,\eta}_b(\bsl{k}) \right]_{n_1^c n_2^c} + :H_{\text{int}}:\ ,
}
where $:H_{\text{int}}:$ above is understood to be restricted to the lowest three conduction bands.
In this work, we only consider the case where the conduction electrons are polarized in valley $\K$ and spin $\uparrow$. Thus we finally study the following Hamiltonian
\eqa{
\label{eq:ED_H_3D_int_ave}
H & = \sum_{\bsl{k} n_1^c n_2^c } c^\dagger_{\K,\bsl{k},n_1^c, \uparrow} c_{\K,\bsl{k},n_2^c, \uparrow}  \left[h_{0}(\bsl{k})+ h^{H}_b(\bsl{k})  + h^{F}_b(\bsl{k}) \right]_{n_1^c n_2^c} \\
& \quad + \frac{1}{2\V} \sum_{\bsl{k}_1 \bsl{k}_2 \bsl{q} }  \sum_{n_1 n_2 n_3 n_4}  V_{n_1 n_2 n_3 n_4}^{\K \K}(\bsl{k}_1,\bsl{k}_2,\bsl{q})c^\dagger_{K,\bsl{k}_1+\bsl{q},n_1 ,\uparrow} c^\dagger_{\K,\bsl{k}_2-\bsl{q},n_2 ,\uparrow}  c_{\K,\bsl{k}_2,n_3 ,\uparrow}  c_{\K,\bsl{k}_1,n_4 ,\uparrow}\\
& = H_0 + H_b + : H_{\text{int},\K\uparrow} : \ ,
}
where
\eq{
h_{0}(\bsl{k}) = h_{0,K}(\bsl{k})\ ;\ h^{H}_b(\bsl{k}) = h^{H,K}_b(\bsl{k})\ ;\ h^{F,K}_b(\bsl{k}) = h^{F}_b(\bsl{k}) \ ,
}
and
\eqa{
\label{eq:ED_H_3D_int_ave_terms}
& H_0 = \sum_{\bsl{k} n_1^c n_2^c } c^\dagger_{\K,\bsl{k},n_1^c, \uparrow} c_{\K,\bsl{k},n_2^c, \uparrow}  \left[h_{0}(\bsl{k}) \right]_{n_1^c n_2^c} \\
& H_b = \sum_{\bsl{k} n_1^c n_2^c } c^\dagger_{\K,\bsl{k},n_1^c, \uparrow} c_{\K,\bsl{k},n_2^c, \uparrow}  \left[ h^{H}_b(\bsl{k})  + h^{F}_b(\bsl{k}) \right]_{n_1^c n_2^c} \\
& : H_{\text{int},\K\uparrow} : = \frac{1}{2\V} \sum_{\bsl{k}_1 \bsl{k}_2 \bsl{q} }  \sum_{n_1 n_2 n_3 n_4}  V_{n_1 n_2 n_3 n_4}^{\K \K}(\bsl{k}_1,\bsl{k}_2,\bsl{q})c^\dagger_{K,\bsl{k}_1+\bsl{q},n_1 ,\uparrow} c^\dagger_{\K,\bsl{k}_2-\bsl{q},n_2 ,\uparrow}  c_{\K,\bsl{k}_2,n_3 ,\uparrow}  c_{\K,\bsl{k}_1,n_4 ,\uparrow}\ .
}

\subsubsection{Charge-Neutrality Scheme}
\label{app:3DintCNP}

For the CN scheme, we simply normal order the Coulomb interaction against the charge neutrality gap below the lowest conduction band. For $V>0$ where the conduction electrons are biased away from the hBN, this amounts to effectively treating the conduction bands as an electron gas (virtually unaffected by the valence bands bound to the moir\'e pattern). The total many-body Hamiltonian, projected to the lowest three conduction bands in valley $\K$ and spin $\uparrow$, reads
\eqa{
\label{eq:ED_H_3D_int_CN}
H & = \sum_{\bsl{k} n_1^c n_2^c } c^\dagger_{\K,\bsl{k},n_1^c, \uparrow} c_{\K,\bsl{k},n_2^c, \uparrow}  \left[h_{0}(\bsl{k})\right]_{n_1^c n_2^c} \\
& \quad + \frac{1}{2\V} \sum_{\bsl{k}_1 \bsl{k}_2 \bsl{q} }  \sum_{n_1 n_2 n_3 n_4}  V_{n_1 n_2 n_3 n_4}^{\K \K}(\bsl{k}_1,\bsl{k}_2,\bsl{q})c^\dagger_{\K,\bsl{k}_1+\bsl{q},n_1 ,\uparrow} c^\dagger_{\K,\bsl{k}_2-\bsl{q},n_2 ,\uparrow}  c_{\K,\bsl{k}_2,n_3 ,\uparrow}  c_{\K,\bsl{k}_1,n_4 ,\uparrow}\\
& = H_0 + : H_{\text{int},\K\uparrow} :\ ,
}
which differs from the 3D Coulomb Hamiltonian in the AVE scheme by to the absence of one-body background terms ($H_b$ in \Eq{eq:ED_H_3D_int_ave}).

\newpage
\clearpage

\section{Multi-Band ED Computational Details}
In this Appendix, we discuss the details of our numerical ED computations. \cref{eq:hilbertspacedim} discusses the novel approximations used to reduce Hilbert space sizes in multi-band ED. \cref{eq:HFbasis} discusses our HF calculations used to obtain a unitarily rotated basis for ED and to investigate finite-size dependence.   \App{app:mommesh} lists the momentum meshes used in ED calculations. \App{app:HFhamiltonianbasis} describes the connection between the HF basis and an alternative normal-ordering of the interaction with respect to the HF band structure.
\App{app:biasing_method} then introduces the biasing terms used to further investigate convergence of the FCI states on finite-size multi-band calculations.

\subsection{Multi-Band ED Hilbert Space Reduction Techniques}
\label{eq:hilbertspacedim}

The main limitation of Exact Diagonalization (ED) is the size of the many-body Hilbert space, which scales exponentially in the number of single-particle states (``orbitals" for short) and becomes computationally intractable at, or before, the Hilbert space dimension reaches $\sim10^9$. Symmetries reduce the Hilbert space size by block diagonalizing the Hamiltonian matrix by symmetry sector. In FCI problems, translation symmetry is used to reduce the Hilbert space dimension by block diagonalizing into different (lattice) momentum sectors. Typically, band projection is used to further reduce the degrees of freedom.
We now review the combinatorics of the 1-band Hilbert space dimension as a warm-up for more involved multi-band schemes.

First we consider a 1-band problem with $N_s=N_x N_y$ total orbitals on an $N_x \times N_y$ torus, according to the convention in \cref{eq:momentum_mesh}. In the $N$-particle sector, the total Hilbert space dimension is $\dim \mathcal{H} = \binom{N_xN_y}{N}$. Momentum conservation divides the total Hilbert space into $N_xN_y$ sectors. Although not every sector has exactly the same dimension for finite sizes, an  accurate estimate is the average dimension in each momentum sector, which is
\bea
\dim \mathcal{H}_{\mbf{k}} \simeq \frac{1}{N_x N_y} \binom{N_xN_y}{N} \ .
\eea
In a multi-band system with $N_b$ bands included, the Hilbert space dimension increases exponentially as $\dim \mathcal{H}_{\mbf{k}} \simeq \frac{1}{N_x N_y} \binom{N_b N_xN_y}{N}$. This quickly makes ED computationally intractable: for 3 bands at $\nu=2/3$, an 18-site calculation already has dimension $1.9\times 10^{10}$. Hence it is imperative to find strategies to reduce the scaling of the Hilbert space dimension. We focus on 3-band ED, but all results can be easily generalized to more (or fewer) bands.

In the pentalayer problem, we encounter a gapless system:  there are 3 nearly ($<0.05$meV) degenerate conduction bands at the $K_M,K'_M$ corners of the moir\'e BZ and 2 nearly degenerate bands along the edges of the moir\'e BZ for the bare kinetic energy. An unbiased ED calculation must hence include at least 3 bands. We index these bands by 0,1,2 in increasing order of kinetic energy. Band $0$ has quasi-flat dispersion for all momenta in the moir\'e BZ. Away from regions of near-degeneracy with band 0, the other conduction bands 1 and 2 quickly disperse to high energies. It is hence physically reasonable (and true upon further testing)  that such high-energy orbitals will not have large weight in the ground state and low-lying excited states of the system at the fillings $\nu\leq 1$ of interest. This allows a reduction of the ED Hilbert space as detailed below.

We first implement an approximation that we refer to as ``band maximum" where we limit the maximum total number of particles in band 1 to $N_{\text{band1}}$ and in band 2 to $N_{\text{band2}}$ in the Hilbert space. We refer to this approximation as band-max $\{N_{\text{band1}},N_{\text{band2}}\}$. We do not restrict the number of particles in band 0. Computationally, this is easily implemented in the generation of the Fock basis. The total Hilbert space dimension is
\bea
\dim \mathcal{H} &= \sum_{i=0}^{N_{\text{band1}}} \sum_{j=0}^{N_{\text{band2}}} \binom{N_xN_y}{N-i-j} \binom{N_xN_y}{i}\binom{N_xN_y}{j}
\eea
corresponding to enumerating the Hilbert space by partitioning the $N$ particles into $N-{N_{\text{band1}}}-{N_{\text{band2}}},{N_{\text{band1}}},{N_{\text{band2}}}$ particles, in bands 0, 1, 2 respectively. Our estimate of the Hilbert space dimension in each momentum sector is $\dim \mathcal{H}_{\mbf{k}} \simeq (N_xN_y)^{-1} \dim \mathcal{H}$. Note that taking $N_{\text{band1}},N_{\text{band2}} \to N$ recovers the full 3-band calculation, and hence band-max is a controlled approximation which can be systematically benchmarked. The approximation is valid when there is small probability to have greater than $N_{\text{band1}}$ and $N_{\text{band2}}$ particles in band 1 and band 2 respectively. Note that this depends on the choice of the single-particle ``band" basis, which we discuss later in \cref{eq:HFbasis}. As an example for the magnitude of Hilbert space reduction, full 3-band ED is impossible for a $\nu=2/3$ an 18-site calculation since the Hilbert space has size $1.9\times 10^{10}$, but implementing band-max $\{3,1\}$ reduces the Hilbert space to a manageable $4.6\times 10^7$.
This approximation would be ideal (and has been used~\cite{Rezayi2011breaking,2017PhRvL.119b6801R,kwan2024abelianFTI}) to investigate mixing from remote bands gapped from the band under consideration.
However, here the bands are nearly degenerate around the edge of the moir\'e BZ, and the momentum area where the bands are nearly degenerate and expected to strongly mix is a fixed fraction of  the moir\'e BZ determined by the single-particle band structure.
Hence, it is clear that both of these numbers $\{N_{\text{band1}},N_{\text{band2}}\}$ should scale with the number of unit cells $N_s$ in the thermodynamic limit for converged results.

To deal with this challenging computational problem, we implement a second approximation in certain calculations. We can also restrict the allowed orbitals in each band based on their momentum, since the degeneracy of the bands is quickly lifted away from the moir\'e BZ edge due to the large Fermi velocity of graphene.  We denote this approximation scheme as $\{N_{\text{band1}} \ (N_{\text{orb1}}),N_{\text{band2}} \ (N_{\text{orb2}})\}$ when we keep a maximum of $N_{\text{band1}}$ and $N_{\text{band2}}$ particles in bands 1 and 2 truncated to $N_{\text{orb1}}$ and $N_{\text{orb2}}$ orbitals respectively. There is a freedom in choosing which orbitals to restrict to, and this shall be specified in all calculations. Typically we remove the orbitals with the largest single-particle energies since physically we expect that  their occupations will be suppressed in the states belonging to the low-lying many-body spectrum. The total Hilbert space dimension is
\bea
\dim \mathcal{H} &= \sum_{i=0}^{N_{\text{band1}}} \sum_{j=0}^{N_{\text{band2}}} \binom{N_xN_y}{N-i-j} \binom{N_{\text{orb1}}}{i}\binom{N_{\text{orb2}}}{j}
\eea
with an estimate of $\dim \mathcal{H}_{\mbf{k}} \simeq  (N_xN_y)^{-1} \dim \mathcal{H}$ for each momentum sector. Orbital restriction (which can be used independently of band-max), is again a controlled approximation that exactly reproduces the 3-band calculation as $N_{\text{orb1}}$ and $N_{\text{orb2}}$ tend to $N_s$. The approximation is valid when the weight of a particular set of orbitals in the low-lying many-body states is small.

Finally, we also have the freedom to define the ``bands" according to a unitary transformation of the Hilbert space. We will refer to the Bloch states of the bare kinetic Hamiltonian $h_0(\bsl{k})$ (\cref{eq:ED_H_2D_int_ave}) as the \emph{bare} basis ($c^\dagger_{\bsl{k},n}$). We will also make use of the diagonal basis  ($d^\dagger_{\bsl{k},n}$) of the entire one-body term in \cref{eq:ED_H_2D_int_ave} and the HF basis ($\gamma^\dagger_{\bsl{k},\alpha}$) (where we use the HF eigenbasis to define orbitals for the many-body ED problem). We will elaborate on these bases in the following sections.

Both Hilbert space reduction methods described here are approximations that reduce the full multi-band ED problem by extracting a principal sub-matrix of the many-body Hamiltonian. By the Cauchy interlacing theorem, the ground state energy of the reduced problem is greater than or equal to that of the full problem (but there is no monotonicity result for the gaps). Note that these approximations will converge to the exact answer as the cutoffs are increased, and in practice their accuracy can be judged from the convergence of the low energy spectrum. Our approximations are justified by the physical nature of our problem --- that of a flat band degenerate with other dispersive bands.

\subsubsection{Restricted Orbital Configurations}
\label{app:restrictedorb}

Restricting the allowed orbitals within the higher bands provides a way to reduce the Hilbert space without compromising the low energy spectrum. To determine which orbitals can be removed, we employ a few approaches. First, on physical grounds, the highest energy orbitals are good candidates for truncation since their weight in the ground state is suppressed if the interaction is smaller than their energy above the flat band. Second, one can check the correlation functions $\braket{\gamma^\dag_{\mbf{k},\al}\gamma_{\mbf{k},\al}}$ for smaller band-max and determine which orbitals have the lowest weight. Where possible, symmetry-related orbitals are removed together, although not all lattices preserve $C_3$. It is also important to check the convergence of the calculations. We give an example below on the $4\times 6$ lattice where energies in two momentum sectors are compared between two calculations where we keep 15 versus 19 orbitals in band 1. The 19 orbital calculation is too costly to compute all sectors with, but serves as a check of the good ($97\%$) convergence between the 15 and 19 orbital calculations.

For the $9\times2$ lattice at $\nu=1$ with band-max $\{5,1\}$, we use the momenta $(3, 0)$, $
(6, 0)$, $
(4, 0)$, $
(2, 0)$, $
(5, 0)$, $
(5, 1)$, $
(7, 0)$, $
(0, 1)$, $
(4, 1)$, $
(1, 1)$, $
(8, 1)$, $
(6, 1)$, $
(2, 1)$, $
(3, 1)$, $
(7, 1)$, $
(1, 0)$ in band 1 and $(6,0)$, $
(7,0)$, $
(3,0)$, $
(2,0)$, $
(3 1)$, $
(4,0)$, $
(5,0)$, $
(1,1)$, $
(6,1)$, $
(8,1)$, $
(5,1)$, $
(4,1)$, $
(7,1)$, $
(0,1)$ in band 2.

For the $9\times2$ lattice with band-max $\{6,0\}$, we use the momenta
$(3, 0)$, $
(6, 0)$, $
(4, 0)$, $
(2, 0)$, $
(5, 0)$, $
(5, 1)$, $
(7, 0)$, $
(0, 1)$, $
(4, 1)$, $
(1, 1)$, $
(8, 1)$, $
(6, 1)$, $
(2, 1)$, $
(3, 1)$, $
(7, 1)$, $
(1, 0)$ in band 1 (and all momenta in band 2).

For the $21\times1$ lattice with band-max $\{5,0\}$ and $\nu=2/3$ and $\{6,0\}$ at $\nu=1$, we use the momenta
$(7 ,0)$, $
(14 ,0)$, $
(8 ,0)$, $
(11 ,0)$, $
(2 ,0)$, $
(12 ,0)$, $
(3 ,0)$, $
(6 ,0)$, $
(19 ,0)$, $
(10 ,0)$, $
(13 ,0)$, $
(15 ,0)$, $
(9 ,0)$, $
(18 ,0)$, $
(16 ,0)$ in band 1 and all momenta in band 2. For band-max $\{5,0\}$ and $\{4,1\}$ at $\nu=1$, we use momenta $(7 ,0)$, $
(14 ,0)$, $
(8 ,0)$, $
(11 ,0)$, $
(2 ,0)$, $
(12 ,0)$, $
(3 ,0)$, $
(6 ,0)$, $
(19 ,0)$, $
(10 ,0)$, $
(13 ,0)$, $
(15 ,0)$, $
(9 ,0)$, $
(18 ,0)$, $
(16 ,0)$ in band 1 and all the momenta in band 2.

For the $4\times6$ lattice with band-max $\{4,0\}$ at $\nu=2/3$, we use the momenta $(0, 2)$, $
(0, 4)$, $
(3, 5)$, $
(1, 1)$, $
(0, 3)$, $
(2, 3)$, $
(2, 0)$, $
(3, 3)$, $
(1, 3)$, $
(3, 4)$, $
(1, 2)$, $
(1, 4)$, $
(3, 2)$, $
(2, 5)$, $
(2, 4)$. We have confirmed the convergence of this calculation at a few $k$ points using the additional momenta $(0, 1)$, $(0, 5)$, $ (2, 2)$, $ (2, 1)$. Specifically, we check that the energy difference between the lowest states in the $(0,0)$ and $(0,1)$ momenta is within $3\%$ agreement between the two calculations. We only check two points because calculations with 19 orbitals (Hilbert space dimension $5.53\times 10^8$) can take hundreds of hours per momentum sector.

\subsubsection{ED Hilbert Space Dimensions at $\nu=1$}

For reference, we include the following tables of approximate Hilbert space dimensions per momentum $\mbf{k}$ with band-max restriction for $N_xN_y = 12,15,18,21,24$ sites at filling $\nu=1$. We omit entries that are larger than $5\times 10^8$, and hence difficult to access computationally.

\bea
\begin{array}{c|ccccccccc}
\mbf{12},\nu=1 & N_{\text{band2}}={0} & N_{\text{band2}}={1} & N_{\text{band2}}={2} & N_{\text{band2}}={3} & N_{\text{band2}}={4} & N_{\text{band2}}={5} & N_{\text{band2}}={6} & N_{\text{band2}}={7} & N_{\text{band2}}={8} \\
\hline
N_{\text{band1}}={0} & 1 & 12 & 3.8\times 10^2 & 4.4\times 10^3 & 2.5\times 10^4 & 7.7\times 10^4 & 1.5\times 10^5 & 2.\times 10^5 & 2.2\times 10^5 \\
N_{\text{band1}}={1} & 12 & 8.2\times 10^2 & 1.6\times 10^4 & 1.3\times 10^5 & 5.4\times 10^5 & 1.3\times 10^6 & 2.1\times 10^6 & 2.6\times 10^6 & 2.7\times 10^6 \\
N_{\text{band1}}={2} & 3.8\times 10^2 & 1.6\times 10^4 & 2.1\times 10^5 & 1.3\times 10^6 & 4.2\times 10^6 & 8.4\times 10^6 & 1.2\times 10^7 & 1.3\times 10^7 & 1.3\times 10^7 \\
N_{\text{band1}}={3} & 4.4\times 10^3 & 1.3\times 10^5 & 1.3\times 10^6 & 6.1\times 10^6 & 1.6\times 10^7 & 2.8\times 10^7 & 3.5\times 10^7 & 3.7\times 10^7 & 3.7\times 10^7 \\
N_{\text{band1}}={4} & 2.5\times 10^4 & 5.4\times 10^5 & 4.2\times 10^6 & 1.6\times 10^7 & 3.6\times 10^7 & 5.5\times 10^7 & 6.5\times 10^7 & 6.7\times 10^7 & 6.8\times 10^7 \\
N_{\text{band1}}={5} & 7.7\times 10^4 & 1.3\times 10^6 & 8.4\times 10^6 & 2.8\times 10^7 & 5.5\times 10^7 & 7.7\times 10^7 & 8.7\times 10^7 & 9.\times 10^7 & 9.1\times 10^7 \\
N_{\text{band1}}={6} & 1.5\times 10^5 & 2.1\times 10^6 & 1.2\times 10^7 & 3.5\times 10^7 & 6.5\times 10^7 & 8.7\times 10^7 & 9.8\times 10^7 & 1.\times 10^8 & 1.\times 10^8 \\
N_{\text{band1}}={7} & 2.\times 10^5 & 2.6\times 10^6 & 1.3\times 10^7 & 3.7\times 10^7 & 6.7\times 10^7 & 9.\times 10^7 & 1.\times 10^8 & 1.\times 10^8 & 1.\times 10^8 \\
N_{\text{band1}}={8} & 2.2\times 10^5 & 2.7\times 10^6 & 1.3\times 10^7 & 3.7\times 10^7 & 6.8\times 10^7 & 9.1\times 10^7 & 1.\times 10^8 & 1.\times 10^8 & 1.\times 10^8 \\
\end{array}
\eea

\bea
\begin{array}{c|ccccccccc}
\mbf{15},\nu=1& N_{\text{band2}}={0} & N_{\text{band2}}={1} & N_{\text{band2}}={2} & N_{\text{band2}}={3} & N_{\text{band2}}={4} & N_{\text{band2}}={5} & N_{\text{band2}}={6} & N_{\text{band2}}={7} & N_{\text{band2}}={8} \\
\hline
N_{\text{band1}}={0} & 1 & 15 & 7.5\times 10^2 & 1.5\times 10^4 & 15 10^5 & 7.4\times 10^5 & 2.4\times 10^6 & 5.2\times 10^6 & 7.9\times 10^6 \\
N_{\text{band1}}={1} & 1.5\times 10^1 & 1.6\times 10^3 & 5.\times 10^4 & 6.8\times 10^5 & 4.9\times 10^6 & 2.1\times 10^7 & 5.4\times 10^7 & 9.9\times 10^7 & 1.3\times 10^8 \\
N_{\text{band1}}={2} & 7.5\times 10^2 & 5.\times 10^4 & 1.1\times 10^6 & 1.1\times 10^7 & 6.3\times 10^7 & 2.1\times 10^8 & 4.7\times 10^8 & \text{} & \text{} \\
N_{\text{band1}}={3} & 1.5\times 10^4 & 6.8\times 10^5 & 1.1\times 10^7 & 9.1\times 10^7 & 4.1\times 10^8 & \text{} & \text{} & \text{} & \text{} \\
N_{\text{band1}}={4} & 1.4\times 10^5 & 4.9\times 10^6 & 6.3\times 10^7 & 4.1\times 10^8 & \text{} & \text{} & \text{} & \text{} & \text{} \\
N_{\text{band1}}={5} & 7.4\times 10^5 & 2.1\times 10^7 & 2.1\times 10^8 & \text{} & \text{} & \text{} & \text{} & \text{} & \text{} \\
N_{\text{band1}}={6} & 2.4\times 10^6 & 5.4\times 10^7 & 4.7\times 10^8 & \text{} & \text{} & \text{} & \text{} & \text{} & \text{} \\
N_{\text{band1}}={7} & 5.2\times 10^6 & 9.9\times 10^7 & \text{} & \text{} & \text{} & \text{} & \text{} & \text{} & \text{} \\
N_{\text{band1}}={8} & 7.9\times 10^6 & 1.3\times 10^8 & \text{} & \text{} & \text{} & \text{} & \text{} & \text{} & \text{} \\
\end{array}
\eea

\bea
\begin{array}{c|ccccccccc}
\mbf{18},\nu=1 & N_{\text{band2}}={0} & N_{\text{band2}}={1} & N_{\text{band2}}={2} & N_{\text{band2}}={3} & N_{\text{band2}}={4} & N_{\text{band2}}={5} & N_{\text{band2}}={6} & N_{\text{band2}}={7} & N_{\text{band2}}={8} \\
\hline
N_{\text{band1}}={0} & 1 & 18 & 1.3\times 10^3 & 3.8\times 10^4 & 5.6\times 10^5 & 4.6\times 10^6 & 2.4\times 10^7 & 8.\times 10^7 & 1.9\times 10^8 \\
N_{\text{band1}}={1} & 18 & 2.8\times 10^3 & 1.3\times 10^5 & 2.7\times 10^6 & 2.9\times 10^7 & 1.9\times 10^8 & \text{} & \text{} & \text{} \\
N_{\text{band1}}={2} & 1.3\times 10^3 & 1.3\times 10^5 & 4.2\times 10^6 & 6.6\times 10^7 & \text{} & \text{} & \text{} & \text{} & \text{} \\
N_{\text{band1}}={3} & 3.8\times 10^4 & 2.7\times 10^6 & 6.6\times 10^7 & \text{} & \text{} & \text{} & \text{} & \text{} & \text{} \\
N_{\text{band1}}={4} & 5.6\times 10^5 & 2.9\times 10^7 & \text{} & \text{} & \text{} & \text{} & \text{} & \text{} & \text{} \\
N_{\text{band1}}={5} & 4.6\times 10^6 & 1.9\times 10^8 & \text{} & \text{} & \text{} & \text{} & \text{} & \text{} & \text{} \\
N_{\text{band1}}={6} & 2.4\times 10^7 & \text{} & \text{} & \text{} & \text{} & \text{} & \text{} & \text{} & \text{} \\
N_{\text{band1}}={7} & 8.\times 10^7 & \text{} & \text{} & \text{} & \text{} & \text{} & \text{} & \text{} & \text{} \\
N_{\text{band1}}={8} & 1.9\times 10^8 & \text{} & \text{} & \text{} & \text{} & \text{} & \text{} & \text{} & \text{} \\
\end{array}
\eea

\bea
\begin{array}{c|ccccccccc}
\mbf{21},\nu=1 & N_{\text{band2}}={0} & N_{\text{band2}}={1} & N_{\text{band2}}={2} & N_{\text{band2}}={3} & N_{\text{band2}}={4} & N_{\text{band2}}={5} & N_{\text{band2}}={6} & N_{\text{band2}}={7} & N_{\text{band2}}={8} \\
\hline
N_{\text{band1}}={0} &1 & 21 & 2.1\times 10^3 & 8.6\times 10^4 & 1.8\times 10^6 & 2.2\times 10^7 & 1.6\times 10^8 & \text{} & \text{} \\
N_{\text{band1}}={1} & 21 & 4.5\times 10^3 & 2.9\times 10^5 & 8.3\times 10^6 & 1.3\times 10^8 & \text{} & \text{} & \text{} & \text{} \\
N_{\text{band1}}={2} & 2.1\times 10^3 & 2.9\times 10^5 & 1.3\times 10^7 & 2.9\times 10^8 & \text{} & \text{} & \text{} & \text{} & \text{} \\
N_{\text{band1}}={3} & 8.6\times 10^4 & 8.3\times 10^6 & 2.9\times 10^8 & \text{} & \text{} & \text{} & \text{} & \text{} & \text{} \\
N_{\text{band1}}={4} & 1.8\times 10^6 & 1.3\times 10^8 & \text{} & \text{} & \text{} & \text{} & \text{} & \text{} & \text{} \\
N_{\text{band1}}={5} & 2.2\times 10^7 & \text{} & \text{} & \text{} & \text{} & \text{} & \text{} & \text{} & \text{} \\
N_{\text{band1}}={6} & 1.6\times 10^8 & \text{} & \text{} & \text{} & \text{} & \text{} & \text{} & \text{} & \text{} \\
N_{\text{band1}}={7} & \text{} & \text{} & \text{} & \text{} & \text{} & \text{} & \text{} & \text{} & \text{} \\
N_{\text{band1}}={8} & \text{} & \text{} & \text{} & \text{} & \text{} & \text{} & \text{} & \text{} & \text{} \\
\end{array}
\eea

\bea
\begin{array}{c|ccccccccc}
\mbf{24},\nu=1 & N_{\text{band2}}={0} & N_{\text{band2}}={1} & N_{\text{band2}}={2} & N_{\text{band2}}={3} & N_{\text{band2}}={4} & N_{\text{band2}}={5} & N_{\text{band2}}={6} & N_{\text{band2}}={7} & N_{\text{band2}}={8} \\
\hline
N_{\text{band1}}={0} &1 & 24 & 3.2\times 10^3 & 1.7\times 10^5 & 4.9\times 10^6 & 8.\times 10^7 & \text{} & \text{} & \text{} \\
N_{\text{band1}}={1} & 24 & 6.7\times 10^3 & 5.7\times 10^5 & 2.2\times 10^7 & 4.8\times 10^8 & \text{} & \text{} & \text{} & \text{} \\
N_{\text{band1}}={2} & 3.2\times 10^3 & 5.7\times 10^5 & 3.5\times 10^7 & \text{} & \text{} & \text{} & \text{} & \text{} & \text{} \\
N_{\text{band1}}={3} & 1.7\times 10^5 & 2.2\times 10^7 & \text{} & \text{} & \text{} & \text{} & \text{} & \text{} & \text{} \\
N_{\text{band1}}={4} & 4.9\times 10^6 & 4.8\times 10^8 & \text{} & \text{} & \text{} & \text{} & \text{} & \text{} & \text{} \\
N_{\text{band1}}={5} & 8.\times 10^7 & \text{} & \text{} & \text{} & \text{} & \text{} & \text{} & \text{} & \text{} \\
N_{\text{band1}}={6} & \text{} & \text{} & \text{} & \text{} & \text{} & \text{} & \text{} & \text{} & \text{} \\
N_{\text{band1}}={7} & \text{} & \text{} & \text{} & \text{} & \text{} & \text{} & \text{} & \text{} & \text{} \\
N_{\text{band1}}={8} & \text{} & \text{} & \text{} & \text{} & \text{} & \text{} & \text{} & \text{} & \text{} \\
\end{array}
\eea

\subsubsection{ED Hilbert Spaces at $\nu=2/3$}

For reference, we include the following tables of approximate Hilbert space dimensions per momentum $\mbf{k}$ with band-max restriction for $N_xN_y = 12,15,18,21,24$ sites at filling $\nu=2/3$. We omit entries that are larger than $5\times 10^8$, and hence difficult to access computationally.

\bea
\begin{array}{c|ccccccc}
\mbf{12},\nu=2/3 & N_{\text{band2}}={0} & N_{\text{band2}}={1} & N_{\text{band2}}={2} & N_{\text{band2}}={3} & N_{\text{band2}}={4} & N_{\text{band2}}={5} & N_{\text{band2}}={6} \\
\hline
N_{\text{band1}}={0} & 4.1\times 10^1 & 8.3\times 10^2 & 5.9\times 10^3 & 2.\times 10^4 & 4.1\times 10^4 & 5.5\times 10^4 & 6.\times 10^4 \\
N_{\text{band1}}={1} & 8.3\times 10^2 & 1.3\times 10^4 & 7.\times 10^4 & 1.9\times 10^5 & 3.2\times 10^5 & 3.9\times 10^5 & 4.1\times 10^5 \\
N_{\text{band1}}={2} & 5.9\times 10^3 & 7.\times 10^4 & 3.1\times 10^5 & 7.\times 10^5 & 1.\times 10^6 & 1.1\times 10^6 & 1.1\times 10^6 \\
N_{\text{band1}}={3} & 2.\times 10^4 & 1.9\times 10^5 & 7.\times 10^5 & 1.4\times 10^6 & 1.8\times 10^6 & 1.9\times 10^6 & 1.9\times 10^6 \\
N_{\text{band1}}={4} & 4.1\times 10^4 & 3.2\times 10^5 & 1.\times 10^6 & 1.8\times 10^6 & 2.2\times 10^6 & 2.3\times 10^6 & 2.4\times 10^6 \\
N_{\text{band1}}={5} & 5.5\times 10^4 & 3.9\times 10^5 & 1.1\times 10^6 & 1.9\times 10^6 & 2.3\times 10^6 & 2.5\times 10^6 & 2.5\times 10^6 \\
N_{\text{band1}}={6} & 6.\times 10^4 & 4.1\times 10^5 & 1.1\times 10^6 & 1.9\times 10^6 & 2.4\times 10^6 & 2.5\times 10^6 & 2.5\times 10^6 \\
\end{array}
\eea

\bea
\begin{array}{c|ccccccc}
\mbf{15},\nu=2/3 & N_{\text{band2}}={0} & N_{\text{band2}}={1} & N_{\text{band2}}={2} & N_{\text{band2}}={3} & N_{\text{band2}}={4} & N_{\text{band2}}={5} & N_{\text{band2}}={6} \\
\hline
N_{\text{band1}}={0} & 2.\times 10^2 & 5.2\times 10^3 & 5.\times 10^4 & 2.5\times 10^5 & 7.\times 10^5 & 1.3\times 10^6 & 1.8\times 10^6 \\
N_{\text{band1}}={1} & 5.2\times 10^3 & 1.1\times 10^5 & 8.3\times 10^5 & 3.3\times 10^6 & 7.9\times 10^6 & 1.3\times 10^7 & 1.5\times 10^7 \\
N_{\text{band1}}={2} & 5.\times 10^4 & 8.3\times 10^5 & 5.2\times 10^6 & 1.7\times 10^7 & 3.5\times 10^7 & 4.9\times 10^7 & 5.6\times 10^7 \\
N_{\text{band1}}={3} & 2.5\times 10^5 & 3.3\times 10^6 & 1.7\times 10^7 & 4.8\times 10^7 & 8.5\times 10^7 & 1.1\times 10^8 & 1.2\times 10^8 \\
N_{\text{band1}}={4} & 7.\times 10^5 & 7.9\times 10^6 & 3.5\times 10^7 & 8.5\times 10^7 & 1.3\times 10^8 & 1.6\times 10^8 & 1.7\times 10^8 \\
N_{\text{band1}}={5} & 1.3\times 10^6 & 1.3\times 10^7 & 4.9\times 10^7 & 1.1\times 10^8 & 1.6\times 10^8 & 1.9\times 10^8 & 2.\times 10^8 \\
N_{\text{band1}}={6} & 1.8\times 10^6 & 1.5\times 10^7 & 5.6\times 10^7 & 1.2\times 10^8 & 1.7\times 10^8 & 2.\times 10^8 & 2.1\times 10^8 \\
\end{array}
\eea

\bea
\begin{array}{c|ccccccc}
\mbf{18},\nu=2/3& N_{\text{band2}}={0} & N_{\text{band2}}={1} & N_{\text{band2}}={2} & N_{\text{band2}}={3} & N_{\text{band2}}={4} & N_{\text{band2}}={5} & N_{\text{band2}}={6} \\
\hline
N_{\text{band1}}={0} & 1.\times 10^3 & 3.3\times 10^4 & 4.\times 10^5 & 2.6\times 10^6 & 1.\times 10^7 & 2.5\times 10^7 & 4.4\times 10^7 \\
N_{\text{band1}}={1} & 3.3\times 10^4 & 8.5\times 10^5 & 8.7\times 10^6 & 4.7\times 10^7 & 1.5\times 10^8 & 3.3\times 10^8 & \text{} \\
N_{\text{band1}}={2} & 4.\times 10^5 & 8.7\times 10^6 & 7.3\times 10^7 & 3.3\times 10^8 & \text{} & \text{} & \text{} \\
N_{\text{band1}}={3} & 2.6\times 10^6 & 4.7\times 10^7 & 3.3\times 10^8 & \text{} & \text{} & \text{} & \text{} \\
N_{\text{band1}}={4} & 1.\times 10^7 & 1.5\times 10^8 & \text{} & \text{} & \text{} & \text{} & \text{} \\
N_{\text{band1}}={5} & 2.5\times 10^7 & 3.3\times 10^8 & \text{} & \text{} & \text{} & \text{} & \text{} \\
N_{\text{band1}}={6} & 4.4\times 10^7 & \text{} & \text{} & \text{} & \text{} & \text{} & \text{} \\
\end{array}
\eea

\bea
\begin{array}{c|ccccccc}
\mbf{21},\nu=2/3 & N_{\text{band2}}={0} & N_{\text{band2}}={1} & N_{\text{band2}}={2} & N_{\text{band2}}={3} & N_{\text{band2}}={4} & N_{\text{band2}}={5} & N_{\text{band2}}={6} \\
\hline
N_{\text{band1}}={0} & 5.5\times 10^3 & 2.1\times 10^5 & 3.1\times 10^6 & 2.5\times 10^7 & 1.3\times 10^8 & 4.1\times 10^8 & \text{} \\
N_{\text{band1}}={1} & 2.1\times 10^5 & 6.6\times 10^6 & 8.4\times 10^7 & \text{} & \text{} & \text{} & \text{} \\
N_{\text{band1}}={2} & 3.1\times 10^6 & 8.4\times 10^7 & \text{} & \text{} & \text{} & \text{} & \text{} \\
N_{\text{band1}}={3} & 2.5\times 10^7 & \text{} & \text{} & \text{} & \text{} & \text{} & \text{} \\
N_{\text{band1}}={4} & 1.3\times 10^8 & \text{} & \text{} & \text{} & \text{} & \text{} & \text{} \\
N_{\text{band1}}={5} & 4.1\times 10^8 & \text{} & \text{} & \text{} & \text{} & \text{} & \text{} \\
N_{\text{band1}}={6} & \text{} & \text{} & \text{} & \text{} & \text{} & \text{} & \text{} \\
\end{array}
\eea

\bea
\begin{array}{c|ccccccc}
\mbf{24},\nu=2/3 & N_{\text{band2}}={0} & N_{\text{band2}}={1} & N_{\text{band2}}={2} & N_{\text{band2}}={3} & N_{\text{band2}}={4} & N_{\text{band2}}={5} & N_{\text{band2}}={6} \\
\hline
N_{\text{band1}}={0} & 3.1\times 10^4 & 1.3\times 10^6 & 2.4\times 10^7 & 2.3\times 10^8 & \text{} & \text{} & \text{} \\
N_{\text{band1}}={1} & 1.3\times 10^6 & 5.\times 10^7 & \text{} & \text{} & \text{} & \text{} & \text{} \\
N_{\text{band1}}={2} & 2.4\times 10^7 & \text{} & \text{} & \text{} & \text{} & \text{} & \text{} \\
N_{\text{band1}}={3} & 2.3\times 10^8 & \text{} & \text{} & \text{} & \text{} & \text{} & \text{} \\
N_{\text{band1}}={4} & \text{} & \text{} & \text{} & \text{} & \text{} & \text{} & \text{} \\
N_{\text{band1}}={5} & \text{} & \text{} & \text{} & \text{} & \text{} & \text{} & \text{} \\
N_{\text{band1}}={6} & \text{} & \text{} & \text{} & \text{} & \text{} & \text{} & \text{} \\
\end{array}
\eea

\subsubsection{ED Hilbert Space at $\nu=1/3$}

For reference, we include the following tables of approximate Hilbert space dimensions per momentum $\mbf{k}$ with band-max restriction. At $N_x N_y = 12, 15, 18$, 3-band ED is easily accessible with Hilbert space dimensions $4.9\times 10^3, 8.1 \times 10^4, 1.43\times10^6$ respectively. Below we include tables for accessible Hilbert spaces at $21$ and $24$ sites.

\bea
\begin{array}{c|ccccccc}
\mbf{21},\nu = 1/3 & N_{\text{band2}}=0 & N_{\text{band2}}=1 & N_{\text{band2}}=2 & N_{\text{band2}}=3 & N_{\text{band2}}=4 &
N_{\text{band2}}=5 & N_{\text{band2}}=6 \\
\hline
N_{\text{band1}}=0 & 5.5\times 10^3 & 6.\times 10^4 & 2.6\times 10^5 & 6.4\times 10^5 & 1.\times 10^6 & 1.2\times 10^6 & 1.3\times 10^6 \\
N_{\text{band1}}=1 & 6.\times 10^4 & 5.4\times 10^5 & 2.\times 10^6 & 4.1\times 10^6 & 5.8\times 10^6 & 6.4\times 10^6 & 6.5\times 10^6 \\
N_{\text{band1}}=2 & 2.6\times 10^5 & 2.\times 10^6 & 6.3\times 10^6 & 1.1\times 10^7 & 1.4\times 10^7 & 1.5\times 10^7 & 1.5\times 10^7 \\
N_{\text{band1}}=3 & 6.4\times 10^5 & 4.1\times 10^6 & 1.1\times 10^7 & 1.8\times 10^7 & 2.1\times 10^7 & 2.2\times 10^7 & 2.2\times 10^7 \\
N_{\text{band1}}=4 & 1.\times 10^6 & 5.8\times 10^6 & 1.4\times 10^7 & 2.1\times 10^7 & 2.4\times 10^7 & 2.5\times 10^7 & 2.5\times 10^7 \\
N_{\text{band1}}=5 & 1.2\times 10^6 & 6.4\times 10^6 & 1.5\times 10^7 & 2.2\times 10^7 & 2.5\times 10^7 & 2.6\times 10^7 & 2.6\times 10^7 \\
N_{\text{band1}}=6 & 1.3\times 10^6 & 6.5\times 10^6 & 1.5\times 10^7 & 2.2\times 10^7 & 2.5\times 10^7 & 2.6\times 10^7 & 2.6\times 10^7 \\
\end{array}
\eea

\bea
\begin{array}{c|ccccccc}
\mbf{24},\nu=1/3 & N_{\text{band2}}=0 & N_{\text{band2}}=1 & N_{\text{band2}}=2 & N_{\text{band2}}=3 & N_{\text{band2}}=4 &
N_{\text{band2}}=5 & N_{\text{band2}}=6 \\
\hline
N_{\text{band1}}=0 & 3.1\times 10^4 & 3.8\times 10^5 & 1.9\times 10^6 & 5.5\times 10^6 & 1.\times 10^7 & 1.4\times 10^7 & 1.5\times 10^7 \\
N_{\text{band1}}=1 & 3.8\times 10^5 & 4.\times 10^6 & 1.7\times 10^7 & 4.2\times 10^7 & 6.9\times 10^7 & 8.4\times 10^7 & 8.9\times 10^7 \\
N_{\text{band1}}=2 & 1.9\times 10^6 & 1.7\times 10^7 & 6.4\times 10^7 & 1.4\times 10^8 & 2.\times 10^8 & 2.2\times 10^8 & 2.3\times 10^8 \\
N_{\text{band1}}=3 & 5.5\times 10^6 & 4.2\times 10^7 & 1.4\times 10^8 & 2.6\times 10^8 & 3.4\times 10^8 & 3.7\times 10^8 & 3.7\times 10^8 \\
N_{\text{band1}}=4 & 1.\times 10^7 & 6.9\times 10^7 & 2.\times 10^8 & 3.4\times 10^8 & 4.2\times 10^8 & 4.5\times 10^8 & 4.6\times 10^8 \\
N_{\text{band1}}=5 & 1.4\times 10^7 & 8.4\times 10^7 & 2.2\times 10^8 & 3.7\times 10^8 & 4.5\times 10^8 & 4.8\times 10^8 & 4.9\times 10^8 \\
N_{\text{band1}}=6 & 1.5\times 10^7 & 8.9\times 10^7 & 2.3\times 10^8 & 3.7\times 10^8 & 4.6\times 10^8 & 4.9\times 10^8 & 5.\times 10^8 \\
\end{array}
\eea

\subsection{HF Basis}
\label{eq:HFbasis}

Pentalayer graphene features a flat conduction band with $\Ch = 5$ separated by a small gap ($<$ 0.1 meV over a range of $V$) to two highly dispersive bands. The screened Coulomb interaction is expected to strongly mix these single-particle bands. At integer filling $\nu=1$, experiments observe a $\Ch = 1$ state, which is captured in HF calculations that project into the four lowest single-particle conduction bands and four highest single-particle valence bands for each spin and valley~\cite{kwan2023MFCI3}.
The single occupied HF band above the charge neutrality gap has $\Ch = 1$. We find that the $\Ch=1$ state is also reproduced in HF calculations that only project into the three lowest conduction bands in one spin-valley flavor (and hence assume that the single-particle valence bands are fully occupied in all flavors).
From \cref{fig:BS}, we clearly observe that for at least $40$\,meV around the flat band there exist only 2 other bands, and hence 3-band spin-valley-polarized HF should be physically  appropriate.

Two questions now arise. First, is HF reliable, especially for the CN scheme where it gives the AHC state as the moir\'e potential is taken to zero~\cite{2024arXiv240307873D}?  For example, it is well known that HF fails to accurately predict the Wigner crystallization transition in the homogeneous 2D spinful electron gas with parabolic dispersion and Coulomb interactions. This system is characterized by a dimensionless parameter $r_s=\frac{m_ee^2}{4\pi^{3/2}\epsilon\hbar^2\sqrt{n}}$, which gives the ratio of the interaction scale $E_U=\frac{e^2}{4\pi\epsilon}\sqrt{\pi n}$ and the kinetic energy scale $E_K=\frac{\hbar^2}{2m_e}(2\pi n)$. For sufficiently small $r_s$ (high density) the ground state is expected to be a paramagnetic Fermi liquid to minimize the kinetic energy, while for sufficiently large $r_s$ (small density) Wigner crystallization is expected to occur to reduce the repulsion energy. In HF calculations~\cite{Trail2003HF,Bernu2011HF}, the ground state is found to be a commensurate Wigner crystal for $r_s\gtrsim 1.2$, where the spatial symmetry and spin ordering of the Wigner crystal depend on $r_s$. On the other hand, quantum Monte Carlo calculations~\cite{tanatar1989_2deg,attaccalite2002_2deg,drummond2009_2deg} find that Wigner crystallization only occurs for $r_s\gtrsim 31$. We also note that several possible intermediate phases have been proposed~\cite{spivak2004phases,reza2005universal,falakshahi2005hybrid,kim2024dynamical,valenti2024nematic}. In the pentalayer setting, we can make a \emph{very} crude lower estimate of $r_s$ by first considering moir\'e-less pentalayer for $V=0$ at an electron density that corresponds to $\nu=1$ in $\theta=0.77^\circ$ R5G/hBN. Using $\epsilon=5\epsilon_0$ and estimating $E_K$ as the kinetic energy of the lowest conduction band at momentum $|\mbf{q}_1|$, we obtain $r_s\sim 2.3$. An externally applied displacement field will flatten the conduction band and increase $r_s$.
We caution that the pentalayer graphene setting differs the standard 2D electron gas in several respects, in that the dispersion at low energies is far from parabolic (and even non-monotonic for sufficiently large $V$), the Bloch wavefunctions carry non-trivial quantum geometry, there is an additional valley degree of freedom, and the Wigner crystal state can be topological with $\Ch=1$. As such, it remains to be seen how well HF captures the physics of the AHC in this platform.

Second, to study the FCIs at fractional filling $\nu<1$, is the projection onto the occupied HF band at $\nu=1$ reliable?
We find that fractionally filling the projected HF band will result in an FCI at \emph{both} $\nu=1/3$ and $2/3$ (in contrast with the experiment) since the HF band is topological, nearly flat, and possesses smooth quantum geometry.
It must be checked whether projection is reliable though, since a full 3-band calculation on 12 sites (where the Hilbert space dimension is still tractable) does not yield FCIs. We access multiband calculations for larger system sizes by developing new methods to systematically enlarge the Hilbert space based on constraining band occupations and selecting orbitals (see \cref{eq:hilbertspacedim}), and by performing rotations to more useful bases where these methods can be fruitfully applied.  The ultimate goal is to understand the physics of full 3-band ED calculations for larger system sizes.

To do so, we start by unitarily transforming the 3-band ED problem to the HF \emph{basis} to study the FCIs at $\nu = 2/3$ and below. We emphasize that this is a unitary transformation that does not affect the many-body spectrum of the 3-band interacting Hamiltonian.  We consider only a single spin-valley flavor ($(\eta,s)=(\K,\uparrow)$), and 3 active conduction bands as the minimal setting for unbiased ED. All other bands are at much higher energies (for the range of displacement field we consider) and hence are irrelevant to the physics of the FCIs and Chern insulators at $\nu\leq 1$.

Explicitly, by performing self-consistent 3-band HF calculations that preserve moir\'e translation invariance, we obtain a converged order parameter $P_{mn}(\bsl{k}) = \langle c^\dag_{K,\mbf{k}m\uparrow}c_{K,\mbf{k}n,\uparrow}  \rangle $ (expectation value taken in the HF state).
Then, the HF basis is the diagonal eigenbasis of the HF Hamiltonian
\bea
\label{eq:hartreefockapp}
H_{\text{HF}}[P] &= H_0 + H_b + H_H[P]+ H_F[P] \\
H_H[P]&= \frac{1}{N} \sum_{\mbf{k}mn} \frac{V(\mbf{G})}{\Omega} M_{mn}(\mbf{k},\mbf{G})\lp\sum_{\mbf{k}'m'n'} M^*_{m'n'}(\mbf{k}',\mbf{G}) P_{m' n'}(\mbf{k}') \rp c^\dag_{K,\mbf{k}m\uparrow}c_{K,\mbf{k}n,\uparrow} \\
H_F[P]&= - \frac{1}{N}\sum_{\mbf{k}mn} \sum_{\mbf{q} \mbf{G} m' n'}\frac{V(\mbf{q}+\mbf{G})}{\Omega} M^*_{n'm}(\mbf{k},\mbf{q}+\mbf{G}) P^*_{m' n'}(\mbf{k}+\mbf{q}) M_{m' n}(\mbf{k},\mbf{q}+\mbf{G}) c^\dag_{K,\mbf{k}m \uparrow}c_{K,\mbf{k}n,\uparrow}
\eea
where $H_0$ and $H_b$ are defined in \cref{eq:ED_H_2D_int_ave_terms} for the AVE scheme, and $H_b$ should be set to zero for the CN scheme. The eigenbasis $ \tilde{U}_{n}(\mbf{k})$ of $H_{\text{HF}}[P]$ defines the HF band operators
\bea
\label{eq:HF_basis}
H_{\text{HF}}[P] &= \sum_{\mbf{k},\alpha} \eps^{\text{HF}}_{\alpha}(\mbf{k})\gamma^\dag_{\mbf{k},\alpha} \gamma_{\mbf{k},\alpha}, \qquad  \gamma^\dag_{\mbf{k},\alpha} = \sum_{n^c}\gamma^\dag_{\K,\bsl{k}, n^c, \uparrow}  \tilde{U}_{n^c \alpha}(\mbf{k}) \ ,
\eea
where $\tilde{U}_{\alpha}(\mbf{k})$ is the $\alpha$th eigenvector of $H^{\text{HF}}[P]$ at $\mbf{k}$, with $\alpha=0,1,2$ ordered by increasing HF band energy. Then, the total many-body Hamiltonian for the three-band ED calculation is equal to the original Hamiltonian, except that we express its matrix elements in the HF basis. This is a unitary transformation within the 3-band Hilbert space. Note that we do not consider moir\'e translation symmetry-breaking and hence the HF basis is a unitary transformation of the single-particle basis \emph{at each $\mbf{k}$}.

Recall that the occupied HF band (which we label as $\alpha=0$) self-consistently reproduces the HF spectrum, and that the unoccupied HF bands capture the single-electron excitation energies via Koopman's theorem~\cite{KOOPMANS1934104}, assuming no further orbital rearrangement in the excited state. The gap between the occupied and unoccupied HF bands \emph{does not} indicate a gap in the many-body spectrum, which is instead determined by the low-lying neutral excitations at fixed particle number. Thus, a sizable charge-1 HF gap does not mean that it is a reliable approximation in ED to project to the lowest occupied HF band. In any case, the self-consistent HF solution is obtained at filling $\nu=1$, and it is not \textit{a priori} justified to rigidly empty this band to access lower fillings such as $\nu=1/3$ and $2/3$.

For ED calculations, we compute the HF Hamiltonian and HF basis  on a large mesh (of at least $12\times 12 $ momentum points) where we can check that the HF ground state has $\Ch = 1$ by computing the Berry curvature on small plaquettes~\cite{fukui2005chern}.
Specifically, we consider 6 momentum meshes in our ED calculations: $(N_x, N_y, \widetilde{n}_{11}, \widetilde{n}_{12}, \widetilde{n}_{21}, \widetilde{n}_{22}) = (2,6,1,0,1,1),(15,1,1,-5,0,1),(9,2,1,-2,0,1),(21,1,1,-5,0,1),(4,6,1,0,1,1)$, which we refer to as $12,15,18,21$, and $24$ sites in short. These meshes each have both $K_M$ and $K_M'$ points, which is crucial since these points in the moir\'e BZ are nearly 3-fold degenerate in the non-interacting band structure, and their LSM momenta~\cite{LSM,PhysRevLett.84.1535,PhysRevLett.127.237204} are all distinct so that finite size level repulsion does not inflate the spread of the FCI ground states.

\subsection{Momentum Mesh}
\label{app:mommesh}

\begin{figure}[t]
\centering
\includegraphics[width=\columnwidth]{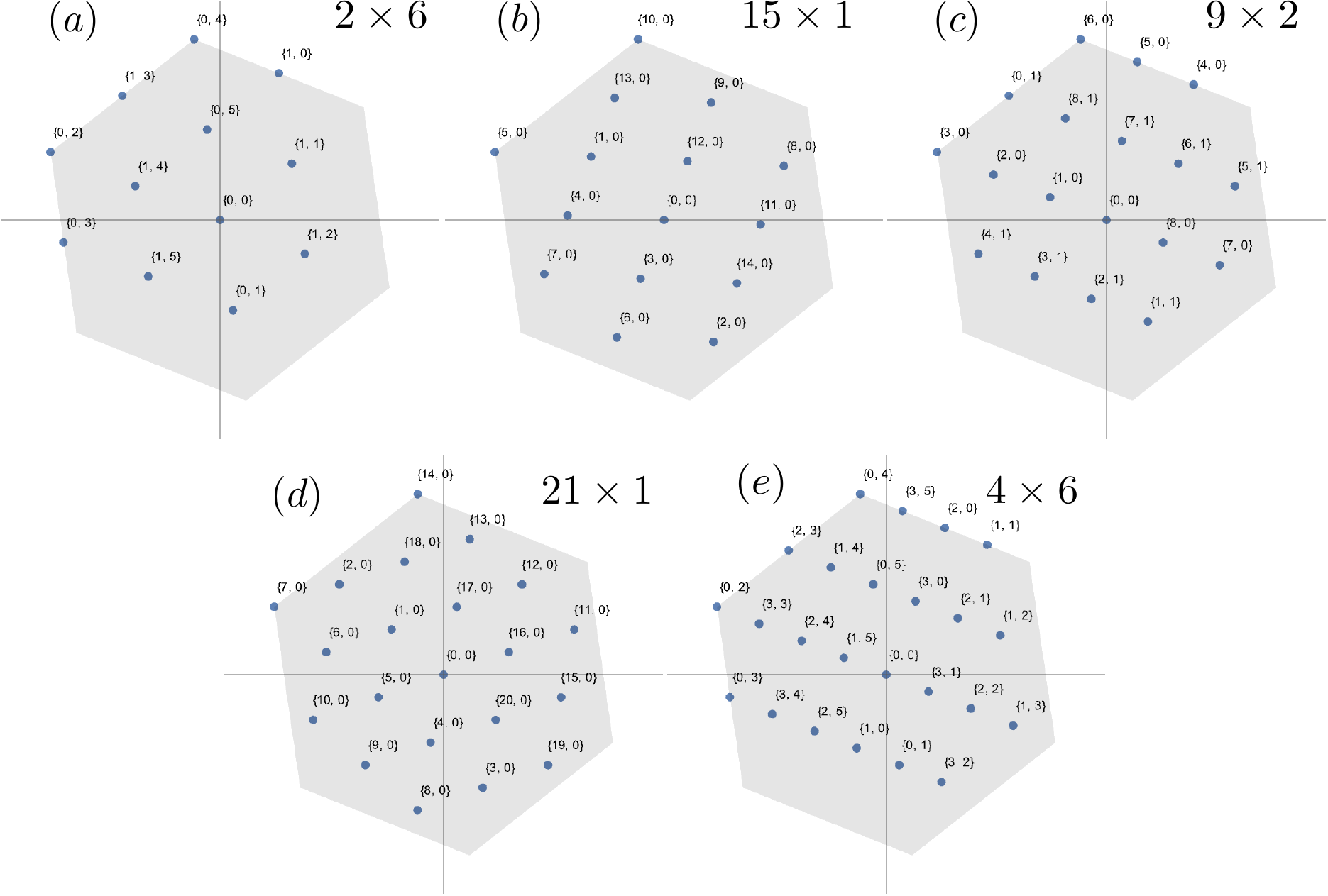}
\caption{We show the momentum meshes (dots) for $(N_x, N_y, \widetilde{n}_{11}, \widetilde{n}_{12}, \widetilde{n}_{21}, \widetilde{n}_{22}) = (2,6,1,0,1,1),(15,1,1,-5,0,1),(9,2,1,-2,0,1),(21,1,1,-5,0,1),
(4,6,1,0,1,1)$ in the moir\'e BZ (grey), according to the convention in \cref{eq:momentum_mesh}.
The corresponding values of $N_x\times N_y$ are indicated in each figure, and  $\{k_x,k_y\}$ values are labeled according to \cref{eq:momentum_mesh}.
Note that we always fold the momentum back to the moir\'e BZ.
}
\label{fig:meshes}
\end{figure}

We now specify the momentum meshes used in this work for ED calculations. They are specified by the form
\eq{
\label{eq:momentum_mesh}
\bsl{k} =\frac{k_x}{N_x} \bsl{f}_1 + \frac{k_y}{N_y} \bsl{f}_2,\qquad k_x = 0, 1, 2, ..., N_x -1, \quad k_y = 0, 1, 2, ..., N_y -1
}
where $\mbf{f}_i$ are moir\'e reciprocal lattice vectors related by $SL(2,\mathds{Z})$ transformations to $\mbf{b}_i = \mbf{q}_3 - \mbf{q}_i$. Explicitly, the $SL(2,\mathds{Z})$ transformations are specified by $\bsl{f}_1 = \widetilde{n}_{11} \bsl{b}_{M,1} + \widetilde{n}_{12} \bsl{b}_{M,2}$ and $\bsl{f}_2 = \widetilde{n}_{21} \bsl{b}_{M,1} + \widetilde{n}_{22} \bsl{b}_{M,2}$. In \cref{fig:meshes}, we show the meshes used in this work labeled by the abbreviations $N_x \times N_y$. The moir\'e lattice vectors used in each case are $\mbf{f}_1 = \mbf{b}_{M,1}, \mbf{f}_{2} = \mbf{b}_{M,1}+\mbf{b}_{M,2}$ for $2\times 6$ and $4\times 6$, $\mbf{f}_1 = \mbf{b}_{M,1} - 2\mbf{b}_{M,2}, \mbf{f}_2 = \mbf{b}_{M,2}$ for $9\times 2$, and $\mbf{f}_1 = \mbf{b}_{M,1} - 5 \mbf{b}_{M,2}, \mbf{f}_2 = \mbf{b}_{M,2}$ for $15\times1, 21 \times 1,$ and $24\times 1$.
Note that if $N_i=1$, then $\bsl{f}_i$ does not affect the momentum mesh, but we still ensure the $\bsl{f}_i$ constitute a $SL(2,\mathds{Z})$ transformation for consistency.
In calculations, we always fold the momentum $\bsl{k}$ in \cref{eq:momentum_mesh} back to the moir\'e BZ.

We choose these lattices to achieve the following properties: (1) we only consider lattices where the total number of sites is divisible by 3 in order to study $\nu=1/3$ and $\nu=2/3$, (2) we must include the $\K_M$ and $\K'_M$ points to resolve the near three-fold degeneracy of the bare bands at the moir\'e BZ corners and (3) we require the LSM momenta of the FCI states to be distinct in order to minimize the effect of level repulsion between the FCI states on small systems.

\begin{figure}[t]
\centering
\includegraphics[width=\columnwidth]{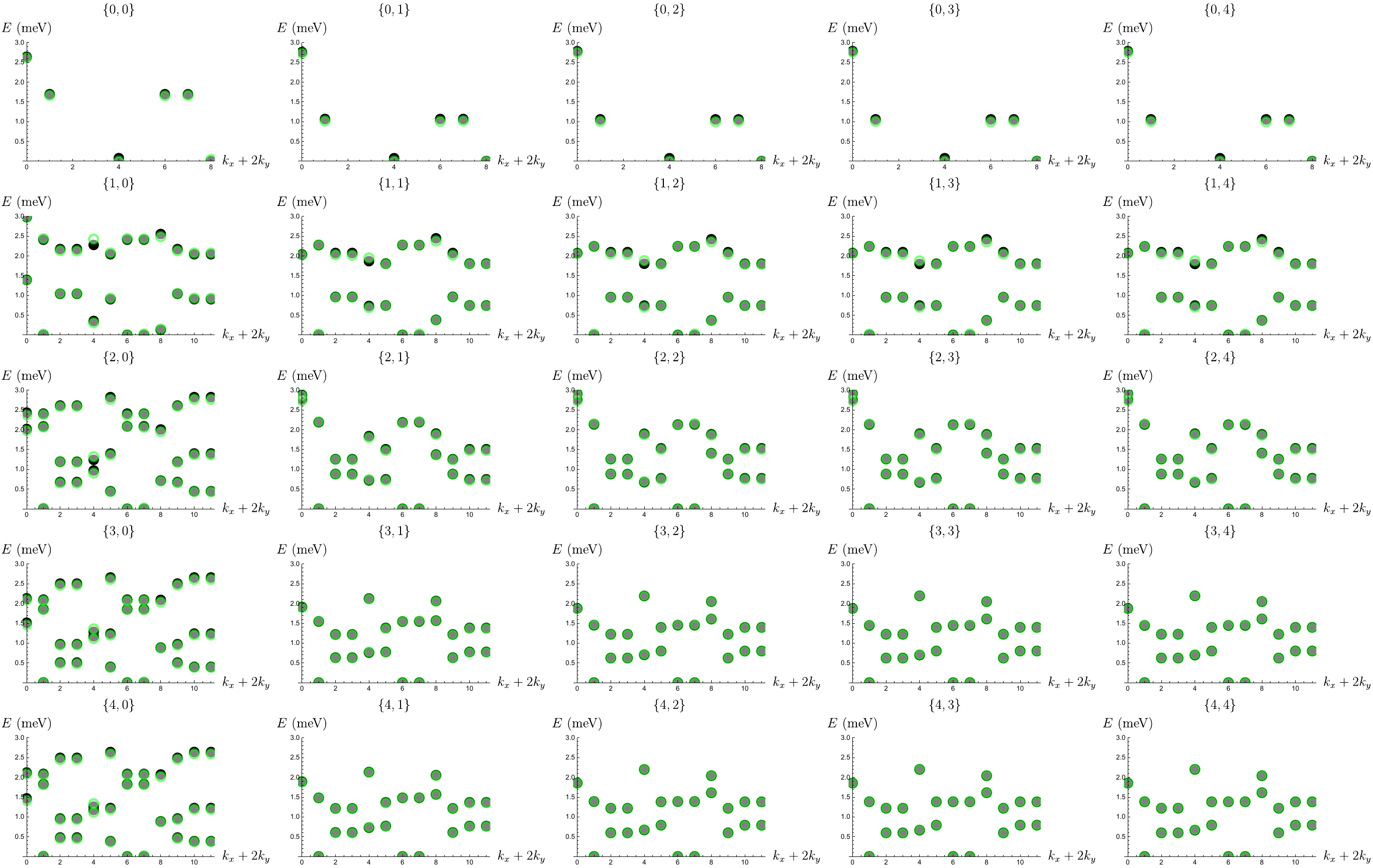}
\caption{We compare multi-band ED spectra for 2D interactions in the AVE scheme at $\nu=1/3$ and $V=22$\,meV on the $2\times 6$ lattice using the HF basis computed on a $6\times 6$ mesh (green) and $12\times12$ mesh (black). All meshes use the same lattice vectors. Very good agreement is obtained.
}
\label{fig:check61213}
\end{figure}

\begin{figure}[t]
\centering
\includegraphics[width=\columnwidth]{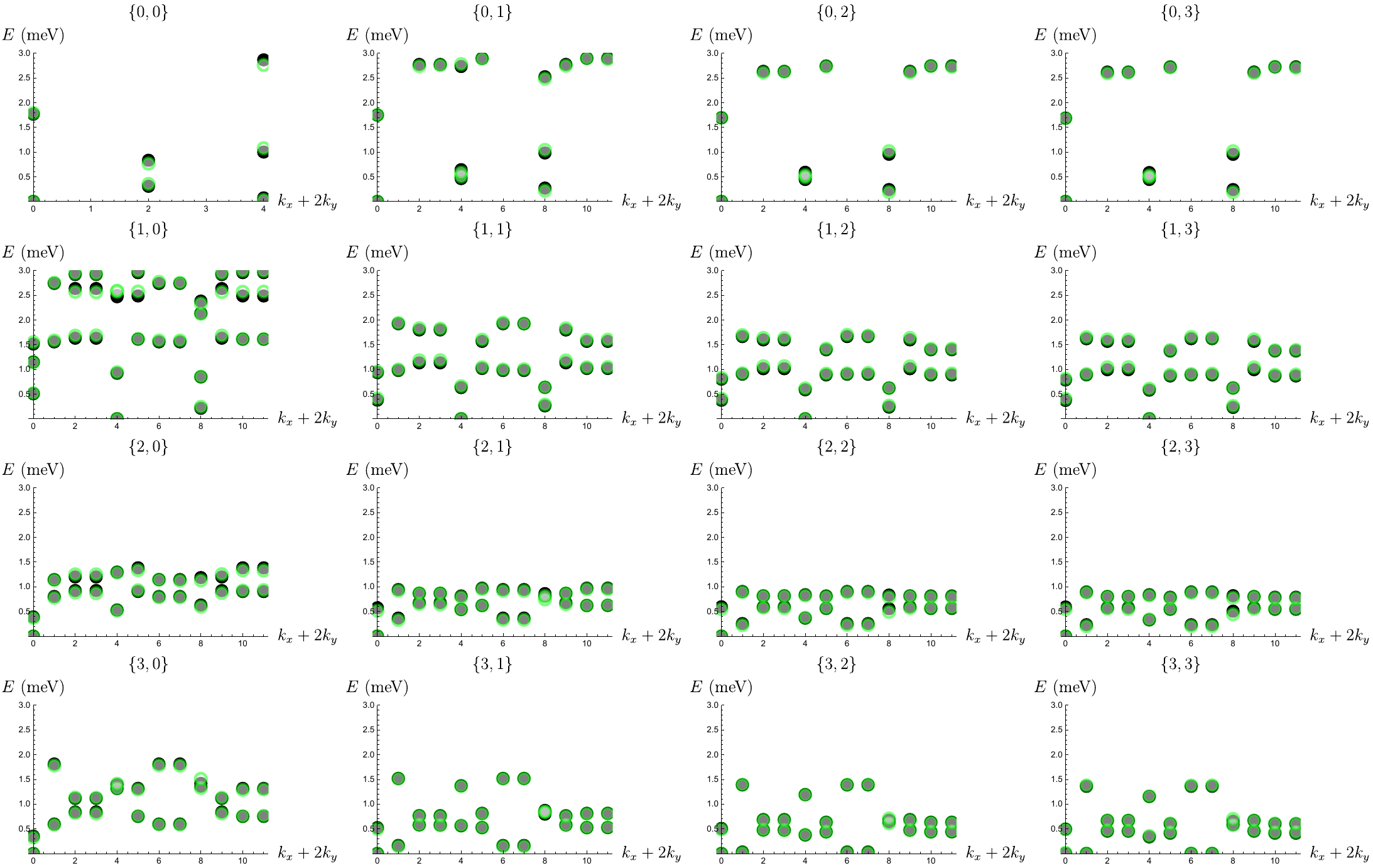}
\caption{We compare multi-band ED spectra for 2D interactions in the AVE scheme at $\nu=2/3$ and $V=22$\,meV on the $2\times 6$ lattice using the HF basis computed on a $6\times 6$ mesh (green) and $12\times12$ mesh (black). All meshes use the same lattice vectors. Very good agreement is obtained.
}
\label{fig:check61223}
\end{figure}

In our HF calculations, we always use a finer momentum mesh (which contains the corresponding ED mesh as a subset) to compute the HF Hamiltonian and $\tilde{U}_n(\mbf{k})$ in \cref{eq:HF_basis} in order to reduce finite size effects. Since we do not allow moir\'e translation symmetry-breaking in these HF calculations, $\tilde{U}_n(\mbf{k})$ is a strictly unitary transformation at each $\mbf{k}$ regardless of the whether the finer mesh is used. For the $2\times 6$ and $4\times 6$ lattices, we compute the HF Hamiltonian on a $12\times12$ mesh with the same lattice vectors. On the $15\times 1, 21 \times 1,$ and $24 \times 1$ meshes, we compute the HF Hamiltonian on $15\times 15, 21 \times 21,$ and $24 \times 24$ meshes, respectively, with the same lattice vectors. \Fig{fig:check61213} and \Fig{fig:check61223} compares ED spectra on a $2\times 6$ lattice using matrix elements in the AVE scheme in the HF basis computed using finer $6\times 6$ and $12 \times 12$ meshes. We find that with an HF convergence tolerance of $10^{-6}$\,meV for the total energy, the kinetic energy matrix elements in the HF basis have relative error $\sim10^{-5}$ between different seeds yielding the same Chern number.

\subsection{Normal Ordering and Equivalence of the HF basis}
\label{app:HFhamiltonianbasis}

In our HF-basis ED calculations, the many-body Hamiltonian expressed in the HF basis takes the form
\eq{
\label{eq:H_HF_basis}
H = \sum_{\bsl{k},\alpha\beta}\gamma^\dagger_{\bsl{k},\alpha} \gamma_{\bsl{k},\beta} t_{\alpha\beta}(\bsl{k}) + \frac{1}{2}\sum_{\bsl{k}\bsl{k}'\bsl{q}} \sum_{\alpha\beta\gamma\delta} V_{\alpha\beta\gamma\delta}(\bsl{k},\bsl{k}',\bsl{q}) \gamma^\dagger_{\bsl{k}+\bsl{q},\alpha} \gamma^\dagger_{\bsl{k}'-\bsl{q},\beta} \gamma_{\bsl{k}',\gamma} \gamma_{\bsl{k},\delta}\ ,
}
where $\alpha=0,1,2$ orders the HF bands by increasing HF band energy. Importantly, $t(\bsl{k})$ is the bare kinetic term $t_{\alpha\beta}(\bsl{k}) = \tilde{U}_{\alpha}^\dagger(\bsl{k}) h_{0}(\bsl{k}) \tilde{U}_{\beta}(\bsl{k})$  in the CN scheme, or the sum of the bare kinetic term and background term $t_{\alpha\beta}(\bsl{k}) = \tilde{U}_{\alpha}^\dagger(\bsl{k})\left[h_{0}(\bsl{k})+ h^{H}_b(\bsl{k})  + h^{F}_b(\bsl{k}) \right]\tilde{U}_{\beta}(\bsl{k})$ in the AVE scheme. $\tilde{U}_{\alpha}(\bsl{k})$ is defined in \cref{eq:HF_basis} and $h_{0}(\bsl{k})$, $h^{H}_b(\bsl{k})$ and $h^{F}_b(\bsl{k})$ are introduced in \cref{eq:ED_H_2D_int_ave}). $V_{\alpha\beta\gamma\delta}(\bsl{k},\bsl{k}',\bsl{q})$ is the bare interaction which satisfies
\eq{
\label{eq:double_exchange}
V_{\alpha\beta\gamma\delta}(\bsl{k},\bsl{k}',\bsl{q}) = V_{\beta\alpha\delta\gamma}(\bsl{k}',\bsl{k},-\bsl{q})\ .
}
\cref{eq:H_HF_basis} is just a unitary transformation of the Hamiltonian in \cref{eq:ED_H_2D_int_CN} for the CN scheme or in \cref{eq:ED_H_2D_int_ave} for the AVE scheme.
When performing the calculation, we will vary the maximum number of allowed particles in band 1 ($\gamma_{\bsl{k},1}^\dagger$) and band 2 ($\gamma_{\bsl{k},2}^\dagger$), which we refer to as band-max $\{N_{\text{band1}},N_{\text{band2}}\}$ (see \cref{eq:hilbertspacedim}). Our ED calculations are performed at $\nu=1/3,2/3$ and $1$.

Refs.~\cite{zhou2023fractional,dong2023anomalous,dong2023theory,guo2023theory} use a different procedure to perform ED for $\nu<1$.
Explicitly, they perform a HF calculation at $\nu=1$, and obtain the lowest occupied HF conduction band $\epsilon_0^{\text{HF}}(\bsl{k})$ and its eigenstates, created by $\gamma^\dagger_{\bsl{k},0}$ in our notation.
Then, they consider the following Hamiltonian
\eq{
\label{eq:H_HFB}
H_{\text{HFB}} =- \sum_{\bsl{k}} \epsilon_{\bsl{k},0}^{\text{HF}} \gamma_{\bsl{k},0} \gamma_{\bsl{k},0}^\dagger + \frac{1}{2}\sum_{\bsl{k}\bsl{k}'\bsl{q}}  V_{0000}(\bsl{k},\bsl{k}',\bsl{q}) \gamma_{\bsl{k},0} \gamma_{\bsl{k}',0} \gamma^\dagger_{\bsl{k}'-\bsl{q},0} \gamma^\dagger_{\bsl{k}+\bsl{q},0}\ ,
}
where HFB stands for ``HF-band projected", which is the name we will use for the method employed in Refs.~\cite{zhou2023fractional,dong2023anomalous,dong2023theory,guo2023theory}.
In our calculation, we can also choose $N_{\text{band1}}=N_{\text{band2}}=0$, thereby also concentrating only on the lowest HF band.
Then, a natural question arises: what is the connection between our HF basis calculation with $N_{\text{band1}}=N_{\text{band2}}=0$, and the HFB ED calculations of Refs.~\cite{zhou2023fractional,dong2023anomalous,dong2023theory,guo2023theory} using \cref{eq:H_HFB}?
In this section, we will answer this question in two different cases, depending on a subtlety for the HF calculation.

We will first show that if the HF calculation (both for defining our HF basis, and in the HFB method) is done on the same momentum mesh as the ED calculation, then our HF basis calculation with $N_{\text{band1}}=N_{\text{band2}}=0$ is \emph{equivalent} to the HFB ED calculation.
For the HF basis calculation with $N_{\text{band1}}$=$N_{\text{band2}}$=0, we are effectively considering the following Hamiltonian
\eq{
\label{eq:H_HF_basis_0_0}
H_{(0,0)} = \sum_{\bsl{k}}\gamma^\dagger_{\bsl{k},0} \gamma_{\bsl{k},0} t_{00}(\bsl{k}) + \frac{1}{2}\sum_{\bsl{k}\bsl{k}'\bsl{q}}  V_{0000}(\bsl{k},\bsl{k}',\bsl{q}) \gamma^\dagger_{\bsl{k}+\bsl{q},0} \gamma^\dagger_{\bsl{k}'-\bsl{q},0} \gamma_{\bsl{k}',0} \gamma_{\bsl{k},0}\ .
}
To compare \cref{eq:H_HF_basis_0_0} to the Hamiltonian used in HFB calculations, we need to normal-order it with respect to the $\nu=1$ HF ground state corresponding to filling all $\gamma^\dagger_{\bsl{k},0}$.
After some algebra, we obtain
\eqa{
H_{(0,0)} & = \sum_{\bsl{k}}t_{00}(\bsl{k})  - \sum_{\bsl{k}} \gamma_{\bsl{k},0} \gamma^\dagger_{\bsl{k},0} t_{00}(\bsl{k}) - \frac{1}{2}\sum_{\bsl{k}\bsl{q}}  V_{0000}(\bsl{k},\bsl{k}+\bsl{q},\bsl{q}) \gamma^\dagger_{\bsl{k}+\bsl{q},0} \gamma_{\bsl{k}+\bsl{q},0}  + \frac{1}{2}\sum_{\bsl{k}\bsl{k}'}  V_{0000}(\bsl{k},\bsl{k}',0)  \gamma^\dagger_{\bsl{k}',0} \gamma_{\bsl{k}',0} \\
& \qquad - \frac{1}{2}\sum_{\bsl{k}\bsl{k}'\bsl{q}}  V_{0000}(\bsl{k},\bsl{k}',0) \gamma_{\bsl{k},0} \gamma^\dagger_{\bsl{k},0} + \frac{1}{2}\sum_{\bsl{k}\bsl{q}}  V_{0000}(\bsl{k},\bsl{k}+\bsl{q},\bsl{q}) \gamma_{\bsl{k},0}  \gamma^\dagger_{\bsl{k},0} + \frac{1}{2}\sum_{\bsl{k}\bsl{k}'\bsl{q}}  V_{0000}(\bsl{k},\bsl{k}',\bsl{q}) \gamma_{\bsl{k},0} \gamma_{\bsl{k}',0} \gamma^\dagger_{\bsl{k}'-\bsl{q},0} \gamma^\dagger_{\bsl{k}+\bsl{q},0}\\
& = \sum_{\bsl{k}}t_{00}(\bsl{k})  - \frac{1}{2}\sum_{\bsl{k}\bsl{q}}  V_{0000}(\bsl{k}-\bsl{q},\bsl{k},\bsl{q}) + \frac{1}{2}\sum_{\bsl{k}\bsl{k}'}  V_{0000}(\bsl{k},\bsl{k}',0)   - \sum_{\bsl{k}} \gamma_{\bsl{k},0} \gamma^\dagger_{\bsl{k},0} t_{00}(\bsl{k})  - \frac{1}{2}\sum_{\bsl{k}\bsl{k}'}  V_{0000}(\bsl{k}',\bsl{k},0)  \gamma_{\bsl{k},0} \gamma^\dagger_{\bsl{k},0}   \\
& \qquad + \frac{1}{2}\sum_{\bsl{k}\bsl{q}}  V_{0000}(\bsl{k}-\bsl{q},\bsl{k},\bsl{q})  \gamma_{\bsl{k},0} \gamma^\dagger_{\bsl{k},0}  - \frac{1}{2}\sum_{\bsl{k}\bsl{k}'\bsl{q}}  V_{0000}(\bsl{k},\bsl{k}',0) \gamma_{\bsl{k},0} \gamma^\dagger_{\bsl{k},0} + \frac{1}{2}\sum_{\bsl{k}\bsl{q}}  V_{0000}(\bsl{k},\bsl{k}+\bsl{q},\bsl{q}) \gamma_{\bsl{k},0}  \gamma^\dagger_{\bsl{k},0} \\
& \qquad  + \frac{1}{2}\sum_{\bsl{k}\bsl{k}'\bsl{q}}  V_{0000}(\bsl{k},\bsl{k}',\bsl{q}) \gamma_{\bsl{k},0} \gamma_{\bsl{k}',0} \gamma^\dagger_{\bsl{k}'-\bsl{q},0} \gamma^\dagger_{\bsl{k}+\bsl{q},0}\\
& = E_0 - \sum_{\bsl{k}} \epsilon_{\bsl{k}} \gamma_{\bsl{k},0} \gamma_{\bsl{k},0}^\dagger + \frac{1}{2}\sum_{\bsl{k}\bsl{k}'\bsl{q}}  V_{0000}(\bsl{k},\bsl{k}',\bsl{q}) \gamma_{\bsl{k},0} \gamma_{\bsl{k}',0} \gamma^\dagger_{\bsl{k}'-\bsl{q},0} \gamma^\dagger_{\bsl{k}+\bsl{q},0}
}
where
\eq{
\label{eq:E_0}
E_0 = \sum_{\bsl{k}}t_{00}(\bsl{k})  - \frac{1}{2}\sum_{\bsl{k}\bsl{q}}  V_{0000}(\bsl{k}-\bsl{q},\bsl{k},\bsl{q}) + \frac{1}{2}\sum_{\bsl{k}\bsl{k}'}  V_{0000}(\bsl{k},\bsl{k}',0) \ ,
}
\eqa{
\label{eq:epsilon_k}
\epsilon_{\bsl{k}} & =  t_{00}(\bsl{k})  + \frac{1}{2}\sum_{\bsl{k}'}  V_{0000}(\bsl{k}',\bsl{k},0)  - \frac{1}{2}\sum_{\bsl{q}}  V_{0000}(\bsl{k}-\bsl{q},\bsl{k},\bsl{q}) + \frac{1}{2}\sum_{\bsl{k}}  V_{0000}(\bsl{k},\bsl{k}',0) - \frac{1}{2}\sum_{\bsl{q}}  V_{0000}(\bsl{k},\bsl{k}+\bsl{q},\bsl{q})  \\
& =  t_{00}(\bsl{k})  + \sum_{\bsl{k}'}  V_{0000}(\bsl{k},\bsl{k}',0) -  \sum_{\bsl{q}}  V_{0000}(\bsl{k}-\bsl{q},\bsl{k},\bsl{q})  \ ,
}
and we have used \eqnref{eq:double_exchange}.

\begin{figure}
\centering
\includegraphics[width=0.6\columnwidth]{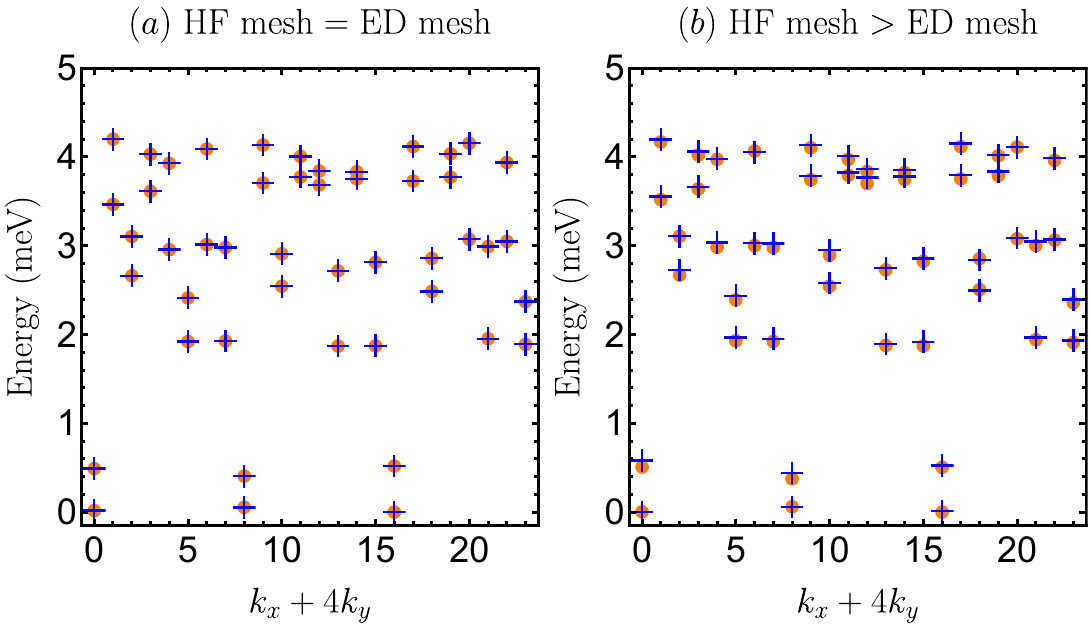}
\caption{Comparison between HF-basis ED calculation with $N_{\text{band1}}=N_{\text{band2}}=0$ in \cref{eq:H_HF_basis_0_0} (orange dots) and the HFB method of \cref{eq:H_HFB} (blue crosses) proposed in Refs.~\cite{zhou2023fractional,dong2023anomalous,dong2023theory,guo2023theory}.
All the ED calculations are done for $4\times 6$ momentum mesh with 16 electrons (or equivalently 8 holes in the occupied HF band) for the 2D interaction in the AVE scheme at $V=22$\,meV.
In (a), the HF calculation is done on the same mesh as the ED calculation, and we have carefully chosen the initial state such that the converged final HF state has $\Ch=1$.
In (b), the HF calculation is done on a much larger momentum mesh $(N_x, N_y, \widetilde{n}_{11}, \widetilde{n}_{12}, \widetilde{n}_{21}, \widetilde{n}_{22}) = (12,12,1,0,0,1)$ (the converged HF state has $\Ch=1$), and extract the corresponding HF states (and also energies for the HFB method) on the $4\times 6$ momentum mesh to perform ED calculation.
}
\label{fig:HFbasis00_HFband_projected_comparison_plot}
\end{figure}

Now we show that $ \epsilon_{\bsl{k}}$ is the HF band energy $\epsilon_{0\mbf{k}}^{\text{HF}}$ of the lowest HF band.
To see that, first note that in the HF basis, the order parameter has the form
\eq{
\langle \gamma_{\bsl{k},\alpha}^\dagger  \gamma_{\bsl{k}',\alpha'} \rangle = \delta_{\alpha,0} \delta_{\alpha',0} \delta_{\bsl{k},\bsl{k}'}\ ,
}
as the HF state only occupies $\gamma^\dagger_{\bsl{k},0}$.
As a result, the HF Hamiltonian reads
\eqa{
\label{eq:HHF_HF_basis}
H_{\text{HF}} & = \sum_{\bsl{k},\alpha\beta}\gamma^\dagger_{\bsl{k},\alpha} \gamma_{\bsl{k},\beta} t_{\alpha\beta}(\bsl{k}) +\sum_{\bsl{k}\bsl{k}'\bsl{q}} \sum_{\alpha\beta\gamma\delta} V_{\alpha\beta\gamma\delta}(\bsl{k},\bsl{k}',\bsl{q}) \gamma^\dagger_{\bsl{k}+\bsl{q},\alpha} \langle \gamma^\dagger_{\bsl{k}'-\bsl{q},\beta} \gamma_{\bsl{k}',\gamma} \rangle \gamma_{\bsl{k},\delta}\\
&\quad - \sum_{\bsl{k}\bsl{k}'\bsl{q}} \sum_{\alpha\beta\gamma\delta} V_{\alpha\beta\gamma\delta}(\bsl{k},\bsl{k}',\bsl{q}) \gamma^\dagger_{\bsl{k}+\bsl{q},\alpha} \langle \gamma^\dagger_{\bsl{k}'-\bsl{q},\beta} \gamma_{\bsl{k},\delta} \rangle \gamma_{\bsl{k}',\gamma} \\
& = \sum_{\bsl{k},\alpha\beta}\gamma^\dagger_{\bsl{k},\alpha} \gamma_{\bsl{k},\beta} t_{\alpha\beta}(\bsl{k}) +\sum_{\bsl{k}\bsl{k}'} \sum_{\alpha\delta} V_{\alpha00\delta}(\bsl{k},\bsl{k}',0) \gamma^\dagger_{\bsl{k},\alpha} \gamma_{\bsl{k},\delta} - \sum_{\bsl{k}\bsl{q}} \sum_{\alpha\gamma} V_{\alpha0\gamma0}(\bsl{k},\bsl{k}+\bsl{q},\bsl{q}) \gamma^\dagger_{\bsl{k}+\bsl{q},\alpha}  \gamma_{\bsl{k}+\bsl{q},\gamma} \\
& = \sum_{\bsl{k},\alpha\beta}\gamma^\dagger_{\bsl{k},\alpha} \gamma_{\bsl{k},\beta} \left[ t_{\alpha\beta}(\bsl{k}) +\sum_{\bsl{k}'}  V_{\alpha00\beta}(\bsl{k},\bsl{k}',0) - \sum_{\bsl{q}} V_{\alpha 0\beta 0}(\bsl{k}-\bsl{q},\bsl{k},\bsl{q})\right]  \\
& = \sum_{\bsl{k},\alpha\beta}\gamma^\dagger_{\bsl{k},\alpha} \gamma_{\bsl{k},\beta} \left[ h_{\text{HF}}(\bsl{k})\right]_{\alpha\beta}\ ,
}
where
\eq{
\left[ h_{\text{HF}}(\bsl{k})\right]_{\alpha\beta} =  t_{\alpha\beta}(\bsl{k}) +\sum_{\bsl{k}'}  V_{\alpha00\beta}(\bsl{k},\bsl{k}',0) - \sum_{\bsl{q}} V_{\alpha 0\beta 0}(\bsl{k}-\bsl{q},\bsl{k},\bsl{q}) \ .
}
As the HF basis diagonlizes the HF Hamiltonian, we have
\eq{
\left[ h_{\text{HF}}(\bsl{k})\right]_{\alpha\beta} = \delta_{\alpha\beta} \epsilon^{\text{HF}}_{\alpha}(\bsl{k}) \ ,
}
leading to
\eq{
t_{\alpha\beta}(\bsl{k}) +\sum_{\bsl{k}'}  V_{\alpha00\beta}(\bsl{k},\bsl{k}',0) - \sum_{\bsl{q}} V_{\alpha 0\beta 0}(\bsl{k}-\bsl{q},\bsl{k},\bsl{q}) = \delta_{\alpha\beta} \epsilon^{\text{HF}}_{\alpha}(\bsl{k}) \ ,
}
where $\epsilon^{\text{HF}}_{\alpha}(\bsl{k})$ is the HF band energy.
Combined with \cref{eq:epsilon_k}, we deduce that $\epsilon(\bsl{k}) =\epsilon^{\text{HF}}_{0}(\bsl{k}) $, which is the lowest HF band.
Similarly, the total HF energy reads
\bea
E_0^{\text{HF}} & = \sum_{\bsl{k}}\gamma^\dagger_{\bsl{k},0} \gamma_{\bsl{k},0} t_{00}(\bsl{k}) + \frac{1}{2}\sum_{\bsl{k}\bsl{k}'\bsl{q}}  V_{0000}(\bsl{k},\bsl{k}',\bsl{q}) \langle \gamma^\dagger_{\bsl{k}+\bsl{q},0}  \gamma_{\bsl{k},0} \rangle \langle \gamma^\dagger_{\bsl{k}'-\bsl{q},0} \gamma_{\bsl{k}',0}\rangle + \frac{1}{2}\sum_{\bsl{k}\bsl{k}'\bsl{q}}  V_{0000}(\bsl{k},\bsl{k}',\bsl{q}) \langle \gamma^\dagger_{\bsl{k}+\bsl{q},0} \gamma_{\bsl{k}',0} \rangle \langle \gamma^\dagger_{\bsl{k}'-\bsl{q},0}  \gamma_{\bsl{k},0} \rangle \\
& = \sum_{\bsl{k}}t_{00}(\bsl{k})  - \frac{1}{2}\sum_{\bsl{k}\bsl{q}}  V_{0000}(\bsl{k}-\bsl{q},\bsl{k},\bsl{q}) + \frac{1}{2}\sum_{\bsl{k}\bsl{k}'}  V_{0000}(\bsl{k},\bsl{k}',0)\ ,
\eea
which means $ E_0 $ in \cref{eq:E_0} is the total HF energy.
Therefore, we arrive at
\eq{
\label{eq:H_HF_basis_0_0_rewriten}
H_{(0,0)} =  E_0^{\text{HF}} - \sum_{\bsl{k}} \epsilon_{\bsl{k},0}^{\text{HF}} \gamma_{\bsl{k},0} \gamma_{\bsl{k},0}^\dagger + \frac{1}{2}\sum_{\bsl{k}\bsl{k}'\bsl{q}}  V_{0000}(\bsl{k},\bsl{k}',\bsl{q}) \gamma_{\bsl{k},0} \gamma_{\bsl{k}',0} \gamma^\dagger_{\bsl{k}'-\bsl{q},0} \gamma^\dagger_{\bsl{k}+\bsl{q},0}\ ,
}
which is exactly the HFB Hamiltonian in \cref{eq:H_HFB} up to the constant shift.
As a numerical verification, we show the ED spectrum with \cref{eq:H_HF_basis_0_0} (dots) and with \cref{eq:H_HFB} (crosses) in \cref{fig:HFbasis00_HFband_projected_comparison_plot}a for $\nu=2/3$ on 24 sites and the 2D interaction in the AVE scheme.

However in practice, for the ED calculation with HF basis, we perform HF on a larger mesh compared to the ED calculations, since $\nu=1$ HF on small sizes can easily converge to the $\Ch=0$ state unless the initial states are carefully chosen in both CN and AVE schemes. Refs.~\cite{zhou2023fractional,dong2023anomalous,dong2023theory,guo2023theory} also used a larger mesh for the HF calculation for their HFB ED calculations.
This subtlety does not significantly affect the comparison between the HF basis ED calculation with $N_{\text{band1}}=N_{\text{band2}}=0$, and the HFB ED calculation. As shown in  \cref{fig:HFbasis00_HFband_projected_comparison_plot}b where the HF calculations are performed on a larger $12\times 12$ lattice, the spectrum HF-basis ED calculation with $N_{\text{band1}}=N_{\text{band2}}=0$ is very close to that of the ED calculation performed using the HFB method proposed in Refs.~\cite{zhou2023fractional,dong2023anomalous,dong2023theory,guo2023theory}.

\subsubsection{General discussion on the relation between the HF basis and the interaction normal-ordered against the HF state}

In this part, we will present a general discussion on the relation between the HF basis and the HFB method.
We only consider the case where HF is done on the same mesh as the ED calculation.
We first specify a completely general Hamiltonian with two-body interactions
\begin{equation}\label{eqapp:H_originalbasis}
\hat{H}=\sum_{ij}\tilde{t}_{ij}\tilde{f}^\dagger_i \tilde{f}_j + \frac{1}{2}\sum_{ijkl}\tilde{V}_{ij,kl}\tilde{f}^\dagger_i \tilde{f}^\dagger_j \tilde{f}_k \tilde{f}_l
\end{equation}
where $i$ is a composite index that could include any quantum numbers such as momentum, band, flavor, etc. Due to fermion antisymmetry, the interaction matrix element satisfies
\begin{equation}
\tilde{V}_{ij,kl}=\tilde{V}_{ji,lk}.
\end{equation}

The HF approximation of restricting the manifold of many-body states to single Slater determinants, which take the general form
\eq{
\ket{\text{HF}} = f^\dagger_1 f^\dagger_2 f^\dagger_3 ...  f^\dagger_N \ket{0}\ ,
}
where $N$ is the particle number, and
\begin{equation}
f^\dagger_A=\sum_{i}v_{A,i}\tilde{f}^\dagger_{i}
\end{equation}
where the $v_{A}$'s are orthonormal vectors.
$\ket{\text{HF}}$ can be uniquely determined by the order parameter
\eq{
\label{eq:defineing_P}
P_{i j} = \bra{\text{HF}} \widetilde{f}^\dagger_{i} \widetilde{f}_{j} \ket{\text{HF}} = \sum_{A=1}^N v_{A,i}^* v_{A,j},
} and thus we usually use $P$ to represent the Slater determinant.
We define the HF Hamiltonian for $P$ as
\eqa{
\label{eq:H_HF_gen}
H_{\text{HF}}[P] & = \sum_{i,j}\widetilde{f}^\dagger_i \widetilde{f}_j t_{ij} + \sum_{ijkl} \tilde{V}_{ij,kl} \left( \widetilde{f}^\dagger_{i} \widetilde{f}_{l} P_{jk} -  \widetilde{f}^\dagger_{j} \widetilde{f}_{l} P_{ik}  \right)- E_{0}\\
&  = \sum_{i,j}\widetilde{f}^\dagger_i \widetilde{f}_j \left(h_{\text{HF}}[P]\right)_{ij}- E_{0}
}
where
\eq{
\left( h_{\text{HF}}[P]\right)_{ij} = t_{ij} + \sum_{kl} (\tilde{V}_{il,kj} -\tilde{V}_{il ,jk})P_{lk}
}
and
\eq{
E_0 = \frac{1}{2}\sum_{i,j,k,l} \tilde{V}_{i j, k l} (P_{i l}P_{j k}-P_{i k}P_{j l})\ .
}
Defined this way, $H_{\text{HF}}$ satisfies $\bra{\text{HF}}H_{\text{HF}}\ket{\text{HF}}=\bra{\text{HF}}\hat{H}\ket{\text{HF}}$.

In practice, we perform self-consistent HF calculations to converge on the order parameter $P$.
The converged order parameter $P$ satisfies the condition that its $N$ vectors $v_A$'s in \eqnref{eq:defineing_P} are the $N$ energetically-lowest eigenvectors of $h_{\text{HF}}[P]$.
These $N$ vectors (corresponding to the  $N$ lowest energy HF orbitals) are occupied.
Then, it is convenient to work using the eigenbasis of $h_{\text{HF}}[P]$, \ie, the HF basis.
Explicitly, we introduce an indexing notation for the eigenvectors of $h_{\text{HF}}[P]$, where uppercase Roman letters ($A,B,C,\ldots$) denote occupied HF orbitals (contained in $\ket{\text{HF}}$), and starting lowercase Roman letters ($a,b,c,\ldots$) denote unoccupied HF orbitals. Greek letters ($\alpha,\beta,\gamma,\ldots$) index all the HF orbitals, independent of whether they are occupied or unoccupied.
Then,
$\hat{H}$ can be exactly rewritten in the HF basis
\begin{gather}\label{eqapp:H_HFbasisstarting}
\hat{H}=\sum_{\alpha\beta}t_{\alpha\beta}f^\dagger_\alpha f_\beta+\frac{1}{2}\sum_{\alpha\beta\gamma\delta}V_{\alpha\beta,\gamma\delta}{f}^\dagger_\alpha {f}^\dagger_\beta {f}_\gamma {f}_\delta\equiv \hat{H}^{(1)}+\hat{H}^{(2)}\\
t_{\alpha\beta}=\sum_{ij}v^*_{\alpha,i}\tilde{t}_{ij}v_{\beta,j}\\
V_{\alpha\beta,\gamma\delta}=\sum_{ijkl}v^*_{\alpha,i}v^*_{\beta,j}\tilde{V}_{ij,kl}v_{\gamma,k}v_{\delta,l}=V_{\beta\alpha,\delta\gamma}
\end{gather}
where $\hat{H}^{(1)}$ ($\hat{H}^{(2)}$) collects all the one-body (two-body) terms.
Furthermore, the converged condition for the order parameter $P$ leads to
\begin{equation}\label{eqapp:HFselfconsistency}
t_{\alpha\beta}+\sum_{A}(V_{\alpha A,A \beta}-V_{\alpha A,\beta A})=E_\alpha \delta_{\alpha\beta}
\end{equation}

Our goal is to show that $\hat{H}$ can be exactly rewritten as
\begin{gather}\label{eqapp:H_HFbasis}
\hat{H}=E^\text{HF}_0+\sum_{\alpha}E_{\alpha}::f^\dagger_{\alpha}f_\alpha::+\frac{1}{2}\sum_{\alpha\beta\gamma\delta}V_{\alpha\beta,\gamma\delta}::{f}^\dagger_\alpha {f}^\dagger_\beta {f}_\gamma {f}_\delta::\ ,
\end{gather}
where
\begin{equation}\label{eqapp:E0HF}
E_0^\text{HF}=\sum_A t_{AA}+\frac{1}{2}\sum_{AB}(V_{AB,BA}-V_{AB,AB})\ .
\end{equation}
is the total HF energy of $\ket{\text{HF}}$.
We have introduced a special normal-ordering notation $::\hat{O}::$ in \cref{eqapp:H_HFbasis}, which means that all $f^\dagger_{a}$ and $f_A$ should be ordered to the left of all $f_a$ and $f^\dagger_{A}$ while keeping track of minus signs from anti-commuting.
As a result, $\bra{\text{HF}}::\hat{O}::\ket{\text{HF}}$ vanishes unless $\hat{O}$ is a number.

We begin with the one-body term $\hat{H}^{(1)}$ of \cref{eqapp:H_HFbasisstarting}:
\begin{align}
\hat{H}^{(1)}&=\sum_{AB}t_{AB}f^\dagger_{A}f_B+
\sum_{Ab}t_{Ab}f^\dagger_{A}f_b+\sum_{aB}t_{aB}f^\dagger_{a}f_B+\sum_{ab}t_{ab}f^\dagger_{a}f_b\\
&=\sum_{A}t_{AA}-\sum_{AB}t_{AB}f_B f^\dagger_A+\sum_{Ab}t_{Ab}f^\dagger_{A}f_b-\sum_{aB}t_{aB}f_Bf^\dagger_a +\sum_{ab}t_{ab}f^\dagger_a f_b\\
&=\sum_{A}t_{AA}+\sum_{\alpha\beta}t_{\alpha\beta}::f^\dagger_\alpha f_\beta::\ .\label{eqapp:H1final}
\end{align}

We now tackle the two-body term $\hat{H}^{(2)}$ of \cref{eqapp:H_originalbasis}. For notational simplicity, we will represent the creation/annihilation operators by just the HF orbital index, represent the interaction matrix element using $[\alpha\beta\gamma\delta]$, and use Einstein summation convention. For example, we make the following replacement of notation
\begin{equation}\label{eqapp:ABcD}
\frac{1}{2}\sum_{ABcD}V_{AB,cD}f^\dagger_A f^\dagger_B f_c f_D \rightarrow \frac{1}{2}[ABcD]A^\dagger B^\dagger cD \equiv \left\{ ooeo\right\}\ ,
\end{equation}
where $\left\{ ooeo\right\}$ means the first, second and fourth indexes are occupied (o for occupied), while the third is unoccupied (e for empty).
If we consider $\frac{1}{2}[\alpha\beta\gamma\delta]\alpha^\dagger\beta^\dagger\gamma\delta$, there are 16 different terms depending on whether $\alpha$ is restricted to occupied $A$ or unoccupied $a$ HF orbitals, and so on for $\beta,\delta,\gamma$.
We write down all 16 terms in $\hat{H}^{(2)}$ and anticommute the constituent operators until the four-fermion term is normal-ordered with respect to $\ket{\text{HF}}$, as in \cref{eqapp:H_HFbasis}:
\begin{align}
\{eeee\}& = \frac{1}{2}[abcd]a^\dagger b^\dagger cd = ::\{eeee\}::\\
\{eeeo\}& = - \frac{1}{2}[abcD]a^\dagger b^\dagger D c = ::\{eeeo\}::\\
\{eeoe\}& = \frac{1}{2}[abCd]b^\dagger a^\dagger
Cd = :: \{eeoe\} :: \\
\{eeoo\} & = \frac{1}{2}[abCD]a^\dagger b^\dagger CD = :: \{eeoo\} :: \\
\{eoee\} & = \frac{1}{2}[aBcd]a^\dagger B^\dagger cd = ::\{eoee\}::\\
\{eoeo\} & = \frac{1}{2}[aBcD](-\delta_{BD}a^\dagger c+a^\dagger DcB^\dagger ) = -\frac{1}{2}[aBcB]a^\dagger c + ::\{eoee\}::\\
\{eooe\}&= \frac{1}{2}[aBCd](\delta_{BC}a^\dagger d-a^\dagger CB^\dagger d) = \frac{1}{2}[aBBd]a^\dagger d + :: \{eooe\} ::  \\
\{eooo\} & = \frac{1}{2}[aBCD](-\delta_{BD}a^\dagger C+\delta_{BC}a^\dagger D+a^\dagger C DB^\dagger ) = \frac{1}{2}(-[aBCB]a^\dagger C+[aBBD] a^\dagger D) + :: \{eooo\} :: \\
\{oeee\}& = \frac{1}{2}[Abcd] A^\dagger b^\dagger cd = ::   \{oeee\} ::\\
\{oeeo\}& = \frac{1}{2}[AbcD](\delta_{AD}b^\dagger c + D b^\dagger A^\dagger c) = \frac{1}{2}[AbcA]b^\dagger c  + ::  \{oeeo\} :: \\
\{oeoe\} &= \frac{1}{2}[AbCd](-\delta_{AC}b^\dagger d- Cb^\dagger A^\dagger d) = \frac{1}{2}(-[AbAd]  b^\dagger d)  + :: \{oeoe\} :: \\
\{oeoo\} & = \frac{1}{2}[AbCD](\delta_{AD}b^\dagger C - \delta_{AC}b^\dagger D-b^\dagger C D A^\dagger ) = \frac{1}{2}([AbCA] b^\dagger C - [AbAD]b^\dagger D ) + :: \{oeoo\} :: \\
\{ooee\}&= \frac{1}{2}[ABcd]A^\dagger B^\dagger  c d = :: \{ooee\} ::  \\
\{ooeo\}& = \frac{1}{2}[ABcD](-\delta_{BD}A^\dagger c+\delta_{AD}B^\dagger c-DA^\dagger B^\dagger c)  = \frac{1}{2}(-[ABcB]A^\dagger c+[ABcA]B^\dagger c) + ::  \{ooeo\} :: \\
\{oooe\}& = \frac{1}{2}[ABCd](\delta_{BC}A^\dagger d-\delta_{AC}B^\dagger d- CA^\dagger B^\dagger d ) =\frac{1}{2}([ACCd]A^\dagger d-[CBCd]B^\dagger d) + :: \{oooe\} ::\\
\{oooo\}& = \frac{1}{2}[ABCD](-\delta_{BD}\delta_{AC}+\delta_{BD}CA^\dagger-\delta_{AD}CB^\dagger +\delta_{AD}\delta_{BC} -\delta_{BC}DA^\dagger +\delta_{AC}DB^\dagger +DC B^\dagger A^\dagger  ) \\
& = \frac{1}{2}(-[ABAB]+[ABCB]CA^\dagger-[ABCA] CB^\dagger +[ABBA] -[ABBD] DA^\dagger +[ABAD] DB^\dagger  )  + :: \{oooo\} ::\ ,
\end{align}
where $\delta_{AB}$ is the Kronecker delta function.
Combining all terms, we obtain the following rewriting for the two-body term $\hat{H}^{(2)}$ of \cref{eqapp:H_HFbasisstarting}:
\eqa{
\label{eqapp:H2final}
\hat{H}^{(2)}  &= -\frac{1}{2}[aBcB]a^\dagger c +  \frac{1}{2}[aBBd]a^\dagger d + \frac{1}{2}(-[AbAd]  b^\dagger d) + \frac{1}{2}[AbcA]b^\dagger c + \frac{1}{2}(-[aBCB]a^\dagger C+[aBBD] a^\dagger D) \\
&  + \frac{1}{2}([AbCA] b^\dagger C - [AbAD]b^\dagger D )
+ \frac{1}{2}(-[ABcB]A^\dagger c+[ABcA]B^\dagger c) + \frac{1}{2}([ACCd]A^\dagger d-[CBCd]B^\dagger d \\
& + \frac{1}{2}(-[ABAB] +[ABBA]  ) + \frac{1}{2}( [ABCB]CA^\dagger - [ABCA] CB^\dagger  -[ABBD] DA^\dagger +[ABAD] DB^\dagger  ) \\
& + ::\frac{1}{2}[\alpha\beta\gamma\delta]\alpha^\dagger\beta^\dagger\gamma\delta  :: \\
& = ([a AA b] - [A a A b]) a^\dagger b + ([a AA B] - [a A B A]) a^\dagger B + ([B AA a ] + [B A a A]) B^\dagger a \\
& - ([B C C A] - [B C A C]) A B^\dagger + \frac{1}{2}(-[ABAB] +[ABBA]  ) + ::\frac{1}{2}[\alpha\beta\gamma\delta]\alpha^\dagger\beta^\dagger\gamma\delta ::\ .
}
Before combining the one-body terms (\cref{eqapp:H1final}) with the two-body terms (\cref{eqapp:H2final}), we note that
\eq{
\sum_{A}t_{AA} + \frac{1}{2}(-[ABAB] +[ABBA]  ) = \sum_{A} t_{AA} + \sum_{A,B}(V_{AB,BA} - V_{AB,AB}) = E_{0}^{\text{HF}}\ ,
}
\eq{
\sum_{ab}t_{ab}f^\dagger_a f_b + ([a AA b] - [A a A b]) a^\dagger b  = \sum_{ab}\left\{ t_{ab} + \sum_{A} (V_{a A,A b} - V_{A a, A b}) \right\} f^\dagger_a f_b = \sum_a E_a f^\dagger_a f_a\ ,
}
\eq{
\sum_{aB}t_{aB}f^\dagger_a f_B + ([a AA B] - [a A B A]) a^\dagger B  = \sum_{aB}\left\{ t_{aB} + \sum_{A} (V_{a A,A B} - V_{a A, B A}) \right\} f^\dagger_a f_B = 0\ ,
}
\eq{
\sum_{Ba}t_{Ba}f^\dagger_B f_a + ([B AA a ] + [B A a A]) B^\dagger a  = \sum_{Ba}\left\{ t_{Ba} + \sum_{A} (V_{B A,A a} - V_{B A, a A}) \right\} f^\dagger_B f_a = 0\ ,
}
and
\eq{
- \sum_{AB}t_{BA} f_A f^\dagger_B- ([B C C A] - [B C A C]) A B^\dagger  = -\sum_{AB}\left\{ t_{BA} + \sum_{C} (V_{B C,C B} - V_{B C, A C}) \right\} f_A f^\dagger_B = - \sum_{A} E_{A} f_A f^\dagger_A\ ,
}
where we have used \cref{eqapp:HFselfconsistency}.
As a result, combining the one-body terms (\cref{eqapp:H1final}) with the two-body terms (\cref{eqapp:H2final}) leads to
\eq{
\hat{H} = E_{0}^{\text{HF}}  + \sum_a E_a f^\dagger_a f_a - \sum_{A} E_{A} f_A f^\dagger_A +\frac{1}{2}\sum_{\alpha\beta\gamma\delta}V_{\alpha\beta,\gamma\delta}::{f}^\dagger_\alpha {f}^\dagger_\beta {f}_\gamma {f}_\delta:: \ ,
}
proving \cref{eqapp:H_HFbasis}.

We can project \cref{eqapp:H_HFbasis} to the many-body Hilbert space generated by $f^\dagger_A$ only, resulting in
\eq{
\bar{H} = E_{0}^{\text{HF}}   - \sum_{A} E_{A} f_A f^\dagger_A +\frac{1}{2}\sum_{\alpha\beta\gamma\delta}V_{AB,CD} {{f}_D f}_C  {f}^\dagger_B  {f}^\dagger_A\ .
}
Combined with \cref{eqapp:H_HFbasisstarting}, we arrive at
\eqa{
\sum_{AB}t_{AB}f^\dagger_A f_B+\frac{1}{2}\sum_{ABCD}V_{AB,CD}{f}^\dagger_A {f}^\dagger_B {f}_C {f}_D
& = E_{0}^{\text{HF}}   - \sum_{A} E_{A} f_A f^\dagger_B +\frac{1}{2}\sum_{\alpha\beta\gamma\delta}V_{AB,CD} {{f}_D f}_C  {f}^\dagger_B  {f}^\dagger_A\ .
}
This means that if we project the many-body Hamiltonian to the Hilbert space generated by the occupied converged HF orbitals, it is the same as the Hamiltonian, restricted to the HF orbitals, with HF dispersion and bare interaction normal-ordered with respect to the HF state.

Returning to pentalayer graphene, we therefore conclude that the HF-basis Hamiltonian \cref{eq:H_HF_basis} is equivalent to
\bea
\label{eq:HnormalorderHF_app}
H &= E_0^{\text{HF}} + \sum_{\mbf{k},\alpha} \eps^{\text{HF}}_\alpha(\mbf{k}) ::\gamma^\dag_{\mbf{k},\alpha} \gamma_{\mbf{k},\alpha}:: + :: H_{\text{int}} ::
\eea
where $E_0^{\text{HF}}$ is the energy of the self-consistent HF ground state and $\eps^{\text{HF}}_\alpha(\mbf{k})$, $\alpha=0,1,2\ldots $ is the dispersion of the self-consistent HF Hamiltonian. Crucially, in \Eq{eq:HnormalorderHF}, the double colons signify normal ordering against the \emph{$\nu=1$ HF ground state}, e.g.~the electron operator $\gamma^\dag_{\mbf{k},0}$ and hole operators $\gamma_{\mbf{k},1}, \gamma_{\mbf{k},2}$ are placed to the right (keeping track of the fermionic signs).
If $\alpha$ is restricted to 0 in all terms, then \cref{eq:HnormalorderHF_app} corresponds to the HF-band-projected (HFB) method used in Refs.~\cite{zhou2023fractional,dong2023anomalous,dong2023theory,guo2023theory} (if the HF calculation were performed on the same mesh as the ED calculation, see discussion at the end of \App{app:HFhamiltonianbasis}). We will perform multi-band ED calculations based on 3-band and 5-band HF calculations. Keeping 3 active bands in ED within the 5-band HF Hilbert space can be thought of as doing a 5-band ED calculation with band-max always set to zero in bands 3 and 4 (and nonzero in bands 0,1,2).

\subsection{Biasing Method in One-Body Diagonal Basis}
\label{app:biasing_method}
\begin{figure}[t]
\centering
\includegraphics[height=0.4\columnwidth]{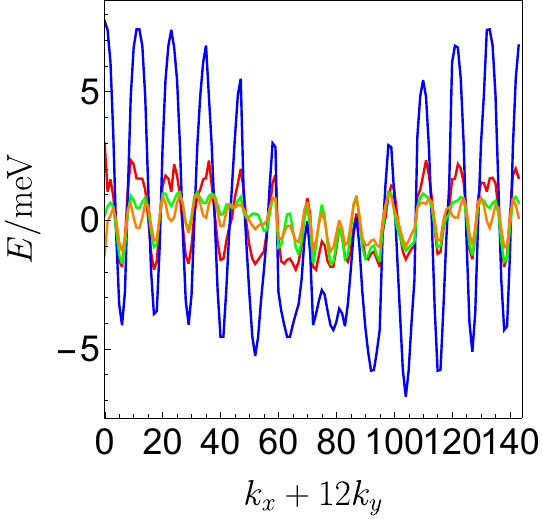}
\caption{ Plot of the different one-body  energies for  system size $12\times 12$, 2D interaction, and $V=22$meV. The momentum mesh is $(N_x, N_y, \widetilde{n}_{11}, \widetilde{n}_{12}, \widetilde{n}_{21}, \widetilde{n}_{22}) = (12,12,1,0,0,1)$ according to \cref{eq:momentum_mesh}. The red, blue, green and orange lines correspond to the (A) lowest HF band at $\nu=1$ with density matrix $P_{\bsl{k}}$ in the AVE scheme, (B) $\Tr[ \left(h_0(\bsl{k}) + h^H(\bsl{k}) + h^F(\bsl{k})\right) P_{\bsl{k}}^*]$, (C) $\Tr[ \left(h_0(\bsl{k}) + h^H(\bsl{k})\right) P_{\bsl{k}}^*]$, and (D) $\Tr[ h_0(\bsl{k}) P_{\bsl{k}}^*]$, respectively. For all plots, we set the mean value of each line to zero in order to compare their dispersions.
}
\label{fig:dispersions}
\end{figure}

In this part, we introduce another multi-band ED method.
We first introduce the method for the AVE scheme, and then extend it to the CN scheme.
In this method, we add an extra artificial one-body term to the Hamiltonian, which is
\eq{
\label{eq:HF_bias}
h_{\text{bias},1}(\bsl{k}) =  E_{\text{bias}} P_{\bsl{k}}^* \ ,
}
where $[P_{\bsl{k}}]_{mn} = \langle \text{HF}|\gamma^\dag_{\mbf{k},m}\gamma_{\mbf{k},n}|\text{HF}\rangle$ is the density matrix for the lowest HF band at $\nu=1$ (see \Eq{eq:hartreefockapp}), and we always choose $E_{\text{bias}} < 0$.
Since we are interested in the parameter region where the HF calculation gives a $\Ch=1$ ground state at $\nu=1$, \cref{eq:HF_bias} will bias the system towards a $\Ch=1$ ground state at $\nu=1$ for $E_{\text{bias}} < 0$.
For $E_{\text{bias}}\rightarrow - \infty$, ED calculations at $\nu\leq 1$ with the extra term in \cref{eq:HF_bias} are equivalent to a 1-band ED calculation within the 1-band Hilbert space specified by $P_{\bsl{k}}^*$.
In other words, in this extreme case, the Hamiltonian effectively reduces to the $N_{\text{band1}}=N_{\text{band2}}=0$ limit of  \cref{eq:H_HF_basis_0_0} up to an overall shift due to $E_\text{bias}$.
We emphasize that the one-body dispersion $t_{00}(\mbf{k})$ in
\cref{eq:H_HF_basis_0_0} is $\Tr[ \left(h_0(\bsl{k}) + h_b^H(\bsl{k}) + h_b^F(\bsl{k})\right) P_{\bsl{k}}^*]$, which is not the same as the HF dispersion of the  lowest $\nu=1$ HF band.
As discussed in \cref{eq:ED_H_2D_int_ave}, $h_0(\bsl{k})$ is the bare kinetic term, and $h_b^H(\bsl{k})$ and $h_b^F(\bsl{k})$ are the Hartree and Fock background terms that result from normal ordering the average scheme interaction.
In fact, the difference is dramatic---the HF dispersion of the lowest $\nu=1$ HF band is much flatter than the effective one-body dispersion for $E_{\text{bias}}\rightarrow - \infty$, as exemplified by the red and blue lines in \cref{fig:dispersions}.

The large effective one-body dispersion can prevent the stabilization of FCIs at fractional fillings for small-size ED calculations.
To address this issue, we introduce a second extra artificial one-body term
\eq{
\label{eq:Fock_bias}
h_{\text{bias},2}(\bsl{k}) =  -(1-\lambda_{\text{Fock}}) h_b^F(\bsl{k}) \ ,
}
where $\lambda_{\text{Fock}} \in [0,1]$.
The motivation for introducing this term is the following.
We observe that the effective one-body dispersion for $E_{\text{bias}}\rightarrow - \infty$ can be dramatically decreased if we drop the Fock background term $h_b^F(\bsl{k})$, as exemplified by the green line in \cref{fig:dispersions}.
Therefore, the effect of \cref{eq:Fock_bias} is to suppress the Fock background term.
In total, the one-body term used in the biased ED calculations reads
\eq{
\label{eq:h_one-body_bias}
h_{\text{one-body}}(\bsl{k}) = h_0(\bsl{k}) + h_b^H(\bsl{k}) + h_b^F(\bsl{k}) + h_{\text{bias},1}(\bsl{k}) + h_{\text{bias},2}(\bsl{k}) = h_0(\bsl{k}) + h_b^H(\bsl{k}) + \lambda_{\text{Fock}} h_b^F(\bsl{k}) + E_{\text{bias}} P_{\bsl{k}}^*\ ,
}
where the unbiased limit is $(E_{\text{bias}},\lambda_{\text{Fock}})=(0,1)$.
The interaction term is still the normal-ordered interaction in \cref{eq:ED_H_2D_int_ave}. Our ED calculations will aim to find the size dependence on the bias, and extrapolate to the non-biased case.

With the Hamiltonian specified, we now discuss the basis used in the calculation.
The choice of basis is  crucial as the computational complexity of ED forces us to truncate the Hilbert space as we increase the system size.
For this biasing method, we work in the diagonal basis of the one-body term \cref{eq:h_one-body_bias}.
We will always keep all orbitals and allow unrestricted particle occupation of the lowest band of \cref{eq:h_one-body_bias} (band 0), while truncating the higher two bands (band 1 and band 2).
The specific truncation of bands 1 and 2 (in terms of particle number and orbitals) will be specified in the discussion of the calculations in \App{app:result_Biasing_2D_ave}.

The same biasing method can also be performed for the CN scheme.
In the CN scheme, we will also introduce the bias term in \cref{eq:HF_bias}.
As shown by the orange line in \figref{fig:dispersions}, the effective one-body dispersion for $E_{\text{bias}}\rightarrow - \infty$ is quite flat in the CN scheme, which means we do not need to introduce any additional biasing terms other than \cref{eq:HF_bias}.
Therefore, the total one-body term for in the CN scheme is
\eq{
\label{eq:h_one-body_bias_CN}
h_{\text{one-body}}(\bsl{k}) = h_0(\bsl{k})  + h_{\text{bias},1}(\bsl{k}) = h_0(\bsl{k}) + E_{\text{bias}} P_{\bsl{k}}^*\ ,
}
where $E_{\text{bias}} = 0 $ is the unbiased limit.
The interaction term is still the normal-ordered interaction, and we will still work in the diagonal basis of the one-body term \cref{eq:h_one-body_bias_CN} to perform further truncation of the Hilbert space.

We will only show results for the biasing method in the AVE scheme in \App{app:result_Biasing_2D_ave}.

\newpage
\clearpage

\section{2D Interaction: Translationally-Invariant Self-Consistent Hartree-Fock Calculations}
\label{app:2D_Int_HF}

\begin{figure}[ht!]
\centering\includegraphics[width=\columnwidth]{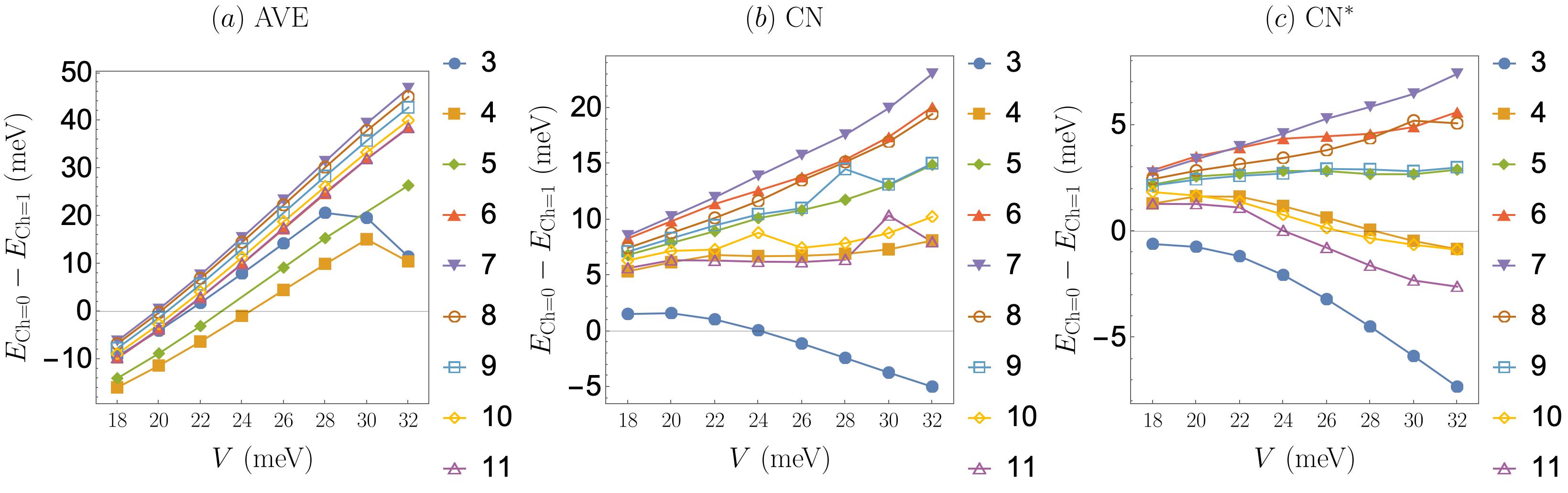}
\caption{Summary of $\nu=1$ HF results with the 2D interaction on the momentum mesh $(N_x, N_y, \widetilde{n}_{11}, \widetilde{n}_{12}, \widetilde{n}_{21}, \widetilde{n}_{22}) = (12,12,1,0,0,1)$  for the (a) AVE, (b) CN, and (c) CN$^*$ schemes.
The CN$^*$ scheme refers to the moir\'e-less limit of the CN scheme.
The number labels the number of lowest conduction bands included in the HF calculations, and we plot the total energy difference between the lowest $\Ch=0$ state and the lowest $\Ch=1$ state.
Moir\'e translation invariance is enforced in the HF calculation.
}
\label{fig:HFSchemes}
\end{figure}

\begin{figure}[ht!]
\centering
\includegraphics[width=0.9\columnwidth]{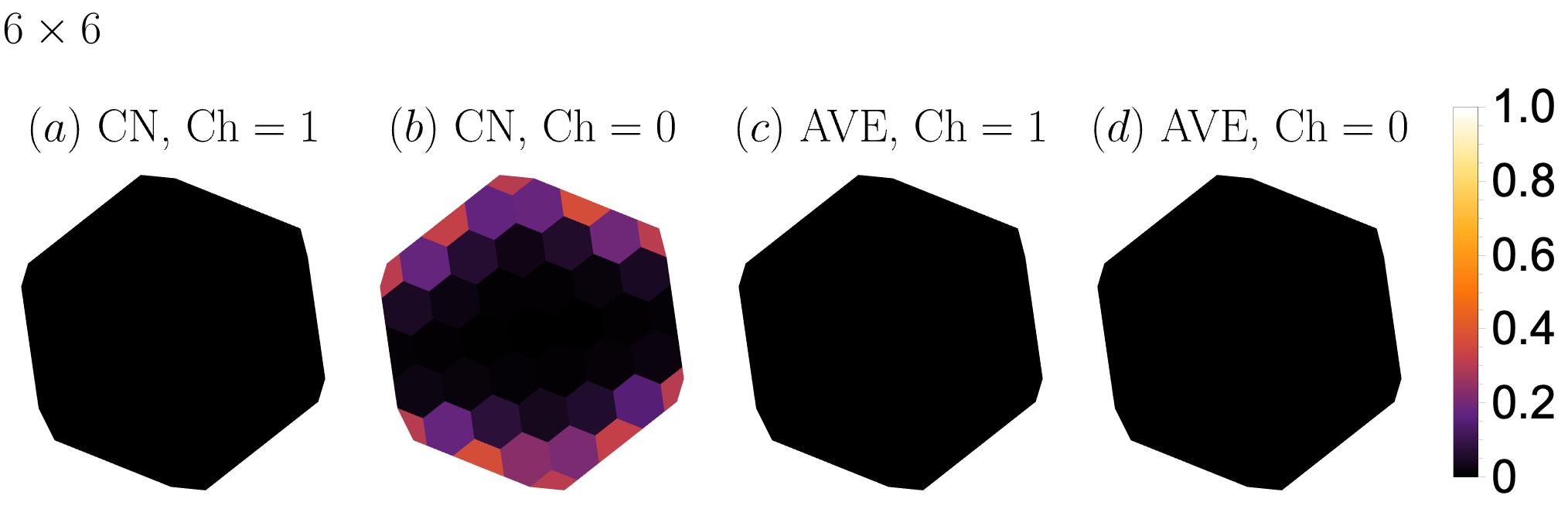}
\caption{Spontaneous $C_3$ breaking in $\nu=1$ HF at $V=22$\,meV for (a) the $\Ch=1$ state in CN scheme, (b) the $\Ch=0$  state in CN scheme, (c)  the $\Ch=1$ state in AVE scheme, and (d) the $\Ch=0$ state in AVE scheme.
The momentum mesh is $(N_x, N_y, \widetilde{n}_{11}, \widetilde{n}_{12}, \widetilde{n}_{21}, \widetilde{n}_{22}) =$ $(6,6,1,0,0,1)$.
In each plot, the color indicates the value of $\lambda_{C_3,\bsl{k}} = 1-|\bra{\text{HF},C_3\bsl{k}} C_3\ket{\text{HF},\bsl{k}}|$, where $\ket{\text{HF},\bsl{k}}$ is HF state.
$C_3$ invariant states have $\lambda_{C_3,\bsl{k}}=0$ for all $\bsl{k}$, while non-zero $\lambda_{C_3,\bsl{k}}$ indicates spontaneous breaking of $C_3$ symmetry.
}
\label{fig:plotC3breaking6x6}
\end{figure}

\begin{table}[ht!]
\centering
\begin{tabular}{|c|c|c|c||c|c|c|c|}
$(N_x, N_y, \widetilde{n}_{11}, \widetilde{n}_{12}, \widetilde{n}_{21}, \widetilde{n}_{22})$ & $E_\text{Ch=0 mod 3}$ & $E_\text{Ch=1 mod 3}$ & $E_\text{Ch=2 mod 3}$ &  $(N_x, N_y, \widetilde{n}_{11}, \widetilde{n}_{12}, \widetilde{n}_{21}, \widetilde{n}_{22})$ & $E_\text{Ch=0 mod 3}$ & $E_\text{Ch=1 mod 3}$ & $E_\text{Ch=2 mod 3}$ \\
\hline
\text{(2,6,1,0,1,1)} & 1547.26 & 1547.89 & \text{N/A} & \text{(78,1,1,10,0,1)} & 10072.86 & 10072.58 & \text{N/A} \\
\text{(15,1,1,-5,0,1)} & 1933.40 & \text{N/A} & \text{N/A} & \text{(81,1,1,10,0,1)} & 10460.33 & 10460.11 & \text{N/A} \\
\text{(9,2,1,-2,0,1)} & 2323.18 & 2323.52 & 2323.09 & \text{(9,9,1,0,0,1)} & 10460.46 & 10460.02 & \text{N/A} \\
\text{(3,6,1,0,0,1)} & 2320.24 & \text{N/A} & \text{N/A} & \text{(84,1,1,10,0,1)} & 10847.96 & 10847.66 & \text{N/A} \\
\text{(21,1,1,-5,0,1)} & 2711.35 & 2711.57 & \text{N/A} & \text{(87,1,1,10,0,1)} & 11235.17 & 11234.87 & \text{N/A} \\
\text{(4,6,1,0,1,1)} & 3097.80 & \text{N/A} & \text{N/A} & \text{(90,1,1,10,0,1)} & 11622.79 & 11622.27 & \text{N/A} \\
\text{(9,3,1,-2,0,1)} & 3486.91 & 3487.11 & \text{N/A} & \text{(93,1,1,10,0,1)} & 12010.01 & 12009.75 & \text{N/A} \\
\text{(3,9,1,0,0,1)} & 3480.39 & \text{N/A} & \text{N/A} & \text{(96,1,1,10,0,1)} & 12397.53 & 12397.14 & \text{N/A} \\
\text{(5,6,0,-1,-1,-1)} & 3873.92 & 3874.05 & \text{N/A} & \text{(99,1,1,10,0,1)} & 12785.04 & 12784.53 & \text{N/A} \\
\text{(33,1,-4,-1,-1,0)} & 4261.31 & 4261.39 & \text{N/A} & \text{(102,1,1,10,0,1)} & 13172.33 & 13171.96 & \text{N/A} \\
\text{(6,6,1,0,0,1)} & 4649.12 & 4649.22 & \text{N/A} & \text{(105,1,1,10,0,1)} & 13559.86 & 13559.39 & \text{N/A} \\
\text{(39,1,1,10,0,1)} & 5036.05 & 5036.08 & \text{N/A} & \text{(108,1,1,10,0,1)} & 13947.11 & 13946.65 & \text{N/A} \\
\text{(42,1,1,10,0,1)} & 5421.25 & 5421.66 & \text{N/A} & \text{(9,12,1,0,0,1)} & 13947.34 & 13946.63 & \text{N/A} \\
\text{(45,1,1,10,0,1)} & 5810.82 & 5810.91 & \text{N/A} & \text{(111,1,1,10,0,1)} & 14334.53 & 14334.12 & \text{N/A} \\
\text{(48,1,1,10,0,1)} & 6198.82 & 6198.70 & \text{N/A} & \text{(114,1,1,10,0,1)} & 14721.90 & 14721.71 & \text{N/A} \\
\text{(51,1,1,10,0,1)} & 6585.58 & 6585.66 & \text{N/A} & \text{(117,1,1,10,0,1)} & 15109.11 & 15109.32 & \text{N/A} \\
\text{(54,1,1,10,0,1)} & 6973.11 & 6973.07 & \text{N/A} & \text{(120,1,1,10,0,1)} & 15496.90 & 15496.35 & \text{N/A} \\
\text{(6,9,1,0,0,1)} & 6973.67 & 6973.54 & \text{N/A} & \text{(123,1,1,10,0,1)} & 15884.32 & 15883.86 & \text{N/A} \\
\text{(57,1,1,10,0,1)} & 7360.91 & 7361.04 & \text{N/A} & \text{(126,1,1,10,0,1)} & 16271.73 & 16271.58 & \text{N/A} \\
\text{(60,1,1,10,0,1)} & 7748.59 & 7748.38 & \text{N/A} & \text{(129,1,1,10,0,1)} & 16659.16 & 16658.67 & \text{N/A} \\
\text{(63,1,1,10,0,1)} & 8135.72 & 8136.17 & \text{N/A} & \text{(132,1,1,10,0,1)} & 17046.57 & 17045.99 & \text{N/A} \\
\text{(66,1,1,10,0,1)} & 8523.49 & 8523.18 & \text{N/A} & \text{(135,1,1,10,0,1)} & 17434.07 & 17433.49 & \text{N/A} \\
\text{(69,1,1,10,0,1)} & 8910.60 & 8910.55 & \text{N/A} & \text{(138,1,1,10,0,1)} & 17821.57 & \text{N/A} & \text{N/A} \\
\text{(72,1,1,10,0,1)} & 9298.14 & 9297.91 & \text{N/A} & \text{(141,1,1,10,0,1)} & 18208.77 & 18208.21 & \text{N/A} \\
\text{(6,12,1,0,0,1)} & 9298.34 & 9297.90 & \text{N/A} & \text{(12,12,1,0,0,1)} & 18596.31 & 18595.27 & \text{N/A} \\
\text{(75,1,1,10,0,1)} & 9685.60 & 9685.32 & \text{N/A} & & & & \\
\end{tabular}
\caption{
The summary of HF results for the 2D interaction in the CN scheme at $V=22$meV.
The first column shows the momentum mesh.
The second, third and fourth columns show the HF energies (in meV) of the lowest state with $\Ch=0 \mod 3$, $\Ch=1 \mod 3$ and $\Ch=2 \mod 3$, respectively.
``N/A" means we have not obtained such states in the HF calculations.
For $N_x N_y \geq 36$, the Chern number is determined to $\Ch=0$ and $\Ch=1$ for $\Ch=0 \mod 3$ and $\Ch=1 \mod 3$ states by the integration of the Berry curvature.
}
\label{tab:CN_HF_results}
\end{table}

\begin{table}[ht!]
\centering
\begin{tabular}{|c|c|c|c||c|c|c|c|}
$(N_x, N_y, \widetilde{n}_{11}, \widetilde{n}_{12}, \widetilde{n}_{21}, \widetilde{n}_{22})$ & $E_\text{Ch=0 mod 3}$ & $E_\text{Ch=1 mod 3}$ & $E_\text{Ch=2 mod 3}$ & $(N_x, N_y, \widetilde{n}_{11}, \widetilde{n}_{12}, \widetilde{n}_{21}, \widetilde{n}_{22})$ & $E_\text{Ch=0 mod 3}$ & $E_\text{Ch=1 mod 3}$ & $E_\text{Ch=2 mod 3}$ \\
\hline
\text{(2,6,1,0,1,1)} & 2042.3 & 2042.77 & \text{N/A} & \text{(78,1,1,10,0,1)} & 13269.5 & 13269.5 & \text{N/A} \\
\text{(15,1,1,-5,0,1)} & 2552.19 & 2552.48 & \text{N/A} & \text{(9,9,1,0,0,1)} & 13785.3 & 13783.5 & \text{N/A} \\
\text{(3,6,1,0,0,1)} & 3064.16 & 3064.16 & \text{N/A} & \text{(81,1,1,10,0,1)} & 13780.9 & 13780.8 & \text{N/A} \\
\text{(9,2,1,-2,0,1)} & 3062.45 & 3062.92 & \text{N/A} & \text{(84,1,1,10,0,1)} & 14292.6 & 14292.7 & \text{N/A} \\
\text{(21,1,1,-5,0,1)} & 3572.7 & 3573.13 & \text{N/A} & \text{(87,1,1,10,0,1)} & 14802.1 & 14802. & \text{N/A} \\
\text{(4,6,1,0,1,1)} & 4084.46 & 4084.8 & \text{N/A} & \text{(90,1,1,10,0,1)} & 15314.2 & 15313.5 & \text{N/A} \\
\text{(3,9,1,0,0,1)} & 4595.79 & 4595.26 & \text{N/A} & \text{(93,1,1,10,0,1)} & 15824.5 & 15823.8 & \text{N/A} \\
\text{(9,3,1,-2,0,1)} & 4594.14 & 4594.26 & \text{N/A} & \text{(96,1,1,10,0,1)} & 16334.6 & 16334.4 & \text{N/A} \\
\text{(5,6,0,-1,-1,-1)} & 5104.68 & 5104.98 & \text{N/A} & \text{(99,1,1,10,0,1)} & 16845.7 & 16844.8 & \text{N/A} \\
\text{(33,1,-4,-1,-1,0)} & 5614.27 & 5614.66 & \text{N/A} & \text{(102,1,1,10,0,1)} & 17354.1 & 17353.8 & \text{N/A} \\
\text{(6,6,1,0,0,1)} & 6128.5 & 6128.15 & \text{N/A} & \text{(105,1,1,10,0,1)} & 17864.5 & 17864.1 & \text{N/A} \\
\text{(39,1,1,10,0,1)} & 6634.19 & 6634.52 & \text{N/A} & \text{(9,12,1,0,0,1)} & 18377.3 & 18376. & \text{N/A} \\
\text{(42,1,1,10,0,1)} & 7145.52 & 7145.93 & \text{N/A} & \text{(108,1,1,10,0,1)} & 18374.2 & 18373.9 & \text{N/A} \\
\text{(45,1,1,10,0,1)} & 7656.15 & 7656.32 & \text{N/A} & \text{(111,1,1,10,0,1)} & 18884.6 & 18884.5 & \text{N/A} \\
\text{(48,1,1,10,0,1)} & 8167.23 & 8167.62 & \text{N/A} & \text{(114,1,1,10,0,1)} & 19395.4 & 19395.9 & \text{N/A} \\
\text{(51,1,1,10,0,1)} & 8676.45 & 8676.84 & \text{N/A} & \text{(117,1,1,10,0,1)} & 19904.2 & 19905.7 & \text{N/A} \\
\text{(6,9,1,0,0,1)} & 9190.39 & 9189.67 & \text{N/A} & \text{(120,1,1,10,0,1)} & 20417.3 & 20416.9 & \text{N/A} \\
\text{(54,1,1,10,0,1)} & 9186.96 & 9187.27 & \text{N/A} & \text{(123,1,1,10,0,1)} & 20926.8 & 20926.6 & \text{N/A} \\
\text{(57,1,1,10,0,1)} & 9697.3 & 9698.52 & \text{N/A} & \text{(126,1,1,10,0,1)} & 21437.1 & 21438. & \text{N/A} \\
\text{(60,1,1,10,0,1)} & 10210.4 & 10210.1 & \text{N/A} & \text{(129,1,1,10,0,1)} & 21947.2 & 21947.2 & \text{N/A} \\
\text{(63,1,1,10,0,1)} & 10718.2 & 10719.3 & \text{N/A} & \text{(132,1,1,10,0,1)} & 22458.6 & 22458.6 & \text{N/A} \\
\text{(66,1,1,10,0,1)} & 11230.9 & 11230.7 & \text{N/A} & \text{(135,1,1,10,0,1)} & 22969.1 & 22968.8 & \text{N/A} \\
\text{(69,1,1,10,0,1)} & 11740.1 & 11740.1 & \text{N/A} & \text{(138,1,1,10,0,1)} & 23479.7 & 23479.5 & \text{N/A} \\
\text{(6,12,1,0,0,1)} & 12253.5 & 12252.6 & \text{N/A} & \text{(141,1,1,10,0,1)} & 23988.5 & 23988.6 & \text{N/A} \\
\text{(72,1,1,10,0,1)} & 12249.8 & 12249.9 & \text{N/A} & \text{(12,12,1,0,0,1)} & 24502.9 & 24501.2 & \text{N/A} \\
\text{(75,1,1,10,0,1)} & \text{N/A} & 12761. & \text{N/A}  & & & &
\end{tabular}
\caption{
The summary of HF results for the 2D interaction in the AVE scheme at $V=22$meV.
The first column shows the momentum mesh.
The second, third and fourth columns show the HF energies (in meV) of the lowest state with $\Ch=0 \mod 3$, $\Ch=1 \mod 3$ and $\Ch=2 \mod 3$, respectively.
``N/A" means we have not obtained such states in the HF calculations.
For $N_x N_y \geq 36$, the Chern number is determined to $\Ch=0$ and $\Ch=1$ for $\Ch=0 \mod 3$ and $\Ch=1 \mod 3$ states by the integration of the Berry curvature.
}
\label{tab:AVE_HF_results}
\end{table}

\begin{figure}[ht!]
\centering
\includegraphics[width=0.4\columnwidth]{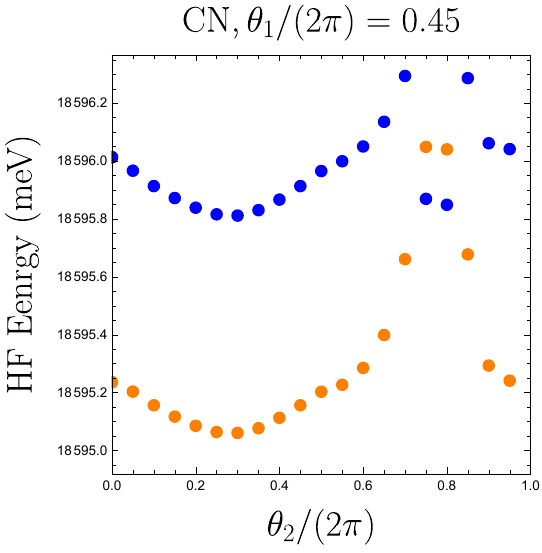}
\includegraphics[width=0.4\columnwidth]{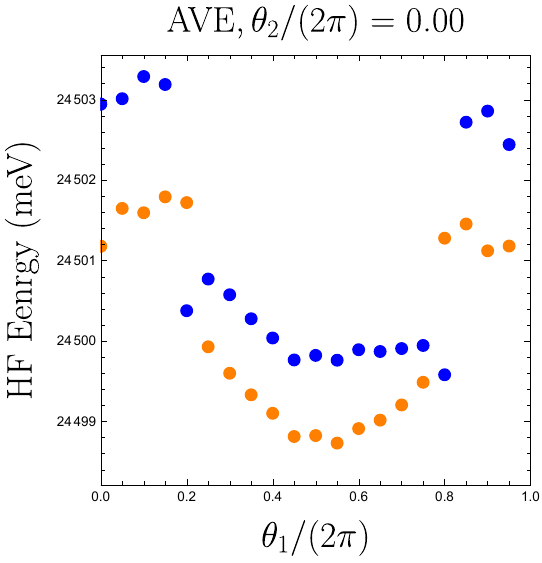}
\caption{
Translationally-invariant self-consistent HF calculations for the 2D interaction with the flux threading in the CN and AVE schemes at $V=22$\,meV.
The momentum mesh is $(N_x, N_y, \widetilde{n}_{11}, \widetilde{n}_{12}, \widetilde{n}_{21}, \widetilde{n}_{22}) =$ $(12,12,1,0,0,1)$, and the flux threading is achieved by shifting the momentum of the single-particle basis: $\bsl{k} \rightarrow  \bsl{k} + \frac{\theta_1}{2\pi} \frac{\bsl{f}_{1}}{N_x} + \frac{\theta_2}{2\pi} \frac{\bsl{f}_{2}}{N_y} $, based on the convention in \cref{eq:momentum_mesh}.
The blue and orange dots are $\Ch=0$ and $\Ch=1$ states, respectively, where $\Ch$ is determined by integrating the Berry curvature.
}
\label{fig:flux_12x12}
\end{figure}

In this section, we discuss the self-consistent HF calculations at $\nu=1$ for the 2D interaction in both the CN and AVE schemes.
We will also discuss the moir\'e-less limit of the CN scheme, referred to as the CN$^*$ scheme.

We require the HF calculations to preserve the moir\'e lattice translation symmetries.
We first perform translationally-invariant HF calculations on the momentum mesh $(N_x, N_y, \widetilde{n}_{11}, \widetilde{n}_{12}, \widetilde{n}_{21}, \widetilde{n}_{22}) = (12,12,1,0,0,1)$ (according to the convention in \cref{eq:momentum_mesh_main}) at various values of $V$ for the AVE, CN and CN$^*$ schemes.
In this calculation, we use the 2D interaction, and perform the HF calculations with $3,4,5,6,7,8,9,10$ and 11 lowest conduction bands, since we wish to study the values of $V$ such that the 3-band calculation is valid and the HF ground state has $\Ch=1$ for large system sizes.
As shown in \cref{fig:HFSchemes}a, the energy difference between $\Ch=0$ and $\Ch=1$ states is very similar between the 3-band results and the $6,7,8,9,10$ and 11-band results in the AVE scheme, which supports the validity of the 3-band calculation here.
Within the 3-band problem, we found that $V=22$\,meV is the best choice since its HF ground state has $\Ch=1$, and it is associated with the strongest FCI for small band mixing as shown in \cref{fig:AVE_V_9x2}.

For the CN scheme, $V=22$\,meV is also acceptable since the HF ground state has $\Ch=1$ for $3,4,5,6,7,8,9,10$ and $11$-band calculations as long as $V\leq 24$\,meV, as shown in \cref{fig:HFSchemes}b.
Therefore, we also choose $V=22$\,meV for 3-band calculations in the CN scheme, unless specified otherwise.

In the CN$^*$ limit, we are less confident about the validity of the 3-band calculation since the HF phase diagram yields different Chern numbers for the ground state for $3$-band versus $5,6,7,8,9$-band calculations as shown in \cref{fig:HFSchemes}c.
Furthermore, the 10 and 11-band results are not consistent with $5,6,7,8$ and 9-band results, suggesting extra caution needs to be taken when performing HF calculations with the CN$^*$ scheme.
We note that the 4 and 10-band results are similar to each other in the CN and CN$^*$ scheme, as well as the 5 and 9-band results and 6 and 8-band results.

At $V=22$meV, we further perform HF calculations on the following momentum meshes specified by $(N_x, N_y, \widetilde{n}_{11}, \widetilde{n}_{12}, \widetilde{n}_{21}, \widetilde{n}_{22}) = (2,6,1,0,1,1)$, $(15,1,1,-5,0,1)$, $(9,2,1,-2,0,1)$, $(3,6,1,0,0,1)$, $(21,1,1,-5,0,1)$, $(4,6,1,0,1,1)$, $(9,3,1,-2,0,1)$, $(5,6,0,-1,-1,-1)$, $(33,1,-4,-1,-1,0)$, $(6,6,1,0,0,1)$, $(6,9,1,0,0,1)$, $(6,12,1,0,0,1)$, $(9,9,1,0,0,1)$, $(9,12,1,0,0,1)$, $(12,12,1,0,0,1)$, and and $(x,1,1,10,0,1)$ with $x=39,42,45,...,141$, which all contain the $K_M$ and $K_M'$ points.
For each mesh and interaction scheme, we use 20 initial random states.
In the AVE scheme, the background term is generated on the same mesh as the HF calculation.
We note that the convergence of the HF calculations in the CN scheme requires substantially more iterations than in the AVE scheme.

We summarize our $V=22$meV HF results in \cref{tab:AVE_HF_results,tab:CN_HF_results}.
The Chern number is determined by $C_3$ eigenvalues at $\Gamma_M$, $\K_M$ and $\K_M'$ (up to mod 3), expect $(N_x,N_y)=(6,6)$ and $(12,12)$ where the Chern number is determined by the integration.
However, there is a subtlety here, since the HF state may spontaneously break the $C_3$ symmetry for small system sizes.
For example, for the $6\times 6$ lattice, we find that the $\Ch=0$ state in the CN scheme spontaneously breaks the $C_3$ symmetry, while the $\Ch=1$ state in the CN scheme and both $\Ch=0,1$ states in the AVE scheme do not have such $C_3$ breaking on the same size, as shown in \cref{fig:plotC3breaking6x6}.
Owing to the $C_3$ breaking, we cannot precisely calculate the $C_3$ eigenvalues of the HF state at $\Gamma_M$, $\K_M$ and $\K_M'$, which makes determining $\Ch\ \mod 3$ from $C_3$ eigenvalues subtle.
To resolve this issue, we use the following method to determine $\Ch\ \mod 3$.
Given a HF state, we can evaluate $\langle c^\dagger_{\K,\bsl{k},n,\uparrow} c_{\K,\bsl{k},n,\uparrow} \rangle$ for $\bsl{k}=\Gamma_M,\K_M,\K_M'$, where $c^\dagger_{\K,\bsl{k},n,\uparrow}$ is the bare basis in \eqnref{eq:ED_H_3D_int_ave} and has definite $C_3$ eigenvalue for $\bsl{k}=\Gamma_M,\K_M,\K_M'$.
At a fixed $\bsl{k}$ (among $\Gamma_M,\K_M,\K_M'$),
the $C_3$ eigenvalue of the HF state is prescribed to be that corresponding to the band $n$ that achieves the largest weight $\langle c^\dagger_{\K,\bsl{k},n,\uparrow} c_{\K,\bsl{k},n,\uparrow} \rangle$ in the HF state.
As a consistency check, we find that $\Ch\ \mod 3$ determined from this method is consistent with $\Ch$ determined from the integration of Berry curvature for all states on $6\times 6$ in both CN and AVE schemes, regardless of whether $C_3$ is broken or not.
We caution that for small sizes, the method of determining $\Ch\ \mod 3$ from $C_3$ eigenvalues might lead to errors if the $C_3$ breaking is strong---the appearance of a $\Ch = 2\ \mod 3 $ state in \cref{tab:CN_HF_results} for $(N_x, N_y, \widetilde{n}_{11}, \widetilde{n}_{12}, \widetilde{n}_{21}, \widetilde{n}_{22})=(9,2,1,-2,0,1)$ is potentially due to this error.

Based on \cref{tab:AVE_HF_results,tab:CN_HF_results}, the HF ground state is mostly $\Ch=0$ ($\Ch=1$) for $N_s < 60$ ($N_s \geq 60$) for both AVE and CN schemes.
We note that the HF calculations on small system sizes are subject to strong finite-size effects. For example, in the AVE scheme, $(N_x, N_y, \widetilde{n}_{11}, \widetilde{n}_{12}, \widetilde{n}_{21}, \widetilde{n}_{22}) =$  $(9,2,1,-2,0,1)$ has the $\Ch\ \mod 3 = 0$ state clearly lower than the $\Ch\ \mod 3 = 1$ state, while  $(N_x, N_y, \widetilde{n}_{11}, \widetilde{n}_{12}, \widetilde{n}_{21}, \widetilde{n}_{22}) =$  $(3,6,1,0,0,1)$ has the $\Ch\ \mod 3 = 0$ state in very close competition with the $\Ch\ \mod 3 = 1$ state.
Even for larger system sizes, the fact that the ground state is not consistently $\Ch=1$ for $N_s$ indicates the strong finite-size effect.
The strong finite-size effect is also reflected by the fact that as we twist the boundary condition, the $\Ch=1$ state does not stay as the ground state even on systems as large as $N_s=144$.
Explicitly, as shown in \cref{fig:flux_12x12} for  $(N_x, N_y, \widetilde{n}_{11}, \widetilde{n}_{12}, \widetilde{n}_{21}, \widetilde{n}_{22}) =$  $(12,12,1,0,0,1)$, changing the boundary condition along $\bsl{b}_{M,1}$ alternates the energetic order of $\Ch=1$ and $\Ch=0$ states in both the CN and AVE schemes.
Nevertheless, our data exhibits a clear trend in HF where small system sizes favor the $\Ch=0$ state, while the $\Ch=1$ state eventually wins as the system size increases.

\newpage

\section{2D Interaction: Charge-Neutrality and Moir\'e-less Schemes}
\label{app:CNscheme}

In this section, we show our ED results in the CN scheme. \App{app:result_2D_int_nu_1} contains a summary of calculations at $\nu=1$, while \App{app:FCI_CN} contains results at $\nu=1/3$ and $\nu=2/3$.

\subsection{ED results at $\nu = 1$}
\label{app:result_2D_int_nu_1}

\begin{figure}[ht!]
\centering
\includegraphics[width=\columnwidth]{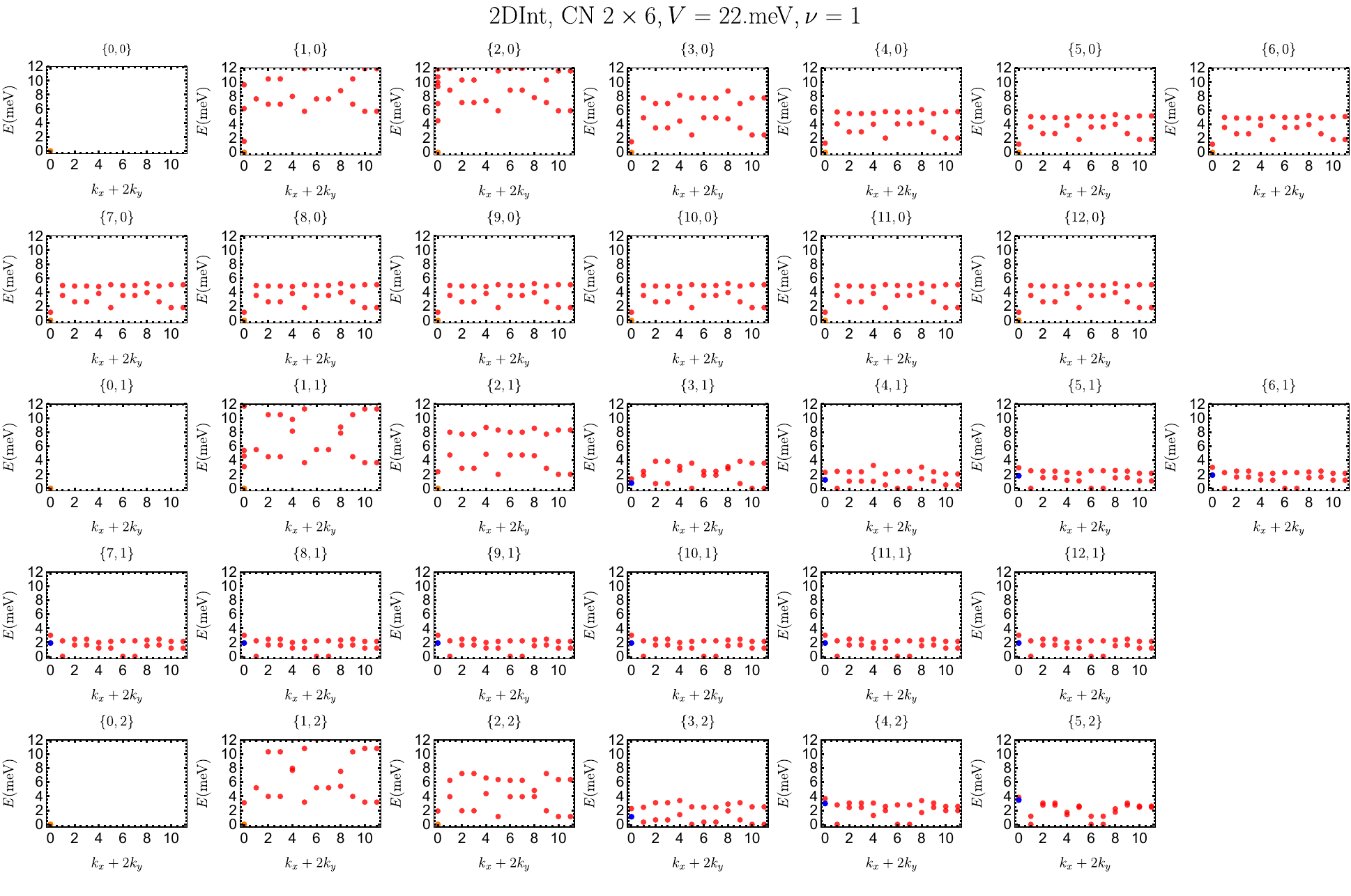}
\caption{Summary of $\nu=1$ ED spectra for 2D interactions in the CN scheme at $V=22$\,meV for $(N_x, N_y, \widetilde{n}_{11}, \widetilde{n}_{12}, \widetilde{n}_{21}, \widetilde{n}_{22}) = (2,6,1,0,1,1)$.
The lowest-energy state at $(k_x, k_y) = (0,0)$ is marked in orange (blue) if it has $\Ch=1$ ($\Ch=0$).
}
\label{fig:nu=1_12_CN}
\end{figure}

\begin{figure}[ht!]
\centering
\includegraphics[width=\columnwidth]{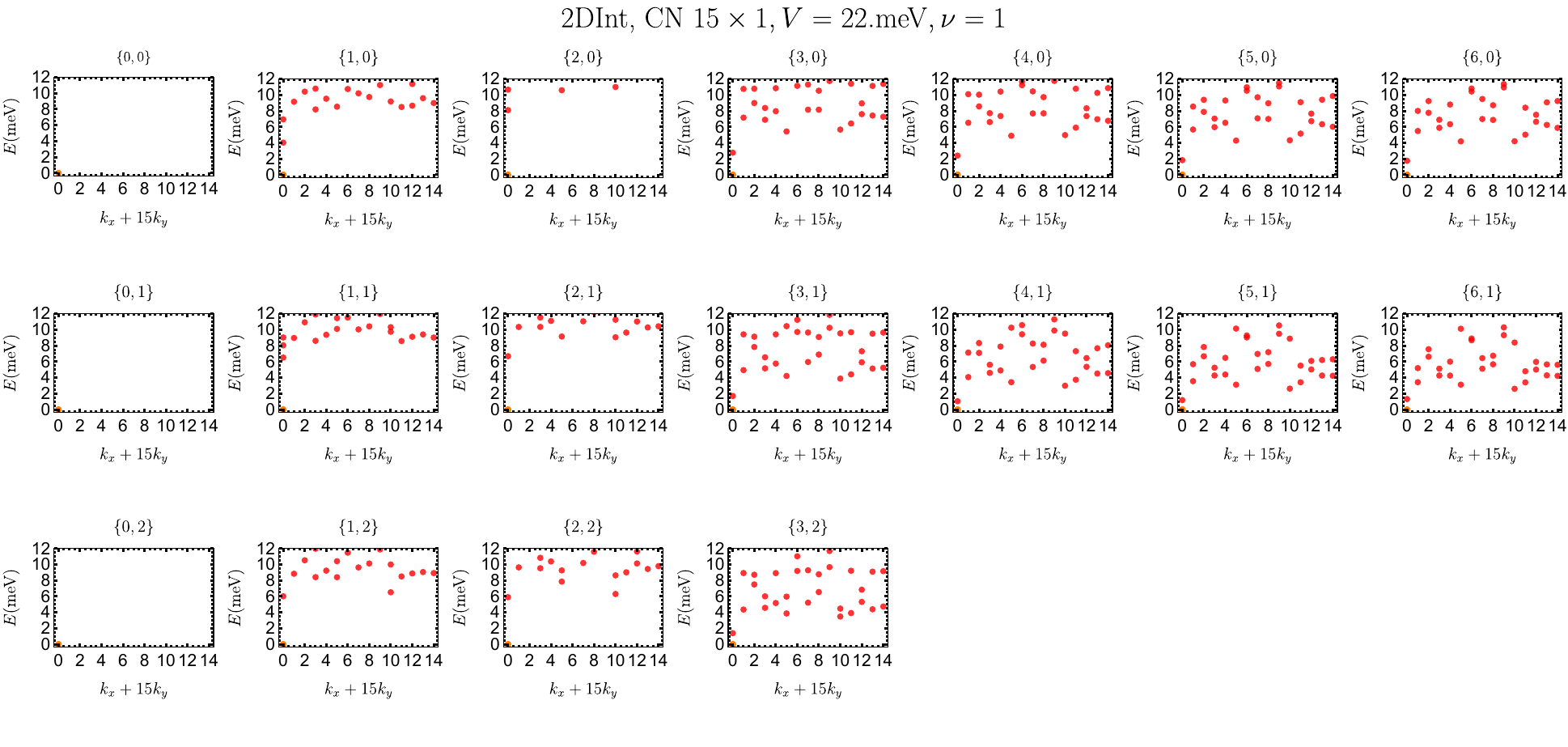}
\caption{Summary of $\nu=1$ ED spectra for 2D interactions in the CN scheme at $V=22$\,meV for $(N_x, N_y, \widetilde{n}_{11}, \widetilde{n}_{12}, \widetilde{n}_{21}, \widetilde{n}_{22}) = (15,1,1,-5,0,1)$.
The lowest-energy state at $(k_x, k_y) = (0,0)$ is marked in orange (blue) if it has $\Ch=1$ ($\Ch=0$).
}
\label{fig:nu=1_15_CN}
\end{figure}

\begin{figure}[ht!]
\centering
\includegraphics[width=\columnwidth]{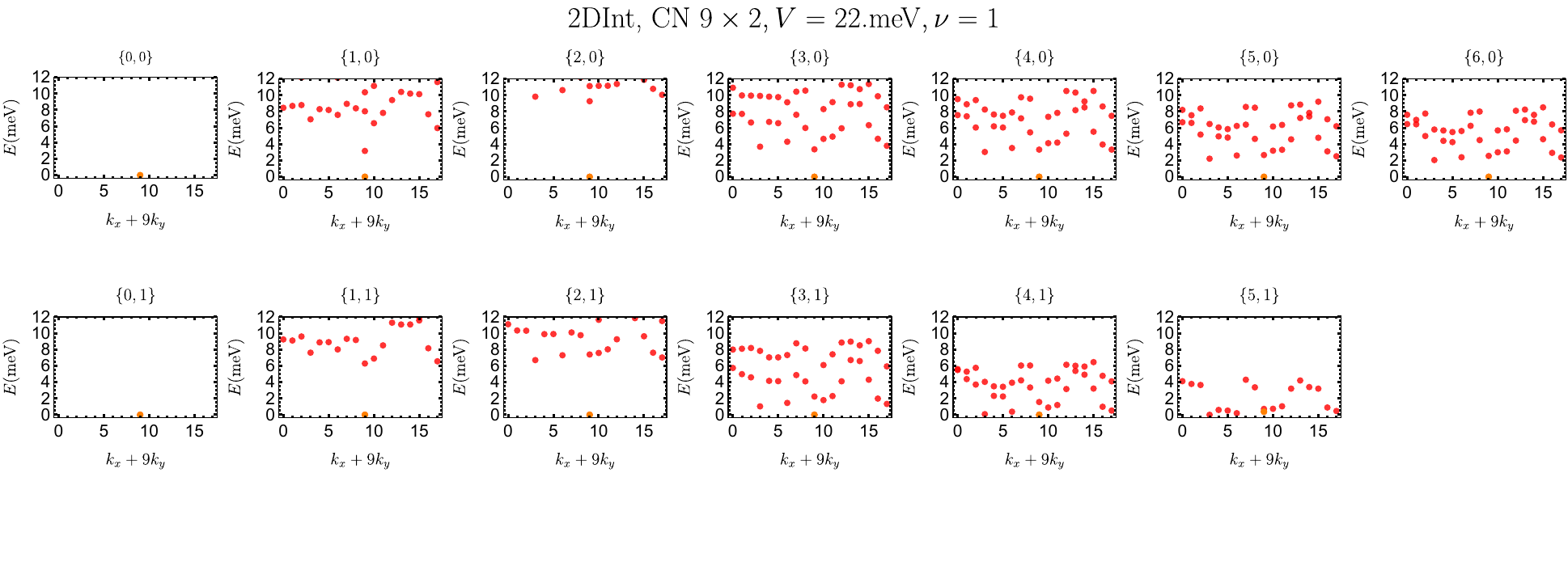}
\caption{Summary of $\nu=1$ ED spectra for 2D interactions in the CN scheme at $V=22$\,meV for $(N_x, N_y, \widetilde{n}_{11}, \widetilde{n}_{12}, \widetilde{n}_{21}, \widetilde{n}_{22}) = (9,2,1,-2,0,1)$.
The lowest-energy state at $(k_x, k_y) = (0,1)$ is marked in orange (blue) if it has $\Ch=1$ ($\Ch=0$).
}
\label{fig:nu=1_18_CN}
\end{figure}

\begin{figure}[ht!]
\centering
\includegraphics[width=\columnwidth]{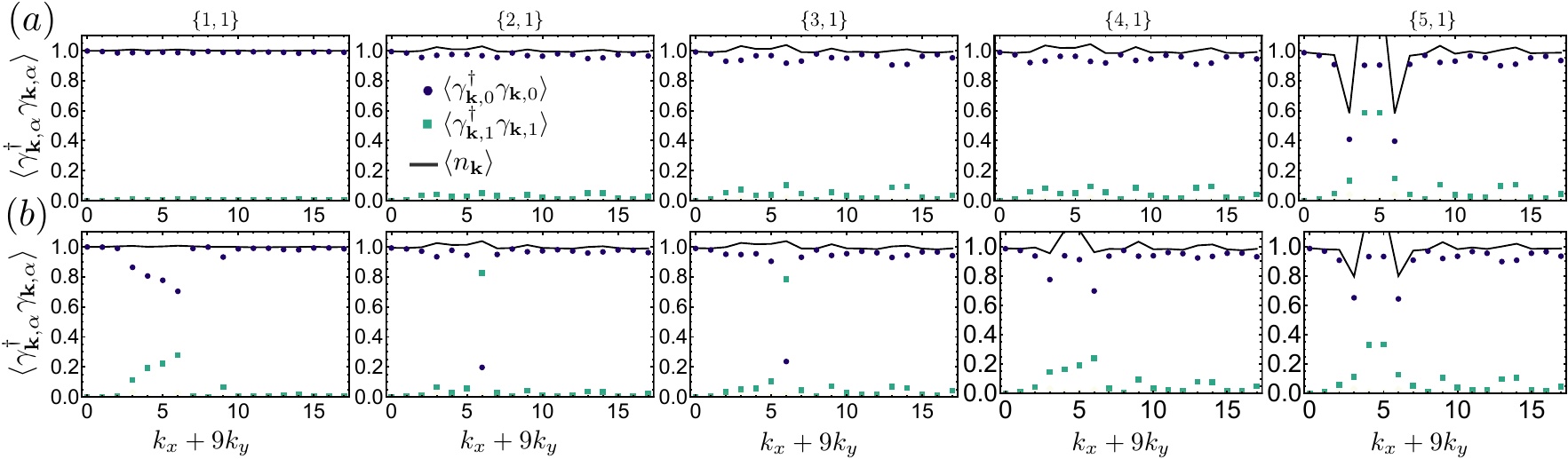}
\caption{Correlation functions for ED calculations in the CN scheme at $\nu=1$ and $V=22$\,meV for $N_{\text{band2}}= 1$ on the $9\times 2$ mesh in $(b)$ (see \Fig{fig:CNintegercollapse}). The lowest $(a)$ and second lowest $(b)$ state at the HF ground state momentum $(k_x,k_y) =(0,1)$ are shown. Note that at band-max $\{5,1\}$ (where the ground state is actually not in this momentum sector, see Main Text), there appears to be an inversion, or strong hybridization, between the $\Ch=1$ and $\Ch=0$ states.
}
\label{fig:nu=1_18_CN_nk}
\end{figure}

\begin{figure}[ht!]
\centering
\includegraphics[width=\columnwidth]{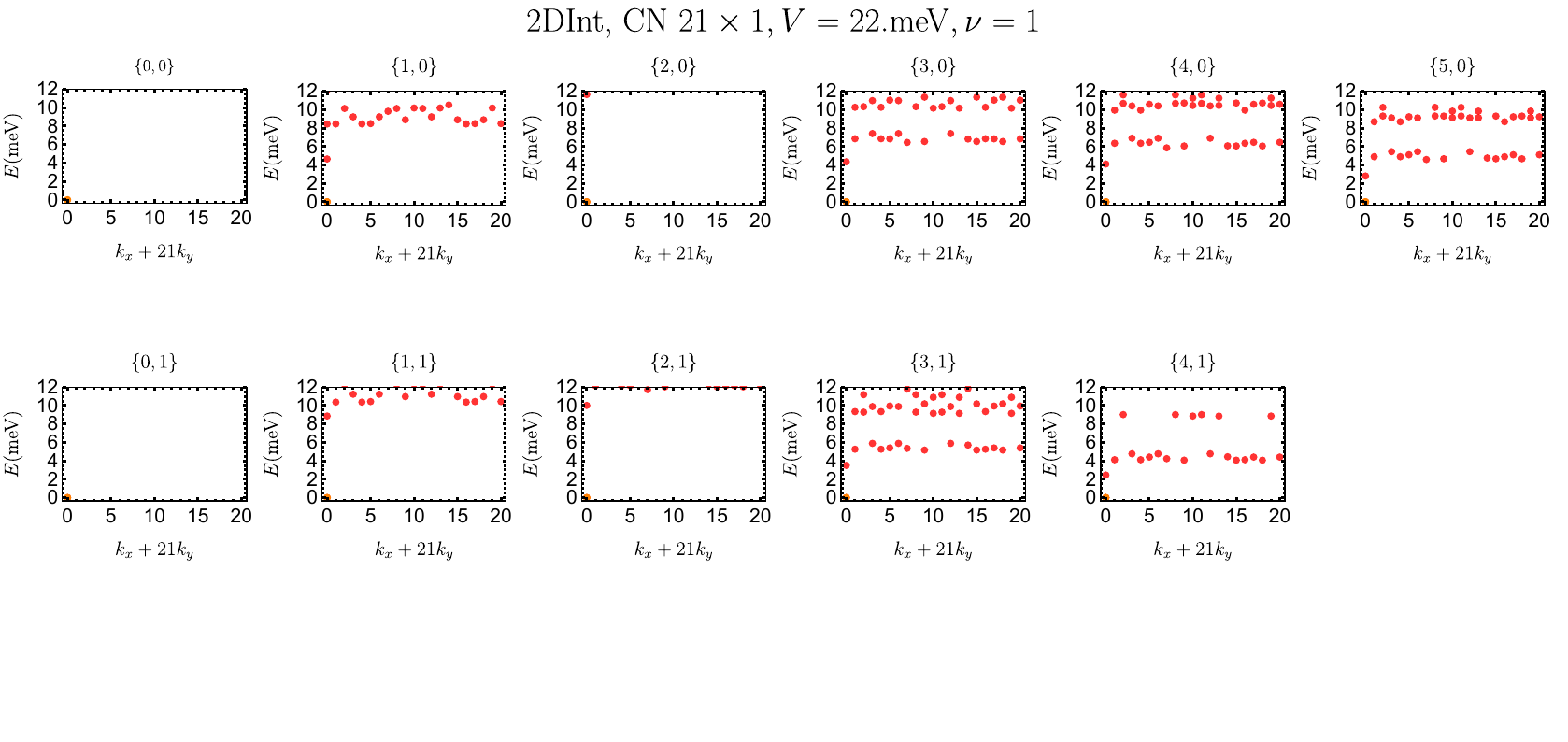}
\caption{Summary of $\nu=1$ ED spectra for 2D interactions in the CN scheme at $V=22$\,meV for $(N_x, N_y, \widetilde{n}_{11}, \widetilde{n}_{12}, \widetilde{n}_{21}, \widetilde{n}_{22}) = (21,1,1,-5,0,1)$.
The lowest-energy state at $(k_x, k_y) = (0,0)$ is marked in orange (blue) if it has $\Ch=1$ ($\Ch=0$).
}
\label{fig:nu=1_21_CN}
\end{figure}
\begin{figure}[ht!]
\centering
\includegraphics[width=\columnwidth]{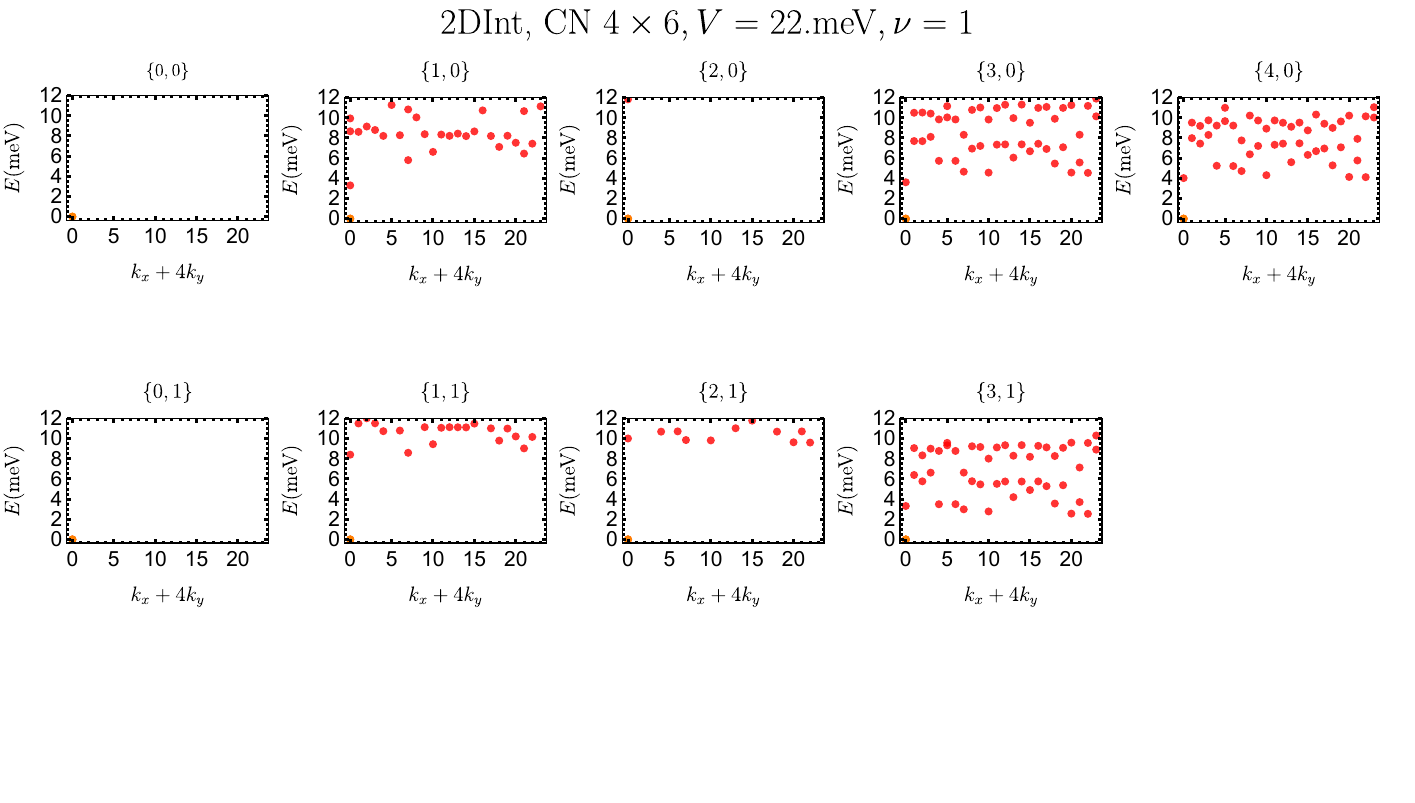}
\caption{Summary of $\nu=1$ ED spectra for 2D interactions in the CN scheme at $V=22$\,meV for $(N_x, N_y, \widetilde{n}_{11}, \widetilde{n}_{12}, \widetilde{n}_{21}, \widetilde{n}_{22}) = (4,6,1,0,1,1)$.
The lowest-energy state at $(k_x, k_y) = (0,0)$ is marked in orange (blue) if it has $\Ch=1$ ($\Ch=0$).
}
\label{fig:nu=1_4x6_CN}
\end{figure}

A summary of our results for $\nu=1$ can be found in \cref{fig:nu=1_12_CN,fig:nu=1_15_CN,fig:nu=1_18_CN,fig:nu=1_21_CN,fig:nu=1_4x6_CN} for $12$, $15$, $18$, $21$ and $24$ sites, respectively.
We find signs for convergence in the $N_{\text{band1}}$ truncation parameter with $N_{\text{band2}}$ = 0.
Recall that for $N_{\text{band1}}=N_{\text{band2}}=0$, the only Fock state in the Hilbert space at $\nu=1$ is the $\Ch=1$ HF state.
For all system sizes, we find that the gap above the $\Ch=1$ state in ED does not close as particles are allowed to populate band 1.
Yet, as shown in \cref{fig:CNintegercollapse}c, we can see that the gap between the continuum (states not at the CI momentum) and the GS is considerably larger at 15 and 18 sites than that at 12,18 and 24 sites.

The $N_{\text{band2}}$=1 sequence appears to indicate a gap closing between the ground state at the CI momentum and the continuum (states not at the CI momentum) for 12 and 18 sites; so does the $N_{\text{band2}}$=2 sequence for 12 sites. \Fig{fig:nu=1_18_CN_nk} shows the $\braket{\gamma^\dag_{\mbf{k},\al}\gamma_{\mbf{k},\al} }$ expectation values that indicate the $\Ch=1$ and $\Ch=0$ competition.
Yet, we have not observed the continuum dropping down on 15 and 21 sites in the $N_{\text{band2}}$=1 sequence.
Such even-odd difference is consistent with the trend that the gap between the continuum (states not at the CI momentum) and the GS is considerably larger at 15 and 21 sites than that at 12 and 18 sites for the $N_{\text{band2}}=0$ sequence.
(One potential reason for such even-odd effect is that the 15- and 21-site meshes miss the $M_M$ points, where band 1 and band 0 is nearly degenerate and band 2 and band 0 are close in energy, as shown in \cref{fig:BS}. We leave a careful study of this even-odd effect for future work.)
As the 24 sites follows the same trend as $12$ and $18$ sites for the $N_{\text{band2}}$=0,1 regarding the gap between the continuum and the GS, we expect the continuum dropping down will happen eventually for the $N_{\text{band2}}$=1 on 24 sites, though we have not reach enough band-mixing on 24 sites to observe it.

There is a close competition between $\Ch=0$ and $\Ch=1$ states on 12 sites with $N_{\text{band2}}=1,2$.
Yet, the $\Ch=0$ state never becomes lower than the $\Ch=1$ state before the continuum drops down, as shown in \cref{fig:nu=1_12_CN}.
In all cases, we diagnose the Chern number by looking at the orbital occupation $\braket{\gamma^\dag_{\mbf{k},\alpha}\gamma_{\mbf{k},\alpha}}$ in the HF basis.
For a $\Ch=1$ state, we require the occupation in band 0 to be greater than that in band 1 and band 2 for all $\mbf{k}$.
The $\Ch=0$ state that we found has a greater occupation in band 1 than in band 0 and band 2 at $K_M'$, while all other $\bsl{k}$ points have maximum occupation in band 0; we identify it as a $\Ch=0$ state because the spinless $C_3$ eigenvalues for band 0 at $\Gamma_M$ ($e^{-\ii 2\pi/3}$), band 0 at $\K_M$ ($1$) and band 1 at $\K_M'$ ($e^{\ii 2\pi/3}$)  gives $\Ch=0\ \mod\ 3$.
The close competition between $\Ch=0$ and $\Ch=1$ states is more clear in the AVE scheme, which is discussed in \cref{app:result_2D_int_nu_1}.

\subsection{ED results at $\nu = 1/3$ and $2/3$}
\label{app:FCI_CN}

\begin{figure}[t]
\centering
\includegraphics[width=\columnwidth]{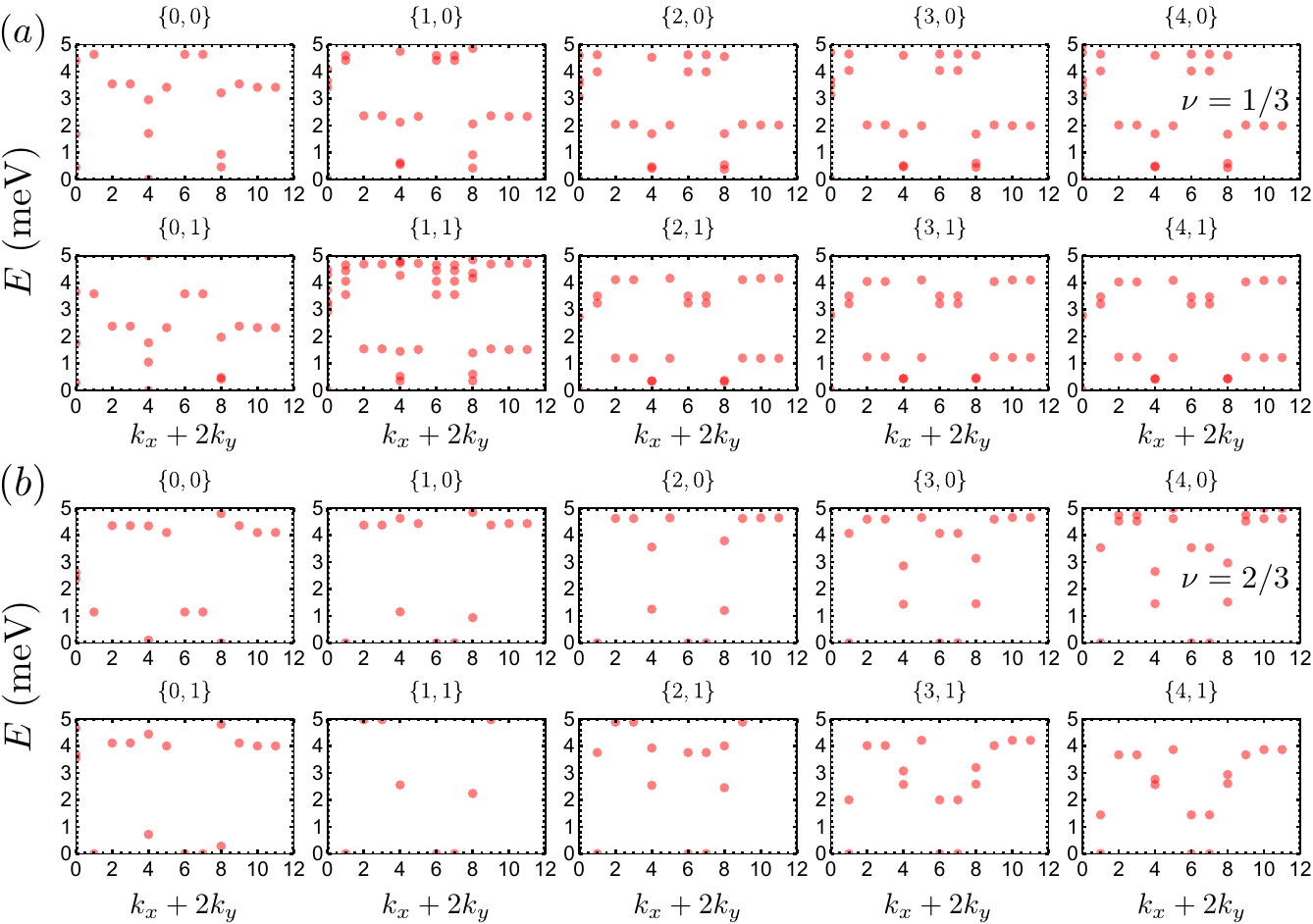}
\caption{
ED spectra for 2D interactions in CN scheme with the moir\'e potential at $V=22$meV using 3-band HF matrix elements on the $2\times 6$ lattice. ED spectra at $\nu=1/3$ are shown in $(a)$  and $\nu=2/3$ in $(b)$.}
\label{fig:AHCCNP12}
\end{figure}
\begin{figure}[t]
\centering
\includegraphics[width=\columnwidth]{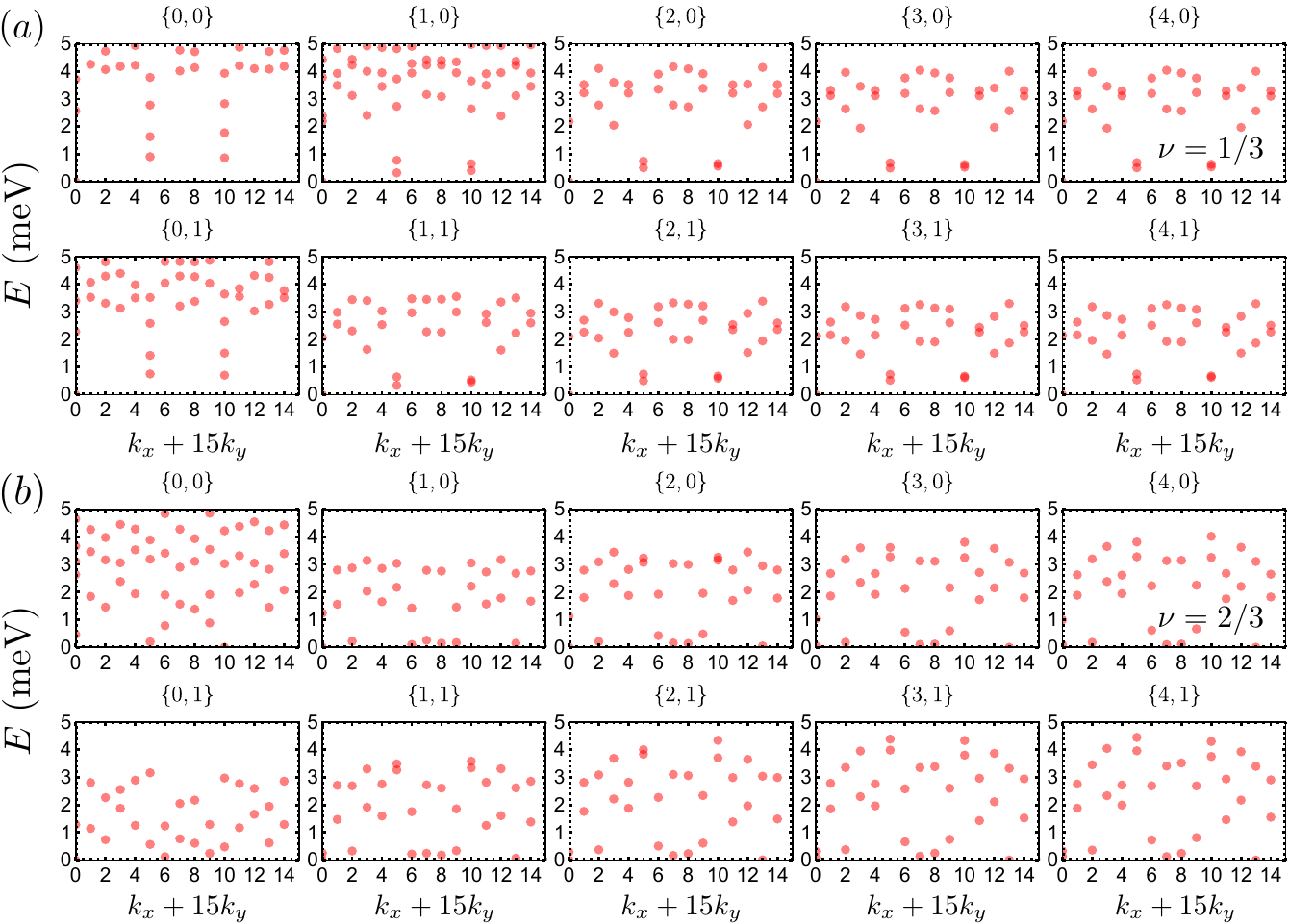}
\caption{
ED spectra for 2D interactions in CN scheme on 15 sites with the moir\'e potential at $V=22$meV using 3-band HF matrix elements. ED spectra at $\nu=1/3$ are shown in $(a)$  and $\nu=2/3$ in $(b)$.}
\label{fig:AHCCNP15}
\end{figure}
\begin{figure}[t]
\centering
\includegraphics[width=\columnwidth]{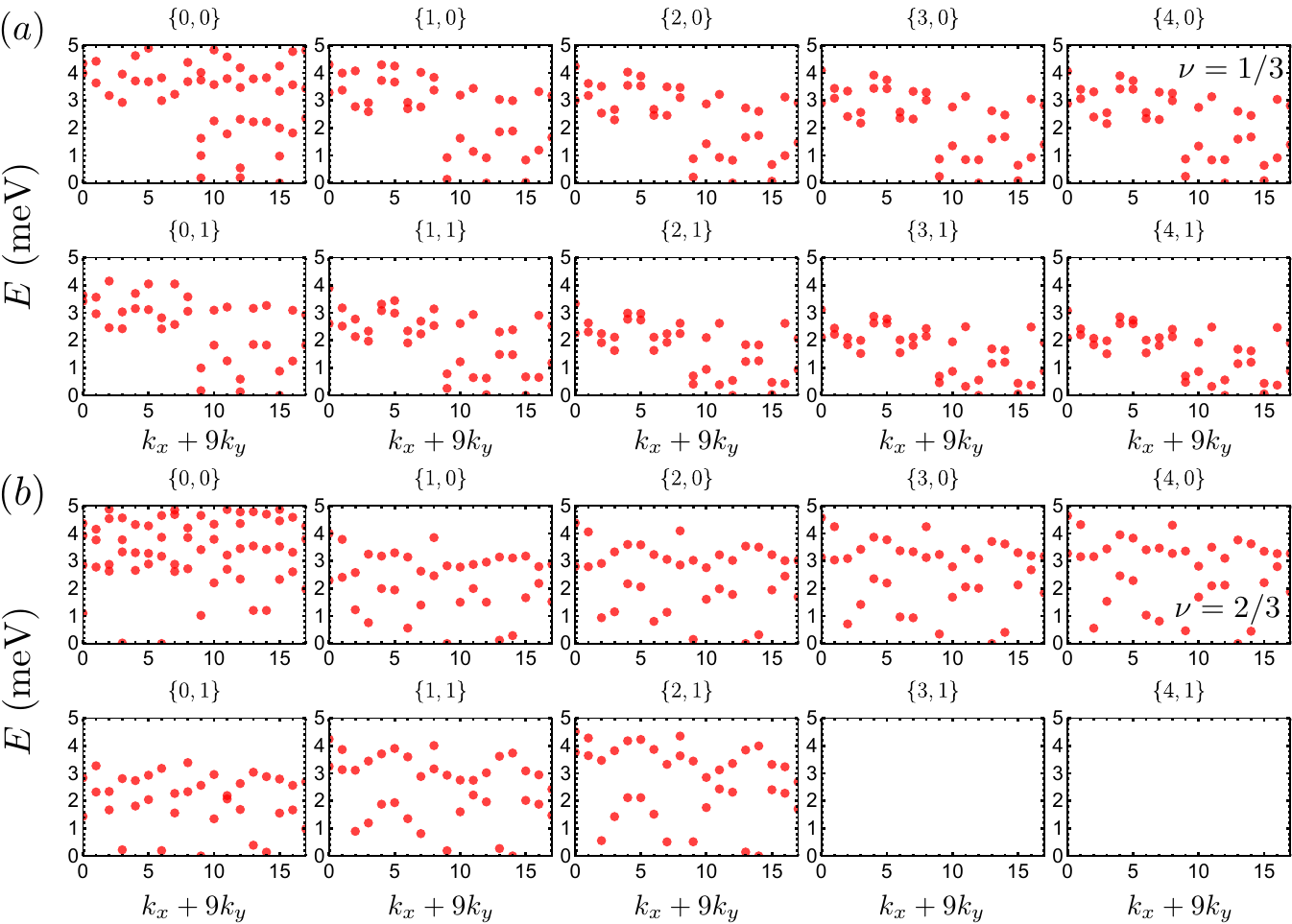}
\caption{
ED spectra for 2D interactions in CN scheme on 18 sites with the moir\'e potential at $V=22$meV using 3-band HF matrix elements. ED spectra at $\nu=1/3$ are shown in $(a)$  and $\nu=2/3$ in $(b)$.}
\label{fig:AHCCNP18}
\end{figure}

\begin{figure}[t]
\centering
\includegraphics[width=\columnwidth]{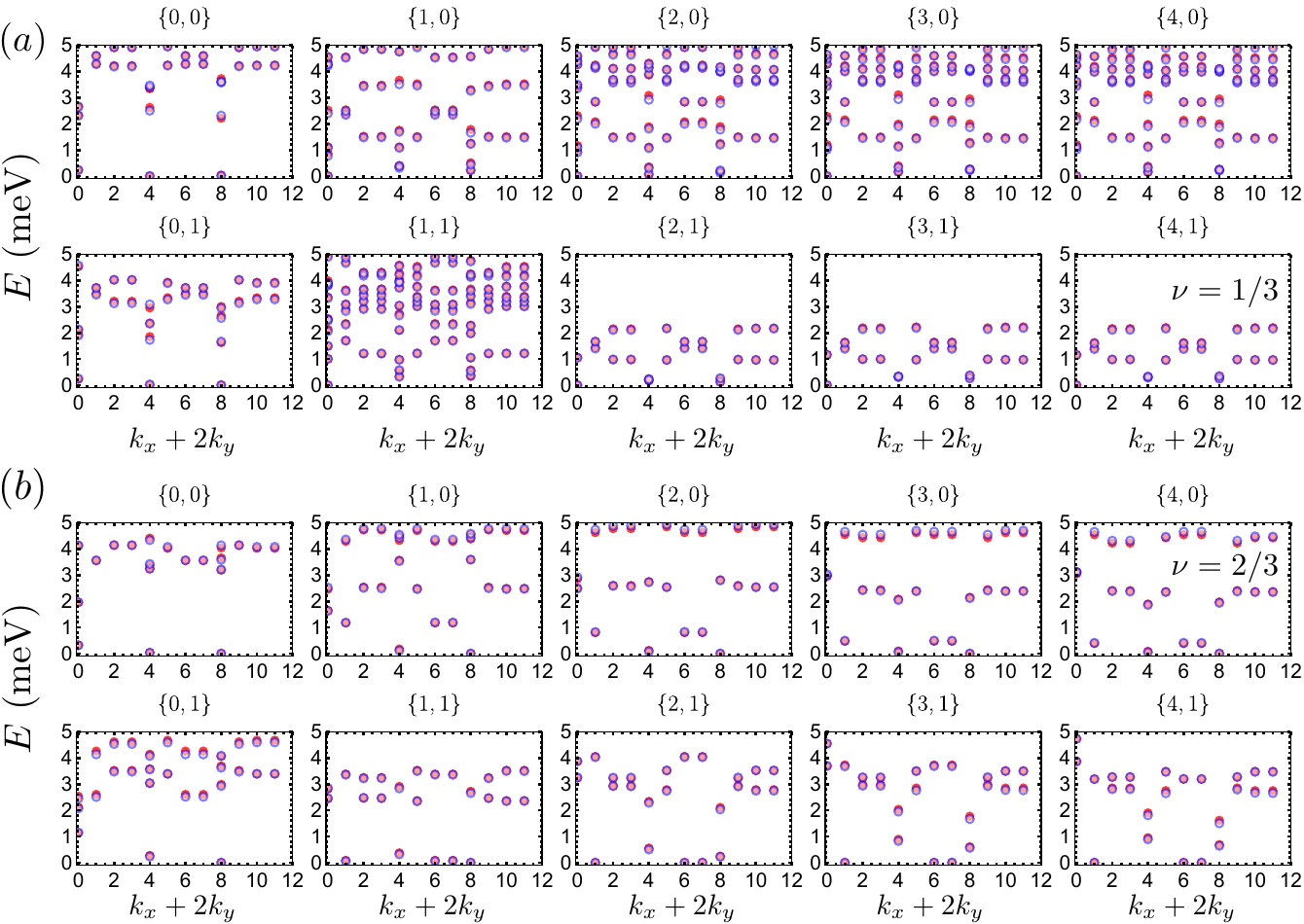}
\caption{
ED spectra on 12 sites for 2D interactions in CN scheme (red) and moir\'e-less limit (blue circles) at $V=28$meV using 5-band HF-basis matrix elements. ED spectra at $\nu=1/3$ are shown in $(a)$  and $\nu=2/3$ in $(b)$.
We forbid particles from populating band 3 and band 4.
}
\label{fig:AHCCNP12_28}
\end{figure}

\begin{figure}[t]
\centering
\includegraphics[width=\columnwidth]{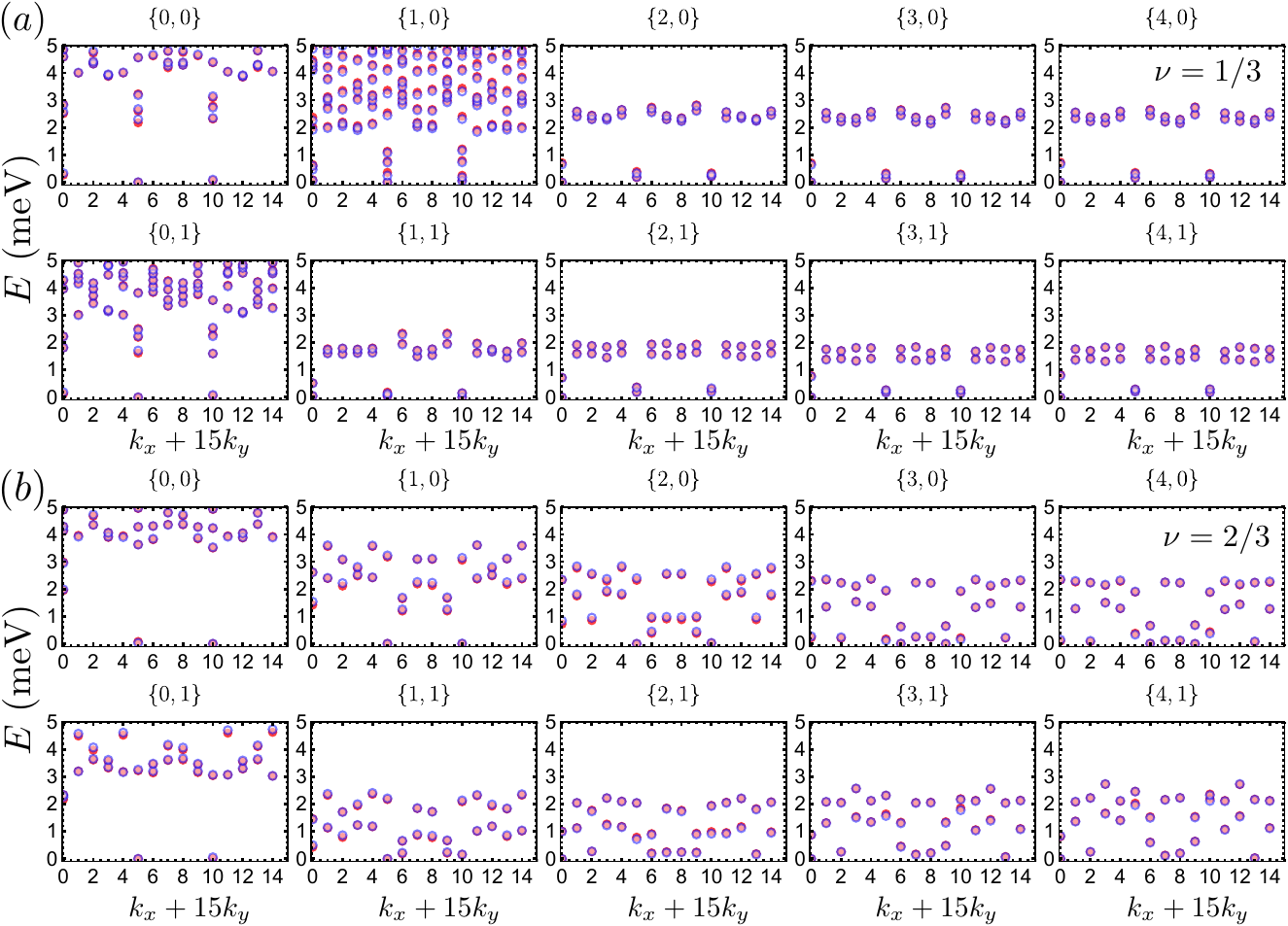}
\caption{
ED spectra on 15 sites for 2D interactions in CN scheme (red) and moir\'e-less limit (blue circles) at $V=28$meV using 5-band HF-basis matrix elements. ED spectra at $\nu=1/3$ are shown in $(a)$  and $\nu=2/3$ in $(b)$.
We forbid particles from populating band 3 and band 4.
}
\label{fig:AHCCNP15_28}
\end{figure}

We now discuss the ED results at $\nu = 1/3$ and $2/3$ in the CN scheme. Figs. \ref{fig:AHCCNP12}, \ref{fig:AHCCNP15}, and \ref{fig:AHCCNP18} show ED spectra at $V=22$\,meV on 12, 15, and 18 sites respectively. At band-max $\{0,0\}$, no clear FCI states are seen, and the spectra undergo significant changes as particles are allowed into the higher bands. In the Main Text, we showed well-developed FCIs at $V=28$\,meV on 18 sites with 5-band HF basis matrix elements restricted to the lowest 3 bands, as 3-band HF is potentially unreliable for $V=28$\,meV in the CN scheme as shown in \cref{fig:HFSchemes}b. We complement these results with 12 and 15 site calculations in \Fig{fig:AHCCNP12_28} and \Fig{fig:AHCCNP15_28} respectively. Both show clear FCIs at $\nu=1/3$ for band-max $\{0,0\}$, but the $\nu=2/3$ FCI on 15 sites has a large spread at band-max $\{0,0\}$ not seen on 12 or 18 sites. Regardless, all FCIs are destroyed by band mixing.


In \Fig{fig:AHCCNP12_28} and \Fig{fig:AHCCNP15_28}, we also compare the results of the CN scheme to those obtained in the fully moir\'e-less limit with continuous translation symmetry where there must be a gapless Goldstone mode in the thermodynamic limit if the ground state possesses Wigner crystalline order. We show that the spectra in the two cases are quantitatively nearly identical.  We find in our CN scheme ED calculations that the $\nu=2/3$ FCI does not survive and its gap collapses as the Hilbert space is enlarged (by increasing band-max $N_{\text{band1}}$ and $N_{\text{band2}}$).
The results pose doubts on the 1-HF-band projected calculations in the CN scheme as well as its moir\'e-less limit.

\newpage
\clearpage

\section{ED results for 2D Interaction in the Average Scheme using the HF basis}
\label{app:result_2D_int}

In this Appendix, we discuss our ED results with the 2D interaction in the HF basis and examine the convergence with respect to band-max. For the largest Hilbert spaces, we also implement orbital restriction (i.e.~keeping only $N_{\text{orb}1}$ and $N_{\text{orb}2}$ orbitals in bands 1 and 2) to further reduce the Hilbert space. We focus on the AVE scheme throughout this section.

\subsection{ED results at $\nu = 1$}

\begin{figure}
\centering
\includegraphics[width=\columnwidth]{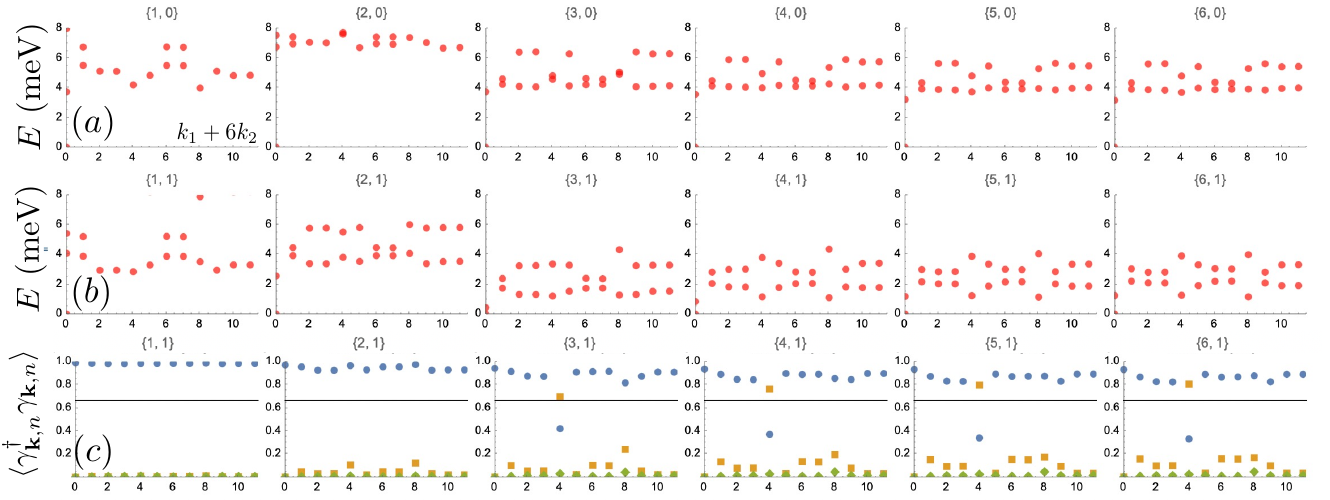}
\caption{Summary of $\nu=1$ ED data for 2D interactions in the AVE scheme at $V=22$\,meV for 12 sites. $(a)$ shows spectra with $N_{\text{band2}}$ = 0, and $(b)$ shows spectra with $N_{\text{band2}}$ = 1. In $(c)$ we show the orbital occupation $\braket{\gamma^\dag_{\mbf{k},n}\gamma_{\mbf{k},n}}$ in the ground states at band-max $\{N_{\text{band1}},1\}$ in panel $(b)$ , where $n=0,1,2$ corresponds to blue, yellow, and green. The spectral eigenvalues close to 1 or 0 indicate an approximately Slater state, allowing us to diagnose a change in the Chern number through a band inversion at the $K_M$ point. At band-max $\{3,1\}$, this band inversion changes the Chern number to $\Ch=0$.}
\label{fig:nu=1_12}
\end{figure}

\begin{figure}
\centering
\includegraphics[width=\columnwidth]{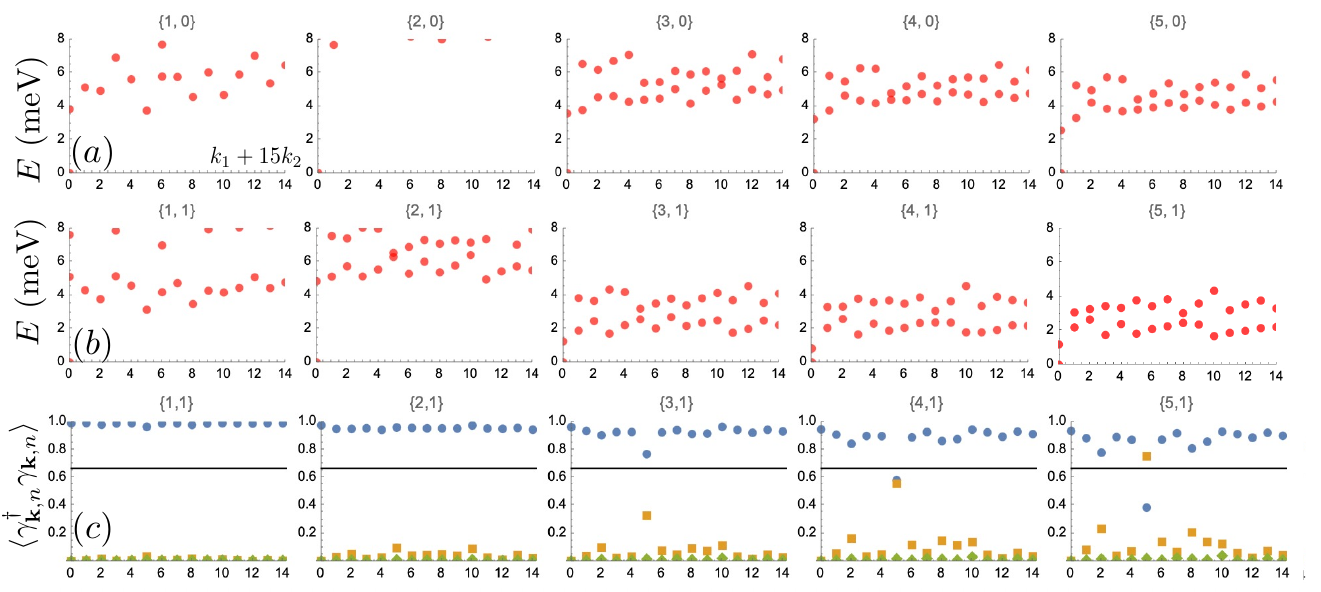}
\caption{Summary of $\nu=1$ ED data for 2D interactions in the AVE scheme at $V=22$\,meV for 15 sites. $(a)$ shows spectra with $N_{\text{band2}}$ = 0, and $(b)$ shows spectra with $N_{\text{band2}}$ = 1. In $(c)$ we show the orbital occupation $\braket{\gamma^\dag_{\mbf{k},n}\gamma_{\mbf{k},n}}$  in the ground states at band-max $\{N_{\text{band1}},1\}$ in panel $(b)$, where $n=0,1,2$ corresponds to blue, yellow, and green. The spectral eigenvalues close to 1 or 0 indicate an approximately Slater state, allowing us to diagnose a change in the Chern number through a band inversion at the $K_M$ point. At band-max $\{4,1\}$, this band inversion changes the Chern number to $\Ch=0$.}
\label{fig:nu=1_15}
\end{figure}

\begin{figure}
\centering
\includegraphics[width=\columnwidth]{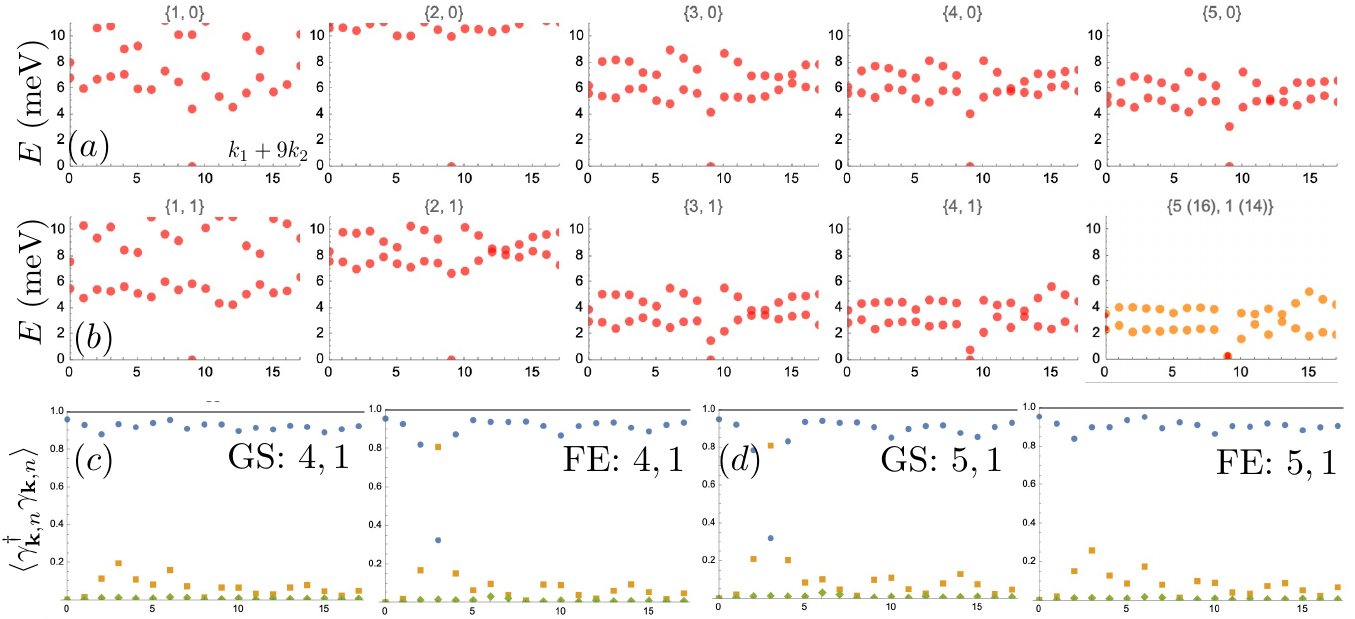}
\caption{Summary of $\nu=1$ ED data for 2D interactions in the AVE scheme at $V=22$\,meV for 18 sites. $(a)$ shows spectra with $N_{\text{band2}}$ = 0, and $(b)$ shows spectra with $N_{\text{band2}}$ = 1. The orange color signifies a spectrum computed with orbital restriction, $\{5\,(16),1 (14)\}$ indicating that 16 out of 18 orbitals in band 1 and 14 out of 18 in band 2 are included in the calculation. To test this approximation, two momentum sectors are computed in the full Hilbert space and shown in red.
In $(c)$ and $(d)$ we show the orbital occupation $\braket{\gamma^\dag_{\mbf{k},n}\gamma_{\mbf{k},n}}$ of the ground state at $\{4,1\}$ and $\{5,1\}$ respectively (no orbital restriction is used). ``GS" denotes the ground state, and ``FE" denotes the first excited state, which is at the same momentum and also has Slater-like orbital occupation. $n=0,1,2$ corresponds to blue, yellow, and green. The spectral eigenvalues close to 1 or 0 indicate an approximately Slater state, allowing us to diagnose a change in the Chern number through a band inversion at the $K_M$ point. At band-max $\{5,1\}$, this band inversion changes the Chern number to $\Ch=0$.}
\label{fig:nu=1_18}
\end{figure}
\begin{figure}
\centering
\includegraphics[width=\columnwidth]{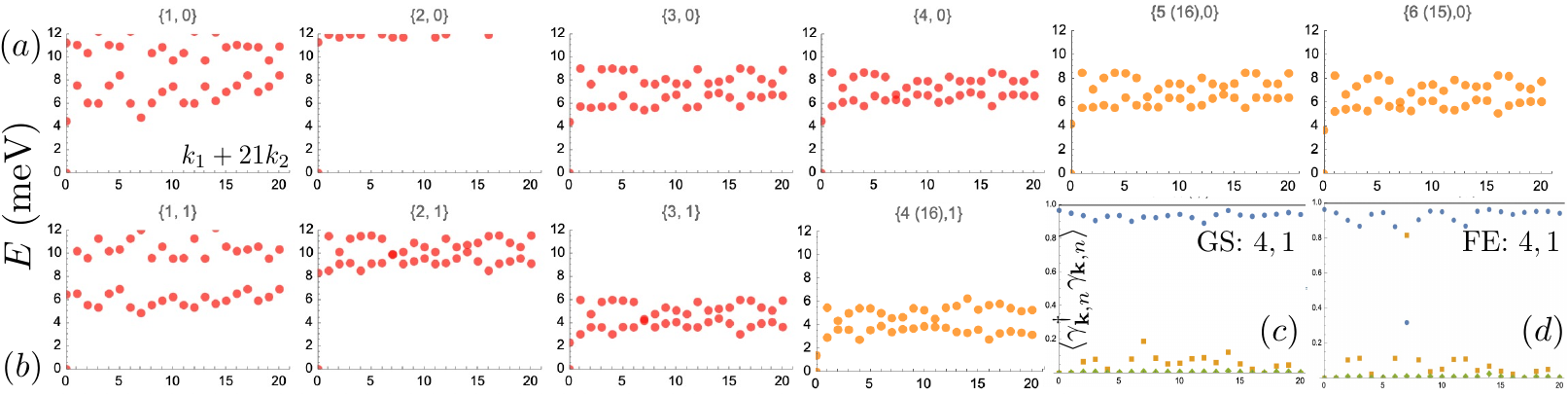}
\caption{Summary of $\nu=1$ ED data for 2D interactions in the AVE scheme at $V=22$\,meV for 21 sites. $(a)$ shows spectra with $N_{\text{band2}}$ = 0, and $(b)$ shows spectra with $N_{\text{band2}}$ = 1. The orange color signifies a spectrum computed with orbital restriction as in \cref{fig:nu=1_18}.
In $(c)$ and $(d)$ we show the orbital occupation $\braket{\gamma^\dag_{\mbf{k},n}\gamma_{\mbf{k},n}}$ of the ground state at $\{4,1\}$ in the ground state ``GS" and first excited state ``FE". $n=0,1,2$ corresponds to blue, yellow, and green. The spectral eigenvalues close to 1 or 0 indicate an approximately Slater state. It is likely that we have not reached a large enough ratio of $N_{\text{band1}}/N_s$ to observe the exchange of $\Ch=1$ and $\Ch=0$. }
\label{fig:nu=1_21}
\end{figure}
\begin{figure}
\centering
\includegraphics[width=\columnwidth]{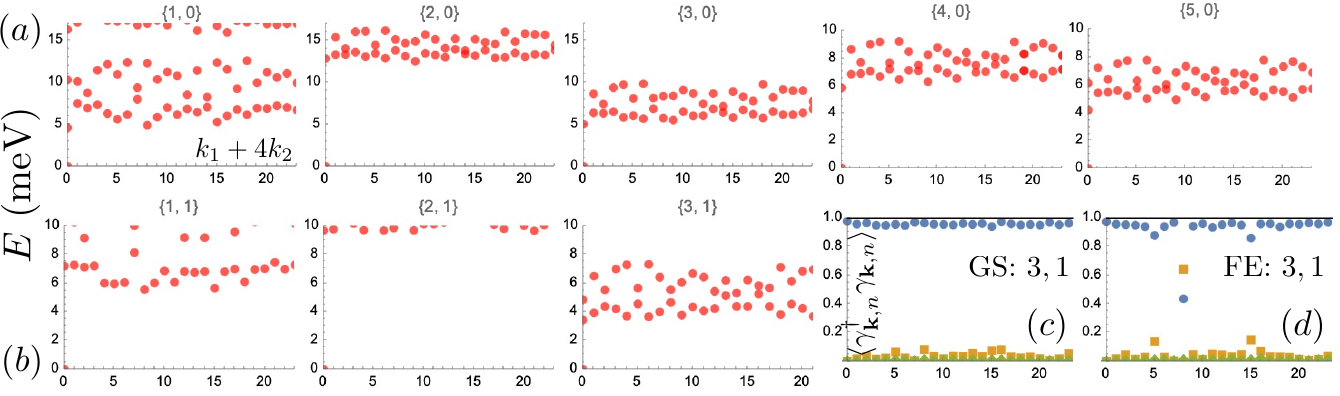}
\caption{Summary of $\nu=1$ ED data for 2D interactions in the AVE scheme at $V=22$\,meV for 24 sites. $(a)$ shows spectra with $N_{\text{band2}}$ = 0, and $(b)$ shows spectra with $N_{\text{band2}}$ = 1.
In $(c)$ and $(d)$ we show the orbital occupation $\braket{\gamma^\dag_{\mbf{k},\alpha}\gamma_{\mbf{k},\alpha}}$ of the ground state at $\{3,1\}$ in the ground state ``GS" and first excited state ``FE". $n=0,1,2$ corresponds to blue, yellow, and green. The spectral eigenvalues close to 1 or 0 indicate an approximately Slater state. It is likely that we have not reached a large enough ratio of $N_{\text{band1}}/N_s$ to observe the exchange of $\Ch=1$ and $\Ch=0$. }
\label{fig:nu=1_24}
\end{figure}

A summary of our results for $\nu=1$ can be found in Figs.~\ref{fig:nu=1_12}$-$\ref{fig:nu=1_24} for $12,15,18$, and $24$ sites respectively. We find evidence for convergence in the $N_{\text{band1}}$ truncation parameter with $N_{\text{band2}}$ = 0. Recall that for $N_{\text{band1}}=N_{\text{band2}}=0$, the only Fock state in the Hilbert space at $\nu=1$ is the $\Ch=1$ HF state. For all system sizes, we find that the gap above the $\Ch=1$ state in ED does not close as particles are allowed to populate band 1.
However, the $N_{\text{band2}}$=1 sequence appears to indicate a gap closing where the ground state transitions from $\Ch=1$ to $\Ch=0$. In all cases, we diagnose the Chern number by looking at the orbital occupation $\braket{\gamma^\dag_{\mbf{k},\alpha}\gamma_{\mbf{k},\alpha}}$ in the HF basis. For a $\Ch=1$ state, we require the occupation in band 0 to be greater than that in band 1 and band 2 for all $\mbf{k}$. If the $K_M$ point has greater occupation in band 1 than that in band 0 and band 2, but the occupation is largest in band 0 for all other $\mbf{k}$, then the state is considered to be $\Ch=0$. As shown in Figs.~\ref{fig:nu=1_12}$-$\ref{fig:nu=1_24}, this is always the case except for the $K_M$ point where a band inversion with HF band 1 can occur. If the $K_M$ point has occupation greater than $66\%$ in HF band 1, but the occupation is greater than $66\%$ for all other $\mbf{k}$, then the state is considered to be $\Ch=0$.

This is a numerically efficient way of determining the Chern number from a single wavefunction which is valid in weakly correlated states. We observe that the $\Ch=1$ to $\Ch=0$ transition occurs for $N_{\text{band1}}/N_s = 3/12 = .25, 4/15=.266, 5/18 = .277$ on $12,15,18$ sites respectively. On 21 and 24 sites, we are not able to reach similar ratios for $N_{\text{band1}}/N_s$.

Close competition between a $\Ch=1$ and $\Ch=0$ state has also observed in earlier HF studies~\cite{guo2023theory,kwan2023MFCI3}, but has not been systemtically investigated. On large sizes ($\geq 6\times 6$ lattices), the $\Ch=1$ state prevails in HF (see discussion in \App{app:2D_Int_HF}). However on small sizes accessible in ED, the $\Ch=0$ state is lower energy.

\subsection{$\nu=2/3$: HF basis and Extrapolation}

\begin{figure}[ht!]
\centering
\includegraphics[width=\columnwidth]{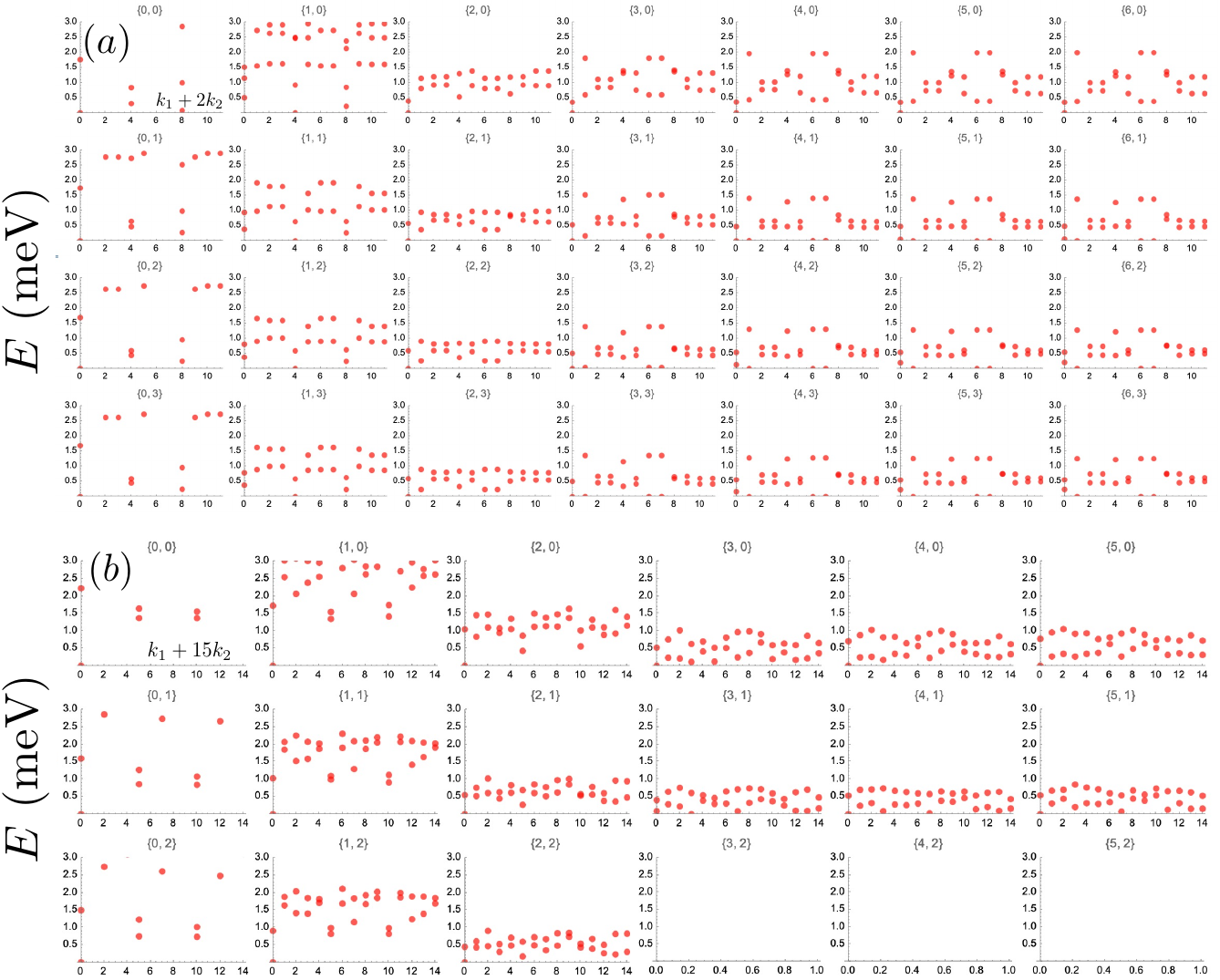}
\caption{We show $\nu=2/3$ ED spectra for the 2D interaction in the AVE scheme for the $2\times 6$ lattice in $(a)$ and $15\times 1$ lattice in $(b)$ at $V=22$meV. No good FCI states develop even at $\{0,0\}$ (although the spread/gap ratio is better on 12 than 15 sites), and the continuum quickly falls yielding a gapless state. The ground state momenta of a gapped FCI are $0,5,10$ corresponding to $\mbf{0},  \mbf{f}_1/3, 2 \mbf{f}_1/3$ respectively, which are the $\Gamma_M$, $K_M$, and $K'_M$ points. At band-max $\{3,1\}$ for instance, the groundstate sectors are $0,3,5$, incompatible with an FCI.}
\label{fig:nu=2/3_12}
\end{figure}

\begin{figure}
\centering
\includegraphics[width=.9\columnwidth]{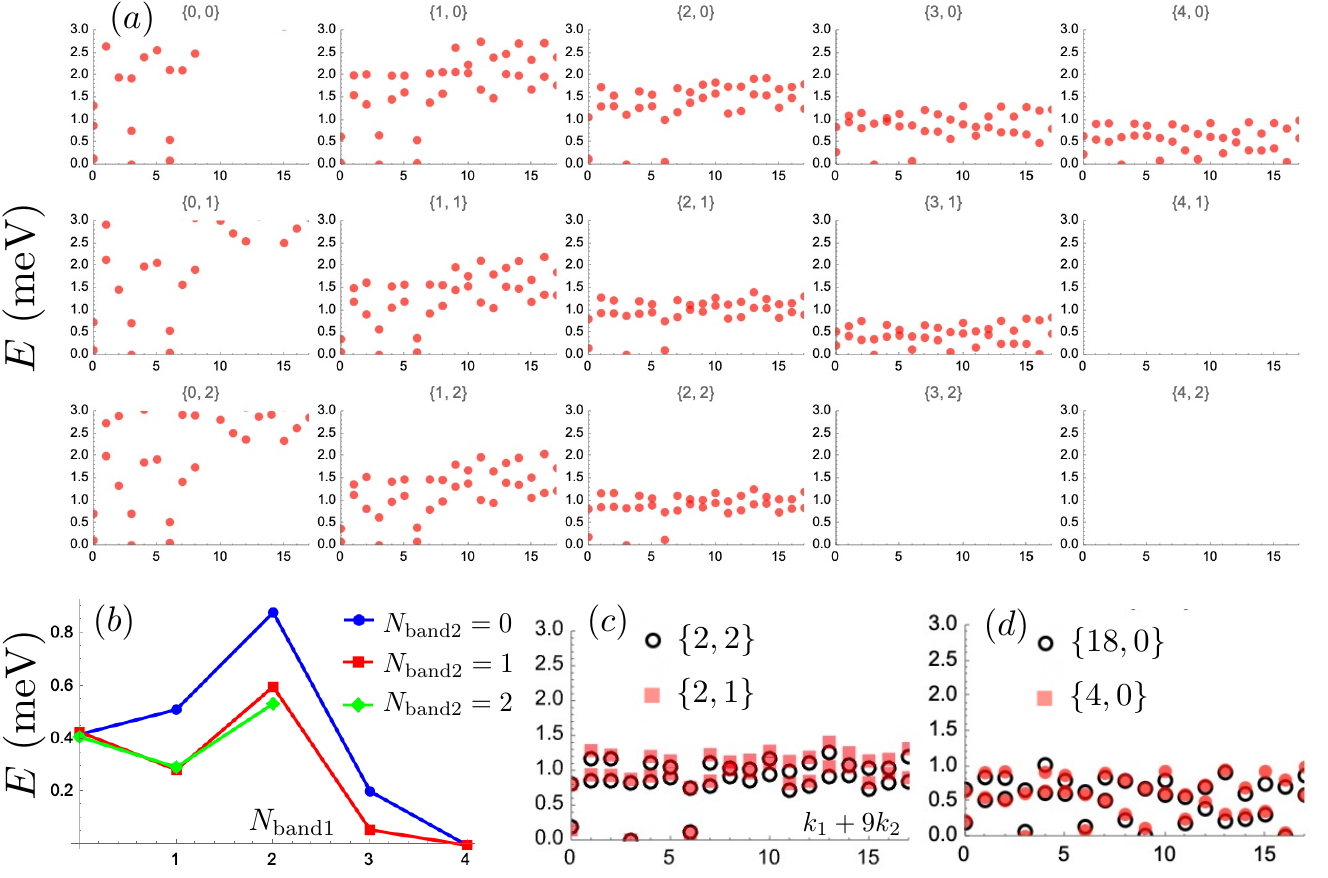}
\caption{Convergence of band-max with the 2D AVE scheme at $V=22$\,meV on $9\times 2$ at $\nu=2/3$. $(a)$ Summary of ED spectra. $(b)$ Behavior of FCI gap with band-max. The $N_{\text{band2}}=0$ (blue) and $N_{\text{band2}}=1$ (red) sequences show the gap closing at  $N_{\text{band1}}=4$. The $N_{\text{band2}}=1$ (red) and $N_{\text{band2}}=2$ (green) sequences converge well for all $N_{\text{band1}}$. This allows us to limit our consideration to $N_{\text{band2}}=1$. $(c)$ Full spectra comparing band-max $\{2,1\}$ and band-max $\{2,2\}$ showing good agreement. $(d)$ Full spectra comparing band-max $\{18,0\}$ and band-max $\{4,0\}$ showing good agreement. }
\label{fig:nu=2/3}
\end{figure}

\begin{figure}
\centering
\includegraphics[width=\columnwidth]{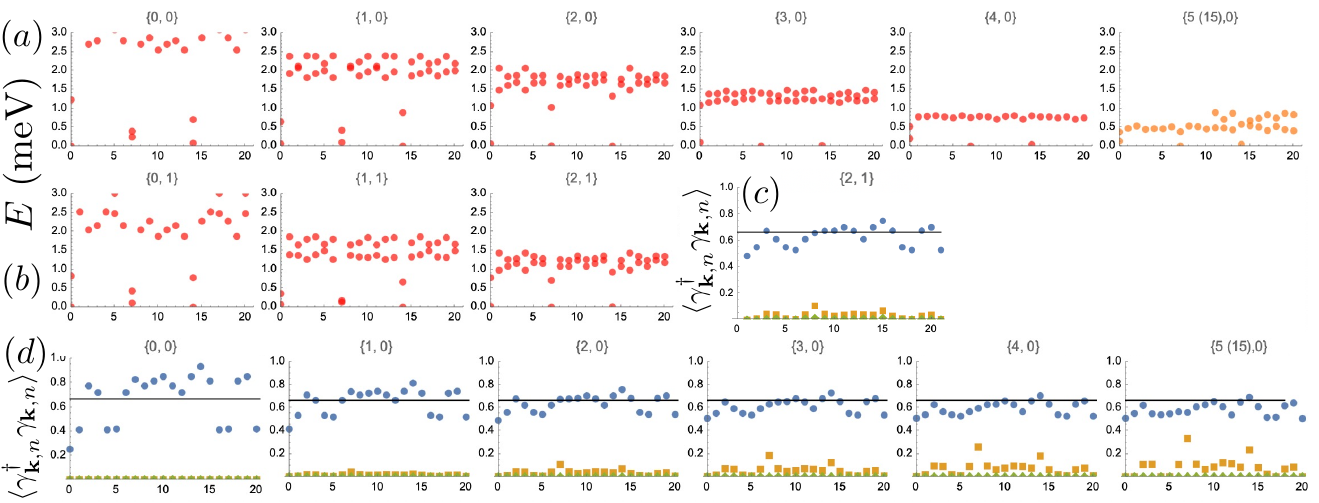}
\caption{Summary of $\nu=2/3$ ED data for the 2D interaction in the AVE scheme at $V=22$\,meV on the $21\times 1$ lattice. $(a)$ and $(b)$ show the accessible $N_{\text{band}2} =0,1$ data respectively. Orange denotes results with largest available orbital restriction, labeled $\{i \, (N),j\}$ for keeping $N$ orbitals in band 1 (we keep all orbitals in band 2). $(c)$ shows the band occupation at $\{2,1\}$ indicating a good FCI after the closing of the gap at $\{1,1\}$. $(d)$ shows the evolution of the band occupation as $N_{\text{band}1}$ is increased for $N_{\text{band2}}$. The particle occpation is most uniform at $\{2,0\}$ consistent with the largest FCI gap in $(a)$.}
\label{fig:nu=2/3_21}
\end{figure}

We summarize our results for $\nu=2/3$ in Figs.~\ref{fig:nu=2/3_12}$-$\ref{fig:nu=2/3_21}. On $N_x N_y=18$ sites, we see the FCI gap go to zero, whereas 21 and 24 sites both show nonzero gaps at the largest Hilbert spaces accessed so far for $N_{\text{band2}} = 0$. Note that results for 18 sites (\cref{fig:nu=2/3}) show very good  convergence in between $N_{\text{band1}} = 4$ and $N_{\text{band1}} = 18$ at fixed $N_{\text{band2}} = 0$. \cref{fig:nu=2/3} also demonstrates convergence between $N_{\text{band2}} = 1$ and $N_{\text{band2}} = 2$. For this reason, we primarily focus on the $N_{\text{band2}} = 0,1$ sequences.


\subsection{$\nu=1/3$: HF basis and Extrapolation}

\begin{figure}
\centering
\includegraphics[width=.9\columnwidth]{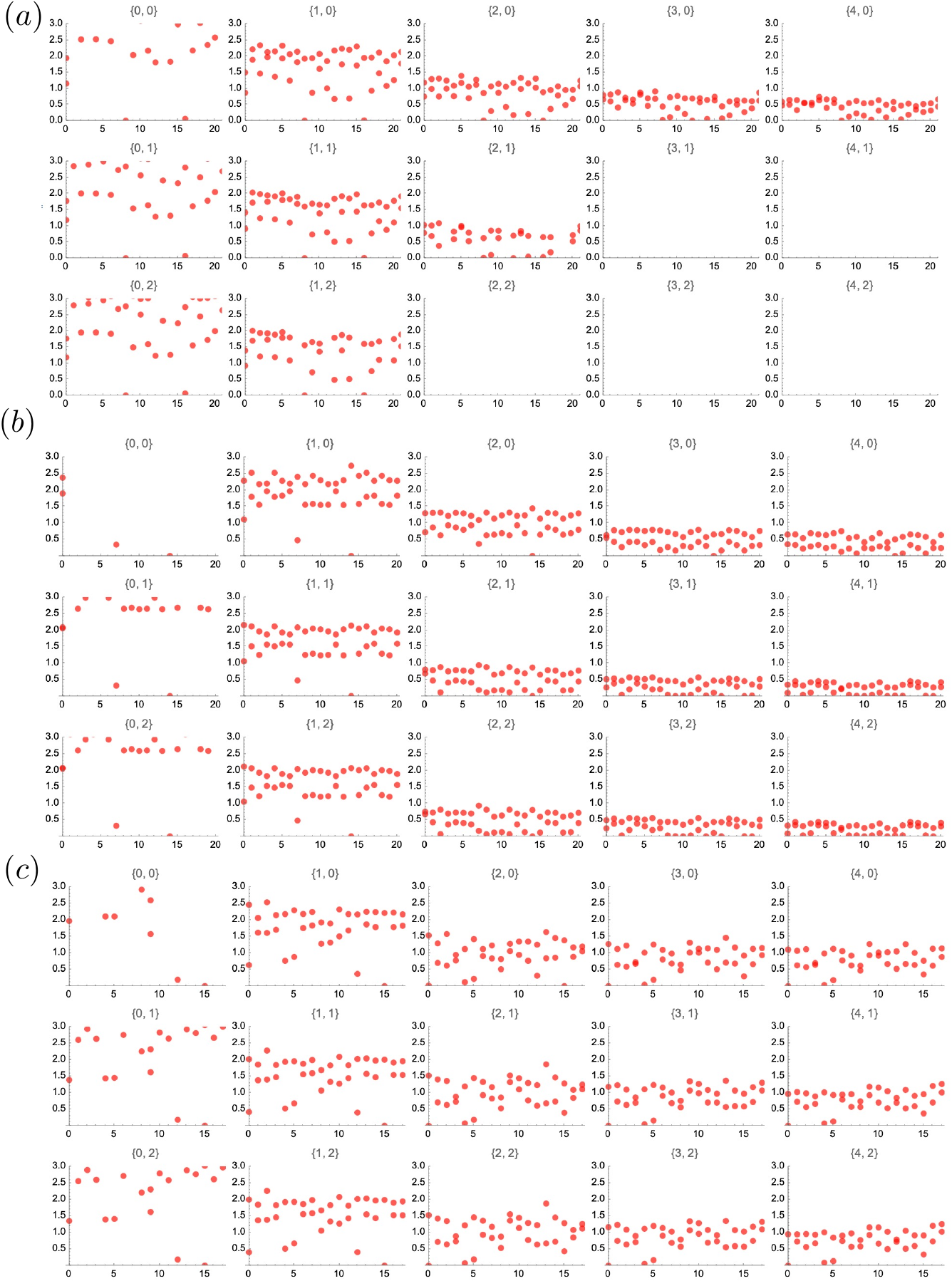}
\caption{$\nu=1/3$ ED spectra for the 2D interaction in the AVE scheme at $V=22$\,meV. $(a)$, $(b)$, $(c)$ show spectra on the $4\times 6$, $21\times 1$, and $9\times 2$ lattices respectively. In all cases, we do not find a well-formed FCI at band-max $\{0,0\}$, and we observe the continuum excitations fall as band-max is increased. }
\label{fig:nu=1/3}
\end{figure}

For $\nu=1/3$, we  find no FCIs as the Hilbert space is increased from band-max $\{0,0\}$, as illustrated in \cref{fig:nu=1/3}. This is consistent with experiments that do not observe an FCI at this filling~\cite{Lu2024fractional}. Even at band-max $\{0,0\}$ in \cref{fig:nu=1/3}, the FCI spread is larger than the gap, indicating the lack of topological degeneracy. The $\nu=1/3$ data on the largest available sizes shows the complete collapse of the continuum already for $N_{\text{band1}}=2$ (see \cref{fig:nu=1/3}a), as is corroborated by calculations on smaller sizes.

\newpage
\clearpage

\section{ED results for the 3D Interaction in the Average scheme with HF basis}

In this Appendix, we discuss our ED results with the 2D interaction in the HF basis and examine the convergence with respect to band-max. We focus on the AVE scheme throughout this section.
Specifically, we focus on the external displacement field $V=64$\,meV corresponding to the the screened displacement field $U(V)=35.554$\,meV.
Although the Hilbert space generated from $U(V)=35.554$\,meV is different from the Hilbert space used for the 2D interaction with $V=22$meV, we will show that the results for the 3D interaction are consistent with those for the 2D interaction.

\begin{figure}[ht!]
\centering
\includegraphics[width=\columnwidth]{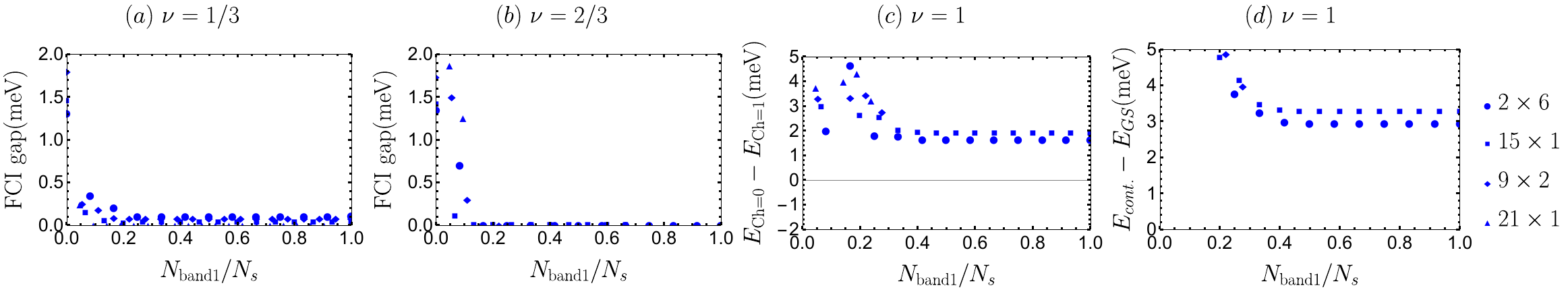}
\caption{
We show the scaling behavior (as a function of $N_{\text{band1}}/N_s$) of the many-body gap for FCI at $\nu=1/3$ in (a), the many-body gap  for FCI at $\nu=2/3$ in (b), and the many-body gap between the $\Ch=0$ and $\Ch=1$ states at $\nu=1$ in the HF momentum sector in $(c)$.
This is for $N_{\text{band2}} = 0$.
The calculation is done for the 3D interaction in the AVE scheme in the HF basis.
At $\nu=1/3$ and $\nu=2/3$, if a system has no FCI, the gap is set to zero.
}
\label{fig:3D_Int_ED_summary}
\end{figure}

\begin{figure}[ht!]
\centering
\includegraphics[width=\columnwidth]{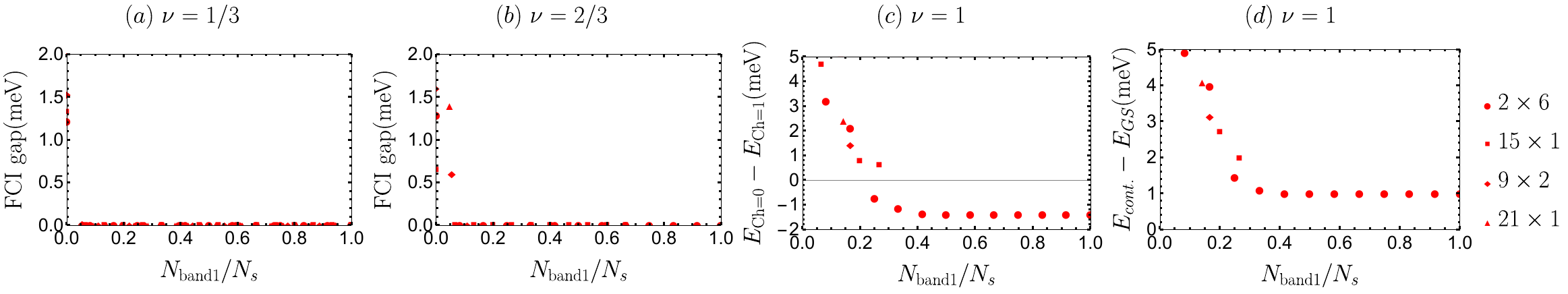}
\caption{
We show the scaling behavior (as a function of $N_{\text{band1}}/N_s$) of the many-body gap for FCI at $\nu=1/3$ in (a), the many-body gap  for FCI at $\nu=2/3$ in (b), and the many-body gap between the $\Ch=0$ and $\Ch=1$ states at $\nu=1$ in the HF momentum sector in $(c)$.
This is for $N_{\text{band2}} = 1$.
The calculation is done for the 3D interaction in the AVE scheme in the HF basis.
At $\nu=1/3$ and $\nu=2/3$, if a system has no FCI, the gap is set to zero.
}
\label{fig:3D_Int_ED_summary2}
\end{figure}

\begin{figure}[ht!]
\centering
\includegraphics[width=\columnwidth]{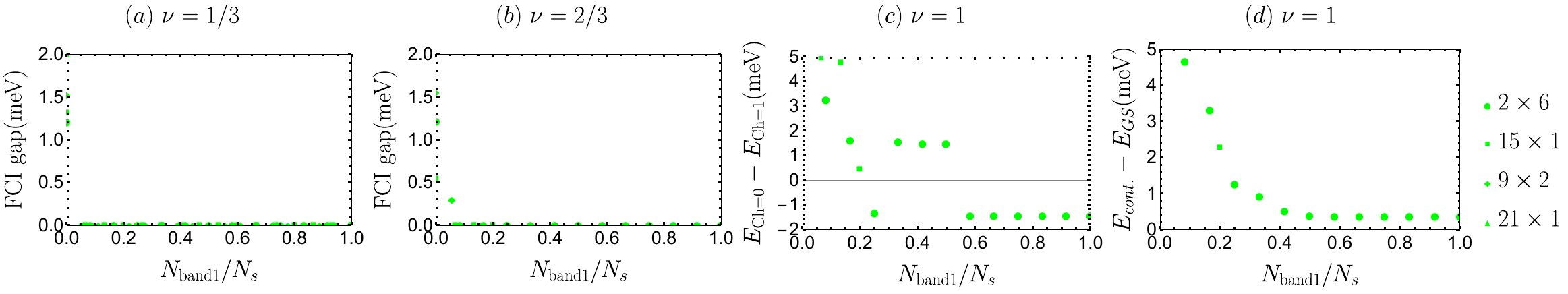}
\caption{
We show the scaling behavior (as a function of $N_{\text{band1}}/N_s$) of the many-body gap for FCI at $\nu=1/3$ in (a), the many-body gap  for FCI at $\nu=2/3$ in (b), and the many-body gap between the $\Ch=0$ and $\Ch=1$ states at $\nu=1$ in the HF momentum sector in $(c)$.
This is for $N_{\text{band2}} = 2$.
The calculation is done for the 3D interaction in the AVE scheme in the HF basis.
At $\nu=1/3$ and $\nu=2/3$, if a system has no FCI, the gap is set to zero.
}
\label{fig:3D_Int_ED_summary3}
\end{figure}

\begin{figure}[ht!]
\centering
\includegraphics[width=\columnwidth]{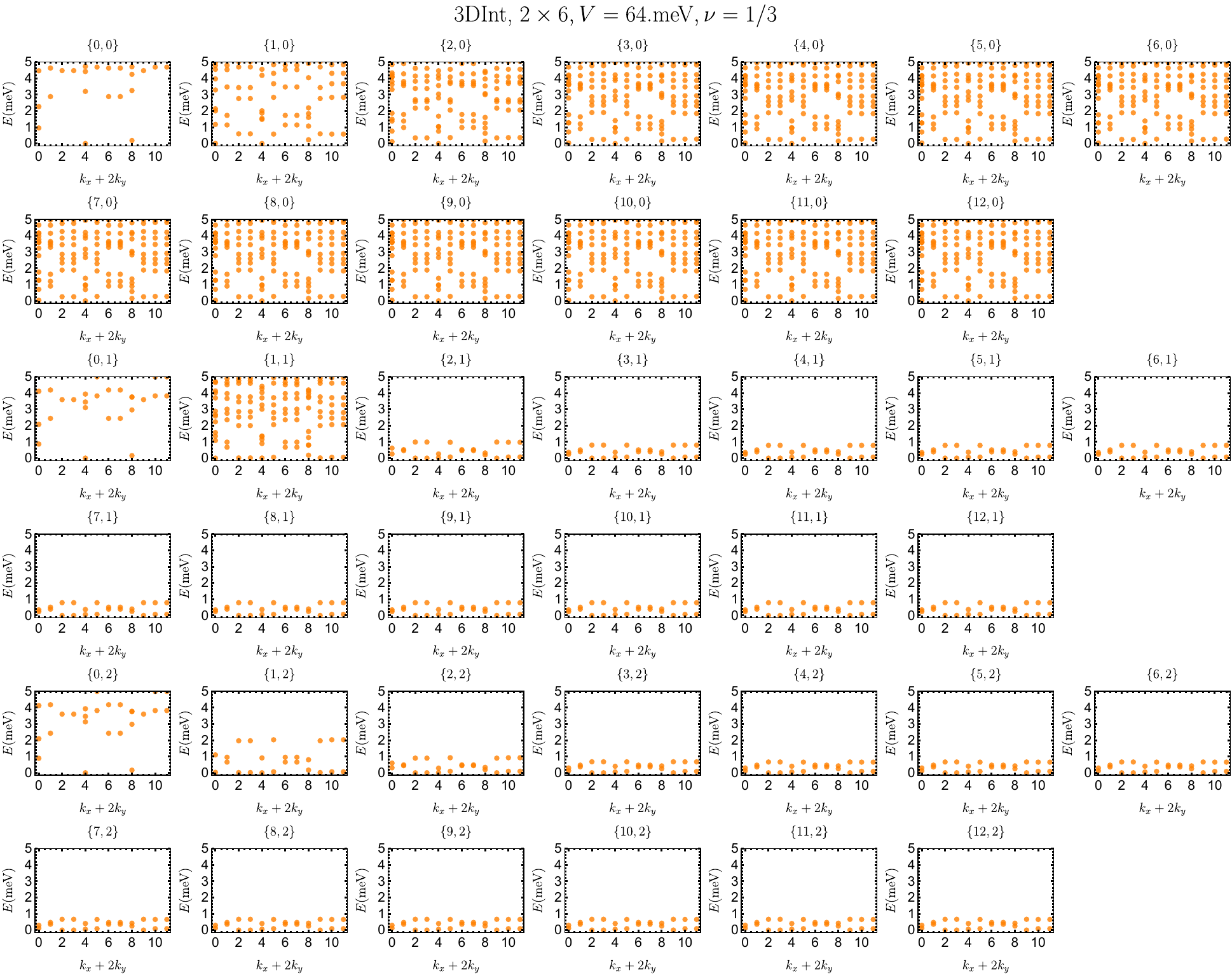}
\caption{Summary of HF-basis  $\nu=1/3$ ED spectrum for 12 sites and for the 3D interaction in the AVE scheme.}
\label{fig:HFbasis_3DInt_2x6_V64_n4}
\end{figure}

\begin{figure}[ht!]
\centering
\includegraphics[width=\columnwidth]{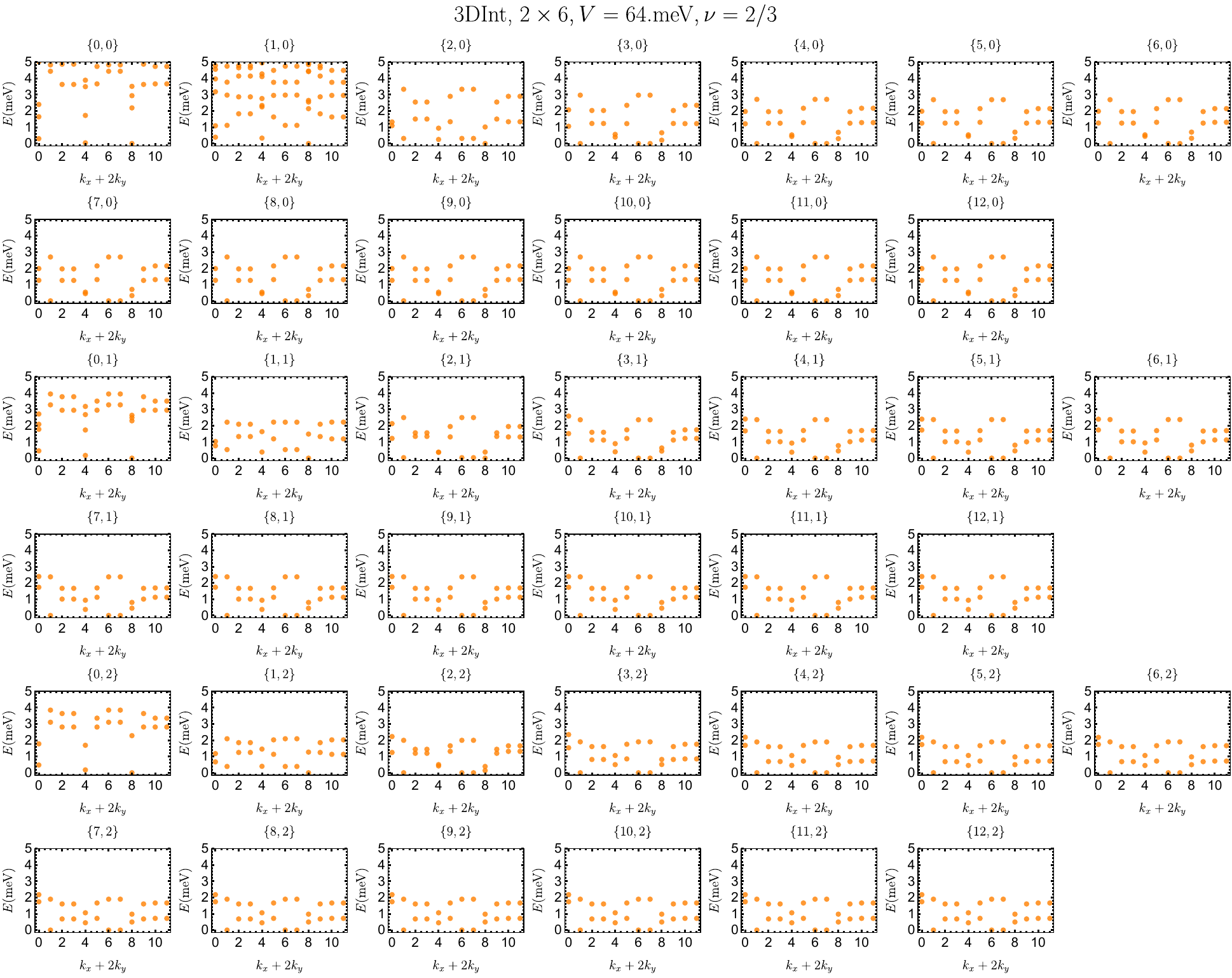}
\caption{Summary of HF-basis  $\nu=2/3$ ED spectrum for 12 sites and for the 3D interaction in the AVE scheme.}
\label{fig:HFbasis_3DInt_2x6_V64_n8}
\end{figure}

\begin{figure}[ht!]
\centering
\includegraphics[width=\columnwidth]{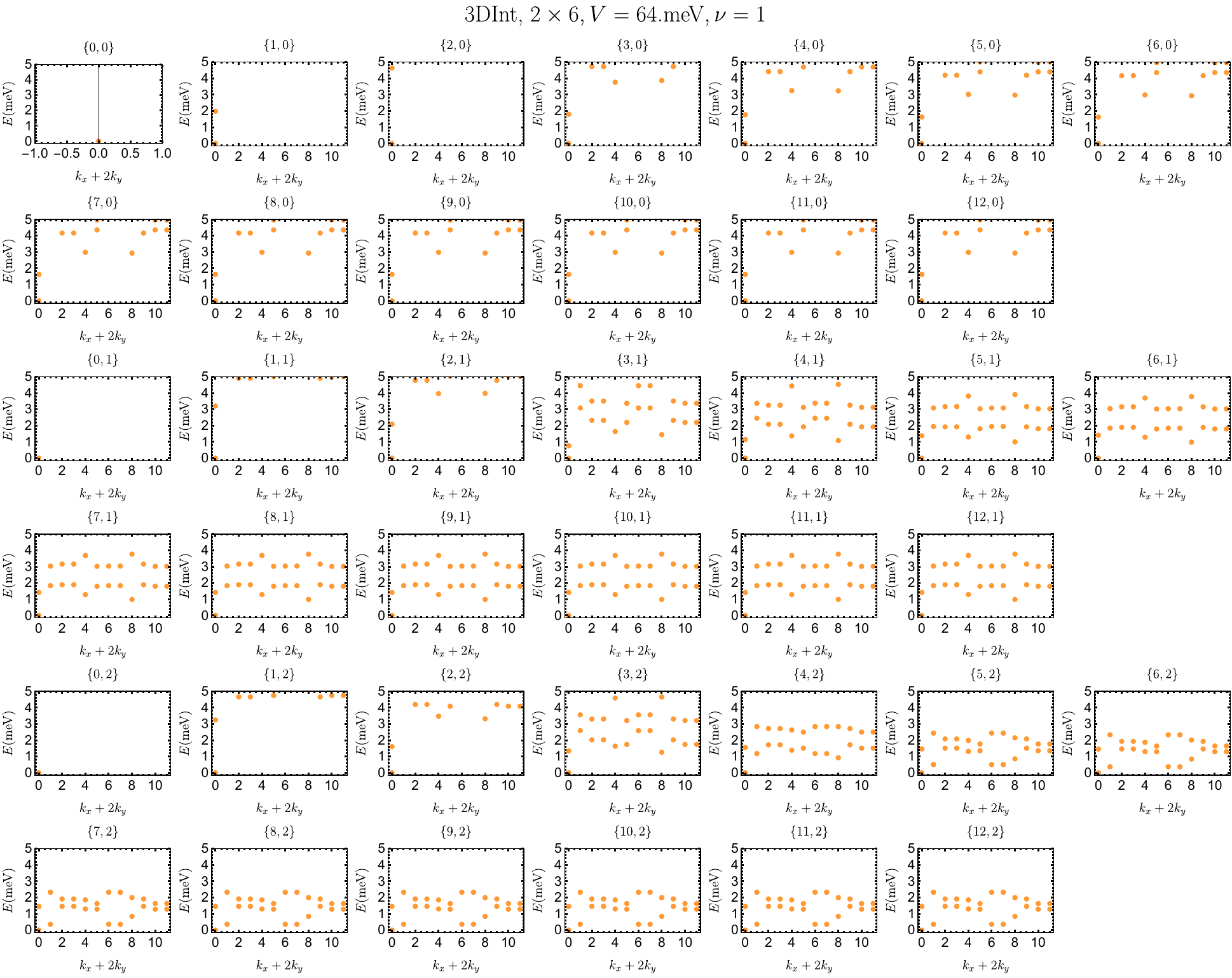}
\caption{Summary of HF-basis  $\nu=1$ ED spectrum for 12 sites and for the 3D interaction in the AVE scheme.}
\label{fig:HFbasis_3DInt_2x6_V64_n12}
\end{figure}

\begin{figure}[ht!]
\centering
\includegraphics[width=\columnwidth]{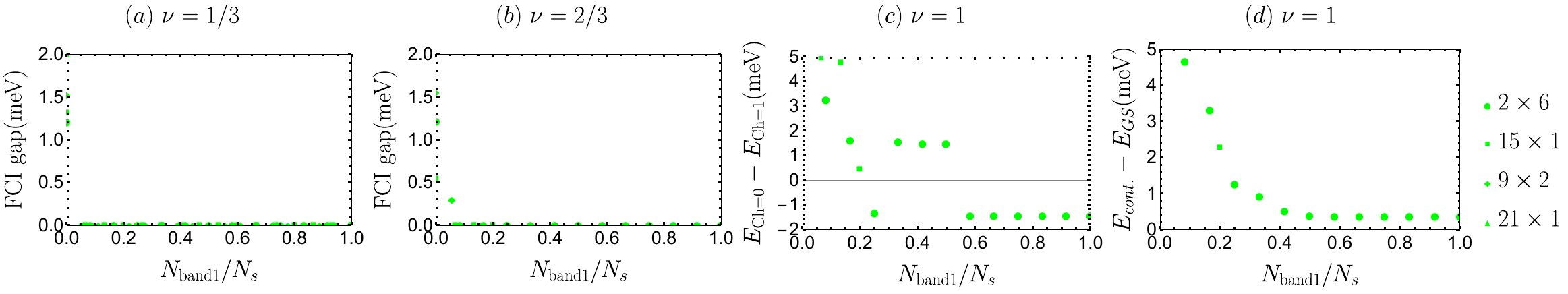}
\caption{
We show the scaling behavior (as a function of $N_{\text{band1}}/N_s$) of the many-body gap for FCI at $\nu=1/3$ in (a), the many-body gap  for FCI at $\nu=2/3$ in (b), and the many-body gap between the $\Ch=0$ and $\Ch=1$ states at $\nu=1$ in the HF momentum sector in $(c)$.
This is for $N_{\text{band2}} = 2$.
The calculation is done for the 3D interaction in the AVE scheme in the HF basis.
At $\nu=1/3$ and $\nu=2/3$, if a system has no FCI, the gap is set to zero.
}
\label{fig:3D_Int_ED_summary4}
\end{figure}

\begin{figure}[ht!]
\centering
\includegraphics[width=\columnwidth]{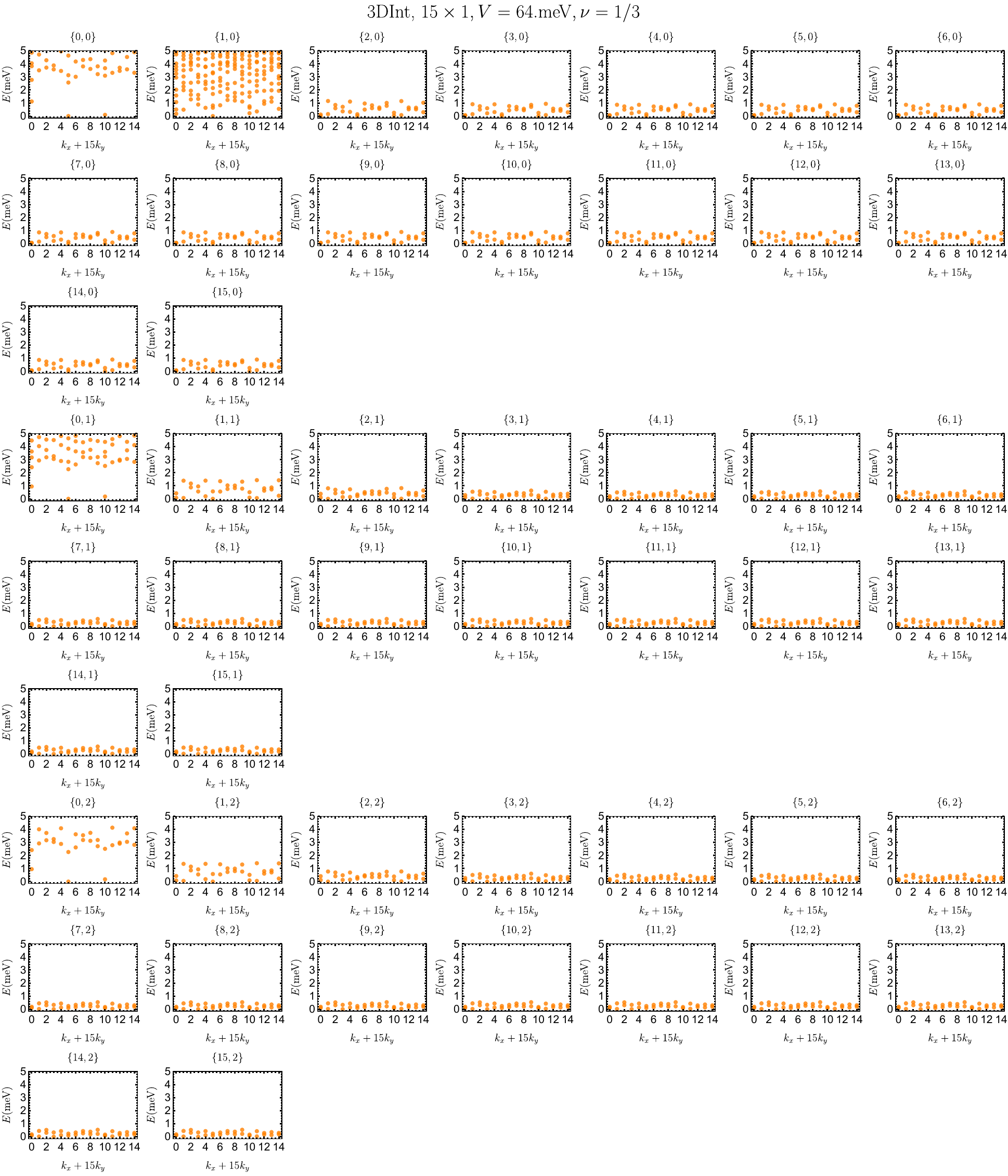}
\caption{Summary of HF-basis  $\nu=1/3$ ED spectrum for 15 sites and for the 3D interaction in the AVE scheme.}
\label{fig:HFbasis_3DInt_15x1_V64_n5}
\end{figure}

\begin{figure}[ht!]
\centering
\includegraphics[width=\columnwidth]{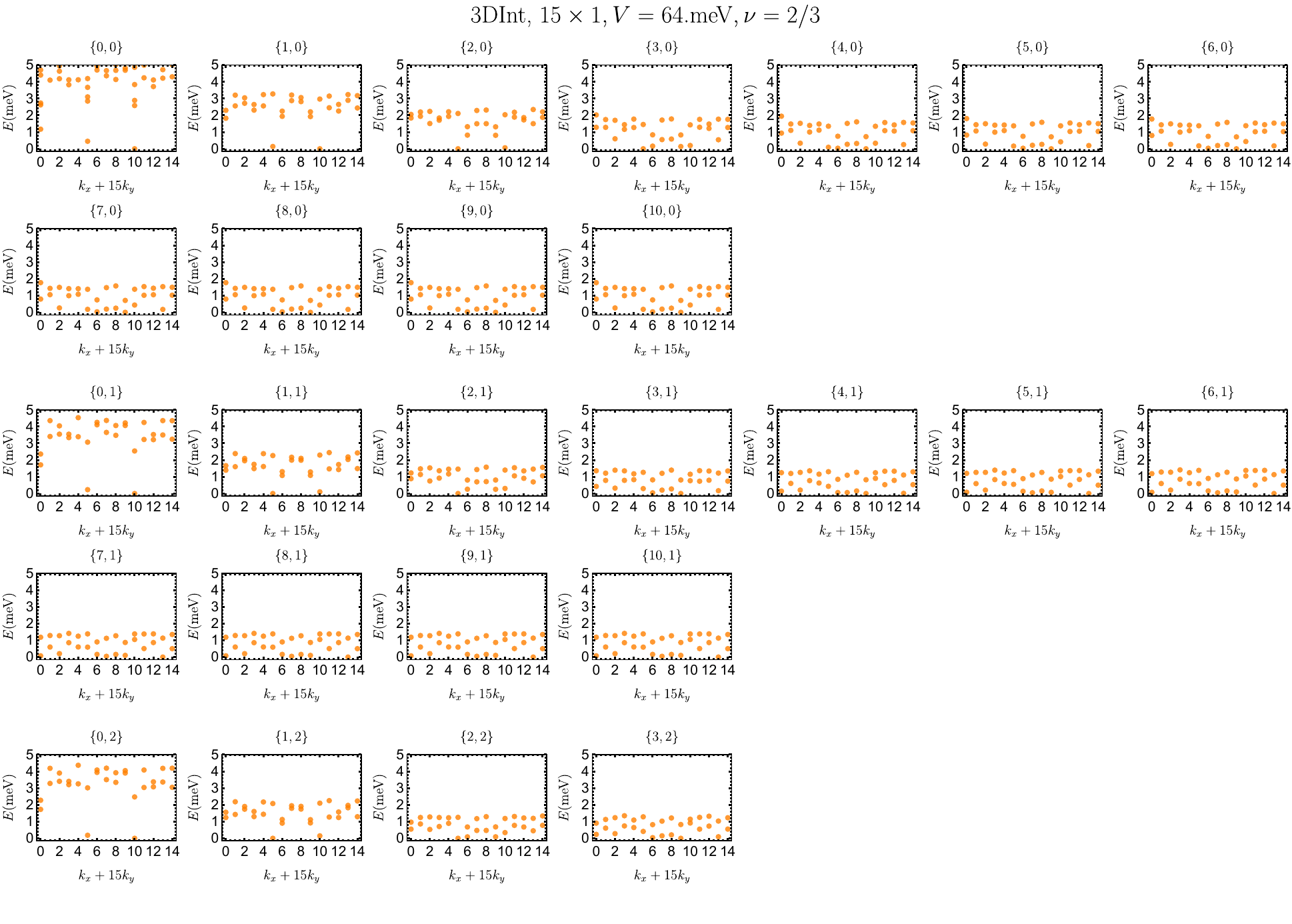}
\caption{Summary of HF-basis  $\nu=2/3$ ED spectrum for 15 sites and for the 3D interaction in the AVE scheme.}
\label{fig:HFbasis_3DInt_15x1_V64_n10}
\end{figure}

\begin{figure}[ht!]
\centering
\includegraphics[width=\columnwidth]{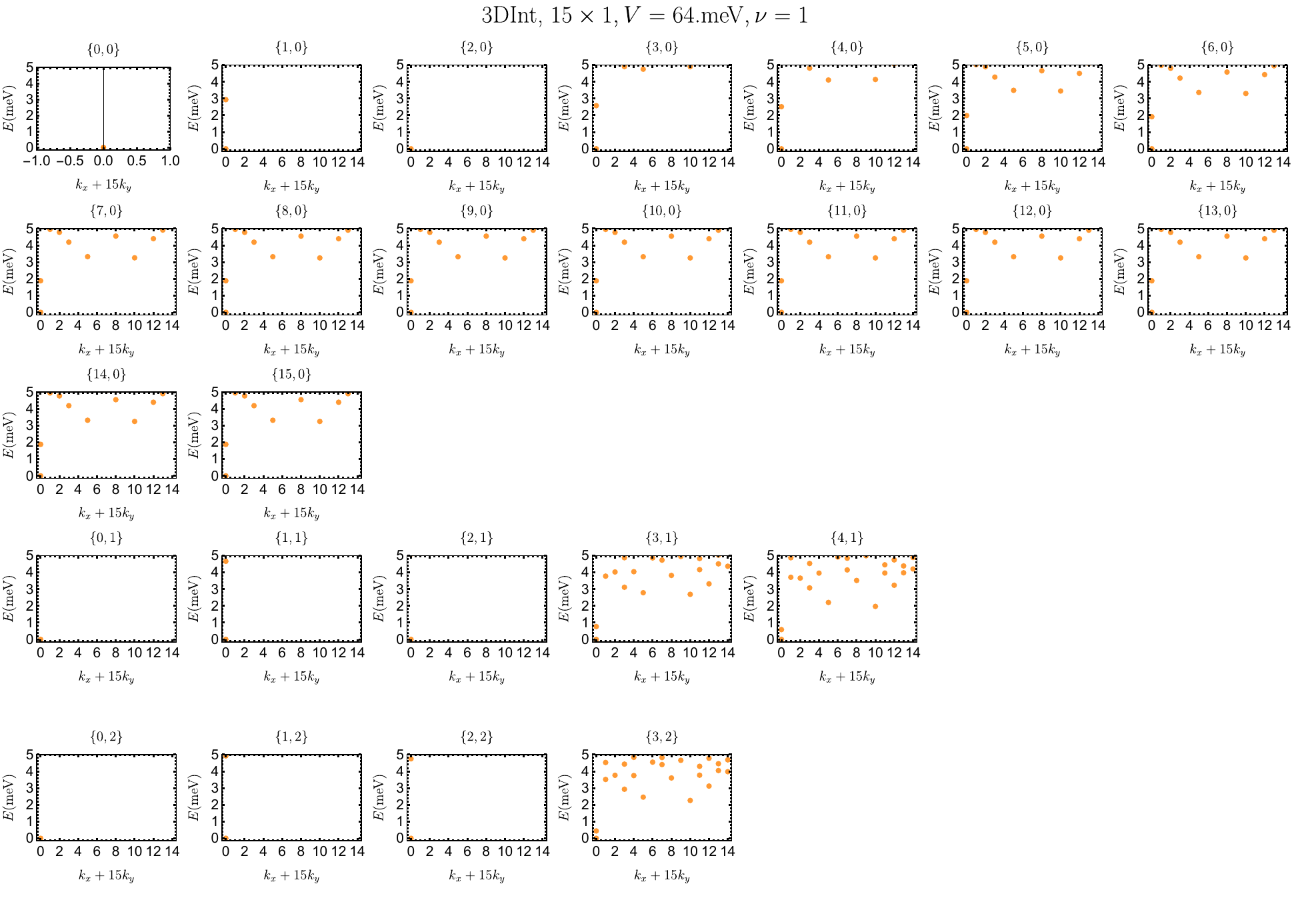}
\caption{Summary of HF-basis  $\nu=1$ ED spectrum for 15 sites and for the 3D interaction in the AVE scheme.}
\label{fig:HFbasis_3DInt_15x1_V64_n15}
\end{figure}

\begin{figure}[ht!]
\centering
\includegraphics[width=\columnwidth]{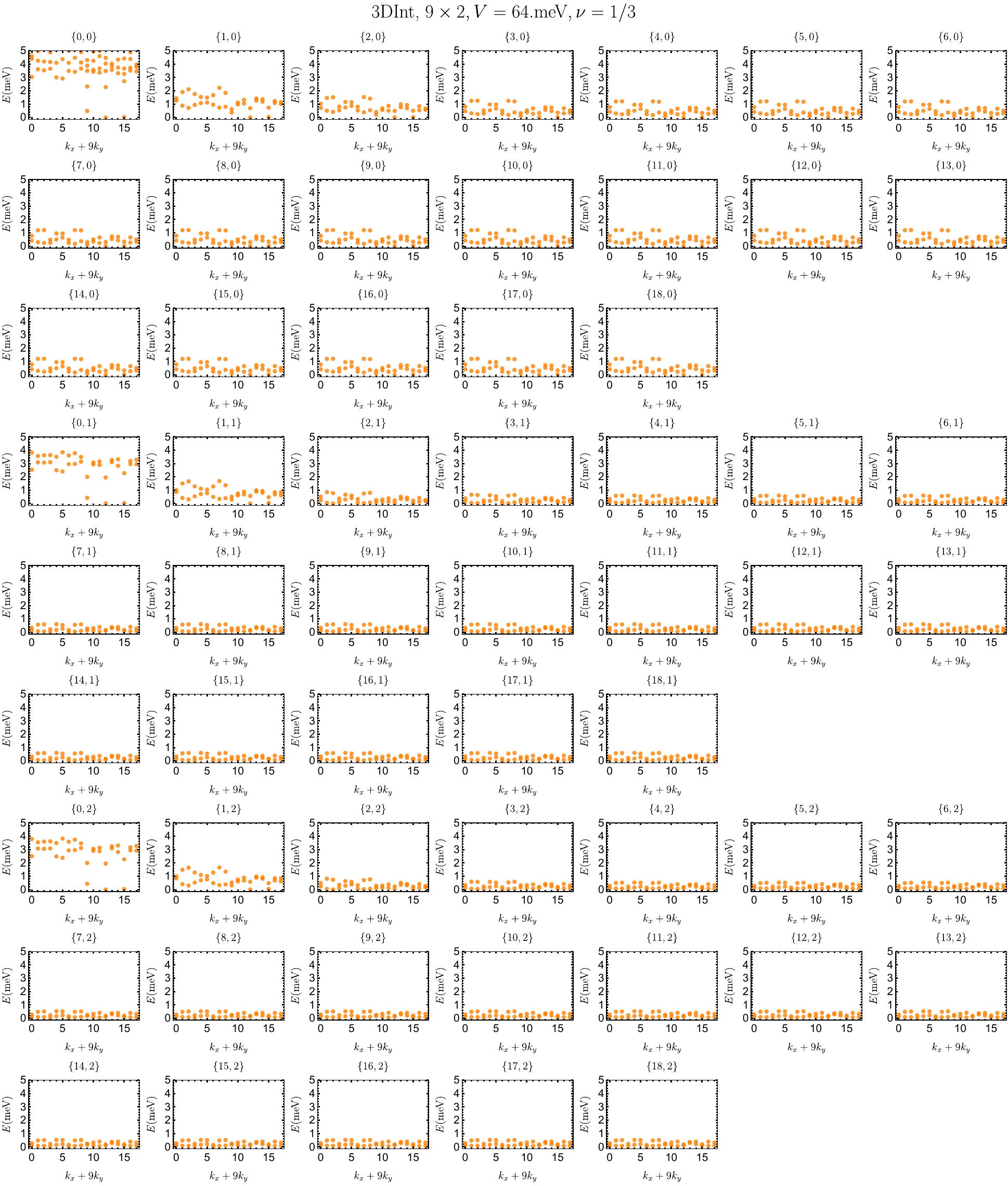}
\caption{Summary of HF-basis $\nu=1/3$ ED spectrum for 18 sites and for the 3D interaction in the AVE scheme.}
\label{fig:HFbasis_3DInt_9x2_V64_n6}
\end{figure}

\begin{figure}[ht!]
\centering
\includegraphics[width=\columnwidth]{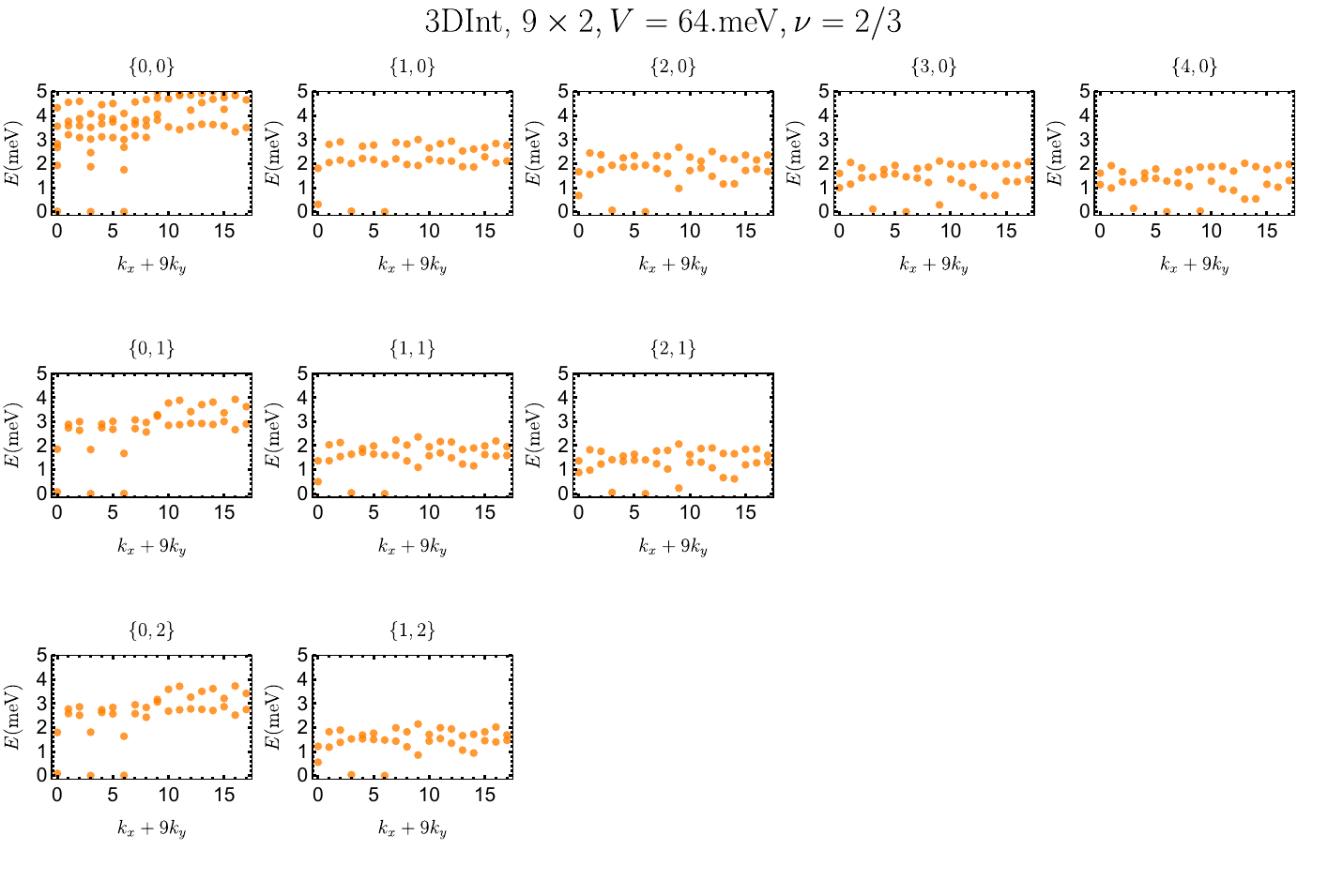}
\caption{Summary of HF-basis $\nu=2/3$ ED spectrum for 18 sites and for the 3D interaction in the AVE scheme.}
\label{fig:HFbasis_3DInt_9x2_V64_n12}
\end{figure}

\begin{figure}[ht!]
\centering
\includegraphics[width=\columnwidth]{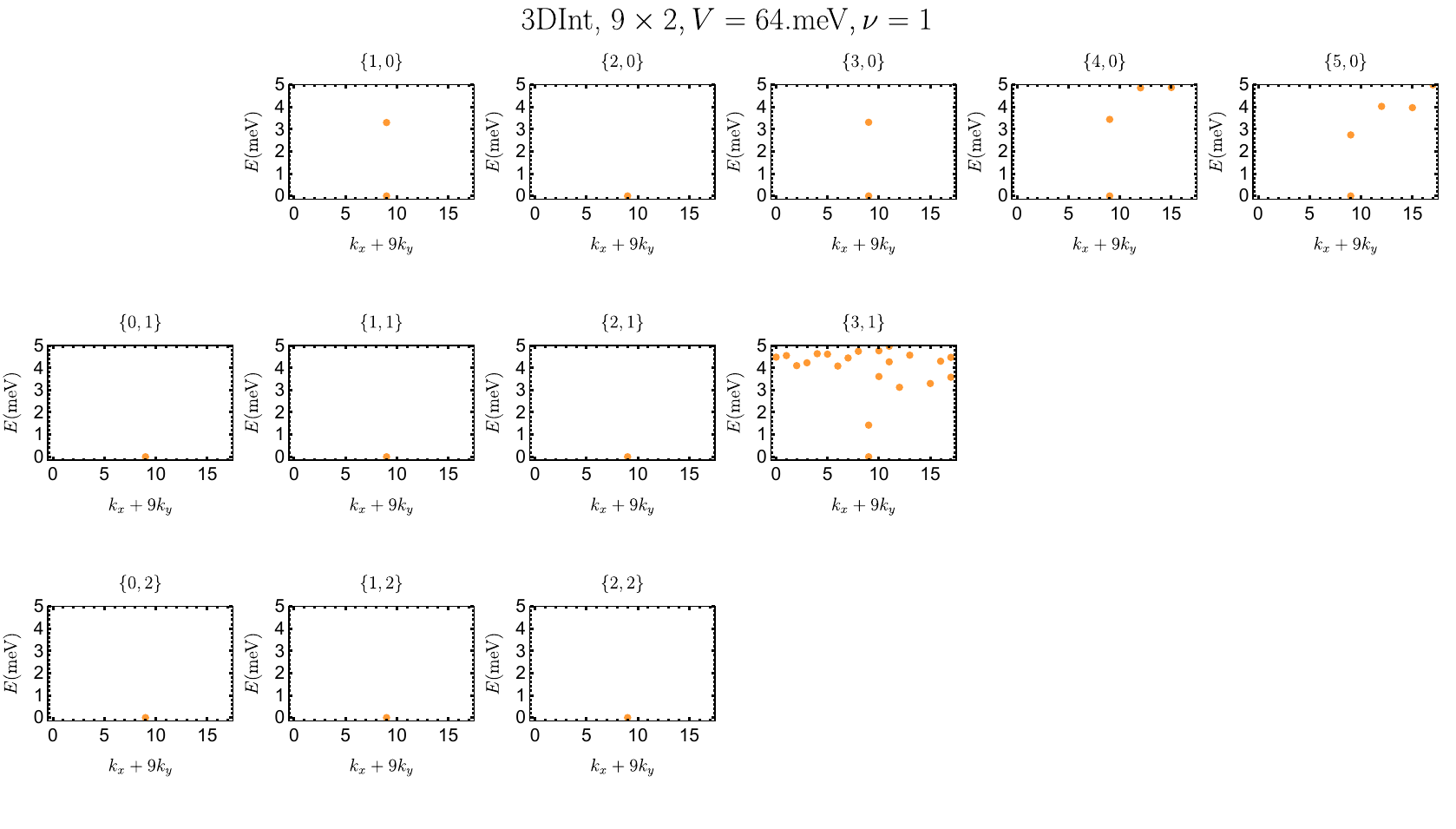}
\caption{Summary of HF-basis  $\nu=1$ ED spectrum for 18 sites and for the 3D interaction in the AVE scheme.}
\label{fig:HFbasis_3DInt_9x2_V64_n18}
\end{figure}

\begin{figure}[ht!]
\centering
\includegraphics[width=\columnwidth]{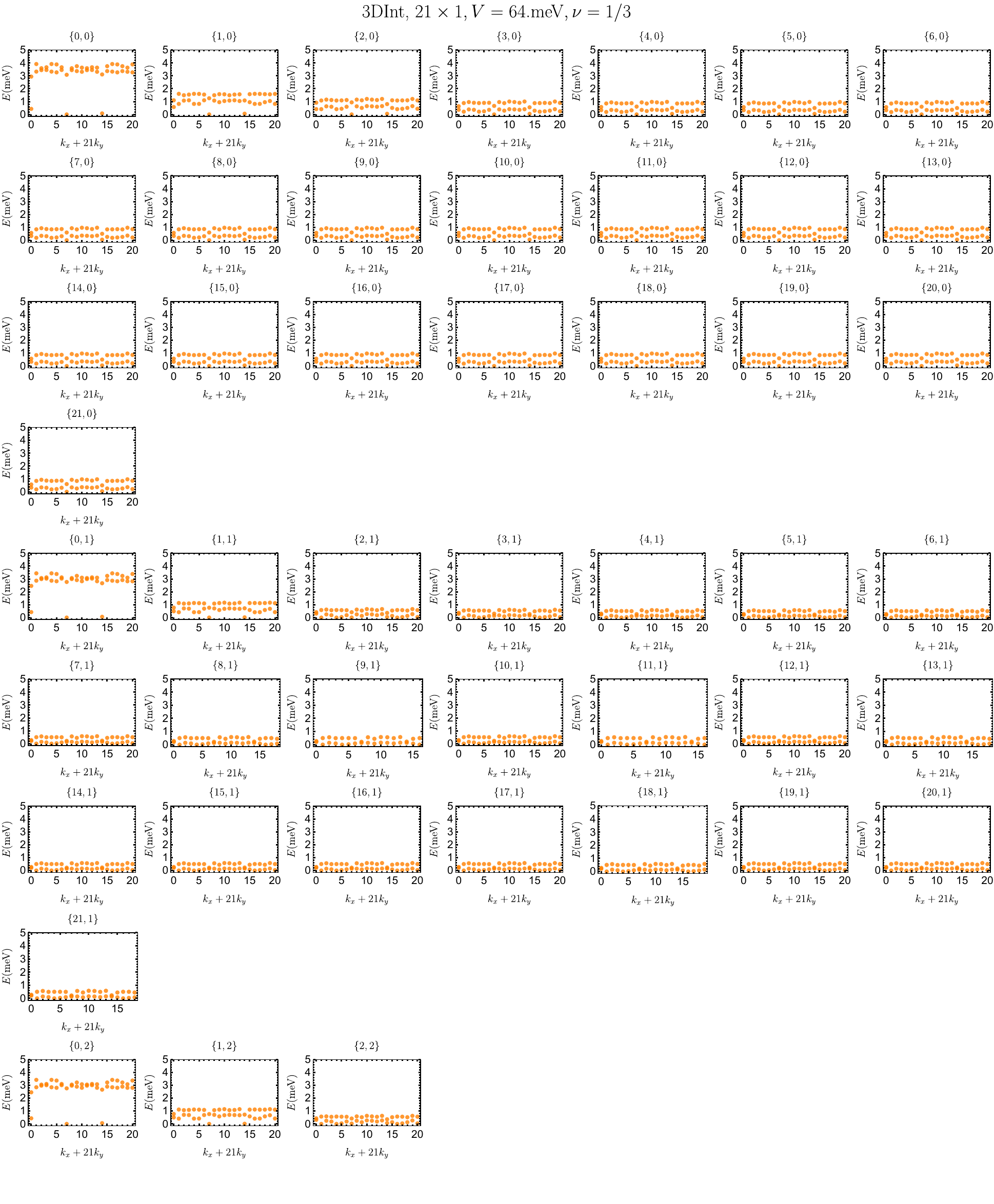}
\caption{Summary of HF-basis $\nu=1/3$ ED spectrum for 21 sites and for the 3D interaction in the AVE scheme.}
\label{fig:HFbasis_3DInt_21x1_V64_n7}
\end{figure}

\begin{figure}[ht!]
\centering
\includegraphics[width=\columnwidth]{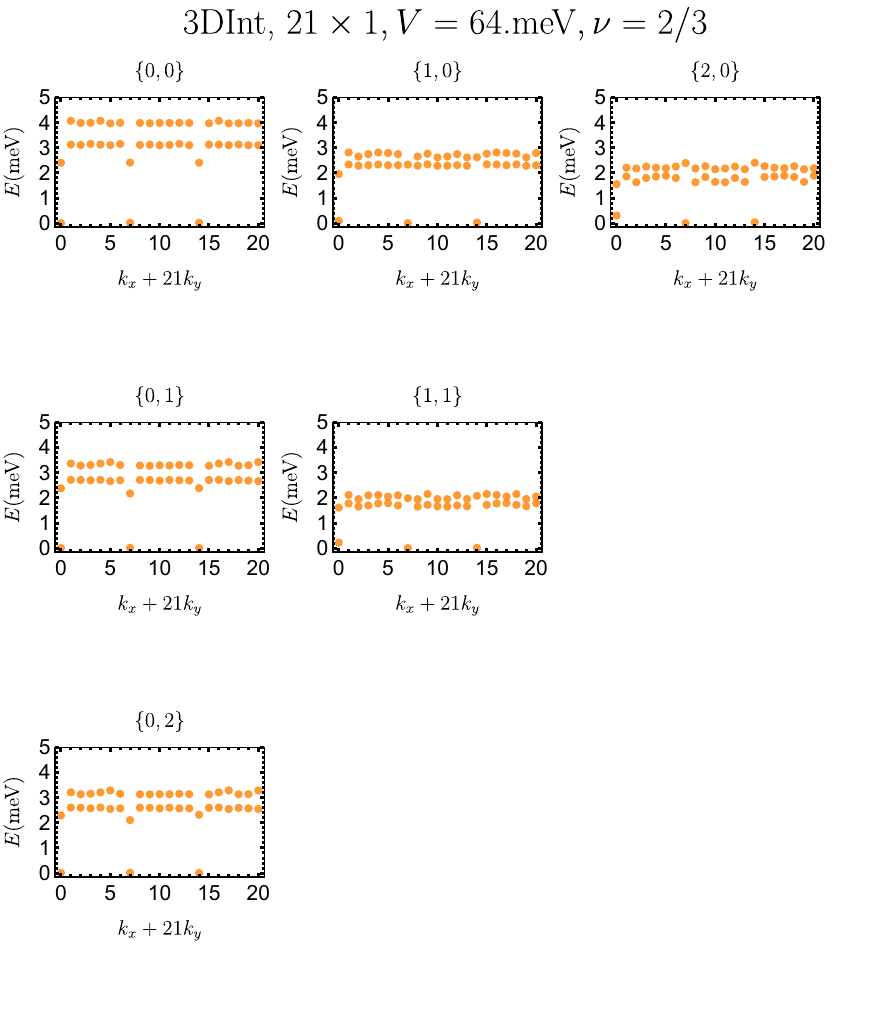}
\caption{Summary of HF-basis $\nu=2/3$ ED spectrum for 21 sites and for the 3D interaction in the AVE scheme.}
\label{fig:HFbasis_3DInt_21x1_V64_n14}
\end{figure}

\begin{figure}[ht!]
\centering
\includegraphics[width=\columnwidth]{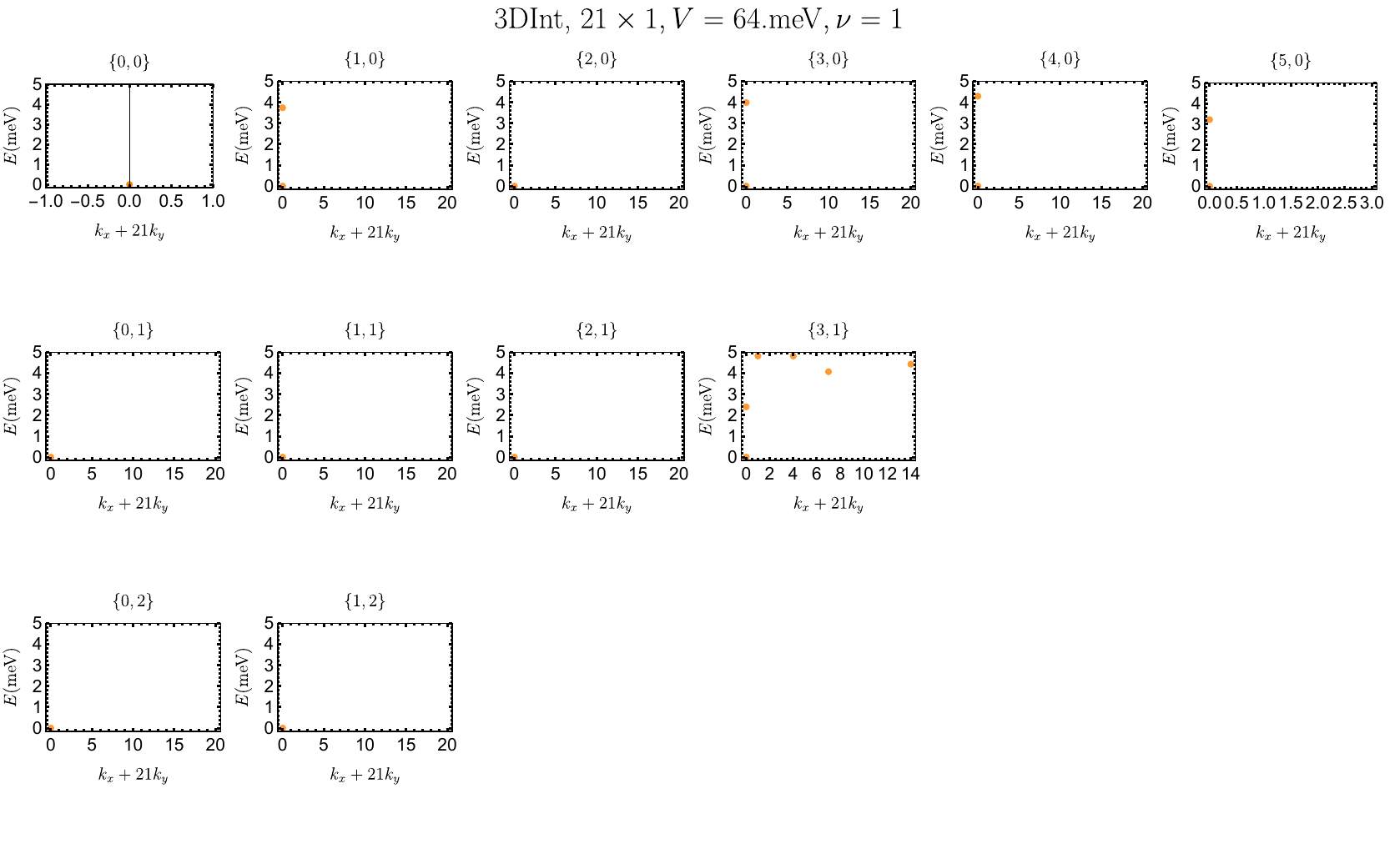}
\caption{Summary of HF-basis  $\nu=1$ ED spectrum for 21 sites and for the 3D interaction in the AVE scheme.}
\label{fig:HFbasis_3DInt_21x1_V64_n21}
\end{figure}

\subsection{ED Results at $\nu = 1$}

A summary of our results at $\nu = 1$ can be found in \cref{fig:3D_Int_ED_summary}c,d and \cref{fig:HFbasis_3DInt_2x6_V64_n12,fig:HFbasis_3DInt_15x1_V64_n15,fig:HFbasis_3DInt_9x2_V64_n18,fig:HFbasis_3DInt_21x1_V64_n21}.
We find evidence for convergence in the $N_{\text{band1}}$ truncation parameter with $N_{\text{band2}}$ = 0, which shows a converged $\Ch=1$ ground state.
However, the $N_{\text{band2}}=1,2$ sequence appears to indicate a gap closing where the ground state transitions from $\Ch=1$ to $\Ch=0$.
We observe the $\Ch=1$ to $\Ch=0$ transition occurring at $N_{\text{band1}}/N_s \in [0.2,0.3]$ on 12 sites at $N_{\text{band2}}=1$, while we have not reached similar ratios at larger system sizes, as shown in \cref{fig:3D_Int_ED_summary}c.
In all cases, we diagnose the Chern number by considering the orbital occupations $\braket{\gamma^\dag_{\mbf{k},\alpha}\gamma_{\mbf{k},\alpha}}$ in the same way as discussed in \App{app:result_2D_int_nu_1}.
Specifically, for a $\Ch=1$ state, we require the occupation in band 0 to be greater than that in band 1 and band 2 for all $\mbf{k}$.
The $\Ch=0$ states that we found have greater occupation in band 1 than that in band 0 and band 2 at $\K_M'$ or $\K_M$, while all other $\bsl{k}$ points have maximum occupation in band 0; we identify them as $\Ch=0$ states, since the spinless $C_3$ eigenvalues for band 0 at $\Gamma_M$ ($e^{-\ii 2\pi/3}$), band 0 at $\K_M$ ($1$) and band 1 at $\K_M'$ ($e^{\ii 2\pi/3}$)  give $\Ch=0\ \mod\ 3$, and so do those for band 0 at $\Gamma_M$ ($e^{-\ii 2\pi/3}$), band 1 at $\K_M$ ($e^{-\ii 2\pi/3}$) and band 0 at $\K_M'$ ($e^{-\ii 2\pi/3}$).
Nevertheless, we have \emph{not} found that the continuum drops down to destroy insulating ground state, as shown in \cref{fig:3D_Int_ED_summary}d.

\subsection{$\nu=2/3$: HF basis and Extrapolation}

We summarize our results at $\nu=2/3$ in \cref{fig:3D_Int_ED_summary}b and \cref{fig:HFbasis_3DInt_2x6_V64_n8,fig:HFbasis_3DInt_15x1_V64_n10,fig:HFbasis_3DInt_9x2_V64_n12,fig:HFbasis_3DInt_21x1_V64_n14}. For $N_{\text{band2}} = 0$, we see for 15 and 18 sites that the FCI gap goes to zero as $N_{\text{band1}}$ is increased, whereas the results for 21 sites show nonzero gaps for the values of $N_{\text{band1}}$ accessed so far.
The collapse of the $\nu=2/3$ FCI on 15 and 18 sites persists to $N_{\text{band2}} = 1,2$.

\subsection{$\nu=1/3$: HF basis and Extrapolation}

We summarize our results at $\nu=1/3$ in \cref{fig:3D_Int_ED_summary}a and \cref{fig:HFbasis_3DInt_15x1_V64_n5,fig:HFbasis_3DInt_9x2_V64_n6,fig:HFbasis_3DInt_21x1_V64_n7}. For $N_{\text{band2}} = 0$, we see for 15 sites that the FCI gap can converge to a small non-zero value, while the results for 18 and 21 sites show a collapse of the FCI as $N_{\text{band1}}$ increases.
The collapse of FCIs for 15, 18 and 21 sites is clear for $N_{\text{band2}} = 1,2$.
Compared to $\nu=2/3$ in \cref{fig:3D_Int_ED_summary}b though, the collapse of FCIs happens at smaller values of $N_{\text{band1}}$ at $\nu=1/3$ given a fixed $N_{\text{band2}}$.

\newpage
\clearpage

\section{ED results for the 2D Interaction in the Average scheme with the Biasing Method}

\label{app:result_Biasing_2D_ave}

\begin{figure}[ht!]
\centering
\includegraphics[width=\columnwidth]{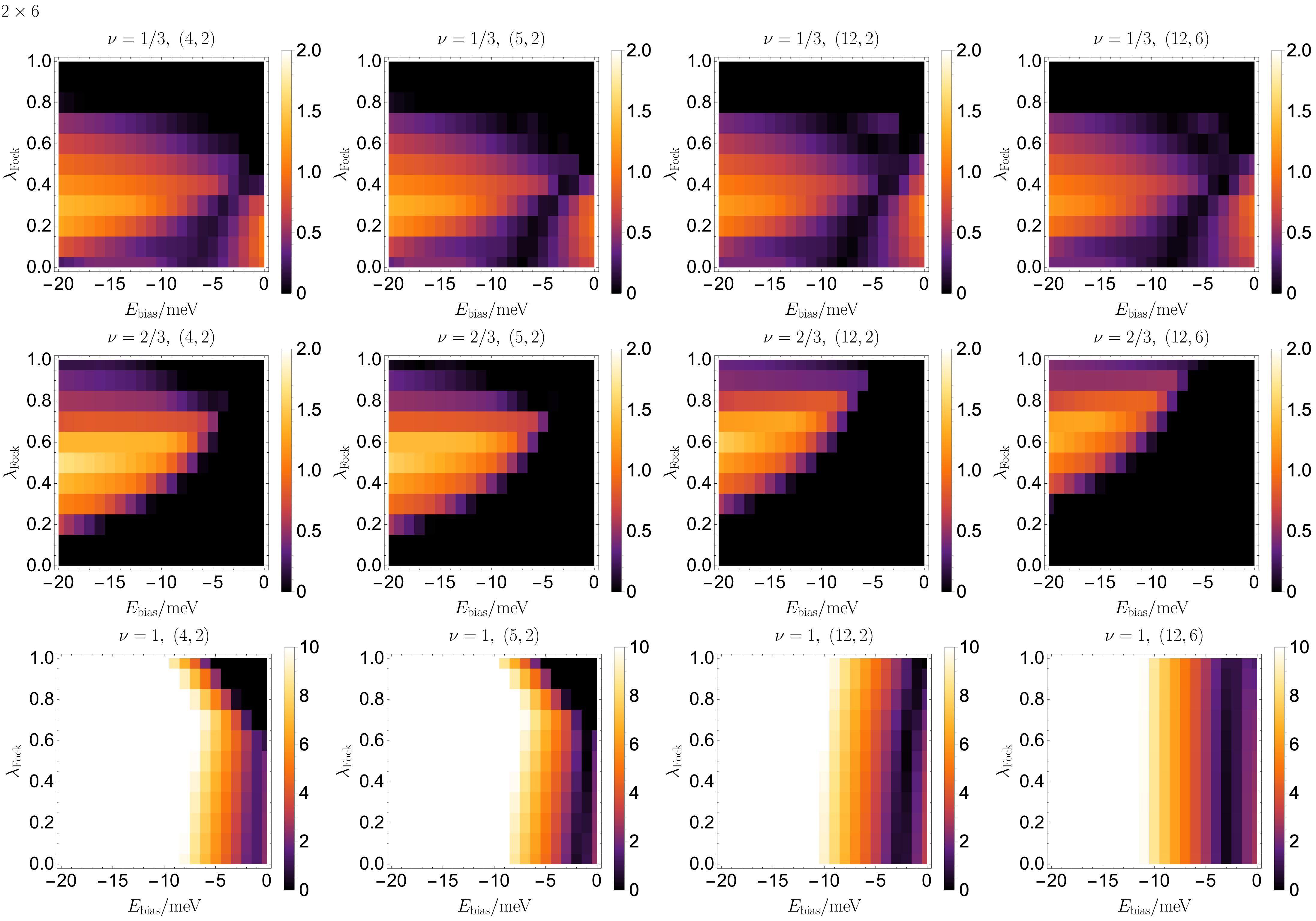}
\caption{Plots of the many-body gap as a function of $\Ebias$ and $\lfock$ (\cref{eq:HF_bias,eq:Fock_bias}) for $\nu=1/3,2/3,1$ and for the displacement field $V=22$meV.
The calculation is done with the 2D interaction and the biasing method in the AVE scheme for 12 sites, \ie, $(N_x, N_y, \widetilde{n}_{11}, \widetilde{n}_{12}, \widetilde{n}_{21}, \widetilde{n}_{22}) = (2,6,1,0,1,1)$, and the color indicates the many body gap.
We label the number of states in band 1 and band 2 in the bracket in the caption of each plot, while we always keep all states in band 0.
Here band 0, band 1 and band 2 refer to the lowest, second lowest and third lowest bands of the one-body term.
The gap at $\nu=1/3$ and $\nu=2/3$ is set to zero if any of the lowest three states are not at the FCI momenta, and the nonzero gap is given by the gap between the third and fourth lowest states in the entire spectrum.
The gap at $\nu=1$ is set to zero if the lowest state is not at the many-body momentum that corresponds to the transitionally invariant gapped state, and the nonzero gap is given by the gap between the lowest and second lowest states.
The system is always FCI at $\nu=1/3,2/3$ and CI at $\nu=1$ for $\Ebias=-20\meV$ and $\lfock=0.5$.
Therefore, the gapped regions that contains $\Ebias=-20\meV$ and $\lfock=0.5$ must correspond to the FCI/CI regions.
}
\label{fig:HFPbias_FockBGScale_Gap_x2y6}
\end{figure}

\begin{figure}[ht!]
\centering
\includegraphics[width=\columnwidth]{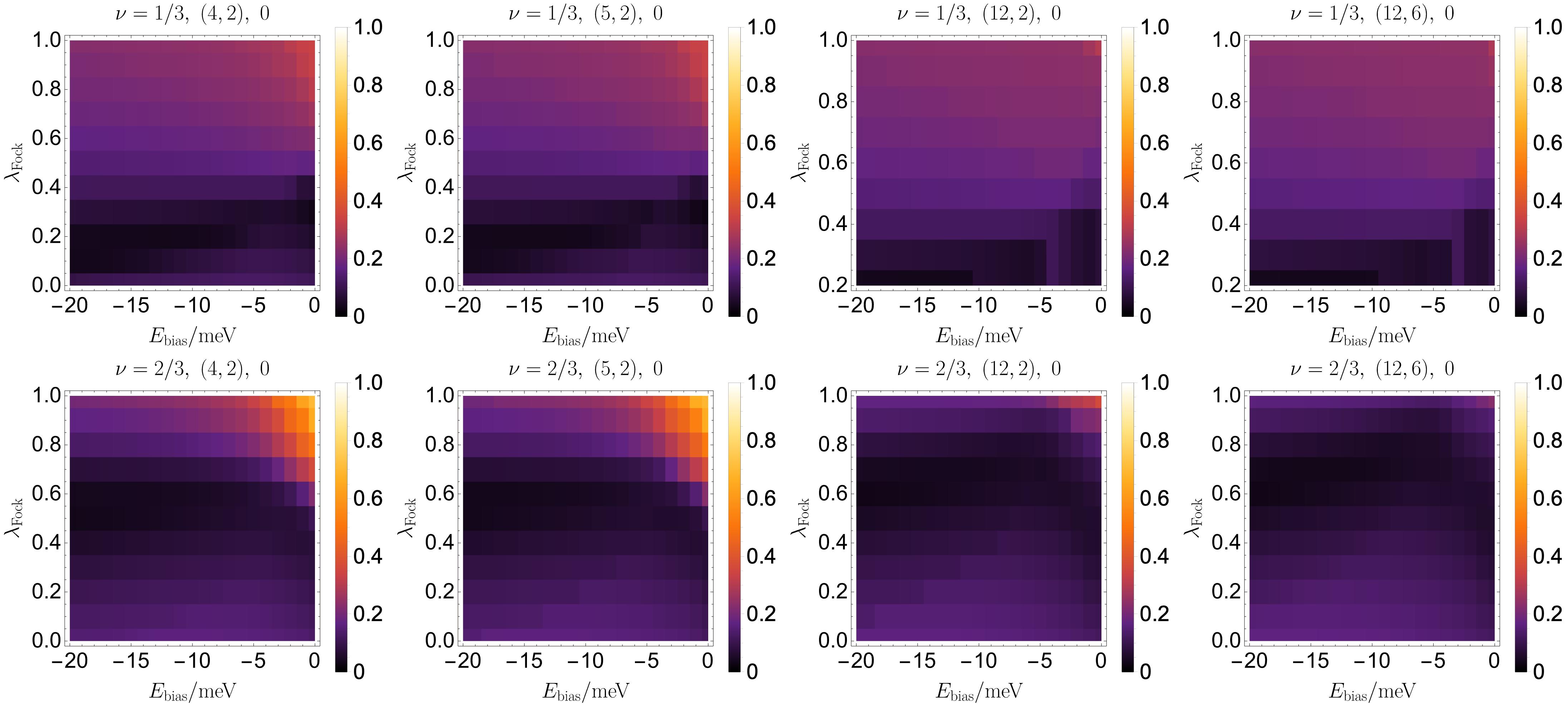}
\caption{
These plots are derived from particle density $\bsl{n}_{\bsl{k}}$ calculated for the 12 sites, \ie, $(N_x, N_y, \widetilde{n}_{11}, \widetilde{n}_{12}, \widetilde{n}_{21}, \widetilde{n}_{22}) = (2,6,1,0,1,1)$ and and for the displacement field $V=22$meV.
The plots for $\nu=1/3$ and $\nu=-2/3$ show the standard deviation of averaged $n_{\bsl{k}}$ for the three lowest (indicated by ``0" in the caption of each plot) states at the FCI momenta $(k_x,k_y)=(0,0),(0,2),(0,4)$.
The caption of each plot specifies (number of states in band 1, number of states in band 2) in parentheses
}
\label{fig:HFPbias_FockBGScale_GSnk_x2y6}
\end{figure}

\begin{figure}[ht!]
\centering
\includegraphics[width=\columnwidth]{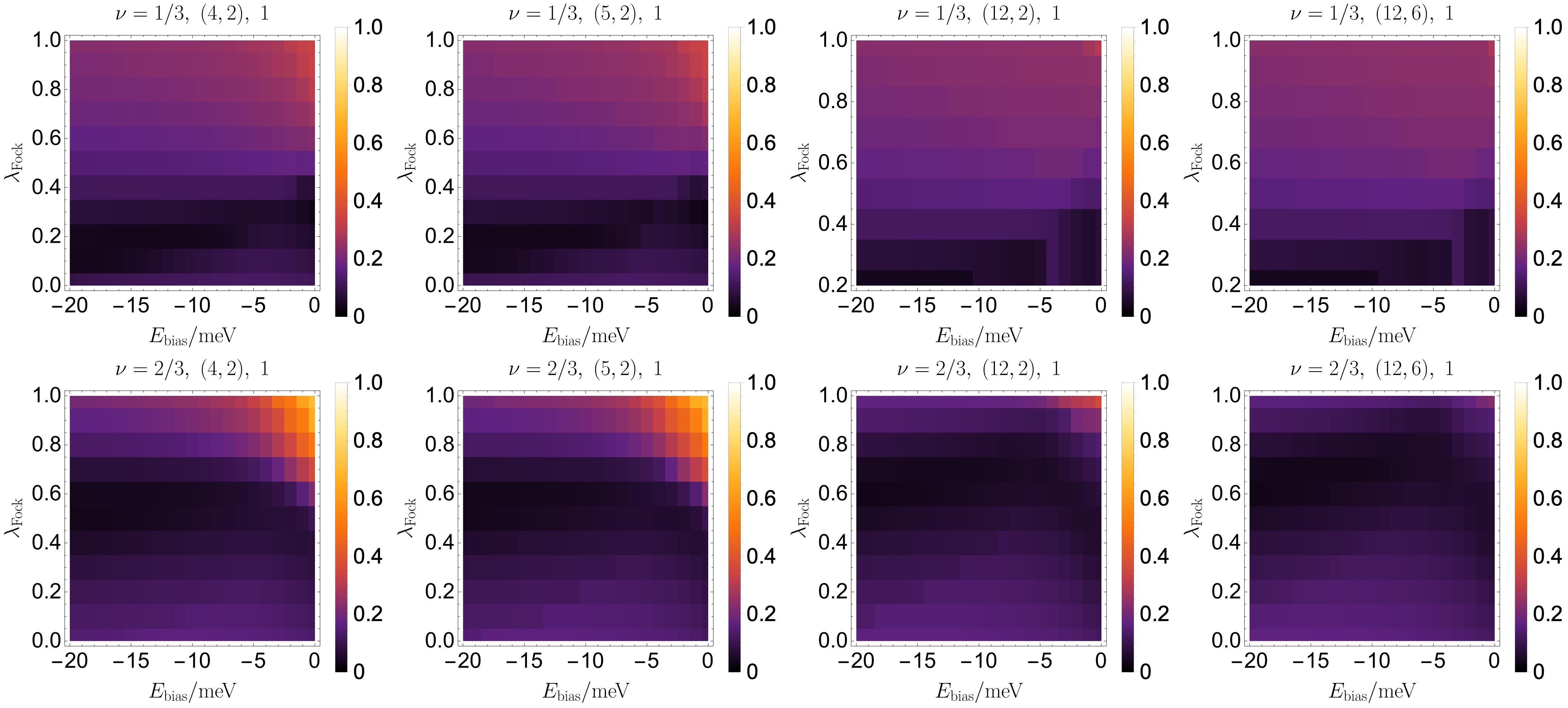}
\caption{
These plots are derived from particle density $\bsl{n}_{\bsl{k}}$ calculated for the 12 sites, \ie, $(N_x, N_y, \widetilde{n}_{11}, \widetilde{n}_{12}, \widetilde{n}_{21}, \widetilde{n}_{22}) = (2,6,1,0,1,1)$ and and for the displacement field $V=22$meV.
The plots for $\nu=1/3$ and $\nu=-2/3$ show the standard deviation of averaged $n_{\bsl{k}}$ for the second lowest set of three (indicated by ``1" in the caption of each plot) states at the FCI momenta $(k_x,k_y)=(0,0),(0,2),(0,4)$.
The caption of each plot specifies (number of states in band 1, number of states in band 2) in parentheses.
}
\label{fig:HFPbias_FockBGScale_LEnk_x2y6}
\end{figure}

\begin{figure}[ht!]
\centering
\includegraphics[width=\columnwidth]{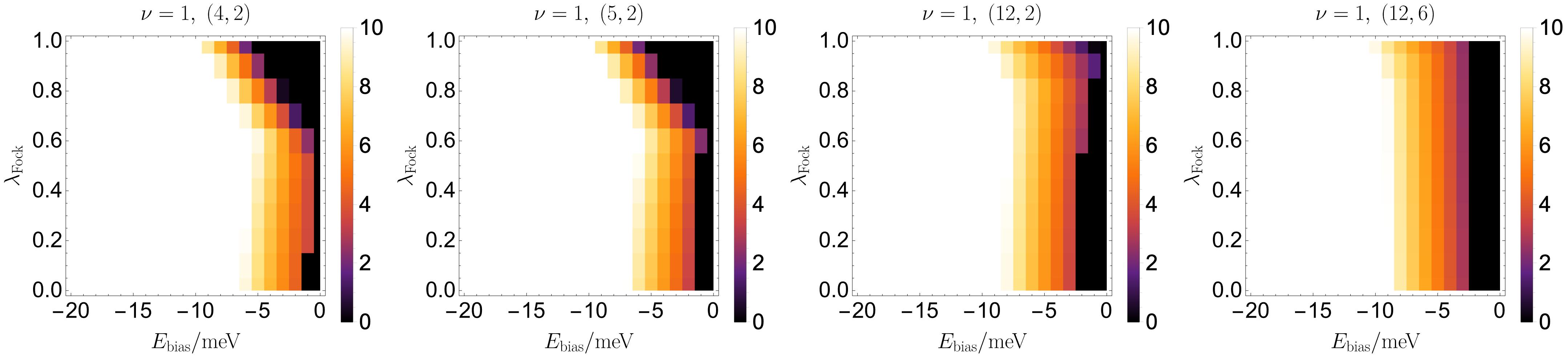}
\caption{$E_{cont.}-E_{\Ch=1}$ as a function of $\Ebias$ and $\lfock$ (\cref{eq:HF_bias,eq:Fock_bias}) for $\nu=1$ and for the displacement field $V=22$meV.
The calculation is done for the 12 sites, \ie, $(N_x, N_y, \widetilde{n}_{11}, \widetilde{n}_{12}, \widetilde{n}_{21}, \widetilde{n}_{22}) = (2,6,1,0,1,1)$, and the color indicates the many body gap.
We set $E_{cont.}-E_{\Ch=1}$ to be zero if it is negative.
}
\label{fig:x2y6n12EcontinuumMinusECh1plots}
\end{figure}

\begin{figure}[ht!]
\centering
\includegraphics[width=\columnwidth]{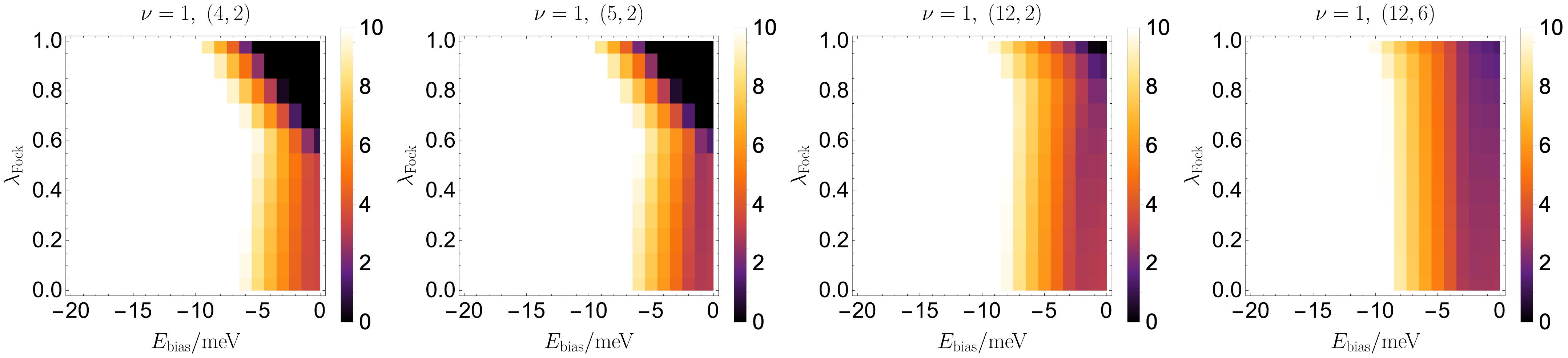}
\caption{$E_{cont.}-E_{GS}$ as a function of $\Ebias$ and $\lfock$ (\cref{eq:HF_bias,eq:Fock_bias}) for $\nu=1$ and for the displacement field $V=22$meV.
The calculation is done for the 12 sites, \ie, $(N_x, N_y, \widetilde{n}_{11}, \widetilde{n}_{12}, \widetilde{n}_{21}, \widetilde{n}_{22}) = (2,6,1,0,1,1)$, and the color indicates the many body gap.
We set $E_{cont.}-E_{GS}$ to be zero if it is negative.
}
\label{fig:x2y6n12EcontinuumMinusEGSplots}
\end{figure}

\begin{figure}
\centering
\includegraphics[width=\linewidth]{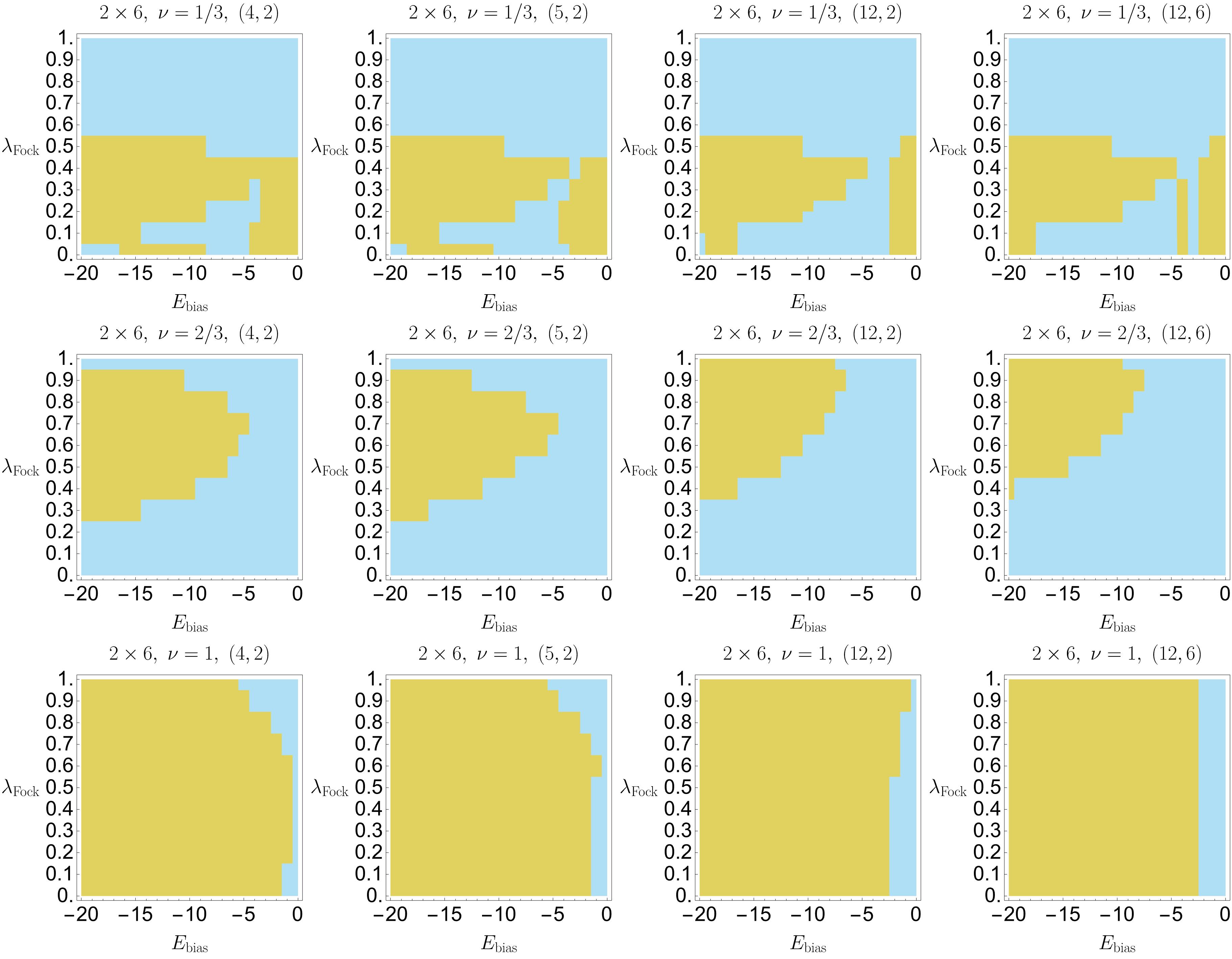}
\caption{
The stability regions (green) for the FCI at $\nu=1/3$ (top row) and $\nu=2/3$ (middle row) and the Chern insulator at $\nu=1$ (bottom row) for 12 sites, \ie, $(N_x, N_y, \widetilde{n}_{11}, \widetilde{n}_{12}, \widetilde{n}_{21}, \widetilde{n}_{22}) = (2,6,1,0,1,1)$. We use the 2D interaction and the biasing method in the AVE scheme with $V=22$meV. We keep all the orbitals in band 0, and the orbital restriction ($N_{\text{orb1}},N_{\text{orb2}}$) for the higher bands is specified in the caption of each plot.}
\label{fig:FCI_Regions_x2y6_plots}
\end{figure}

\begin{figure}[ht!]
\centering
\includegraphics[width=0.81\columnwidth]{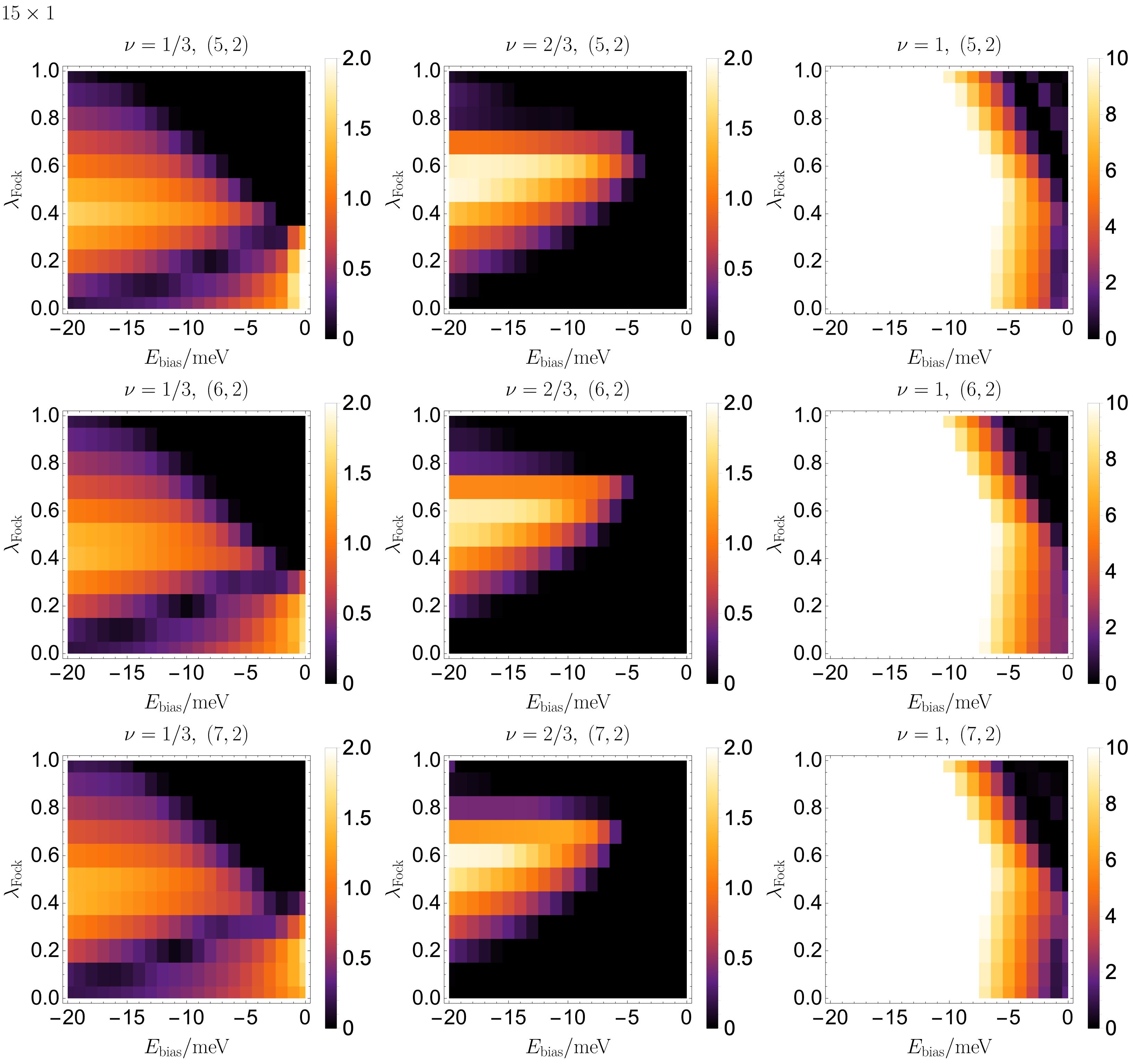}
\caption{Plots of the many-body gap as a function of $\Ebias$ and $\lfock$ (\cref{eq:HF_bias,eq:Fock_bias}) for $\nu=1/3,2/3,1$ and for the displacement field $V=22$meV.
The calculation is done for the 15 sites, \ie, $(N_x, N_y, \widetilde{n}_{11}, \widetilde{n}_{12}, \widetilde{n}_{21}, \widetilde{n}_{22}) = (15,1,1,-5,0,1)$, and the color indicates the many body gap.
The meaning of other labels are the same as \cref{fig:HFPbias_FockBGScale_Gap_x2y6}.
}
\label{fig:HFPbias_FockBGScale_Gap_x15y1}
\end{figure}

\begin{figure}[ht!]
\centering
\includegraphics[width=\columnwidth]{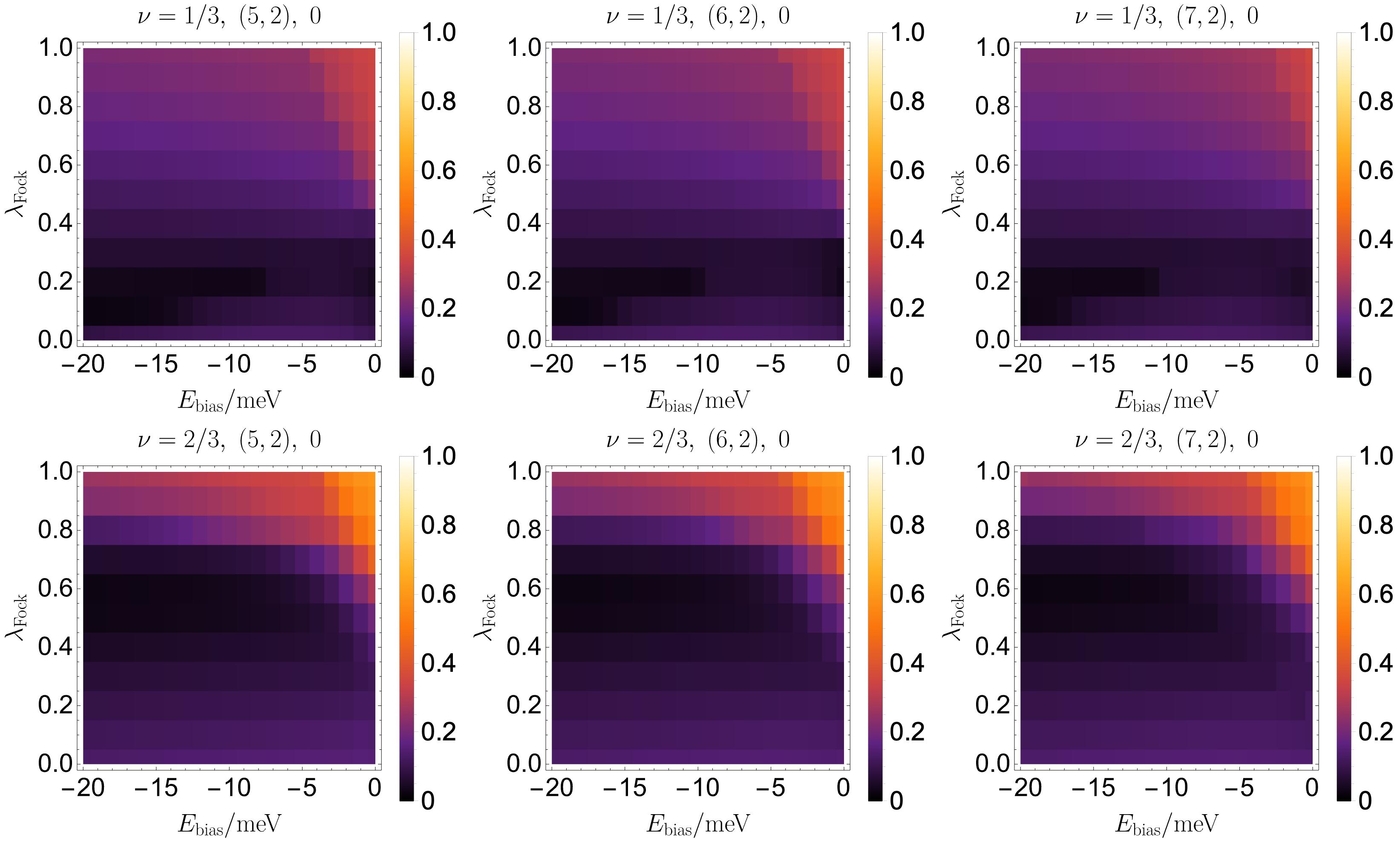}
\caption{
These plots are derived from particle density $\bsl{n}_{\bsl{k}}$ calculated for the 15 sites and for the displacement field $V=22$meV.
The plots for $\nu=1/3$ and $\nu=-2/3$ show the standard deviation of averaged $n_{\bsl{k}}$ for the three lowest (indicated by ``0" in the caption of each plot) states at the FCI momenta $(k_x,k_y)=(0,0),(5,0),(10,0)$.
The caption of each plot specifies (number of states in band 1, number of states in band 2) in parentheses.
}
\label{fig:HFPbias_FockBGScale_GSnk_x15y1}
\end{figure}

\begin{figure}[ht!]
\centering
\includegraphics[width=\columnwidth]{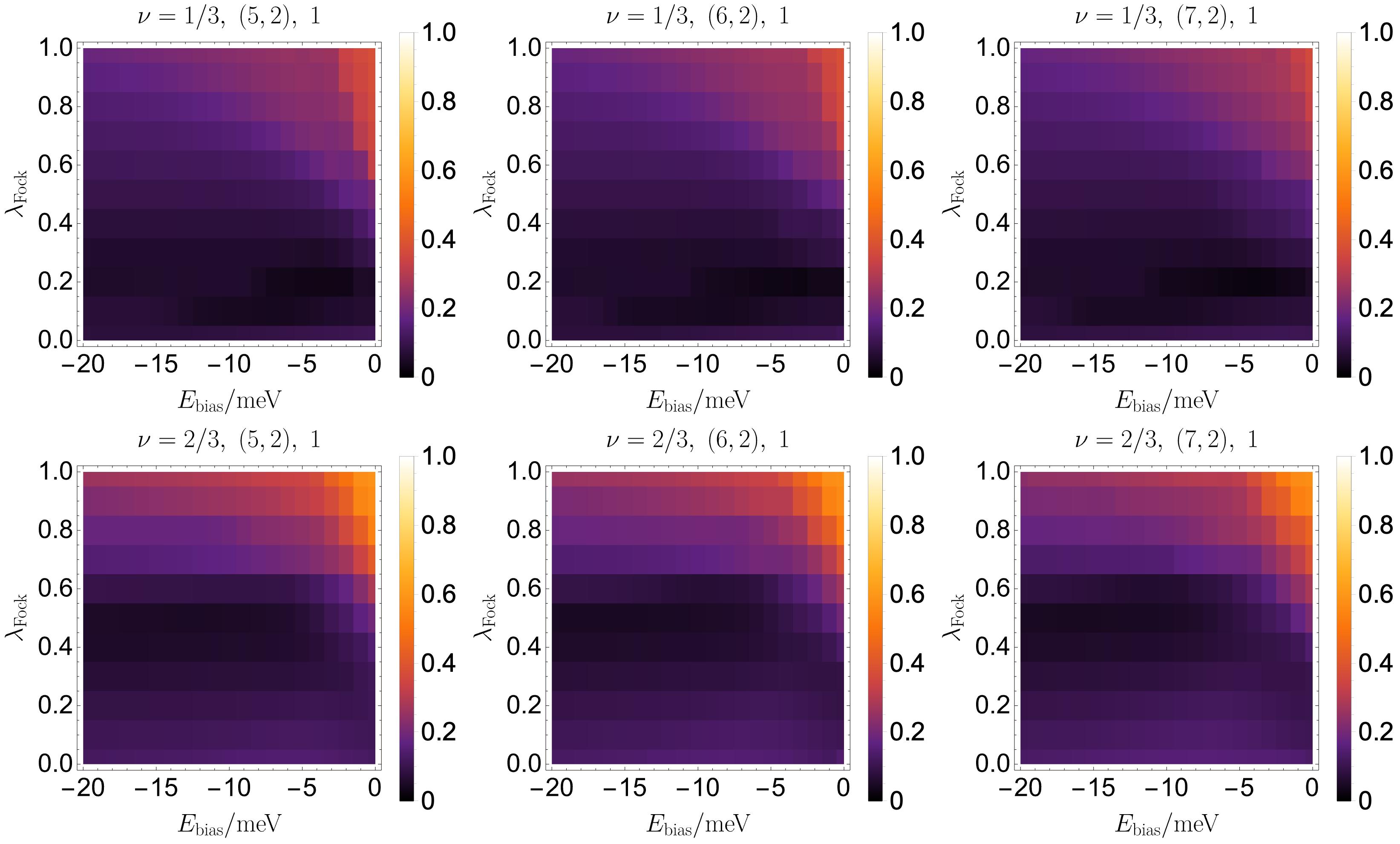}
\caption{
These plots are derived from particle density $n_{\bsl{k}}$ calculated for the 15 sites and for the displacement field $V=22$meV.
The plots for $\nu=1/3$ and $\nu=-2/3$ show the standard deviation of averaged $n_{\bsl{k}}$ for second lowest set of three (indicated by ``1" in the caption of each plot) states at the FCI momenta $(k_x,k_y)=(0,0),(5,0),(10,0)$.
The plot for $\nu=1$ shows the minimum value of particle density in band 0 of the lowest state at the CI momentum $(k_x,k_y)=(0,0)$.
The caption of each plot specifies (number of states in band 1, number of states in band 2) in parentheses.
}
\label{fig:HFPbias_FockBGScale_LEnk_x15y1}
\end{figure}

\begin{figure}[ht!]
\centering
\includegraphics[width=\columnwidth]{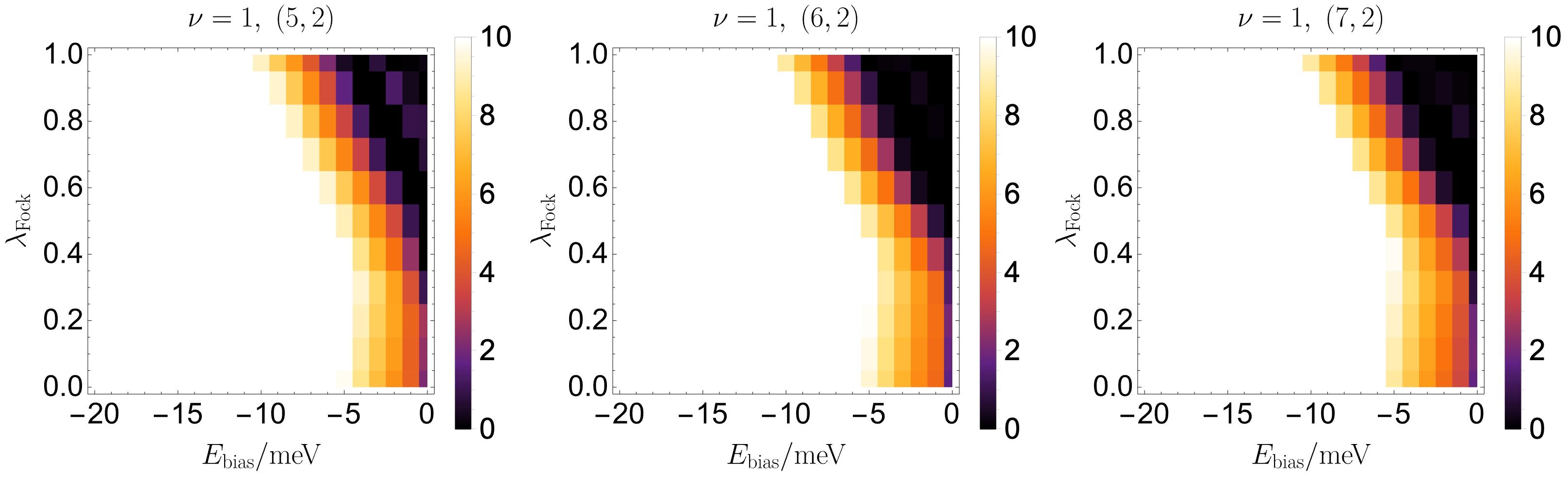}
\caption{$E_{cont.}-E_{\Ch=1}$ as a function of $\Ebias$ and $\lfock$ (\cref{eq:HF_bias,eq:Fock_bias}) for $\nu=1$ and for the displacement field $V=22$meV.
The calculation is done for the 15 sites, \ie, $(N_x, N_y, \widetilde{n}_{11}, \widetilde{n}_{12}, \widetilde{n}_{21}, \widetilde{n}_{22}) = (15,1,1,-5,0,1)$, and the color indicates the many body gap.
We set $E_{cont.}-E_{\Ch=1}$ to be zero if it is negative.
}
\label{fig:x15y1n15EcontinuumMinusECh1plots}
\end{figure}

\begin{figure}[ht!]
\centering
\includegraphics[width=\columnwidth]{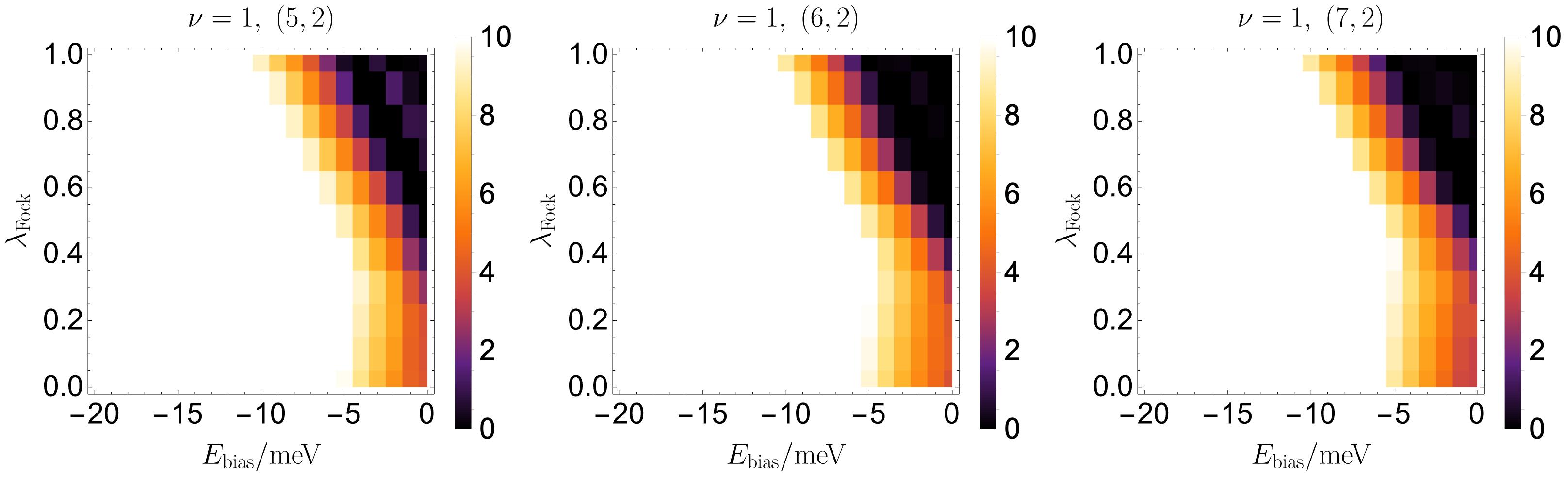}
\caption{$E_{cont.}-E_{GS}$ as a function of $\Ebias$ and $\lfock$ (\cref{eq:HF_bias,eq:Fock_bias}) for $\nu=1$ and for the displacement field $V=22$meV.
The calculation is done for the 15 sites, \ie, $(N_x, N_y, \widetilde{n}_{11}, \widetilde{n}_{12}, \widetilde{n}_{21}, \widetilde{n}_{22}) = (15,1,1,-5,0,1)$, and the color indicates the many body gap.
We set $E_{cont.}-E_{GS}$ to be zero if it is negative.
}
\label{fig:x15y1n15EcontinuumMinusEGSplots}
\end{figure}

\begin{figure}
\centering
\includegraphics[width=\linewidth]{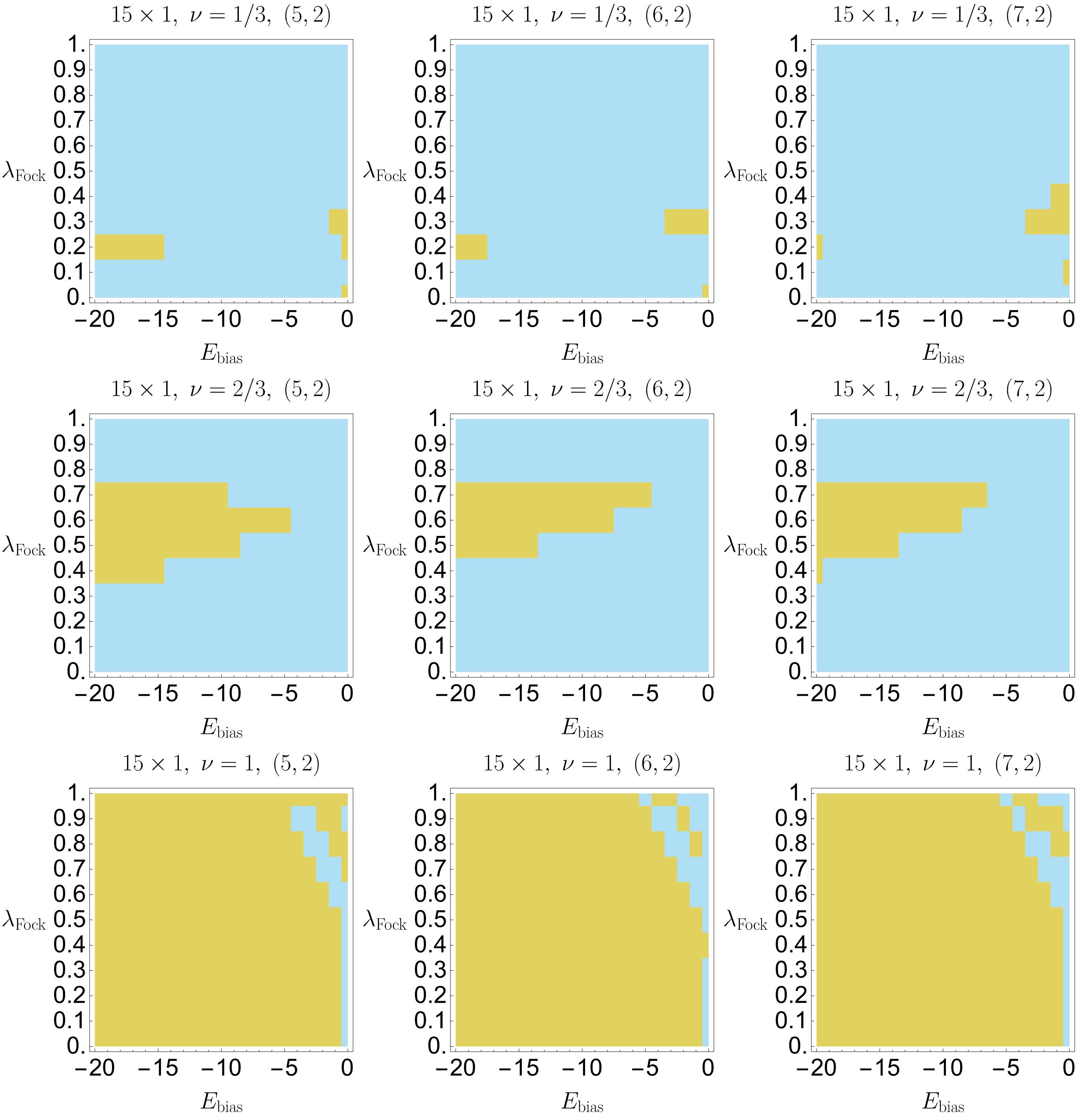}
\caption{
The stability regions (green) for the FCI at $\nu=1/3$ (top row) and $\nu=2/3$ (middle row) and the Chern insulator at $\nu=1$ (bottom row) for 15 sites, \ie, $(N_x, N_y, \widetilde{n}_{11}, \widetilde{n}_{12}, \widetilde{n}_{21}, \widetilde{n}_{22}) = (15,1,1,-5,0,1)$. We use the 2D interaction and the biasing method in the AVE scheme with $V=22$meV. We keep all the orbitals in band 0, and the orbital restriction ($N_{\text{orb1}},N_{\text{orb2}}$) for the higher bands is specified in the caption of each plot.
}
\label{fig:FCI_Regions_x15y1_plots}
\end{figure}

\begin{figure}[ht!]
\centering
\includegraphics[width=\columnwidth]{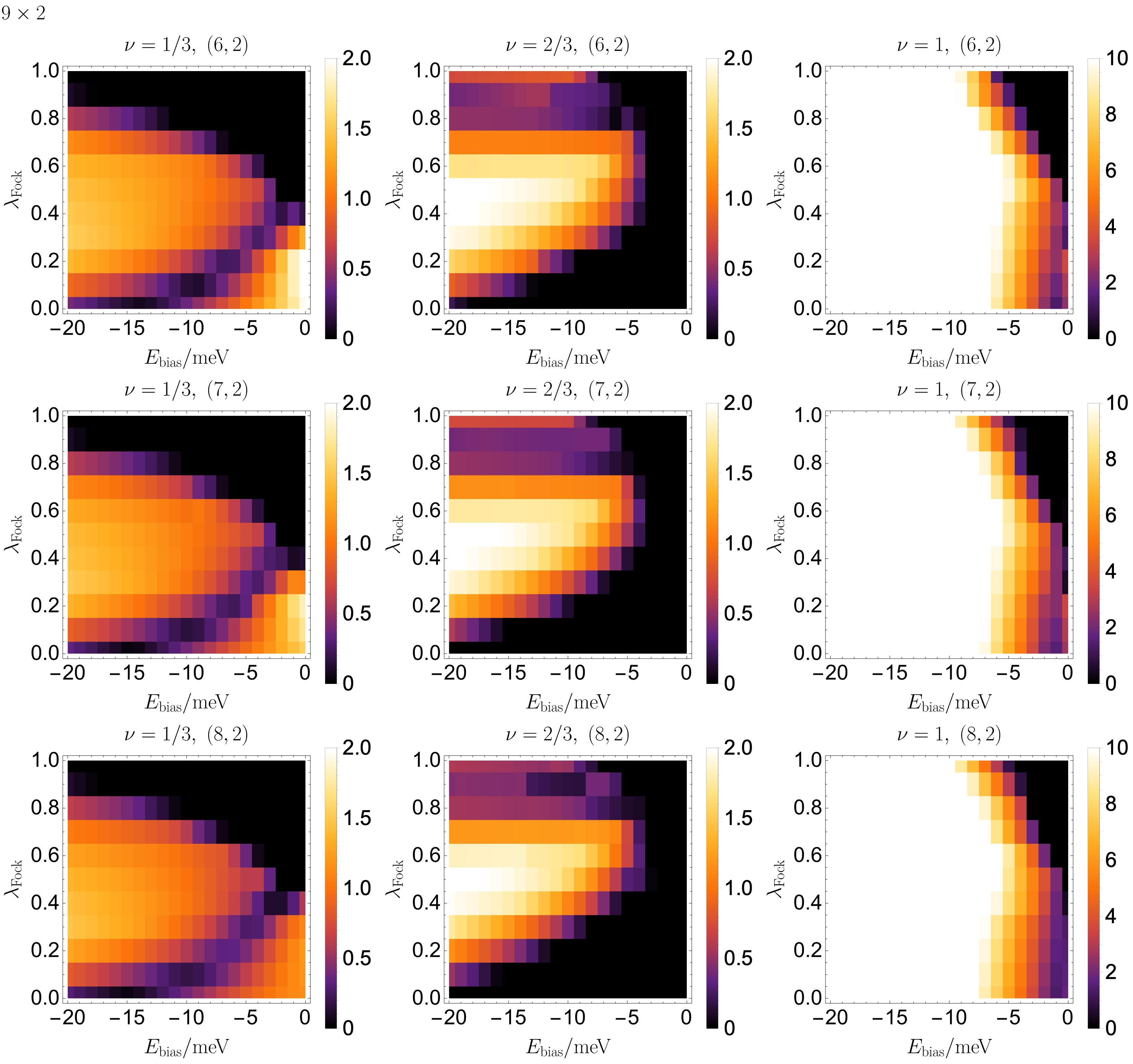}
\caption{Plots of the many-body gap as a function of $\Ebias$ and $\lfock$ (\cref{eq:HF_bias,eq:Fock_bias}) for $\nu=1/3,2/3,1$ with displacement field $V=22$meV.
The calculation is performed for the 18 sites, \ie, $(N_x, N_y, \widetilde{n}_{11}, \widetilde{n}_{12}, \widetilde{n}_{21}, \widetilde{n}_{22}) = (9,2,1,-2,0,1)$, and the color labels the many body gap.
The meaning of other labels are the same as \cref{fig:HFPbias_FockBGScale_Gap_x2y6}.
}
\label{fig:HFPbias_FockBGScale_Gap_x9y2}
\end{figure}

\begin{figure}[ht!]
\centering
\includegraphics[width=\columnwidth]{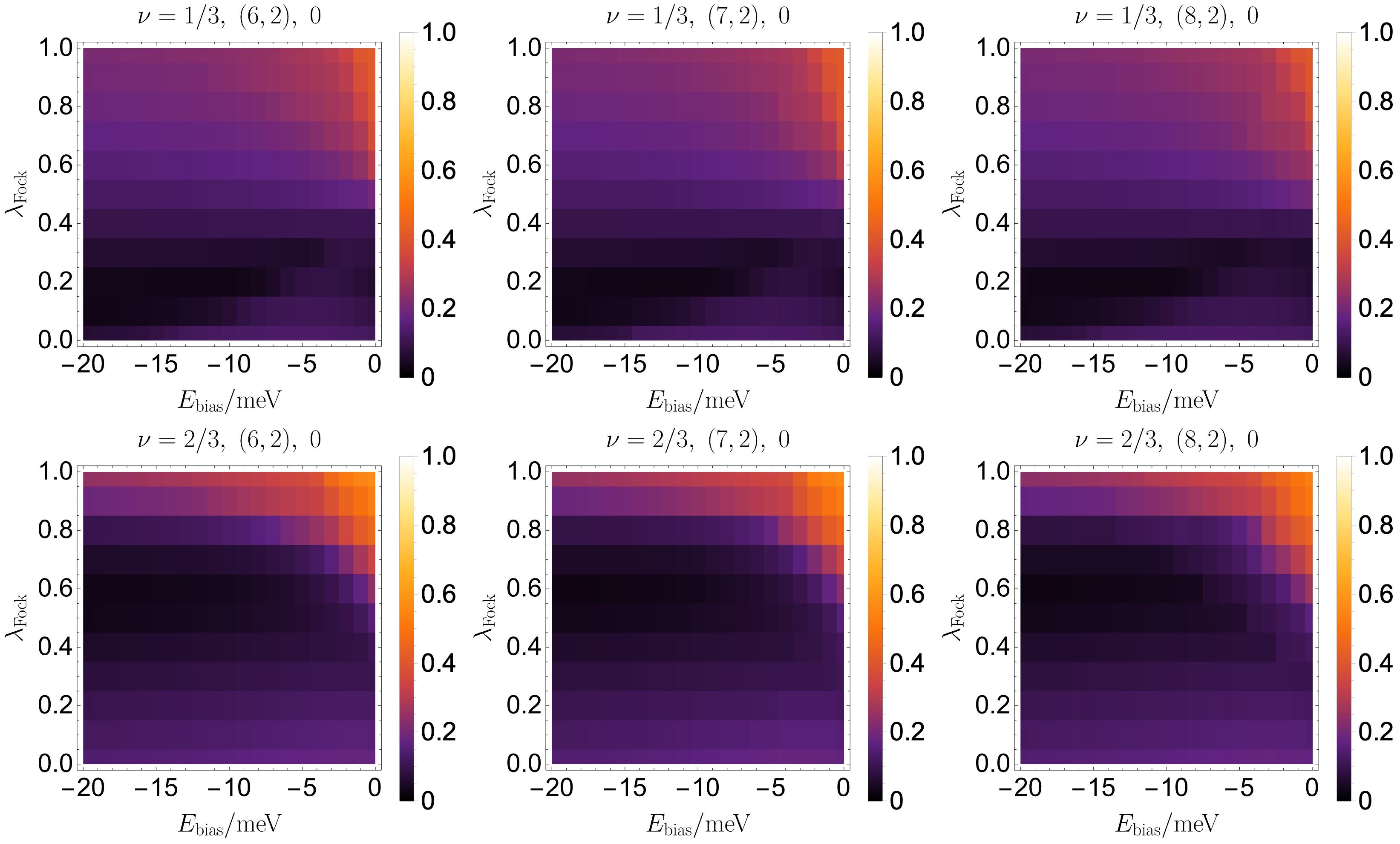}
\caption{
These plots are derived from particle density $\bsl{n}_{\bsl{k}}$ calculated for the 18 sites and for the displacement field $V=22$meV.
The plots for $\nu=1/3$ and $\nu=-2/3$ show the standard deviation of averaged $n_{\bsl{k}}$ for the three lowest (indicated by ``0" in the caption of each plot) states at the FCI momenta $(k_x,k_y)=(0,1),(3,1),(6,1)$ for $\nu=1/3$ and $(k_x,k_y)=(0,0),(3,0),(6,0)$ for $\nu=2/3$.
The caption of each plot specifies (number of states in band 1, number of states in band 2) in parentheses.
}
\label{fig:HFPbias_FockBGScale_GSnk_x9y2}
\end{figure}

\begin{figure}[ht!]
\centering
\includegraphics[width=\columnwidth]{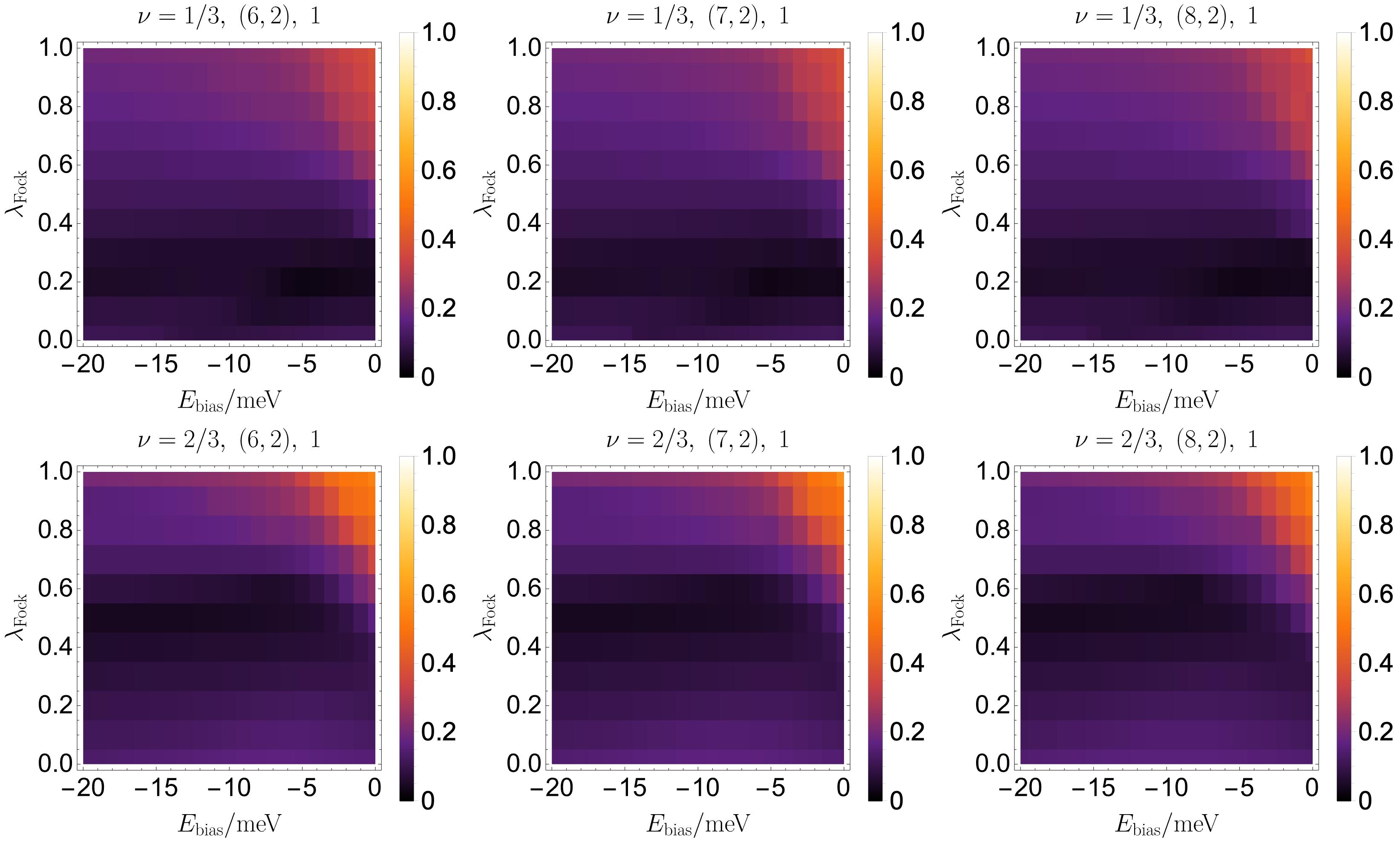}
\caption{
These plots are derived from particle density $\bsl{n}_{\bsl{k}}$ calculated for the 18 sites and for the displacement field $V=22$meV.
The plots for $\nu=1/3$ and $\nu=-2/3$ show the standard deviation of averaged $n_{\bsl{k}}$ for the second lowest set of three (indicated by ``1" in the caption of each plot) states at the FCI momenta $(k_x,k_y)=(0,1),(3,1),(6,1)$ for $\nu=1/3$ and $(k_x,k_y)=(0,0),(3,0),(6,0)$ for $\nu=2/3$.
The caption of each plot specifies (number of states in band 1, number of states in band 2) in parentheses.
}
\label{fig:HFPbias_FockBGScale_LEnk_x9y2}
\end{figure}

\begin{figure}[ht!]
\centering
\includegraphics[width=\columnwidth]{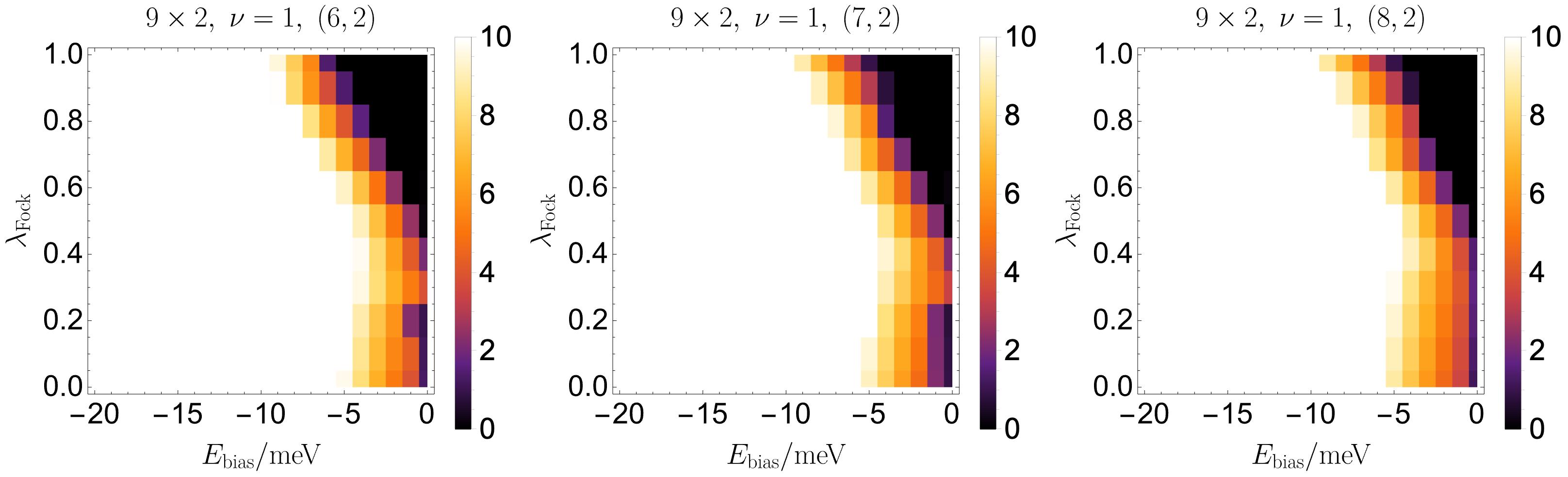}
\caption{$E_{cont.}-E_{\Ch=1}$ as a function of $\Ebias$ and $\lfock$ (\cref{eq:HF_bias,eq:Fock_bias}) for $\nu=1$ and for the displacement field $V=22$meV.
The calculation is performed for the 18 sites, \ie, $(N_x, N_y, \widetilde{n}_{11}, \widetilde{n}_{12}, \widetilde{n}_{21}, \widetilde{n}_{22}) = (9,2,1,-2,0,1)$, and the color indicates the many body gap.
We set $E_{cont.}-E_{\Ch=1}$ to be zero if it is negative.
}
\label{fig:x9y2n18EcontinuumMinusECh1plots}
\end{figure}

\begin{figure}[ht!]
\centering
\includegraphics[width=\columnwidth]{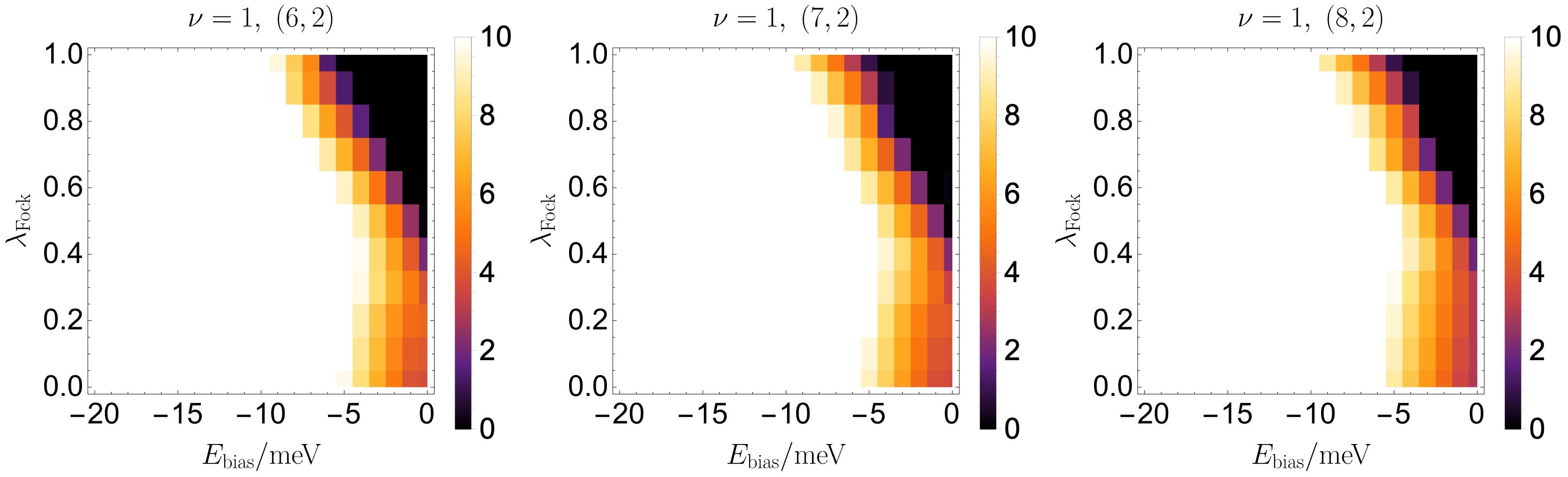}
\caption{$E_{cont.}-E_{GS}$ as a function of $\Ebias$ and $\lfock$ (\cref{eq:HF_bias,eq:Fock_bias}) for $\nu=1$ and for the displacement field $V=22$meV.
The calculation is performed for the 18 sites, \ie, $(N_x, N_y, \widetilde{n}_{11}, \widetilde{n}_{12}, \widetilde{n}_{21}, \widetilde{n}_{22}) = (9,2,1,-2,0,1)$, and the color indicates the many body gap.
We set $E_{cont.}-E_{GS}$ to be zero if it is negative.
}
\label{fig:x9y2n18EcontinuumMinusEGSplots}
\end{figure}

\begin{figure}
\centering
\includegraphics[width=\linewidth]{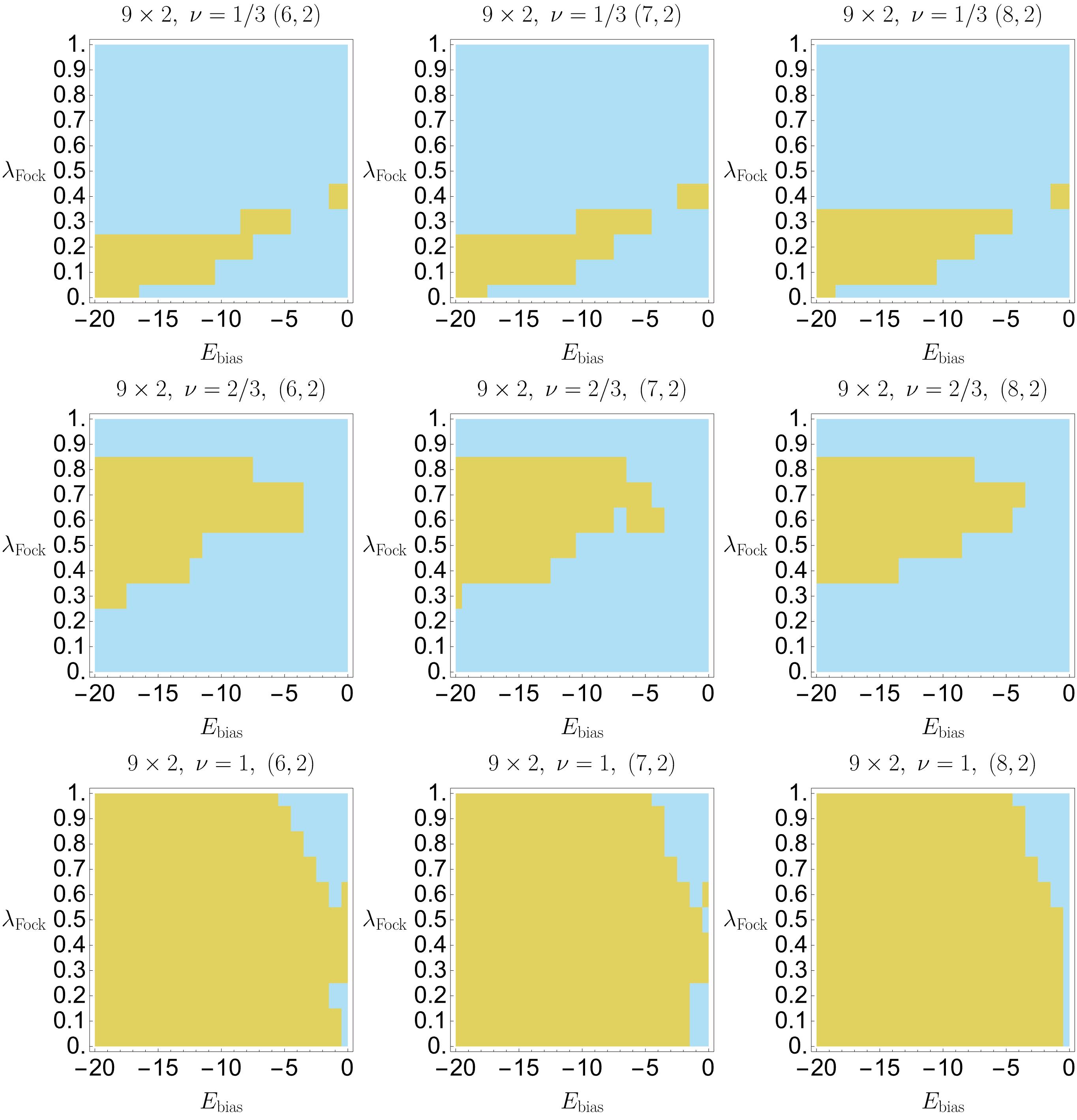}
\caption{The stability regions (green) for the FCI at $\nu=1/3$ (top row) and $\nu=2/3$ (middle row) and the Chern insulator at $\nu=1$ (bottom row) for 18 sites, \ie, $(N_x, N_y, \widetilde{n}_{11}, \widetilde{n}_{12}, \widetilde{n}_{21}, \widetilde{n}_{22}) = (9,2,1,-2,0,1)$. We use the 2D interaction and the biasing method in the  AVE scheme with $V=22$meV.
We keep all the orbitals in band 0, and the orbital restriction ($N_{\text{orb1}},N_{\text{orb2}}$) for the higher bands is specified in the caption of each plot.}
\label{fig:FCI_Regions_x9y2_plots}
\end{figure}

In this appendix, we discuss our ED results performed with the biasing method in \App{app:biasing_method} for the 2D interaction with the AVE scheme and $V=22$meV.
We perform calculations for three system sizes as parameterized by \cref{eq:momentum_mesh}: $(N_x, N_y, \widetilde{n}_{11}, \widetilde{n}_{12}, \widetilde{n}_{21}, \widetilde{n}_{22}) = (2,6,1,0,1,1),(15,1,1,-5,0,1),(9,2,1,-2,0,1)$, which we call 12, 15 and 18 sites in short, respectively.
As discussed in \App{app:biasing_method}, we consider two biasing parameters $E_{\text{bias}}<0$ and $\lambda_{\text{Fock}}\in[0,1]$ with the unbiased limit being $\Ebias=0$ and $\lfock=1$. We work in the diagonal basis for the one-body term in \cref{eq:h_one-body_bias}.
We are interested in three fillings $\nu=1/3,2/3$ and $1$. We impose truncations on the Hilbert space in order to be able to perform calculations at  all three fillings on all system sizes of interest.
As discussed in \App{app:biasing_method}, we do so by limiting the number of single-particle eigenstates (orbital restriction) of the one-body term in \cref{eq:h_one-body_bias}.
Specifically, we keep all states from band 0, but only retain the  lowest $N_{\text{orb1}}$ energy states in band 1, and the  lowest $N_{\text{orb2}}$ energy states in band 2, where bands 0, 1 and 2 are the lowest, middle and highest bands of the one-body term in \cref{eq:h_one-body_bias} respectively.
The values of $N_{\text{orb1}}$ and $N_{\text{orb2}}$ will be specified in the following discussions for each system size.

At $\nu=1/3$ and $2/3$, we consider the system to be in the FCI phase if (1) its lowest three states are have the correct momenta for an FCI, (2) the spread of the $3$ lowest states is smaller than the gap between the 3rd and 4th lowest states, and (3) the averaged particle density $n_{\bsl{k}}$ for the 3 lowest states does not have a standard deviation that is larger than the second set of three lowest states at the FCI momenta.
The third condition is motivated from the fact that the fractional quantum Hall states have zero standard deviation in their $n_{\bsl{k}}$.
At $\nu=1$, the system is considered to be in the $\Ch=1$ phase if (1) its lowest state is at the $\Ch=1$ momentum and (2) its band-resolved $n_{\bsl{k},\alpha}$ at $\Gamma_M$, $\K_M$ and $\K_M'$ is consistent with the HF $\Ch=1$ state.
By consistent, we mean that value of $\alpha$ for which $n_{\bsl{k},\alpha}$ is maximal at $\Gamma_M$, $\K_M$ and $\K_M'$ is the same as that for the HF $\Ch=1$ state at $\Gamma_M$, $\K_M$ and $\K_M'$, which in the HF $\Ch=1$ state determines $\Ch\ \mod\ 3$.

Based on the numerical criteria above, we have obtained the FCI and Chern insulator regions for 12, 15 and 18 sites.
The calculations for 12 sites are summarized in \cref{fig:HFPbias_FockBGScale_Gap_x2y6} (for the many-body gap), \cref{fig:HFPbias_FockBGScale_GSnk_x2y6,fig:HFPbias_FockBGScale_LEnk_x2y6} (for $n_{\bsl{k}}$), and \cref{fig:FCI_Regions_x2y6_plots,fig:x2y6n12EcontinuumMinusECh1plots} (for phase diagrams showing the FCI and Chern insulator regions).
As shown in \cref{fig:FCI_Regions_x2y6_plots}, we can see that as we increase the states included in band 1, the FCI region at $\nu=2/3$ requires less biasing on the Fock background term, eventually requiring zero biasing on the Fock background term $(\lambda_\text{Fock}=1)$. On the other hand, there are minimal changes for the FCI region at $\nu=1/3$ and the Chern insulator region at $\nu=1$.
We note that the calculations in the limit of large negative $\Ebias$ are equivalent to using the HF-band-projected (HFB) method employed in Refs.~\cite{dong2023theory,zhou2023fractional,dong2023anomalous} (though they do not use the AVE scheme).
From \cref{fig:FCI_Regions_x2y6_plots}, we can clearly see that stabilization of the FCI at $\nu=2/3$ requires much less biasing than at $\nu=1/3$, which is consistent with the results using the HF basis and band-max truncation (see \App{app:result_2D_int}).
The Chern insulator at $\nu=1$ needs the least biasing to stabilize in ED, though some biasing is still needed.
As shown in \cref{fig:x2y6n12EcontinuumMinusECh1plots}, the destabilization of the Chern insulator in the unbiased limit not only comes from competition with the $\Ch=0$ state, but also the dropping down of the continuum for 12 sites.
We also note that we do not observe considerable improvement by increasing the number of states in the band 2 from $N_{\text{orb2}}=2$ to $N_{\text{orb2}}=6$, implying that including 2 states in band 2 is enough for acceptable convergence the ED calculation.
We note that at minimum $N_{\text{orb2}}=2$ is required since the two lowest states in band 2 are at $K_M$ and $K_M'$ points where band 2 is close in energy to band 1.
Therefore, for 15 and 18 sites, we fix $N_{\text{orb2}}=2$.

The calculations for 15 sites are summarized in \cref{fig:HFPbias_FockBGScale_Gap_x15y1} (for the many-body gap), \cref{fig:HFPbias_FockBGScale_GSnk_x15y1,fig:HFPbias_FockBGScale_LEnk_x15y1}  (for $n_{\bsl{k}}$), and \cref{fig:FCI_Regions_x15y1_plots,fig:x15y1n15EcontinuumMinusECh1plots} (for FCI and CI regions), and those for 18 sites are summarized in \cref{fig:HFPbias_FockBGScale_Gap_x9y2} (for many-body gap), \cref{fig:HFPbias_FockBGScale_GSnk_x9y2,fig:HFPbias_FockBGScale_LEnk_x9y2}  (for $n_{\bsl{k}}$), and \cref{fig:FCI_Regions_x9y2_plots,fig:x9y2n18EcontinuumMinusECh1plots} (for the phase diagram of the FCI and Chern insulator regions).
In particular, we have chosen the ratio $N_{\text{orb1}}/(N_x N_y)\in [0.33333,0.5]$ to be similar to the corresponding ratios for $(N_{\text{orb1}},N_{\text{orb2}})=(4,2)$ and $(5,2)$ on 12 sites.
By comparing \cref{fig:FCI_Regions_x15y1_plots,fig:FCI_Regions_x9y2_plots} to \cref{fig:FCI_Regions_x2y6_plots} (for $N_{\text{orb1}}=4$ and $5$ on 12 sites), we have not clearly observed that less biasing is required to stabilize the FCI and Chern insulator when increasing the system size.

Yet, we argue that the Chern insulator can be the ground state in the thermodynamic limit at $\nu=1$.
As shown in \cref{fig:x2y6n12EcontinuumMinusEGSplots}, when we include enough extra states in the remote bands, the gap between the ground state and the continuum will stay nonzero even in the physical limit of zero biasing.
Moreover, by comparing $(N_{\text{orb1}},N_{\text{orb2}})=(4,2)$ and $(5,2)$ in \cref{fig:x2y6n12EcontinuumMinusEGSplots} with \cref{fig:x15y1n15EcontinuumMinusEGSplots,fig:x9y2n18EcontinuumMinusEGSplots}, we find that the gap between the ground state and the continuum does not vary much as we change the size.
Therefore, we may conjecture that the gap between the ground state and the continuum will stay nonzero in the thermodynamic limit as long as enough states in band 1 and band 2 are included, \ie, the continuum will not destroy the insulating ground state.
Then, the remaining question becomes what the nature of the insulating ground state in the thermodynamic limit is.
The self-consistent HF calculations of \App{app:2D_Int_HF} show that as the size increases, the $\Ch=1$ insulating state eventually wins over $\Ch=0$ insulating state for sufficiently large system sizes, although the HF calculations performed in this work do not probe the continuum excitations.
Therefore, combining our ED and HF results, we can conjecture that the $\Ch=1$ state is the ground state in the thermodynamic limit for the physical limit of zero biasing.
Such arguments do not generalize to the FCIs here, since they are destroyed by the continuum in the ED spectrum for the realistic parameter values as shown in the HF basis ED calculation.

\end{document}